\documentclass[structabstract]{aa}
\usepackage{natbib}
\bibpunct{(}{)}{;}{a}{}{,}
\usepackage{graphicx}
\usepackage{lscape}
\usepackage{aalongtable}
\usepackage{subfigure}
\usepackage{txfonts}
\newcommand{\Ha}{\mbox{H$\alpha$}}
\newcommand{\Hb}{\mbox{H$\beta$}}
\newcommand{\Hd}{\mbox{H$\delta$}}
\newcommand{\Hg}{\mbox{H$\gamma$}}
\newcommand{\CHbeta}{\mbox{$C$(H$\beta$)}}
\newcommand{\EBV}{\mbox{$E(B-V)$}}

\newcommand{\commentout}[1]{}

\begin{document}
\title{Mapping the properties of blue compact dwarf galaxies:}

\subtitle{integral field spectroscopy with PMAS}

\author{L.~M. Cair{\'o}s \inst{1}
        \and
        N. Caon \inst{2}
	\and	  
	C. Zurita \inst{2}
	\and
	C. Kehrig \inst{1}
	\and
	M. Roth \inst{1}
	\and
	P. Weilbacher \inst{1}
        }
	
\institute{Astrophysikalisches Institut Potsdam, 
           An der Sternwarte 16, D-14482 Potsdam, Germany\\
           \email{luzma; ckehrig; mmroth; pweilbacher@aip.de}
	   \and
           Instituto de Astrof{\'\i}sica de Canarias, 
	   E-38200 La Laguna, Tenerife, Spain and 
	   Departamento de Astrof{\'\i}sica, Universidad de la Laguna, 
	   E-38205, La Laguna, Tenerife, Spain\\ 
           \email{nicola.caon; czurita@iac.es}
	   }

\date{Received 7 January 2010; accepted 29 March 2010}

 
\abstract 
{Blue compact dwarf (BCD) galaxies are low-luminosity, low-metal content dwarf
systems undergoing violent bursts of star formation. They present a unique
opportunity to probe galaxy formation and evolution and to investigate the 
process of star formation in a relatively simple scenario. Spectrophotometric
studies of BCDs are essential to disentangle and characterize their stellar
populations.} 
{We perform integral field spectroscopy of a sample of BCDs with the aim of
analyzing their morphology, the spatial distribution of some of their physical 
properties (excitation, extinction, and electron density) and their 
relationship with the distribution and evolutionary state of the stellar 
populations.} 
{Integral field spectroscopy observations of the sample galaxies were carried 
out with the Potsdam  Multi-Aperture Spectrophotometer (PMAS) at the 3.5 m 
telescope at Calar Alto Observatory.  An area $16\arcsec\times 16\arcsec$ in 
size was mapped with a spatial sampling of  $1\arcsec\times 1\arcsec$.  We
obtained data  in the 3590--6996 \AA\ spectral range, with a linear 
dispersion of 3.2 \AA\  per pixel. \textnormal{From these data we built
two-dimensional maps of the flux of the most prominent emission lines, of two
continuum bands, of the most relevant line ratios, and of the gas velocity
field. Integrated spectra of the most prominent star-forming regions and of
whole objects within the FOV were used to derive their physical parameters and
the gas metal abundances.}  
} 
{Six galaxies display the same morphology both in emission line and in 
continuum maps; only in two objects, Mrk~32 and Tololo~1434+032, the 
distributions of the ionized gas and of the stars differ considerably. In
general the different excitation maps for a same object display the same
pattern and trace the star-forming regions, as expected for objects ionized by
hot stars; only the outer regions of Mrk~32, I~Zw~123 and I~Zw~159 display
higher [\ion{S}{ii}]/\Ha\ values, suggestive of shocks. Six galaxies display 
an inhomogeneous dust distribution. Regarding the kinematics, Mrk~750, Mrk~206
and I~Zw~159 display a clear rotation pattern, while in Mrk~32, Mrk~475 and
I~Zw~123 the velocity fields are flat.}
{}


   
\maketitle
%

\section{Introduction}
\label{Section:Introduction}

Blue compact dwarf (BCD) galaxies are narrow emission-line objects, which
undergo at the present time violent bursts of star formation
\citep{Sargent1970}. They are compact and low-luminosity objects (starburst
diameter $\leq 1$ kpc; $M_{B} \geq -18$ mag), with a low-metal content 
($Z_{\sun}/50 \leq Z \leq Z_{\sun}/2$) and high star-forming (SF) rates, 
able to exhaust their gas content on a time scale much shorter than the age 
of the Universe. Initially it was hypothesized that BCDs were truly young 
galaxies, forming their first generation of stars \citep{Sargent1970, 
Lequeux1980,Kunth1988}, but the subsequent detection of an extended redder
stellar host galaxy in the vast majority of them has shown that most BCDs
are actually old systems
\citep{Loose1987,Telles1995,Papaderos1996a,Cairos2001II,
Cairos2001I,Cairos2002,Cairos2003} undergoing recurrent star-formation 
episodes \citep{Thuan1991,MasHesse1999}.

These galaxies present a unique opportunity to gain insights on central issues 
in contemporary galaxy research. Chemically unevolved nearby SF
dwarfs like BCDs are an important link to the early Universe and the
epoch of galaxy formation, as they have been regarded as the local
counterparts of the distant subgalactic units (building blocks) from which
larger systems are created at high {\em redshifts}
\citep{Kauffmann1993,Lowenthal1997}; the study of these systems hence provides
important insights into the star-formation process of distant
galaxies. Moreover, even though most BCDs are not genuinely young galaxies,
their metal deficiency makes them useful objects to constrain the primordial
$^{4}$He abundance and to monitor the synthesis and dispersal of heavy 
elements in a nearly pristine environment
\citep{Pagel1992,Masegosa1994,Izotov1997,Kunth2000}. 
Blue compact dwarfs are also ideal laboratories for the study of the starburst
phenomenon: as they are smaller and less massive than normal galaxies, they
cannot sustain  a spiral density wave and do not suffer from disk
instabilities, which considerably simplifies the study of the star formation
process. Besides, the radiation emitted by their SF regions is less diluted by
the stellar  continuum than in giant spiral galaxies, allowing for more
precise  studies of element abundance ratios.

However, and in spite of the great effort done during the last two decades on
the field of BCDs, fundamental questions like the mechanisms responsible 
for the ignition of their starburst, their evolutionary status or their SF
histories are still far from well understood.

To answer these questions it is of paramount importance to first disentangle 
and characterize the different components that make up a BCD galaxy. This is
a demanding and difficult task. At any location in the galaxy, the emitted
flux is the sum of the emission from the local starburst, the flux produced 
by the nebula surrounding the young stars, and the emission from the 
underlying, old stellar population, all possibly modulated by dust
\citep{Cairos2002,Cairos2003,Cairos2007}. Substantial work in the field has
shown that photometry alone does not allow us to distinguish the different
components in BCDs (see \citealp{Kunth2000,Cairos2002}). The properties of
the SF knots in the same galaxy may vary widely: accounting for the flux
in emission lines through broad-band filters and for the contribution of the
stellar host is fundamental to derive the actual broad-band colors of the
knots \citep{Cairos2002,Cairos2007}. On the other hand, the dust content
(usually assumed to be negligible in BCDs) turned out to be quite significant
in several objects \citep{Hunt2001,Cairos2003,VanziSauvage2004}.
\commentout{We think the correction is wrong, as "accounting" here   is the
subject of "is fundamental". We rewrote as: "... may vary
widely: accounting for the flux ..."}

The few spectrophotometric studies performed so far have shown indeed that 
they are the right way to tackle the problem: combining high resolution broad-
and narrow-band images with high-quality spatially resolved spectra does allow
us to distinguish the young stars from the older stars, derive the history of 
the SF knots and constrain the evolutionary status of BCDs
\citep{Cairos2002,Guseva2003SBS1129,Guseva2003HS1442,Guseva2003SBS1415,Cairos2007}. That very
few spectrophotometric analyses can be found in the literature, and virtually
all of them focused on one single object, is essentially due to the large
amount of observing time that conventional observational techniques require. 
Acquiring images in several broad-band and narrow-band filters, plus a 
sequence of long-slit spectra sweeping the region of interest translates into
observing times of two or more nights per galaxy.  Thus comprehensive
analysis of a statistically meaningful sample of BCDs based on traditional
imaging and spectroscopic techniques are in terms of observing time just
not feasible. Moreover, these observations usually suffer from varying
instrumental and atmospheric conditions, which makes combining all these data
complicated. Long-slit spectroscopy has also the additional problem
of the uncertainty on the exact location of the slit.

On the other hand, it has been recently shown
\citep{Izotov2006,GarciaLorenzo2008,Kehrig2008,Vanzi2008,Lagos2009,James2009} 
that the state-of-the-art observational technique of integral field
spectroscopy (IFS) offers an alternative way to approach BCDs studies in a
highly effective manner. IFS provides simultaneous spectra of each spatial
resolution element under identical instrumental and atmospheric conditions.
This is not only a more efficient way of observing,  but it also guarantees 
the homogeneity of the dataset. In terms of observing time, IFS observations
of BCDs are one order of magnitude more efficient than traditional observing
techniques. This implies that now, for the first time, spectrophotometric
studies of substantial samples of BCD galaxies have become feasible.

Consequently, we have undertaken a long-term project, which aims to map an
extensive and representative sample of BCDs by means of IFS. This galaxy
sample, composed of about 40 objects, has been chosen so as to span the large
range in luminosities and morphologies found among the galaxies classified as
BCDs. The analysis of such a dataset will allow us to get insights into basic
questions of BCDs research, i.e. how to effectively disentangle the old and
young stellar populations, set constraints of the age and SF history of the
galaxies, study the triggering and propagation mechanisms of the star
formation and investigate the metal abundance patterns.

In the first two papers of this series
\citep{Cairos2009Mrk409, Cairos2009Mrk1418}, we illustrated the full potential
of this study by showing results on two representatives BCDs, Mrk~1418 and
Mrk~409, both observed with the Potsdam multi-aperture spectrophotometer
(PMAS), attached at the 3.5m telescope at Calar Alto Observatory. In this
paper, the remaining objects observed with PMAS are studied. The whole sample
will be analyzed in a series of future publications.

This paper is structured as follows: In Sect.~\ref{Section:Observations} we
describe the observations, the data reduction process and the method employed
to build the maps. In Sect.~\ref{Section:Results} we present the main
results of the work, that is, the flux, emission line and velocity maps, as
well as the results derived from the analysis of the integrated spectra of the
selected galaxy regions. These results are discussed in
Sect.~\ref{Section:Discussion} and summarized in
Sect.~\ref{Section:Conclusions}.

\section{Observations and data reduction}
\label{Section:Observations}

\begin{figure*}
\centering
\includegraphics[width=0.8\textwidth]{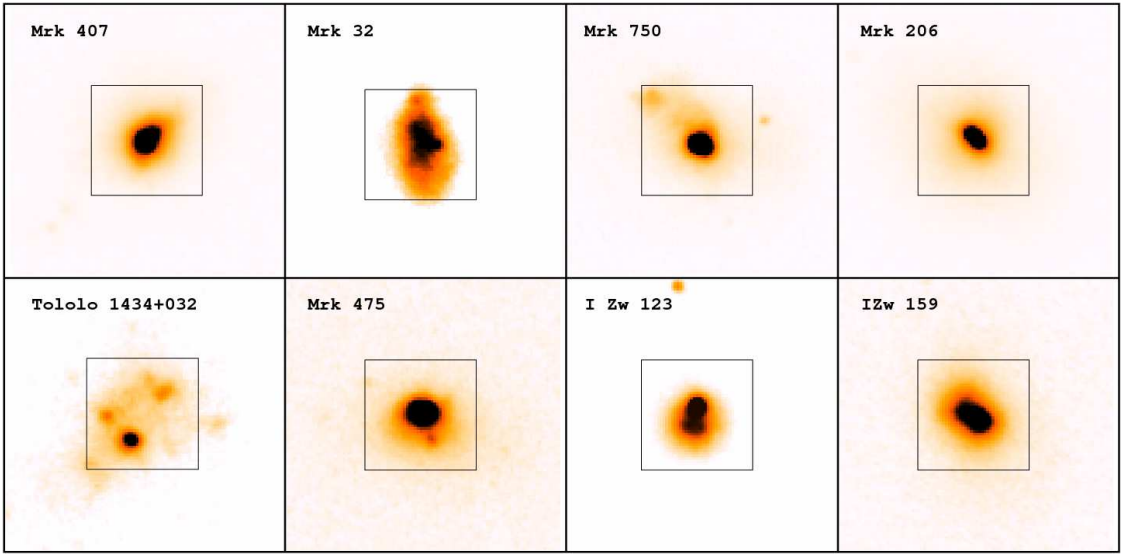}
\caption{Sloan Digital Sky Survey (SDSS) images of our sample of galaxies in 
the g-band; the field of view is 40 arcsec and the central boxes indicate the 
field of view covered by PMAS. North is up, east to the left.}
\label{Figure:mosaic}
\end{figure*}

\subsection{The galaxy sample}
\label{SubSection:GalaxySample}

We present and analyze data of eight galaxies, all of them
previously classified as BCDs. All objects except
Tololo~1434+032 are included in the \cite{ThuanMartin1981} BCD list, while
Tololo~1434+032 appears cataloged as a BCD in \cite{GildePaz2003}.  

Although strictly speaking one of the criteria that a galaxy has to
fulfill to be classified as BCD is to have $M_{B}\geq-18$, 
it is worth mentioning that most BCDs studies
\citep{ThuanMartin1981,Papaderos1996a,Cairos2001II,Cairos2001I,Bergvall2002,Kong2002spectra,GildePaz2003}
include a relatively high fraction of galaxies with luminosities higher
than this limit. Indeed, in practice the term BCD designates a set of
objects that have a very wide range in properties as luminosities ($-13 \geq
M_{B}\geq -21$), morphologies or chemical abundances
\citep{Kunth2000,Cairos2001I}.

We focus on  objects that fall strictly within the dwarf galaxy
regime; these systems, which tend to be also the more compact and more regular
BCDs, are best suited to be observed with PMAS, as their small size makes it
possible to map the whole starburst region in only one exposure. 

The basic data for the sample galaxies are listed in
Table~\ref{Table:sample}. The g-band SDSS images of the sample of galaxies 
are shown in Fig.~\ref{Figure:mosaic}.

\begin{table*}
\caption{The galaxy sample\label{Table:sample}}
\begin{center}
\begin{tabular}{lccccccc} 
\hline\hline 
Galaxy          & R.A.       & Decl.      & $m_{B}$         & $D$    & $M_{B}$ & $A_{B}$  \\
                &  (J2000)   & (J2000)    &	            & (Mpc)  & (mag)    &  (mag)   \\
\hline
Mrk~407         & 09 47 47.6 & +39 05 04  & $15.39\pm0.49$  & 27.2   & $-16.78$ & 0.069   \\[3pt]
Mrk~32          & 10 27 02.0 & +56 16 14  & $16.08\pm0.11$  & 16.4   & $-14.99$ & 0.030   \\[3pt]
Mrk~750         & 11 50 02.7 & +15 01 24  & $15.76\pm0.54$  & 05.2   & $-12.82$ & 0.175   \\[3pt]
Mrk~206         & 12 24 17.0 & +67 26 24  & $15.40\pm0.07$  & 24.3   & $-16.53$ & 0.071   \\[3pt]
Tololo~1434+032 & 14 37 08.9 & +03 02 50  & $16.91\pm0.86$  & 29.2   & $-15.42$ & 0.149   \\[3pt]
Mrk~475         & 14 39 05.4 & +36 48 22  & $16.41\pm0.31$  & 11.9   & $-13.97$ & 0.052   \\[3pt]
I~Zw~123        & 15 37 04.2 & +55 15 48  & $15.44\pm0.08$  & 15.4   & $-15.50$ & 0.062   \\[3pt]
I~Zw~159        & 16 35 21.0 & +52 12 53  & $15.65\pm0.28$  & 43.8   & $-17.56$ & 0.125   \\[3pt]
\hline
\end{tabular}
\end{center}
Notes: R.A., Decl., $D$ and  $A_{B}$ taken from the NED 
(http://nedwww.ipac.caltech.edu/).
Distances were computed using a Hubble constant of 73 km s$^{-1}$ 
Mpc$^{-1}$ and taking into account the influence of the Virgo Cluster, 
the Great Attractor and the Shapley supercluster. $m_{B}$ taken from 
HyperLeda (http://leda.univ-lyon1.fr/; \citealp{Paturel2003}) and $M_{B}$  
computed from the tabulated values of $m_{B}$ and $D$. 
\end{table*}

\subsection{Observations} 
\label{SubSection:Observations} 

Observations were carried out in 2007 March with the PMAS instrument, attached
at the 3.5m telescope in the Observatorio Astron{\'o}mico Hispano Alem{\'a}n
Calar Alto (CAHA). PMAS is an integral field spectrograph, with a lens array
of $16\arcsec\times 16\arcsec$ square elements, \textnormal{each $1\arcsec\times
1\arcsec$ in size} in the configuration used, \textnormal{connected} to a bundle
of 256 optical  fibers; the fibers are re-arranged to form a pseudoslit in the
focal plane of the spectrograph. \textnormal{The final spectrum is thus composed
of 256 spaxels, where by ``spaxel'' we refer to each element of the 16 x 16
fiber matrix}.  For a  detailed description of the instrument see
\cite{Roth2005} and  \cite{Kelz2006}. 

A grating with 300 grooves per mm was used during the observations in
combination with a SITe ST002A 2K $\times$ 4K CCD detector. This setup
provides a spectral range of 3590--6996 \AA, with a linear dispersion of 3.2
\AA\ per pixel (the CCD was binned $2\times2$ in both the spatial and the
spectral directions). 

Calibration frames were taken before and after the exposures of each galaxy.
The calibrations consist of spectra of emission line lamps (HgNe lamp),
which are required to perform the wavelength calibration, and spectra of a
continuum lamp, necessary to locate the 256 individual spectra on the CCD and
to perform the flat-fielding correction. Sky exposures were also obtained,
moving the telescope typically several arcmin from the target position. Bias
and sky-flats exposures were taken at the beginning and at the end of every
night. The spectrophotometric standard stars BD+75325 and BD+332642 were
also observed every night. The seeing ranged between 1.2 and 2 arcsec.

A complete log of the observations is provided in Table~\ref{Table:log}.

\begin{table*}
\caption{Log of the observations\label{Table:log}}
\centering
\begin{tabular}{lccc} 
\hline\hline 
Galaxy          & Exposure time  & airmass  & Seeing     \\
                &  (s)           &          & (arcsec)   \\
\hline
Mrk~407         & 6900           & 1.14--1.01 & 1.4--1.8 \\[3pt]
Mrk~32          & 6600           & 1.06--1.09 & 1.4--1.5 \\[3pt]
Mrk~750         & 5400           & 1.20--1.09 & 1.9--2.0 \\[3pt]
Mrk~206         & 6900           & 1.25--1.16 & 1.5--2.0 \\[3pt]
Tololo~1434+032 & 6600           & 1.33--1.21 & 1.5--1.8 \\[3pt]
Mrk~475         & 8400           & 1.03--1.18 & 1.5--1.9 \\[3pt]
I~Zw~123        & 3900           & 1.23--1.05 & 1.8--2.0 \\[3pt]
I~Zw~159        & 5100           & 1.09--1.04 & 1.6--1.8 \\
\hline  
\end{tabular}
\end{table*}

\subsection{Data reduction}
\label{SubSection:DataReduction}

Although several dedicated software packages have been developed in the last
years to reduce 3D-spectroscopic data \citep{Becker2002,Sanchez2006}, we
decided to process our data using standard IRAF\footnote{IRAF is distributed
by the National Optical Astronomy Observatories, which are operated by the
Association of Universities for Research in Astronomy, Inc., under cooperative
agreement with the National Science Foundation.} tasks. While using 
IRAF routines has, with respect to the use of dedicated pipelines, the main
drawback of requiring a considerable amount of interactive work, which makes
the whole process somewhat slower, it has on the other hand the advantage of 
allowing a complete and precise control of all the parameters involved in 
each data-reduction step. 

The reduction procedure includes the bias subtraction, image trimming, 
tracing and extraction of the individual spectra, wavelength and distortion 
calibrations, flat-fielding, combination of the individual galaxy frames, 
sky-subtraction and flux calibration. 

The first step in the data reduction was the bias subtraction.  All the bias
exposures were averaged to obtain a master bias, which was then
subtracted from the rest of the frames. Bad columns were interpolated with 
the IRAF task \emph{fixpix}.

Next, apertures were defined and traced on the detector. Defining the
apertures means to identify on the detector the spectra produced by the
different fibers (that is, to find out how many and which pixels on the
detector correspond to each fiber). The apertures are affected by the
field distortions and/or by the optics of the system, and therefore each
aperture does not line up along the dispersion axis, but has a clear
curvature. Hence, after we have defined the apertures at a given spectral
position, each of these loci must be traced along the spectral direction.

Apertures were defined on well exposed continuum frames with 
the IRAF task \emph{apall}; the task first finds the centers of each fiber 
(the emission peaks) along the spatial axis at some specified position, and 
then asks for the size of the extraction window, which we set to 6.4 pixels
(the best compromise between including as much signal as possible without
contamination by nearby fibers).  
The apertures  were then traced  by fitting a polynomial to the centroid 
along the dispersion axis. A fifth degree Legendre polynomial was found to
provide good fits, with a typical RMS of about 0.01 pixels.

Once the apertures were defined and traced in the continuum frames, we again 
used \emph{apall} to extract them in all the images. The extraction consists
of summing  the pixels  along the  spatial direction into a final
one-dimensional spectrum.  After that, we had the so-called ``collapse'' or
``row-stacked'' spectra: an image  $M \times N$, where  $M$ is the number of 
pixels in  the dispersion direction, and $N$ is the number of fibers (256 for
PMAS).

Afterwards we performed the  wavelength  calibration and the dispersion
correction. In order to calibrate in wavelength we used the tasks 
\emph{identify} and \emph{reidentify}: i)  first, in the comparison spectra 
(arc) we identified several  emission features of a known wavelength in a
reference fiber; ii) second, a  polynomial was fitted  across the  dispersion
direction;  the standard deviation (RMS) of the polynomial fit gives an
estimate of the uncertainty  in the wavelength calibration. We  obtained
typical RMS of about 0.01 \AA\ by fitting a fifth degree polynomial.  iii)
next, with \emph{reidentify}  we identified the emission lines in all the
remaining fibers of the arc frame, using the selected one as a reference.  

Because of instrumental flexures, there are significant shifts (by up to two
pixels) along both the spatial and the spectral directions, even among a
sequence of consecutive exposures of the same object.  Spatial shifts can be
taken care of by measuring the offset between the brightest fiber spectra in
the galaxy spectrum and in the corresponding continuum, and applying the
correction in \emph{apall}.

As for the wavelength shifts, first we determined that these shifts were
independent of the wavelength itself by comparing arc spectra taken at
different times and telescope positions. We also assessed that the relation
between pixel coordinate and wavelength were essentially linear, with
negligible deviations from linearity.

Then, in each sequence of spectra of a same target we measured the shift in
pixel on the spectral axis of the bright sky line at 5577 \AA, relative to 
the first spectrum of the sequence. The transformation in wavelength was done
with the task \emph{dispcor}, by slightly modifying the starting and ending
wavelengths so that $w_{\rm s} = w_{\rm s,0} + \delta X\cdot D$, where
$w_{\rm s}$ is the starting wavelength, $w_{\rm s,0}$ is the starting 
wavelength of the reference spectrum, $\delta X$ is the shift along the
spectral direction, in pixels, and $D$ is the actual dispersion of the 
spectra. In this way the sky line ends up at exactly the same position (within
a few hundredths of pixels), which ensures that all the wavelength-calibrated
spectra of the same object are at the same zeropoint.

\textnormal{The wavelength calibrated data were corrected for  response (detector
pixel sensitivity variations as well as wavelength-dependent variations in the
fibers transmission curves) by using the wavelength-calibrated continuum
frames, and for throughput (variations in the whole responsivity of the
lenslets and fibers) by using the sky-flat exposures. Both steps were carried
out simultaneously by running the task \emph{msrep1}.} 


After that the individual galaxy frames were corrected for atmospheric
extinction (adopting the ``summer extinction coefficients''  published by
\citealp{Sanchezetal2007}) and combined with the task \emph{imcombine}. The
sky was subtracted from the final combined frame. For each object we took an
offset sky exposure of a shorter duration (typically 5 minutes). Sky spectra
were processed in the same way as galaxy spectra. A one-dimensional sky
spectrum was produced by averaging the signal along the spatial direction with
a sigma-clipping algorithm. The flux of the three to four brightest sky
lines was measured in both the final sky spectrum and in the final galaxy
spectrum to determine the appropriate scaling factor (with an accuracy of a
few percents) by which to multiply the sky spectrum before subtracting it from
the galaxy spectrum.

Because the relative intensity of different sky lines varies noticeably on
short time scales \citep{Patat2003}, it is very difficult to find a scaling
factor that applies equally well to all the sky lines and the sky continuum,
and some fine tuning is required.  The final scale factor was found by trial
and error; we aimed at a value that minimized overall residuals in the sky
lines (especially those close to galaxy emission or absorption lines) even if
it left large residuals in sky bright lines that were not affecting any
interesting spectral features. 

We must say here that minimizing sky-line residuals does not necessarily
imply the best match between the sky background in the sky exposure and the
sky background in the galaxy exposure.  For this reason the uncertainties on
the sky-subtracted galaxy continuum may be relatively large (and difficult to
estimate), especially in the outer regions of galaxies. While this does not
at all affect the emission-line parameters (flux, width, redshift), it can
clearly have a significant impact on the equivalent widths of the less
luminous SF knots and on the outer, fainter spaxels in the continuum maps.

Spectra of spectrophotometric standards were reduced in the same way, except
that the sky spectrum to be subtracted was computed with the median of the
outermost fibers. The integrated spectrum of the spectrophotometric standards
was obtained by summing all the fibers within a radius of about 2 FWHM
(typically 3 to 4 arcsec) from the fiber with the highest signal. The IRAF
tasks \emph{standard} and \emph{sensfunc} were used to derive the sensitivity
curves, after combining the data for the different spectrophotometric stars
observed in the same night.

By comparing the sensitivity curves for different nights and stars in this
same observing run, we can estimate that the relative uncertainty on the
calibration factor is generally equal or less than 2\%, except blueward of
4000 \AA, where the curve shows a marked change of slope and the uncertainty
increases to about 8\%.

No corrections for differential atmospheric refraction (DAR) have been 
applied to our data. For the object observed at the highest airmass 
(Tololo~1434+032, airmass $\simeq 1.3$), the differential shift between the 
bluest and the reddest wavelength in our spectral range, measured on our
data, is less than 1 arcsec (the shift between the \Ha\ and the \Hb\ lines is
about 0.3 arcsec). Given the PMAS spaxel size of 1 square arcsec, the seeing
$\ge 1\farcs2$, and the fact that the diagnostic line ratios we compute
involve emission lines very close to each other in wavelength, we can safely 
ignore DAR effects.

\subsection{Emission line fit}
\label{SubSection:LineFit}

In order to measure the relevant parameters of the emission lines (position,
flux and width), they were fitted by a single Gaussian. The fit was carried
out by the chi square minimization algorithm implemented by 
C.~B.~Markwardt in the \emph{mpfitexpr} IDL library\footnote{URL:
http://cow.physics.wisc.edu/~craigm/idl/idl.html}. The \Hb\ line, where
an absorption component was present, was fitted by two Gaussians.

The continuum (typically 30--50 \AA\ on both sides) was fitted by a straight
line. Lines in a doublet were fitted imposing that they have the same redshift
and width.

Criteria like flux, error on flux, velocity and width were used to do a
first automatic assessment of whether to accept or reject a fit. For
instance, lines with too small (less than the instrumental width) or too
large widths were flagged as rejected, as well as lines with a relative error 
on the flux of more than about 10\% (the exact limits depend on the
specific line and on the overall quality of the spectrum). These criteria
were complemented by a visual inspection of all fits, which led to
override in a few cases the automated criteria decision (either accept a
fit flagged as rejected, or viceversa).

\subsection{Creating the 2D maps}
\label{SubSection:CreatingMaps}

Emission-line maps were constructed in the following way: the emission-line
fit-procedure gives for each line and for each line parameter (for instance
flux) a table with the fiber ID number, the measured value and the
acceptance/rejection flag. This table was then used to produce a 2D map, by
using an IRAF script that takes advantage of the fact that PMAS fibers are
arranged in a regular 16x16 matrix. The script automatically converts ADU
counts into flux (erg s$^{-1}$ cm$^{-2}$) by multiplying by a
wavelength dependent conversion factor computed by using the sensitivity
curve described above.

Continuum maps were obtained by summing the flux within specific wavelength
intervals, selected so as to avoid emission lines or strong residuals from the
sky spectrum subtraction.

Line ratio maps were simply derived by dividing the corresponding flux maps.


\section{Results}
\label{Section:Results}

\subsection{Intensity maps}
\label{SubSection:IntensityMaps}

IFS data provide within the field of view (FOV) of the instrument a
simultaneous mapping of the galaxy emission in a broad wavelength range.
Therefore we can retrieve monochromatic maps at specific wavelengths or
co-added maps equivalent to broad- or narrow-band images. We constructed
emission-line intensity and continuum maps for the observed galaxies. Results
are shown in Figs.~\ref{Figure:mrk407}--\ref{Figure:izw159}.

\subsubsection{Continuum maps}
\label{SubSubSection:ContinuumMaps}

\begin{figure*}[h]
\mbox{
\centerline{
\hspace*{0.0cm}\subfigure{\includegraphics[width=0.24\textwidth]{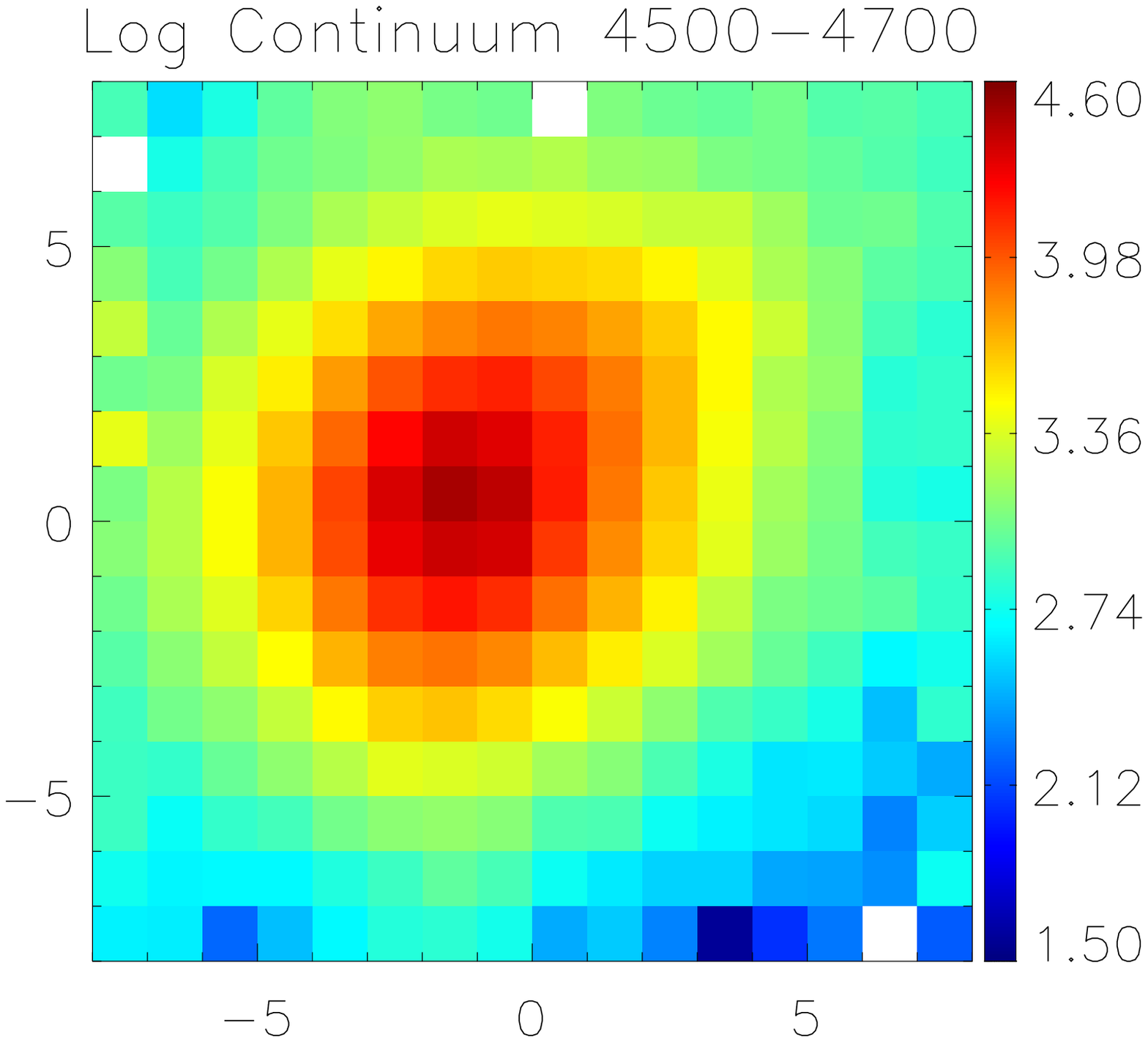}}
\hspace*{0.0cm}\subfigure{\includegraphics[width=0.24\textwidth]{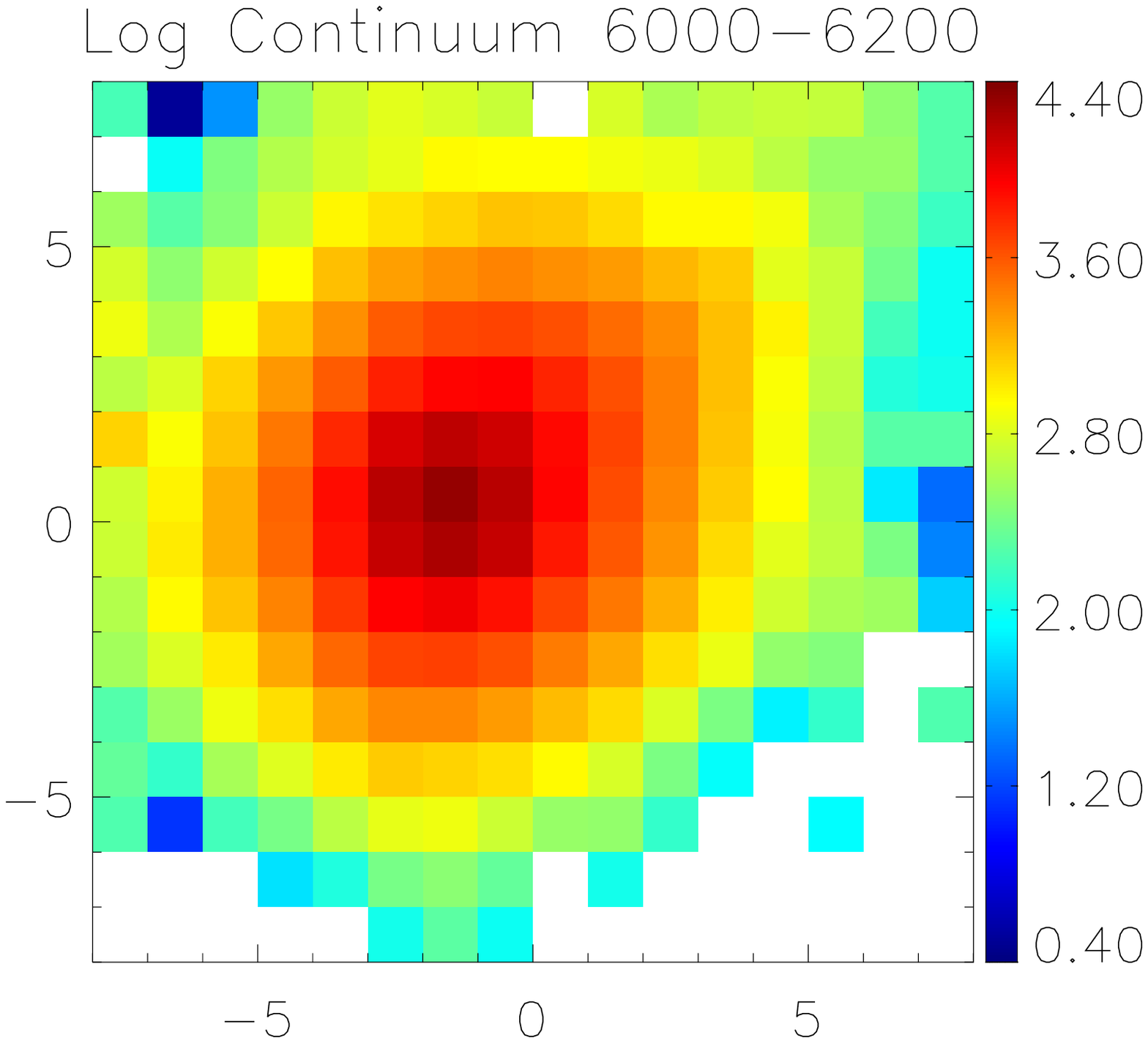}}
\hspace*{0.0cm}\subfigure{\includegraphics[width=0.24\textwidth]{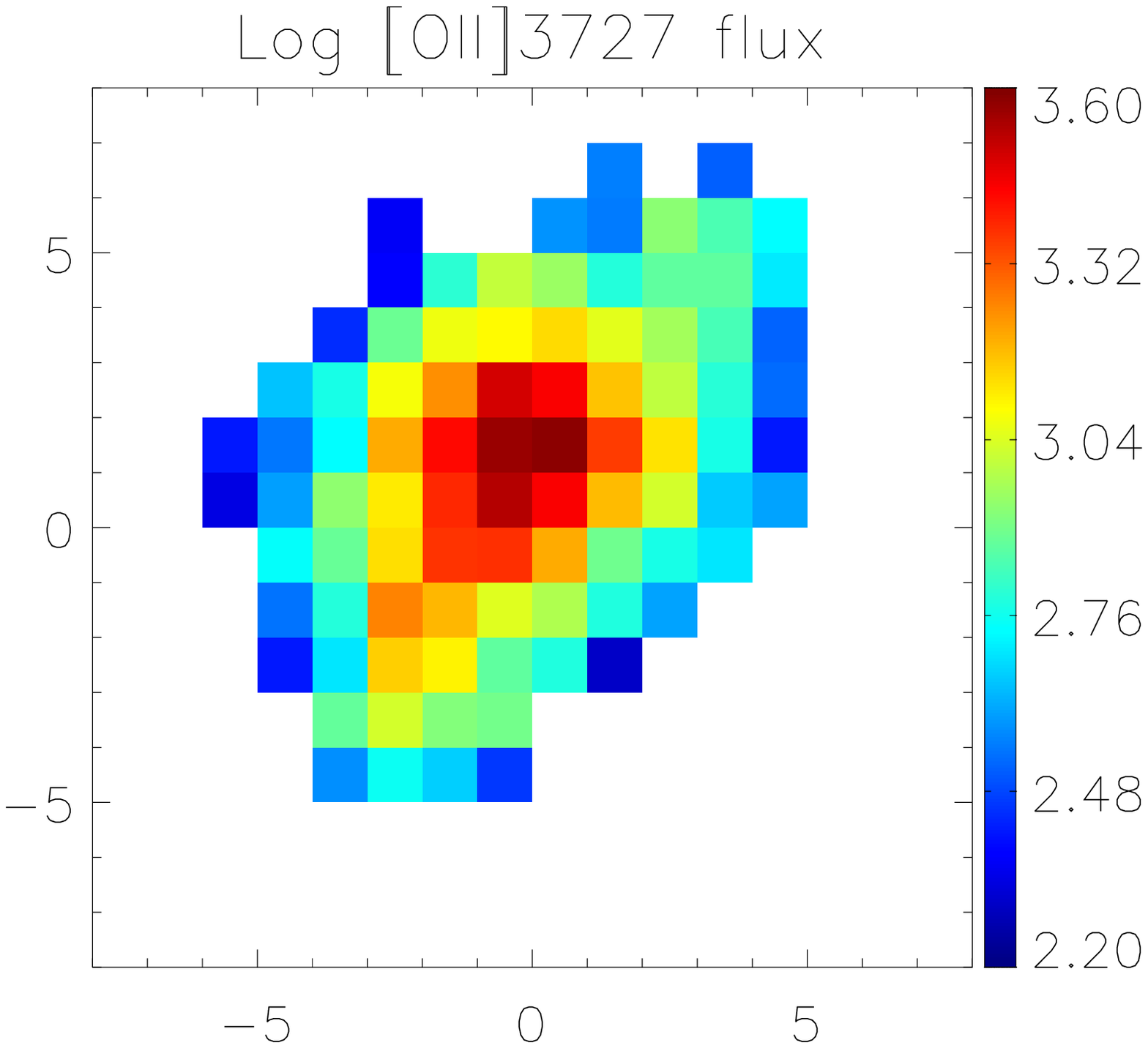}}
\hspace*{0.0cm}\subfigure{\includegraphics[width=0.24\textwidth]{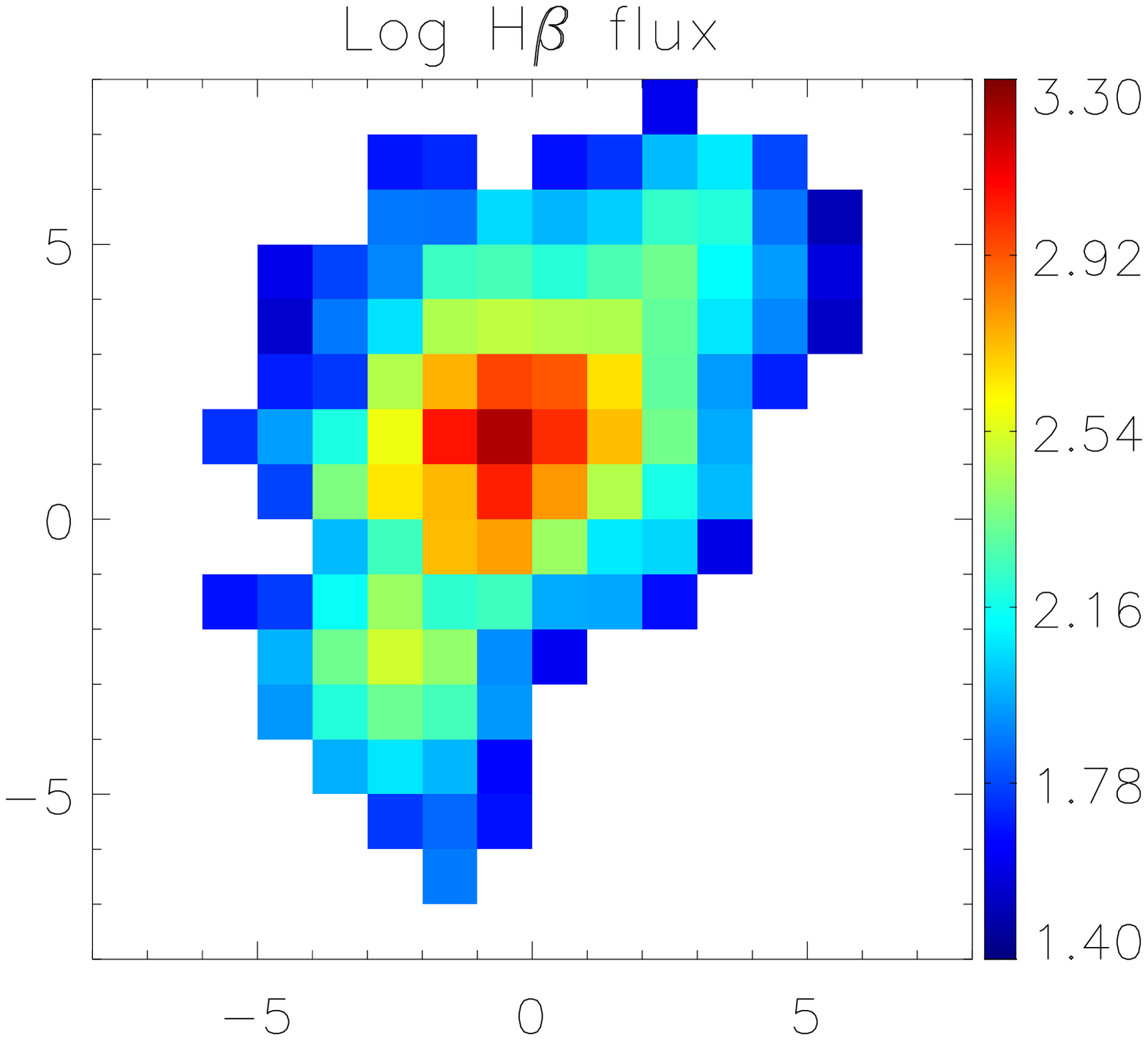}}
}}   
\mbox{
\centerline{
\hspace*{0.0cm}\subfigure{\includegraphics[width=0.24\textwidth]{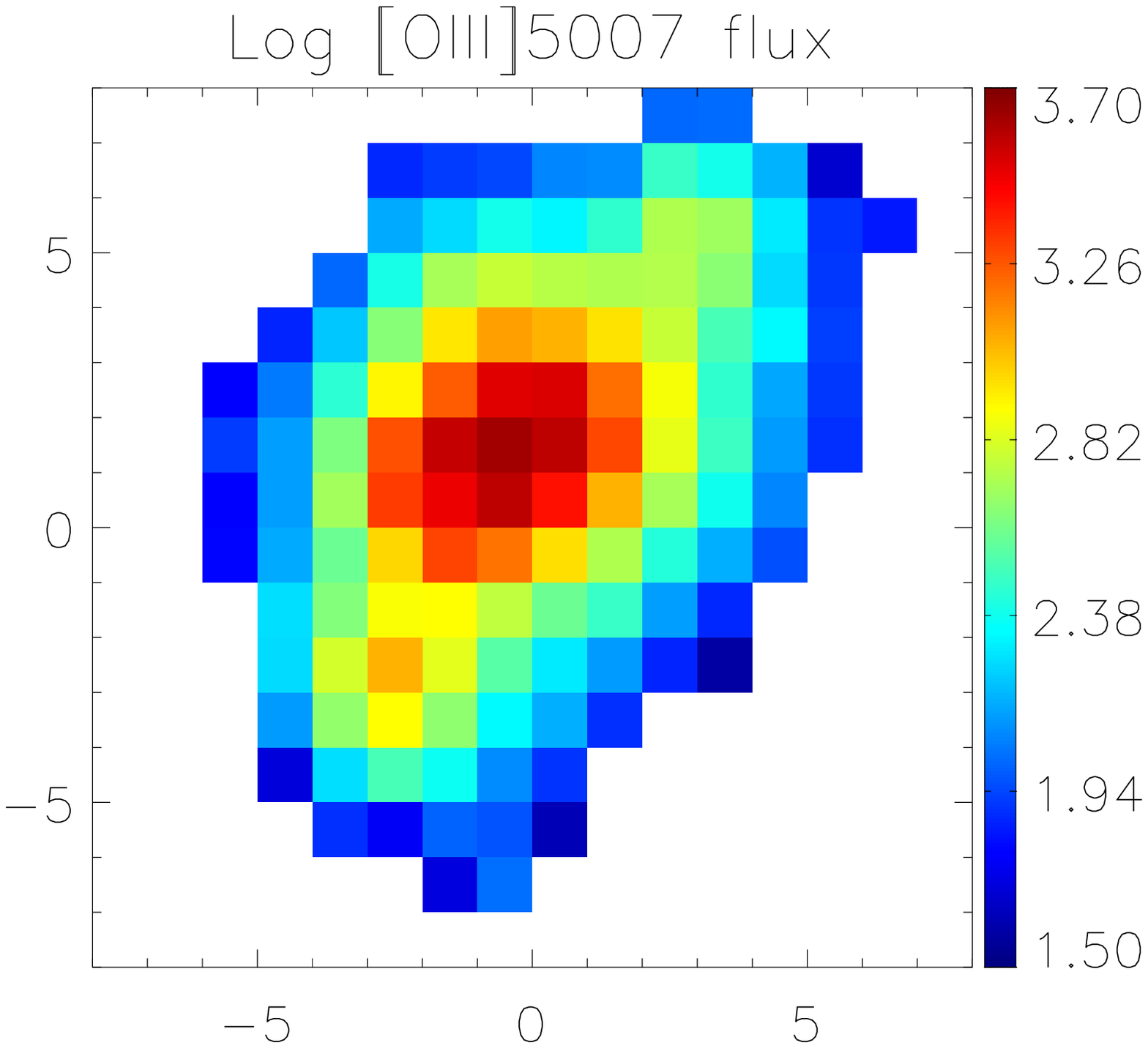}}
\hspace*{0.0cm}\subfigure{\includegraphics[width=0.24\textwidth]{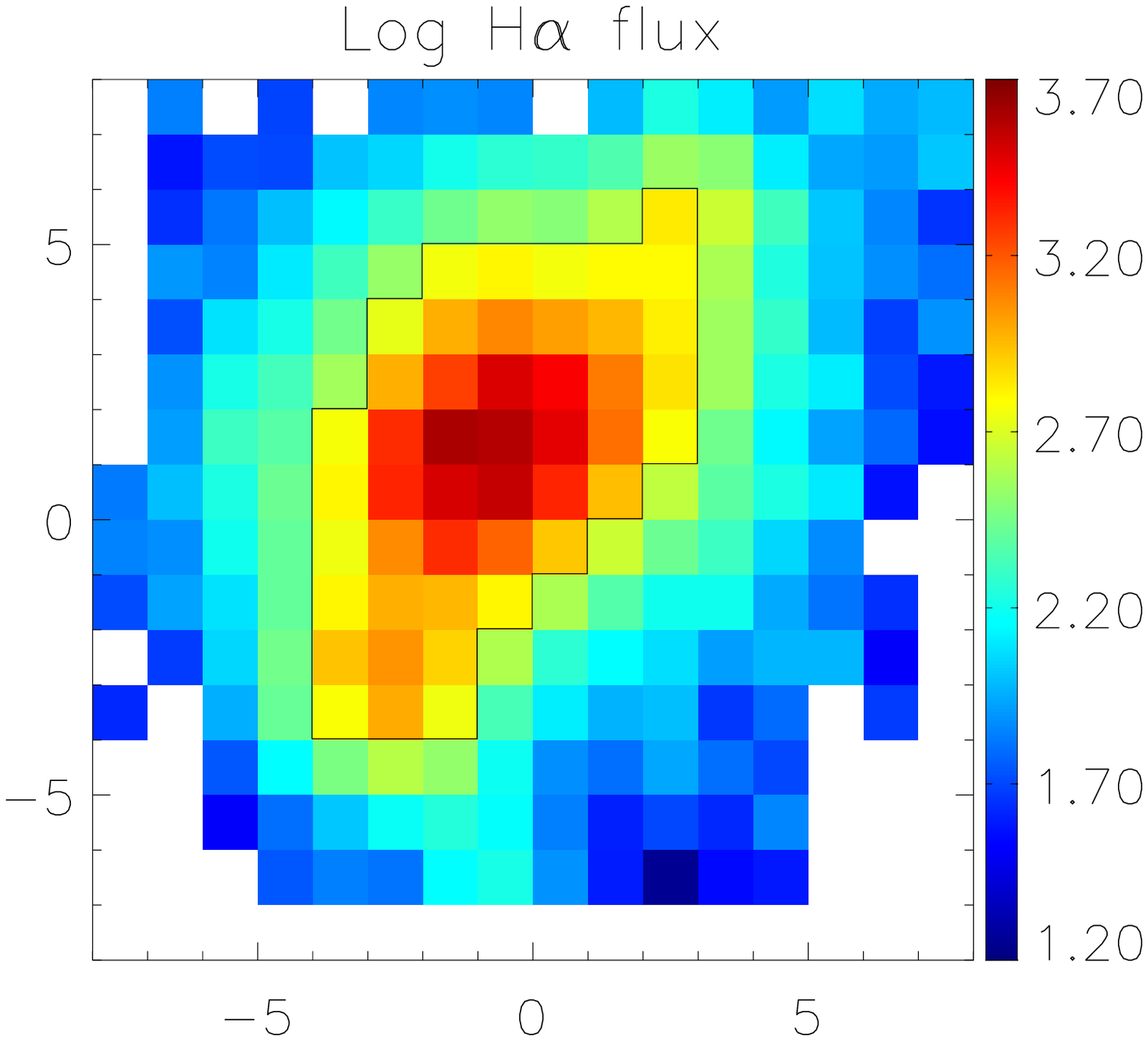}}
\hspace*{0.0cm}\subfigure{\includegraphics[width=0.24\textwidth]{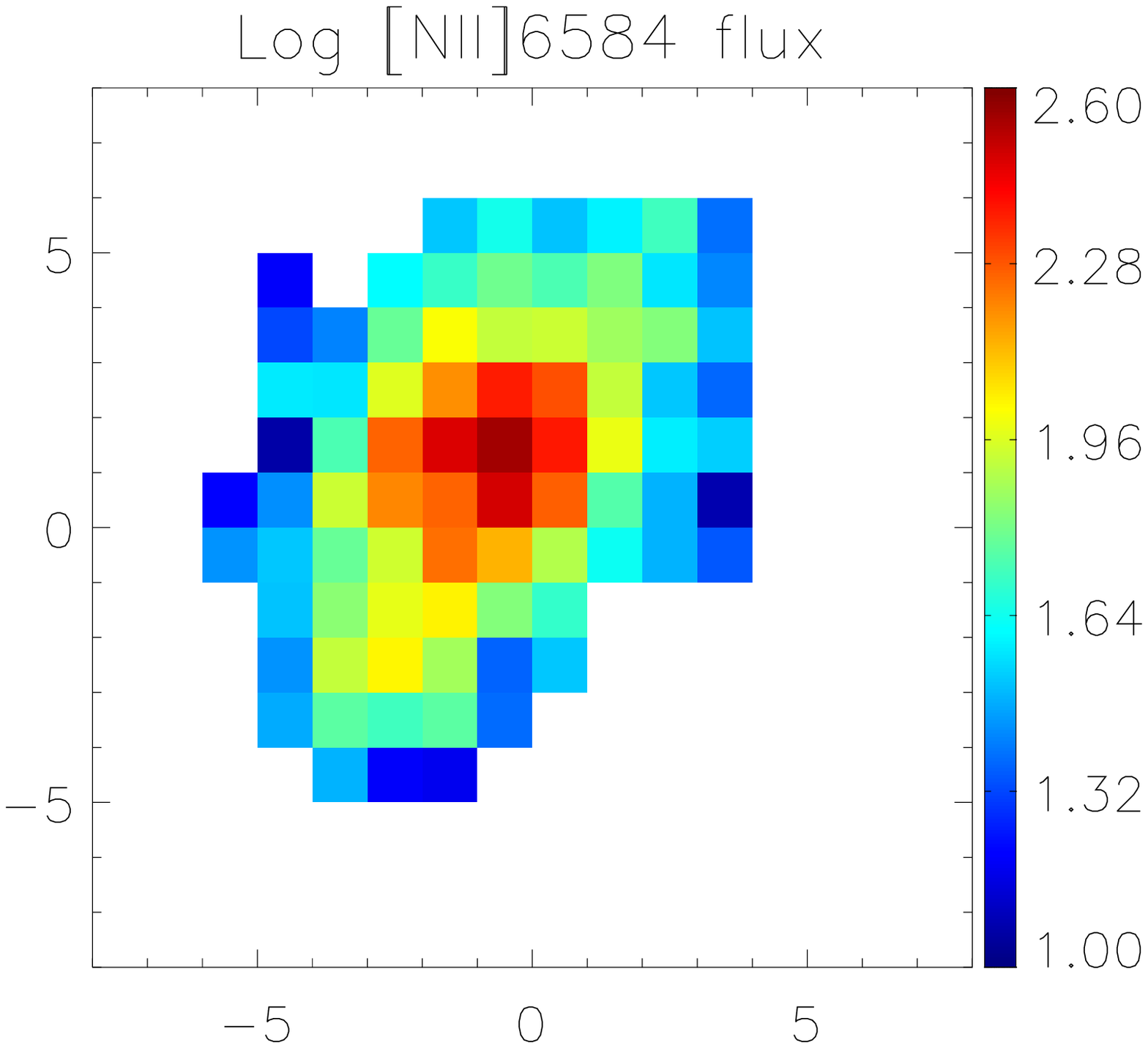}}
\hspace*{0.0cm}\subfigure{\includegraphics[width=0.24\textwidth]{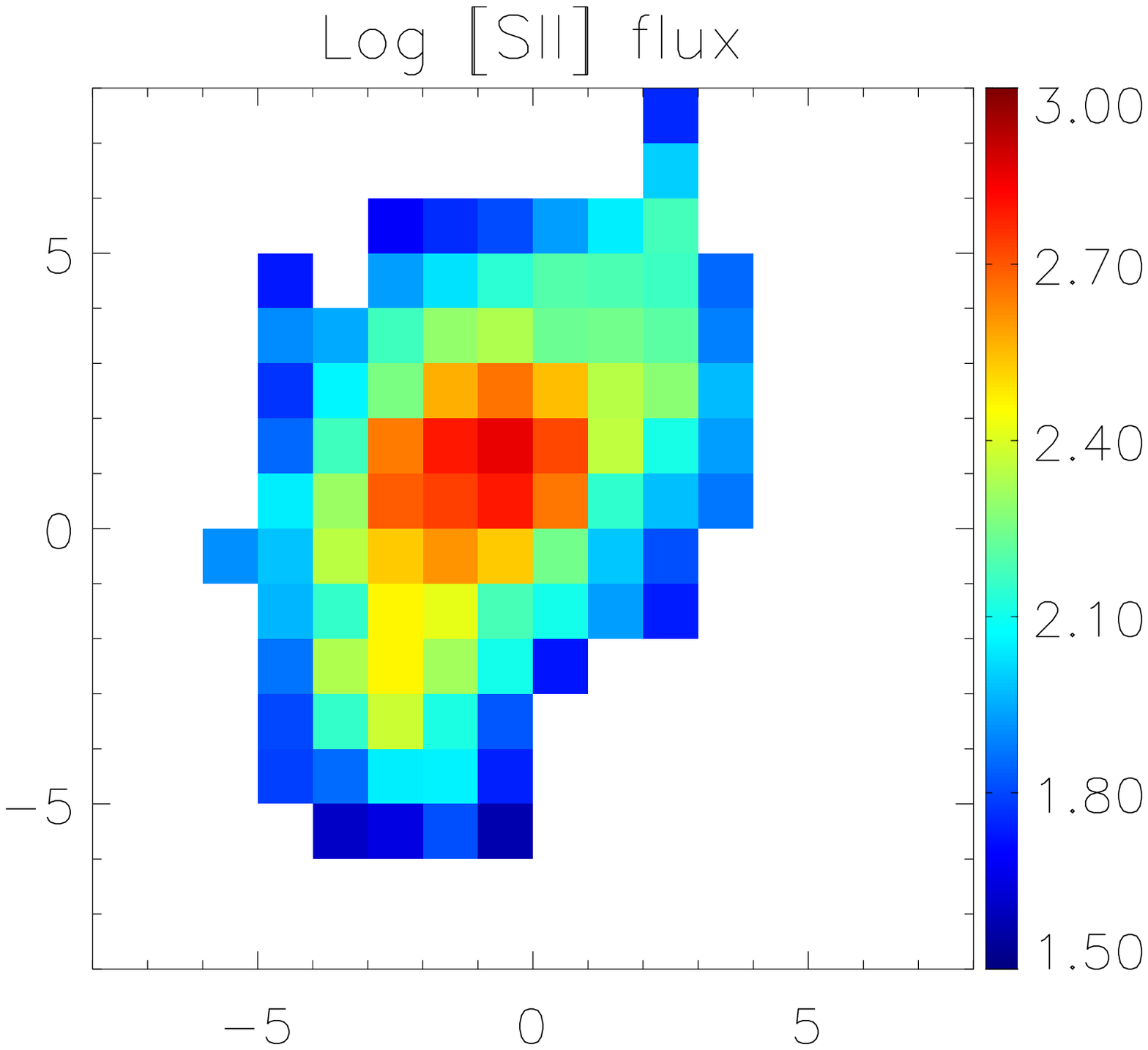}}
}} 
\mbox{
\centerline{
\hspace*{0.0cm}\subfigure{\includegraphics[width=0.24\textwidth]{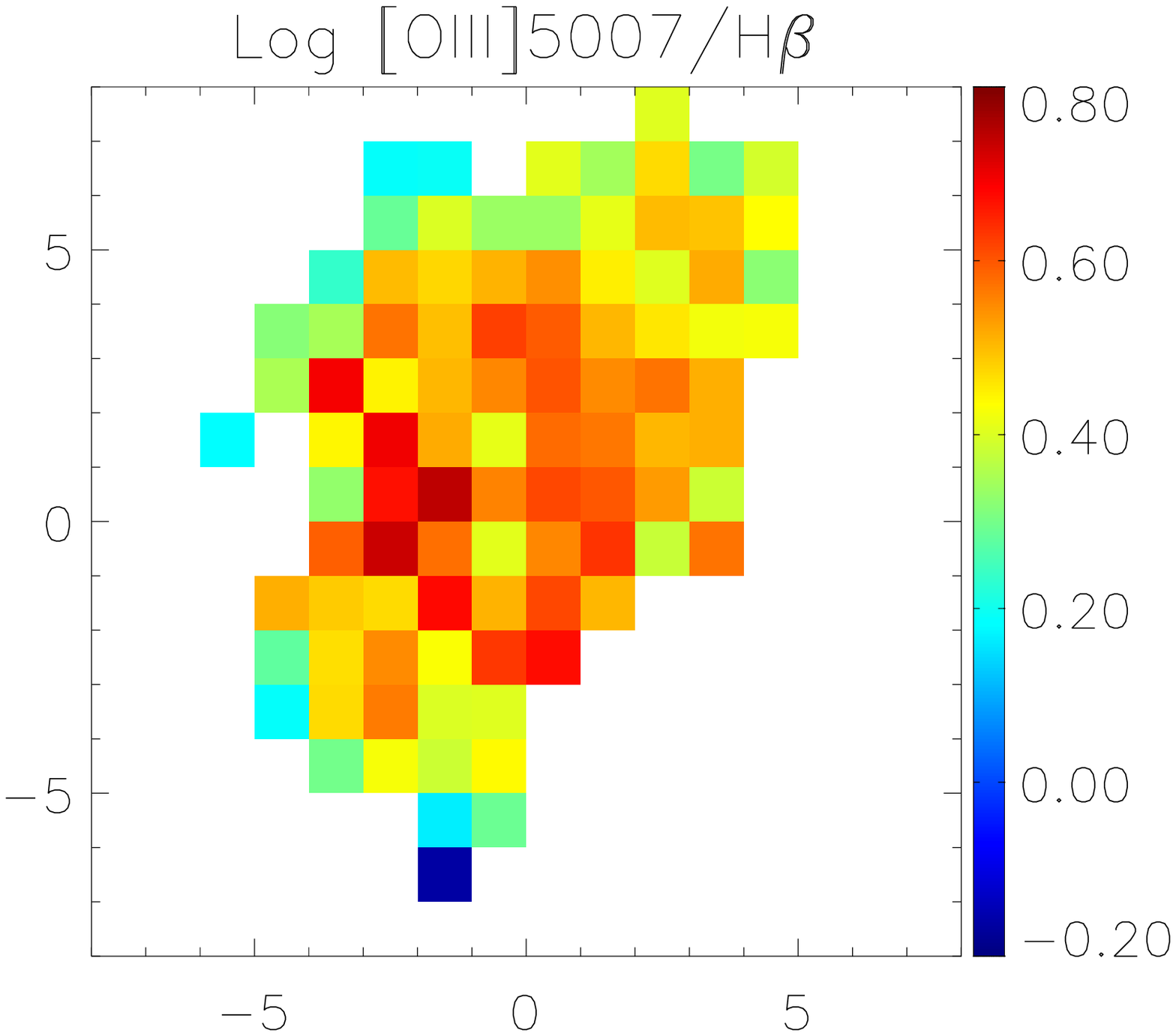}}
\hspace*{0.0cm}\subfigure{\includegraphics[width=0.24\textwidth]{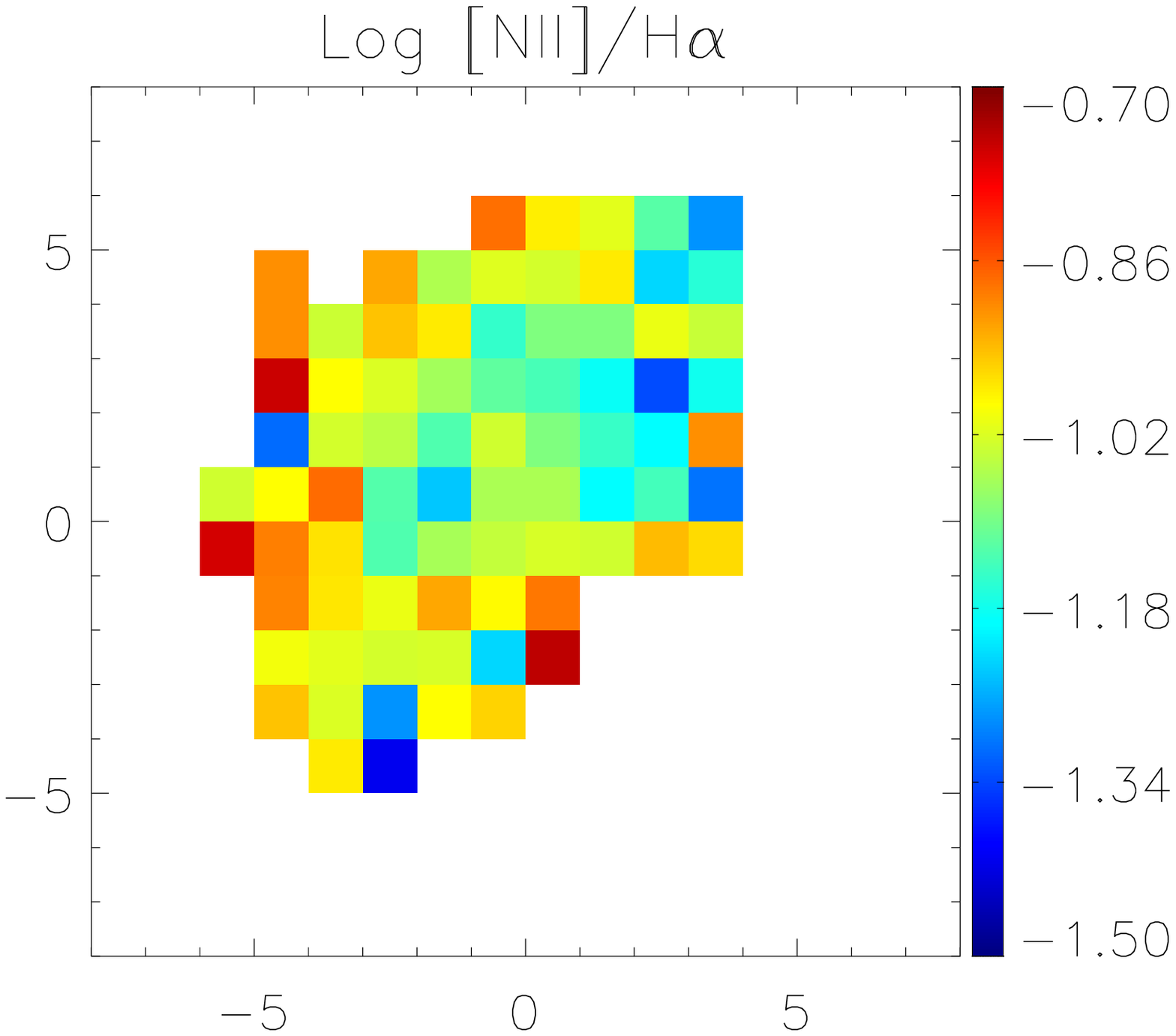}}
\hspace*{0.0cm}\subfigure{\includegraphics[width=0.24\textwidth]{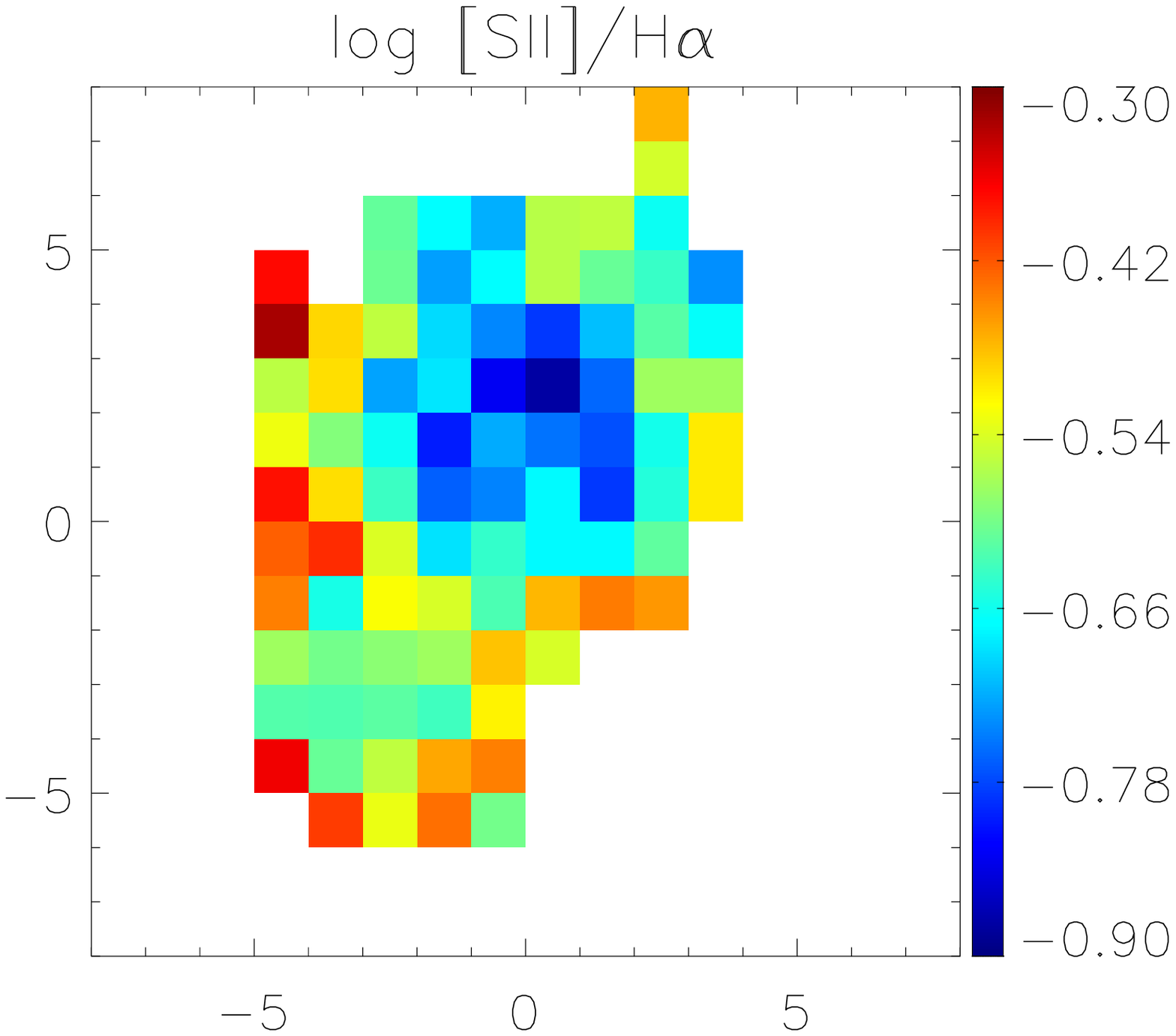}}
\hspace*{0.0cm}\subfigure{\includegraphics[width=0.24\textwidth]{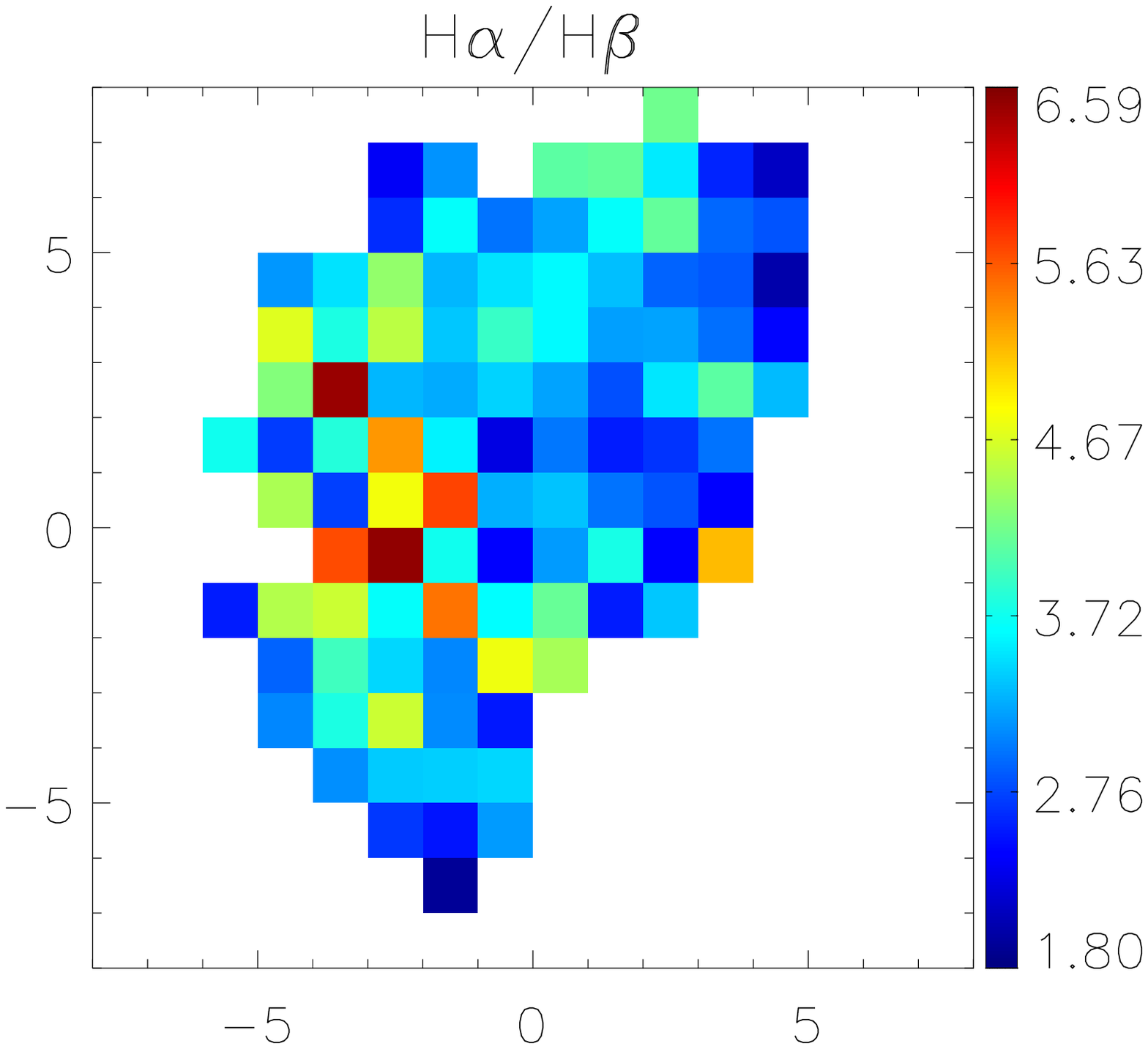}}
}}
\mbox{
\centerline{
\hspace*{0.0cm}\subfigure{\includegraphics[width=0.24\textwidth]{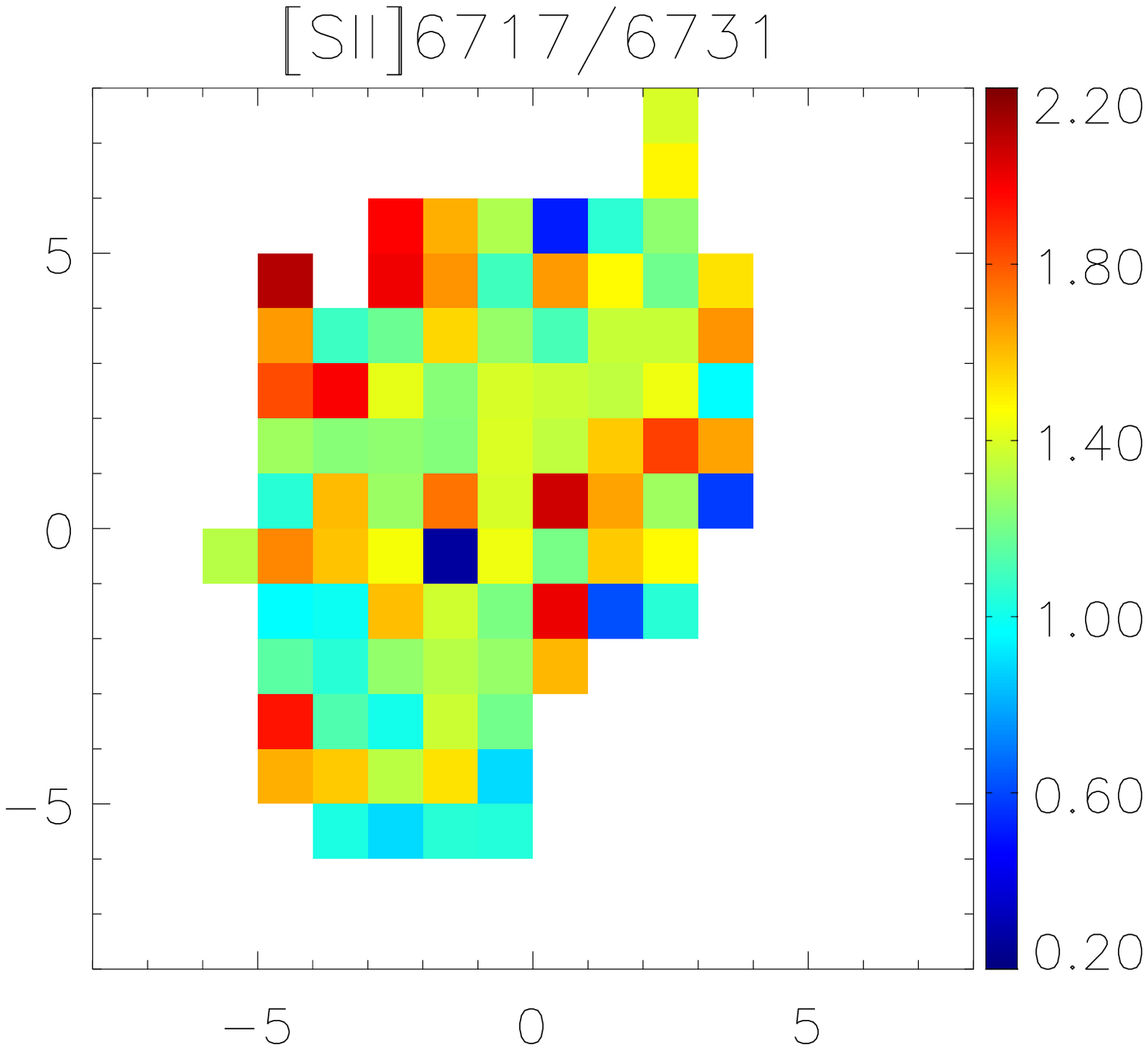}}
\hspace*{0.0cm}\subfigure{\includegraphics[width=0.24\textwidth]{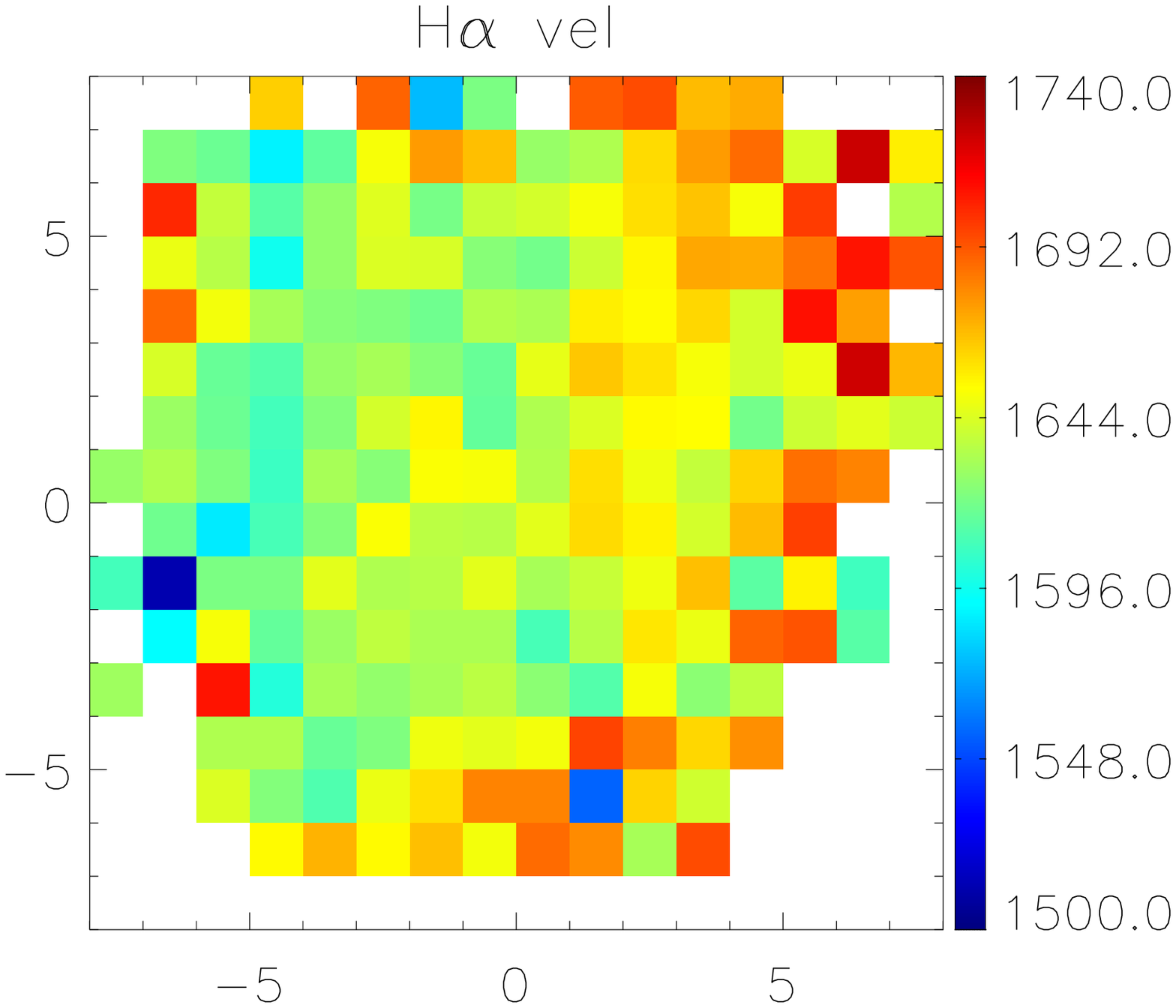}}
\hspace*{0.0cm}\subfigure{\includegraphics[width=0.24\textwidth]{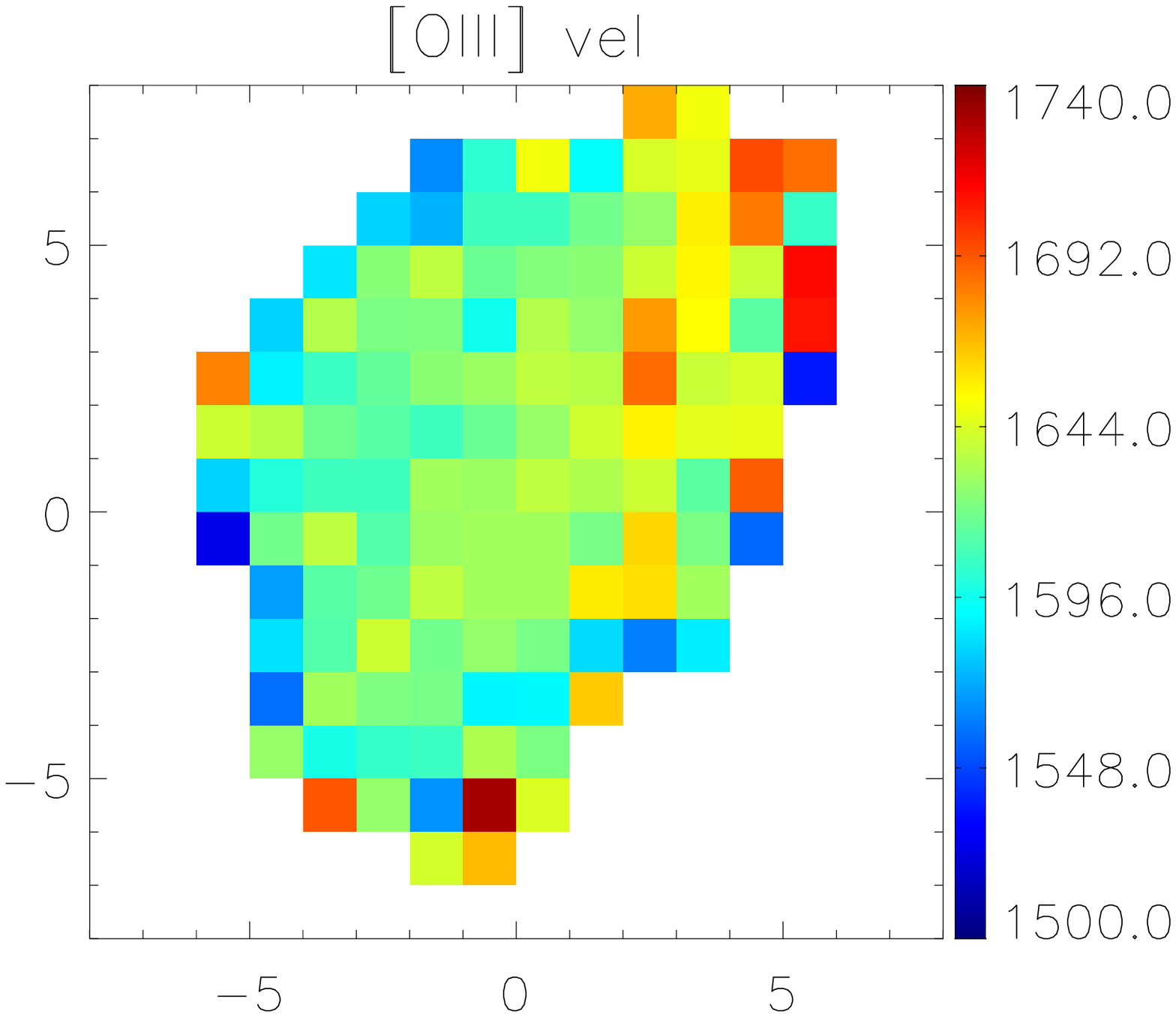}}
}}  
\caption{Two-dimensional maps for Mrk~407. 
Continuum maps in the ``emission line free'' intervals 
4500--4700 \AA\ (blue) and 6000-6200 \AA\ (red); emission line flux maps:   
[\ion{O}{ii}]~$\lambda$3727, \Hb, [\ion{O}{iii}]~$\lambda5007$, \Ha, 
[\ion{N}{ii}]~$\lambda6584$ and [\ion{S}{ii}]~$\lambda\lambda6717,\;6731$;
lines ratio maps:
[\ion{O}{iii}]~$\lambda5007$/\Hb, [\ion{N}{ii}]~$\lambda6584$/\Ha,
[\ion{S}{ii}]~$\lambda\lambda6717,\;6731$/\Ha\ (ionization ratios), 
\Ha/\Hb\ (interstellar extinction) and 
[\ion{S}{ii}]~$\lambda6717$/[\ion{S}{ii}]~$\lambda6731$ (electron density);
velocity fields of the ionized gas in the \Ha\ and 
[\ion{O}{iii}]~$\lambda5007$ lines. 
Axis units are arcseconds; north is up, east to the left. 
All the maps except extinction, electron density and velocities are in  
logarithmic scale. Flux units are $10^{-18}$ ergs cm$^{-2}$ s$^{-1}$.
The outline of the region within which the integrated nuclear spectrum 
was obtained (see 
Sect.~\ref{SubSection:IntegratedSpectroscopy}) is shown in the \Ha\ map.
}
\label{Figure:mrk407}
\end{figure*}

\begin{figure*}
\mbox{
\centerline{
\hspace*{0.0cm}\subfigure{\includegraphics[width=0.24\textwidth]{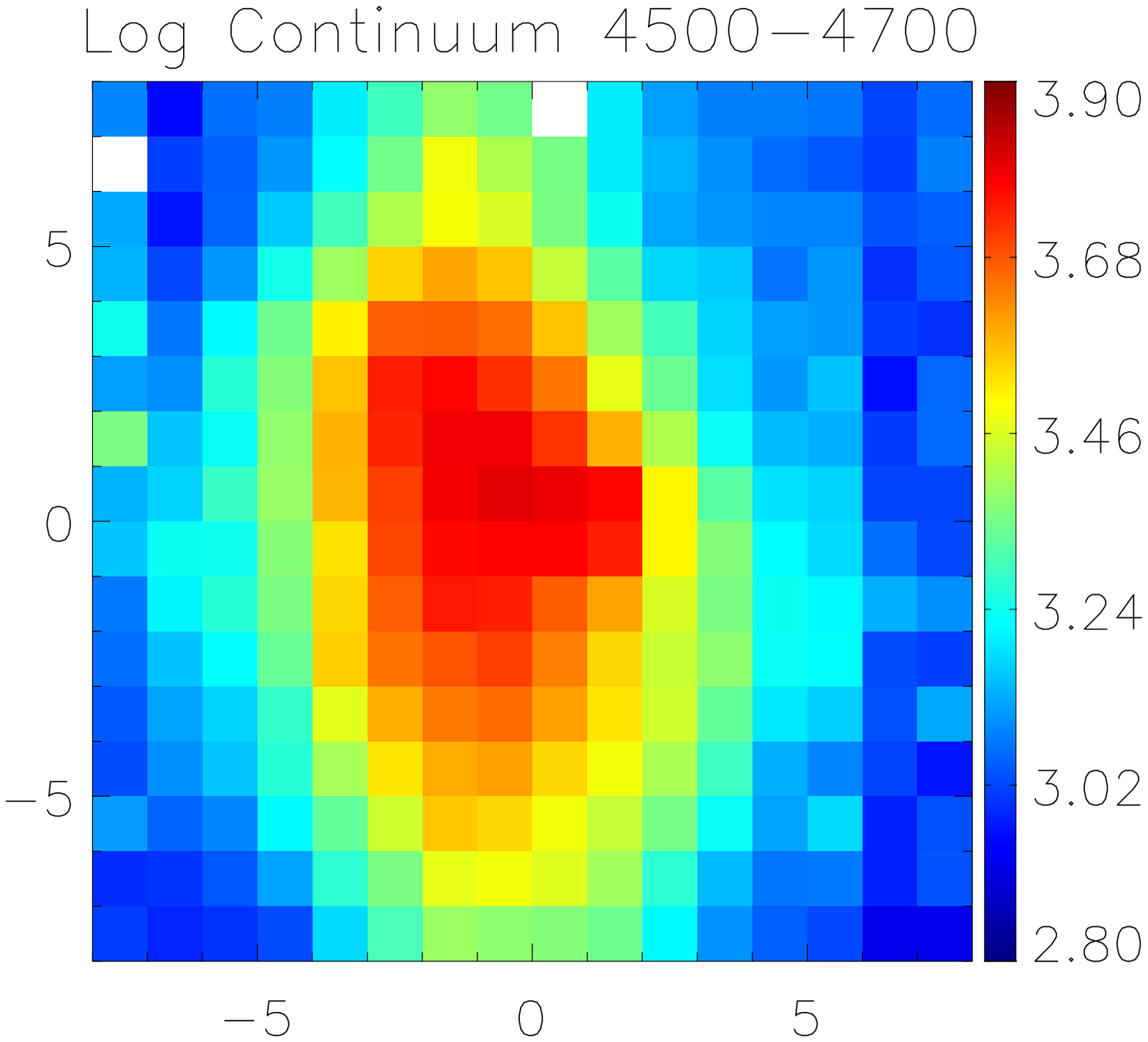}}
\hspace*{0.0cm}\subfigure{\includegraphics[width=0.24\textwidth]{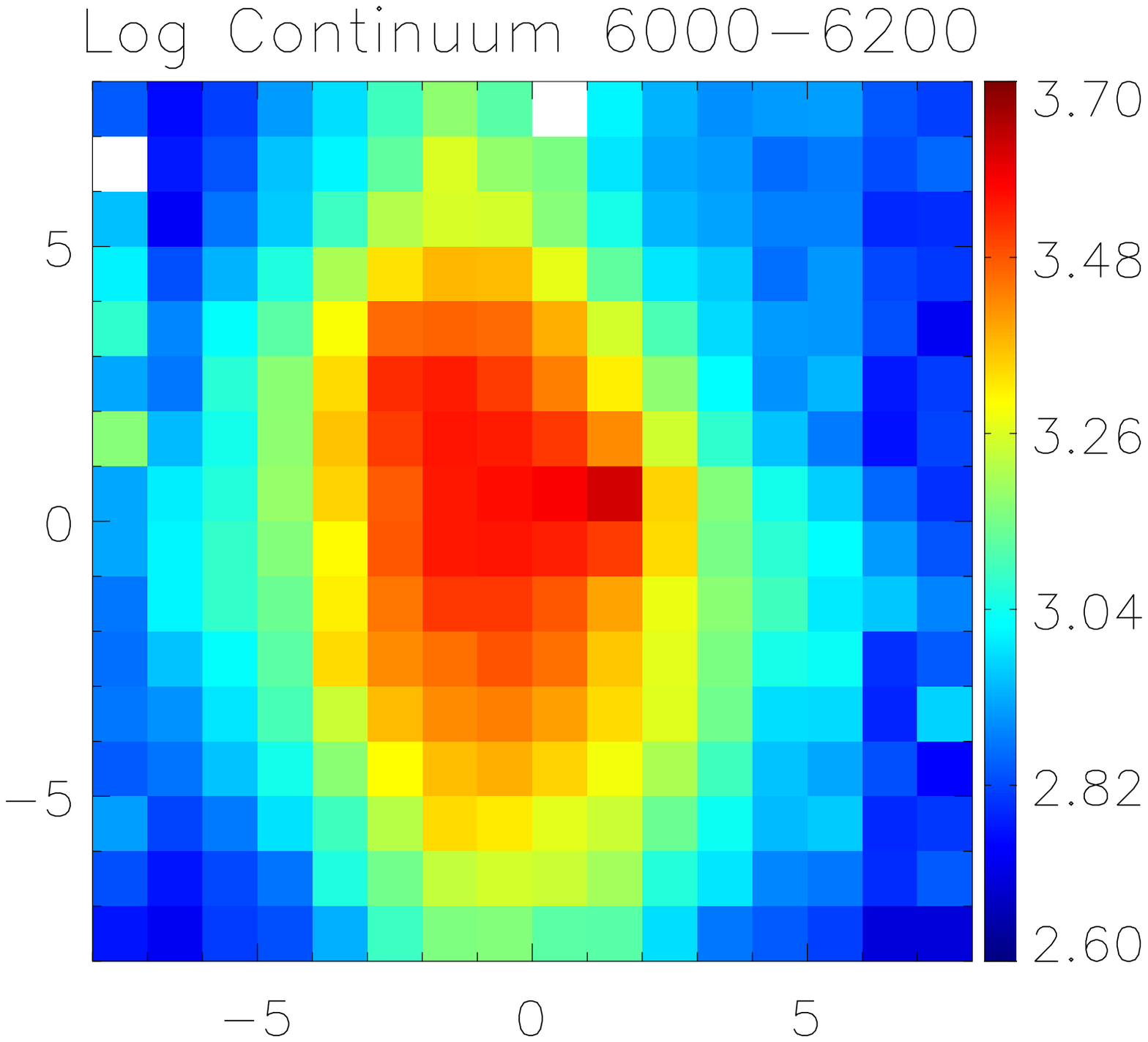}}
\hspace*{0.0cm}\subfigure{\includegraphics[width=0.24\textwidth]{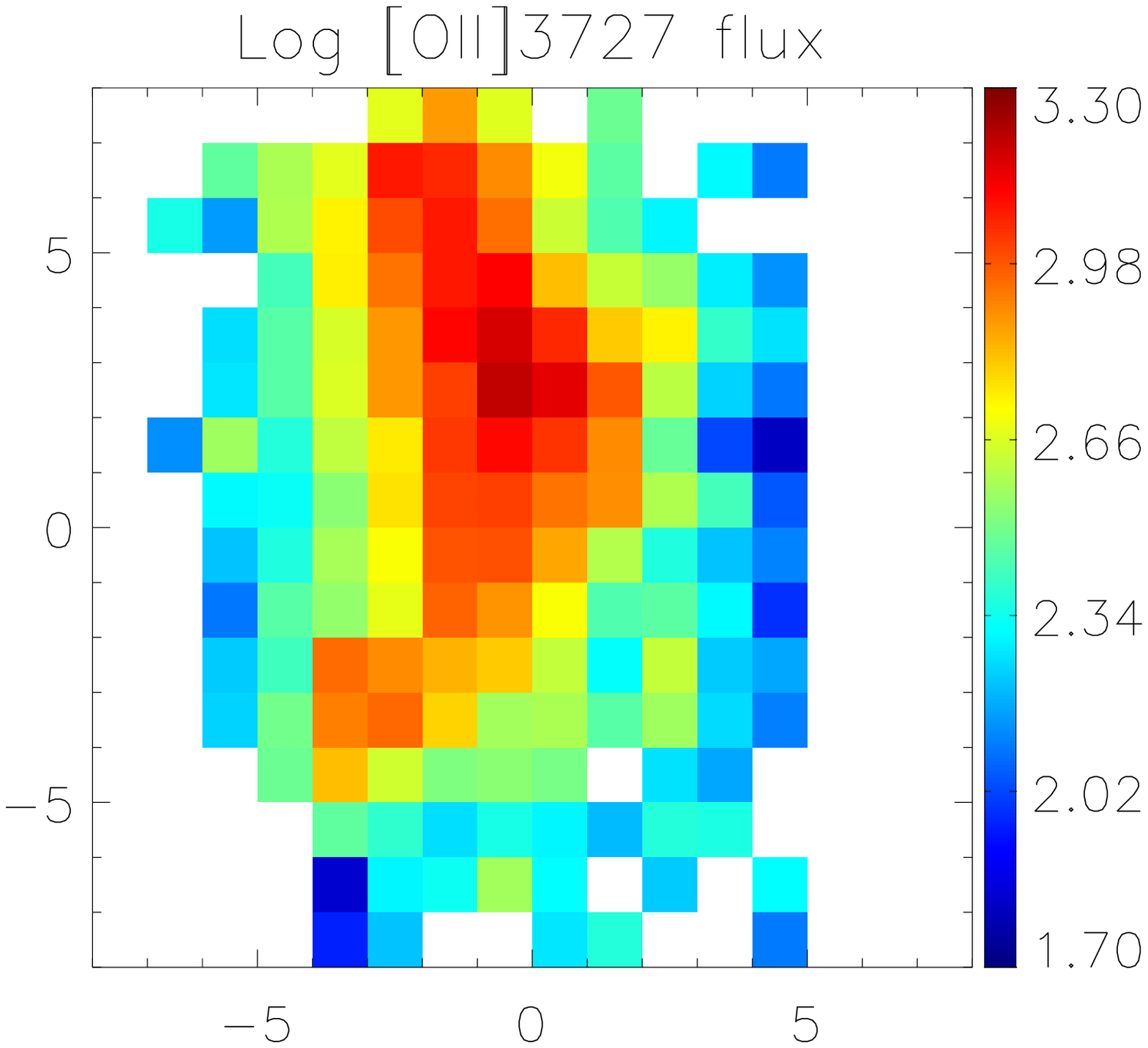}}
\hspace*{0.0cm}\subfigure{\includegraphics[width=0.24\textwidth]{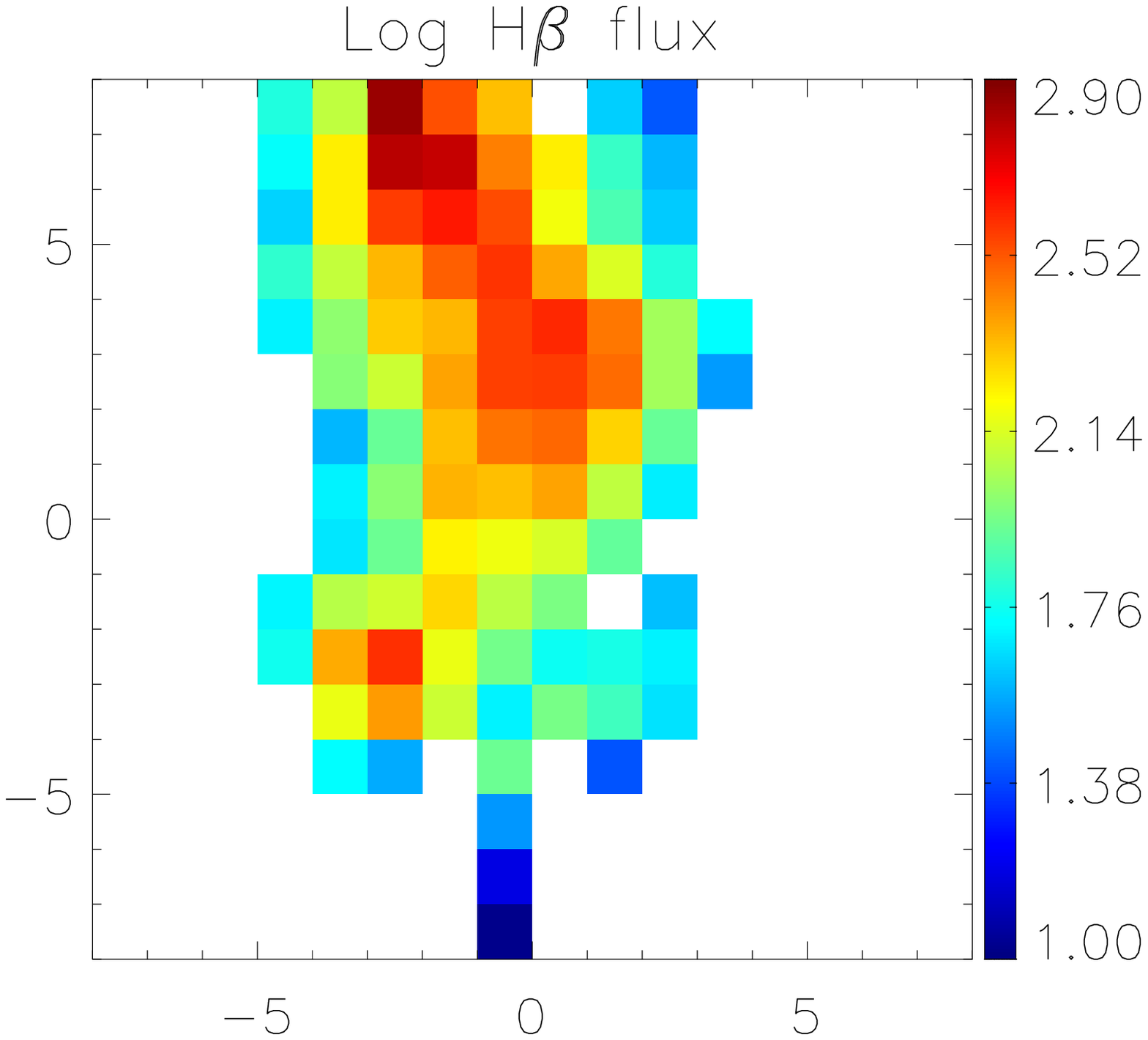}}
}}   
\mbox{
\centerline{
\hspace*{0.0cm}\subfigure{\includegraphics[width=0.24\textwidth]{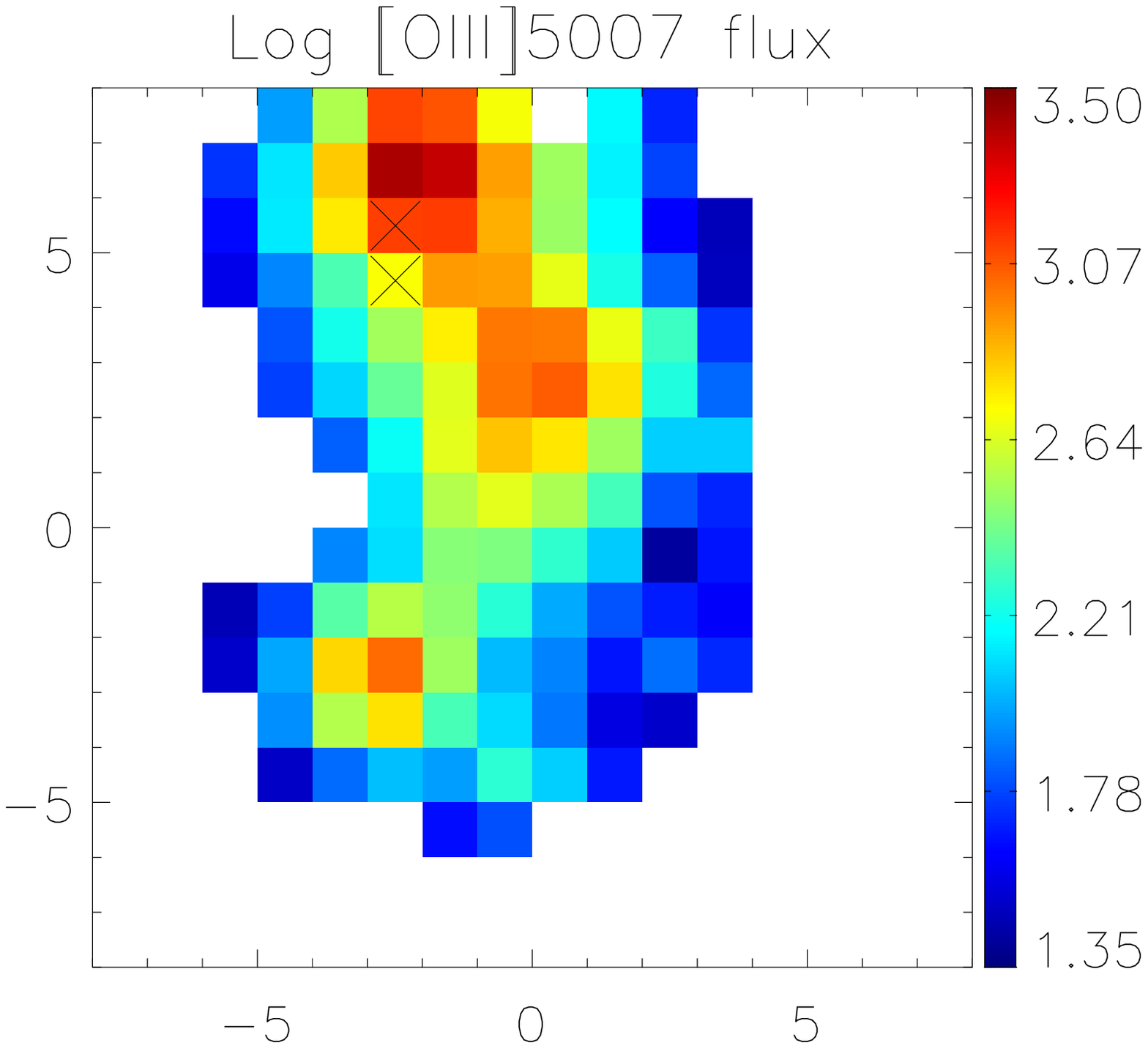}}
\hspace*{0.0cm}\subfigure{\includegraphics[width=0.24\textwidth]{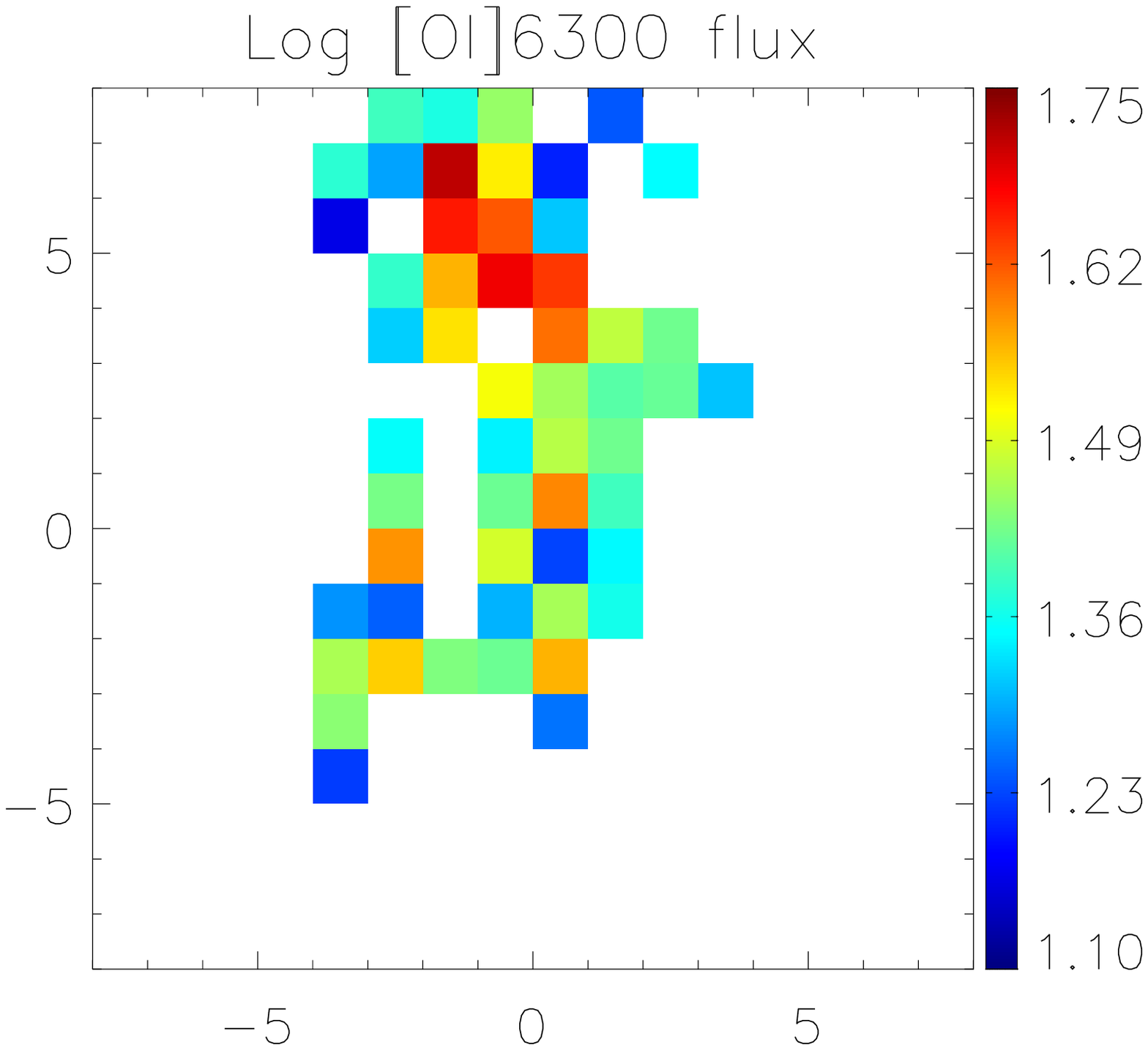}}
\hspace*{0.0cm}\subfigure{\includegraphics[width=0.24\textwidth]{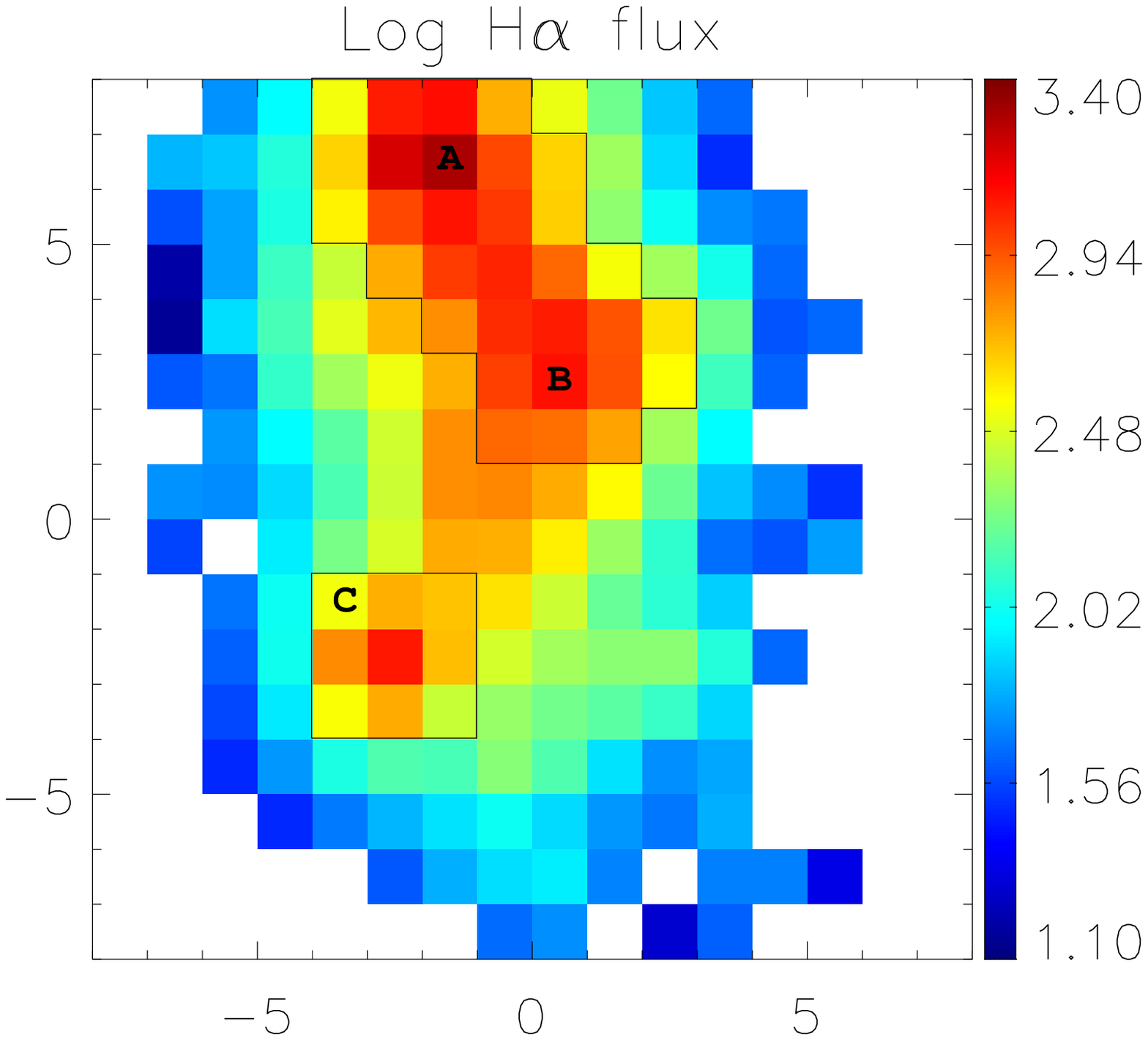}}
\hspace*{0.0cm}\subfigure{\includegraphics[width=0.24\textwidth]{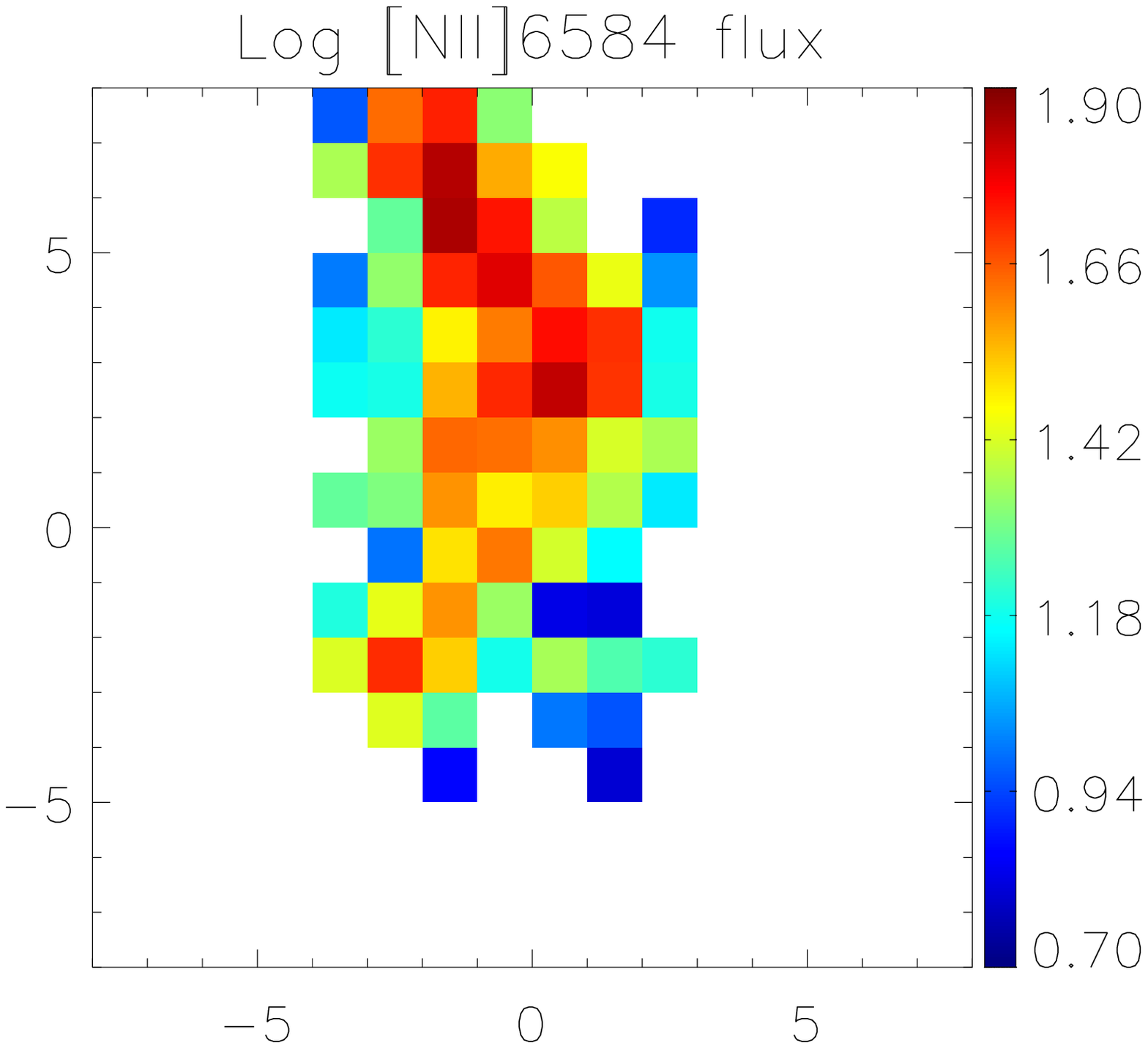}}
}} 
\mbox{
\centerline{
\hspace*{0.0cm}\subfigure{\includegraphics[width=0.24\textwidth]{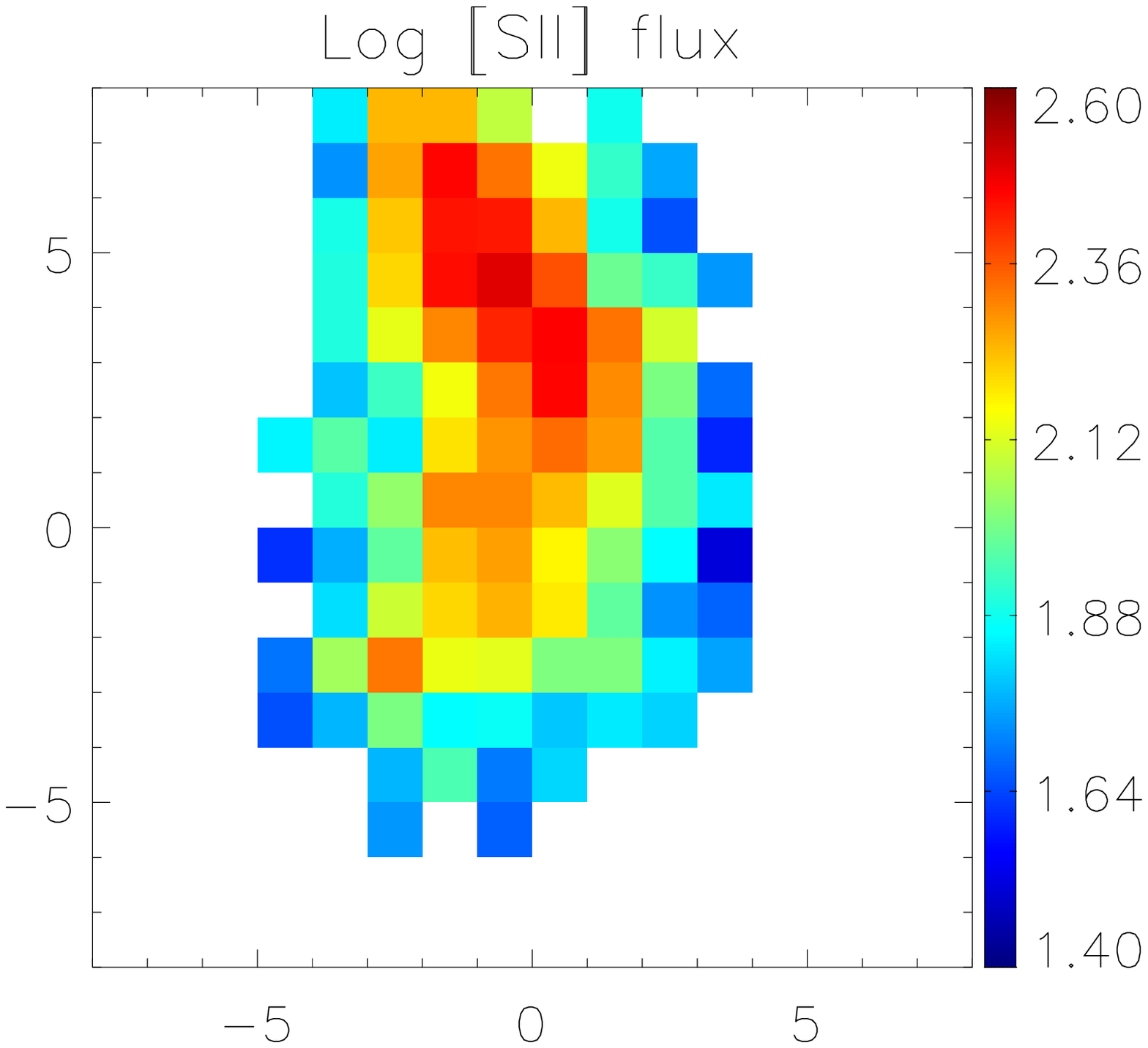}}
\hspace*{0.0cm}\subfigure{\includegraphics[width=0.24\textwidth]{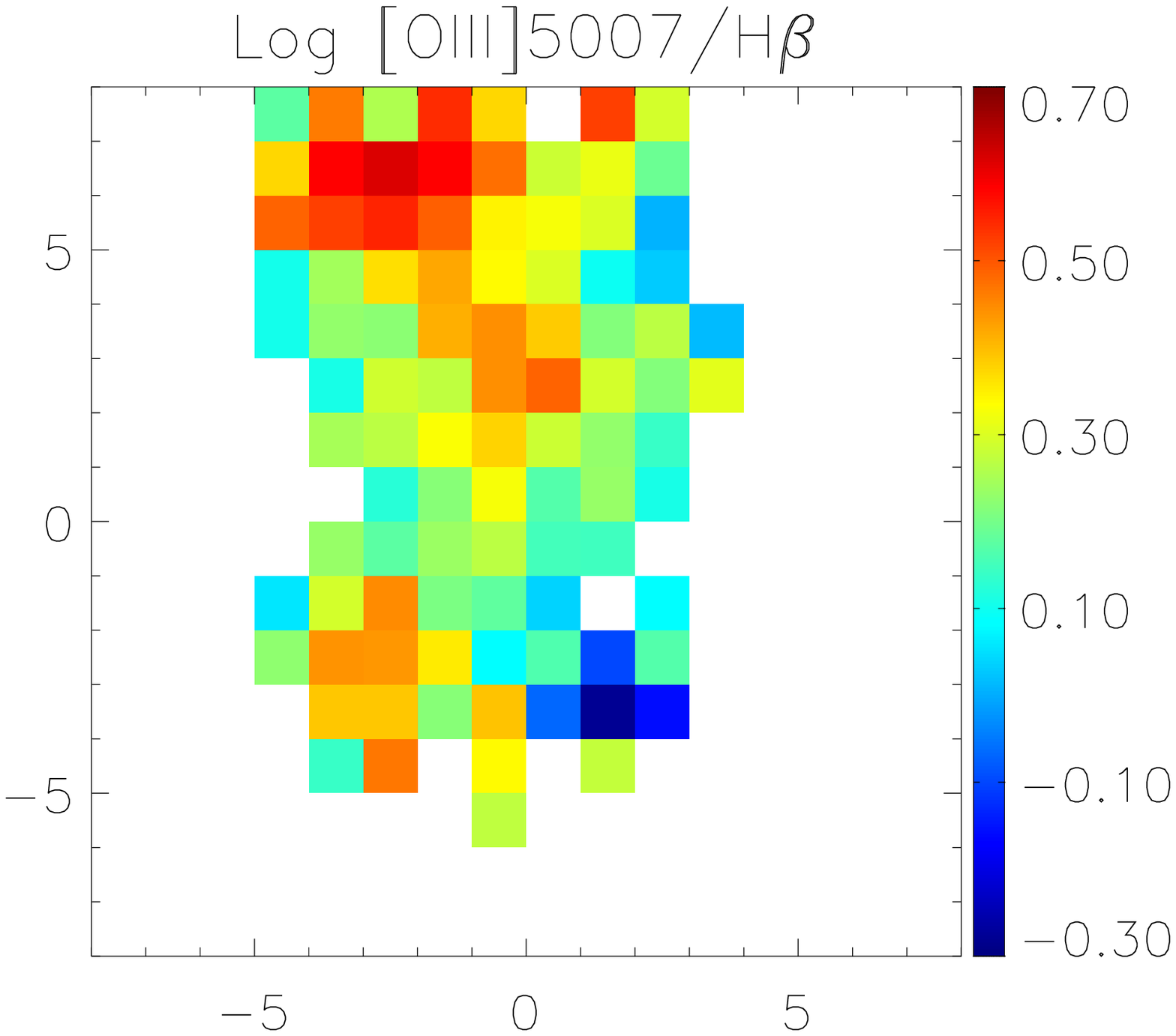}}
\hspace*{0.0cm}\subfigure{\includegraphics[width=0.24\textwidth]{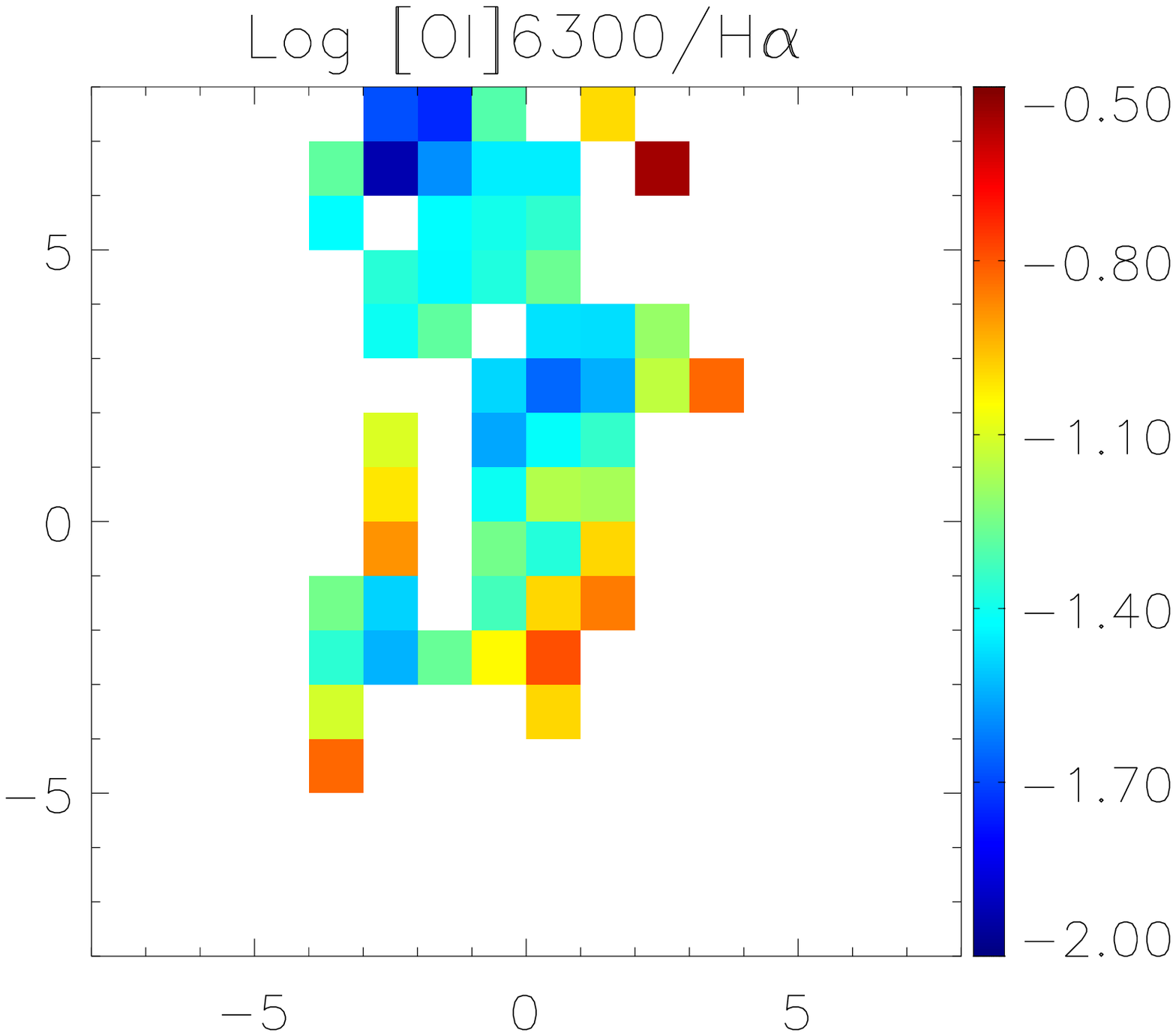}}
\hspace*{0.0cm}\subfigure{\includegraphics[width=0.24\textwidth]{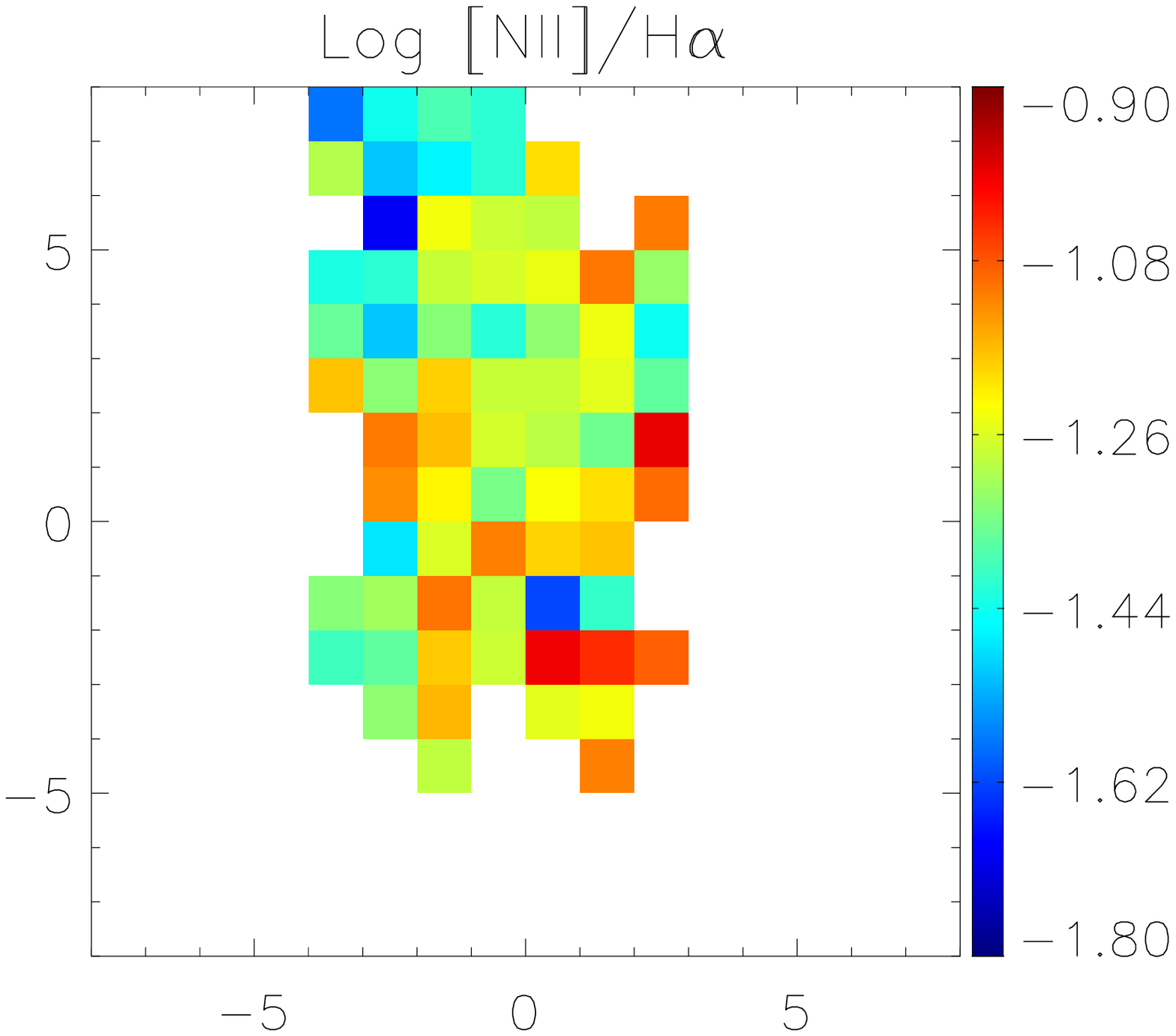}}
}}
\mbox{
\centerline{
\hspace*{0.0cm}\subfigure{\includegraphics[width=0.24\textwidth]{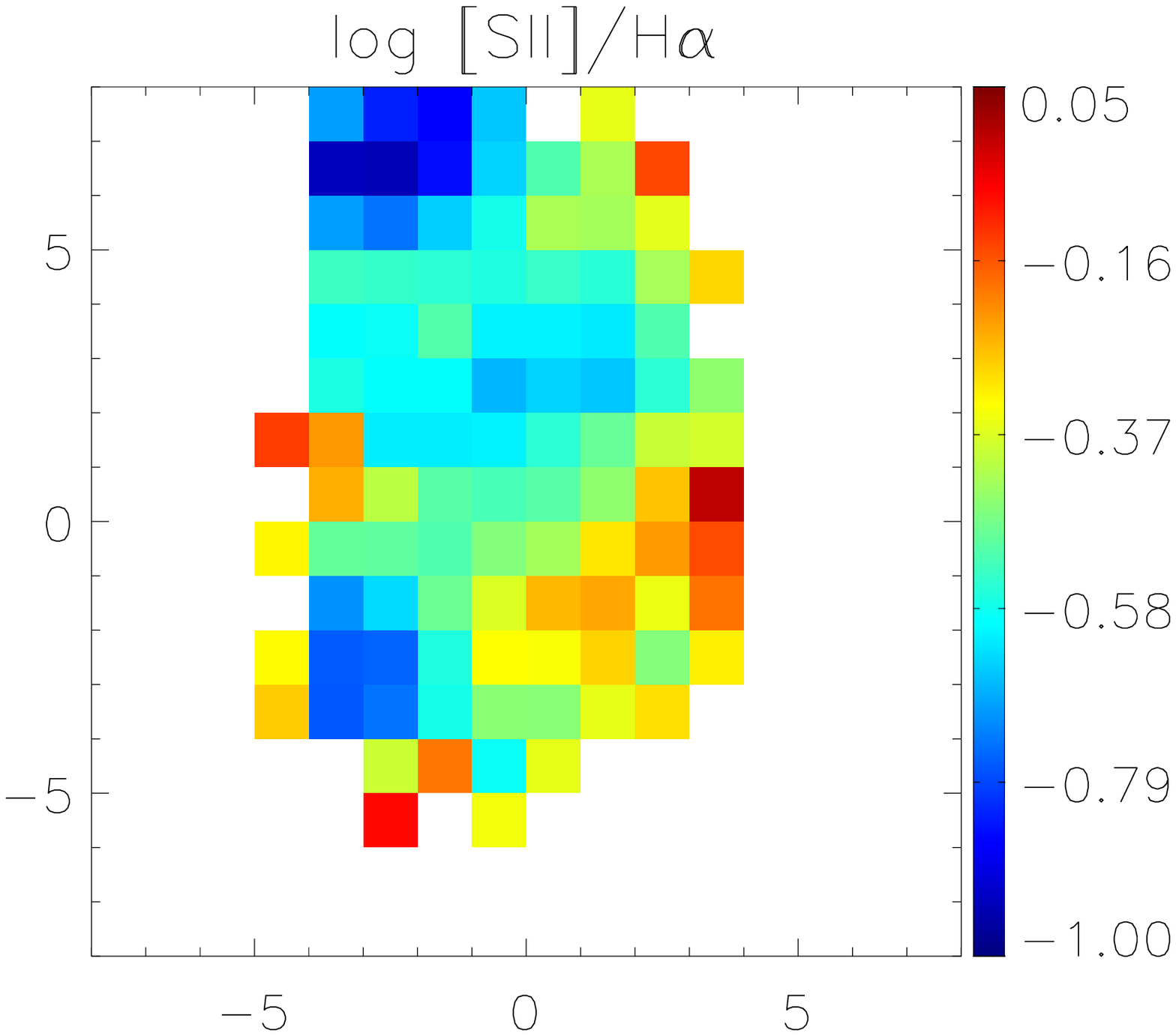}}
\hspace*{0.0cm}\subfigure{\includegraphics[width=0.24\textwidth]{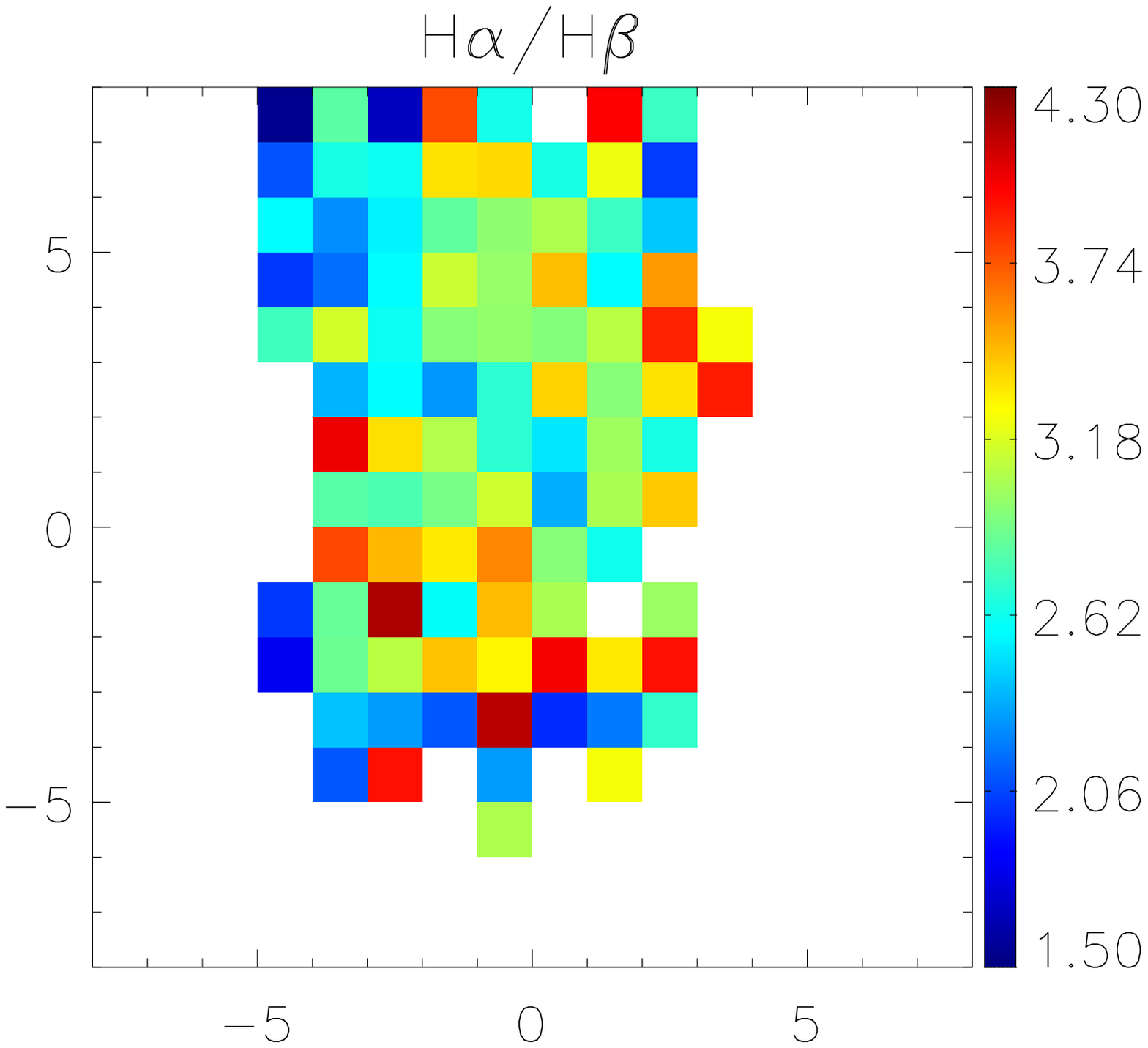}}
\hspace*{0.0cm}\subfigure{\includegraphics[width=0.24\textwidth]{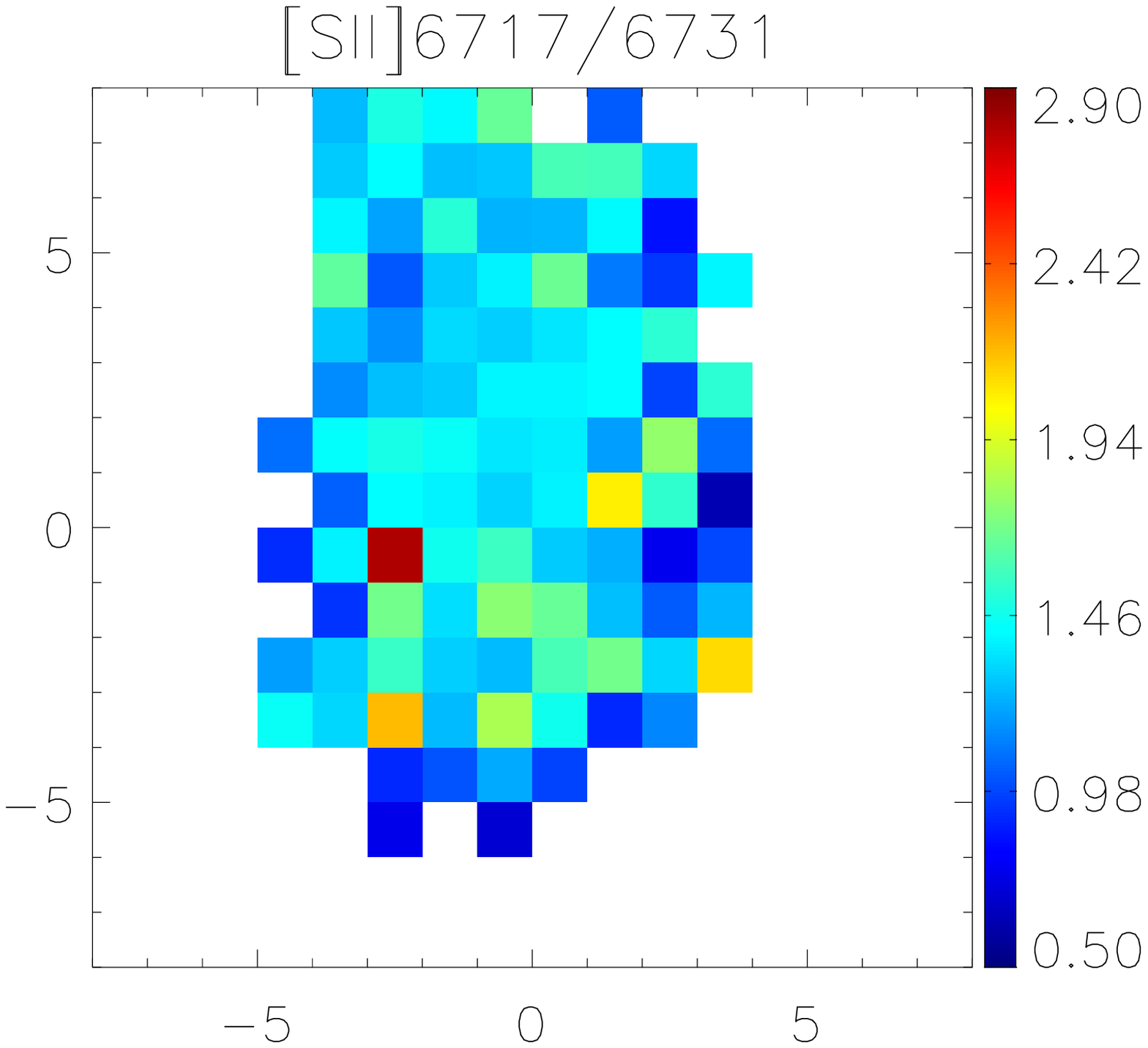}}
}}
\mbox{
\centerline{
\hspace*{0.0cm}\subfigure{\includegraphics[width=0.24\textwidth]{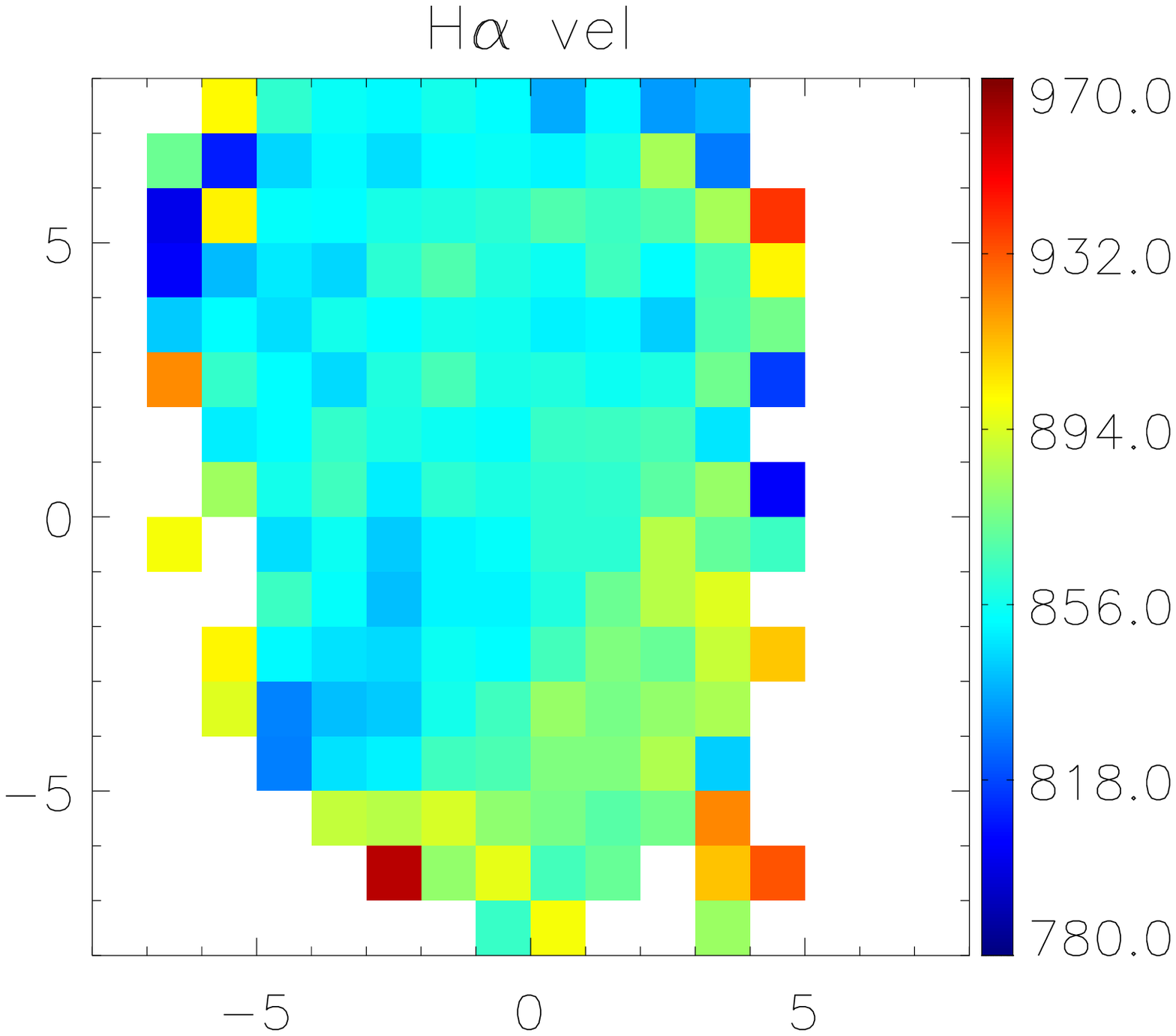}}
\hspace*{0.0cm}\subfigure{\includegraphics[width=0.24\textwidth]{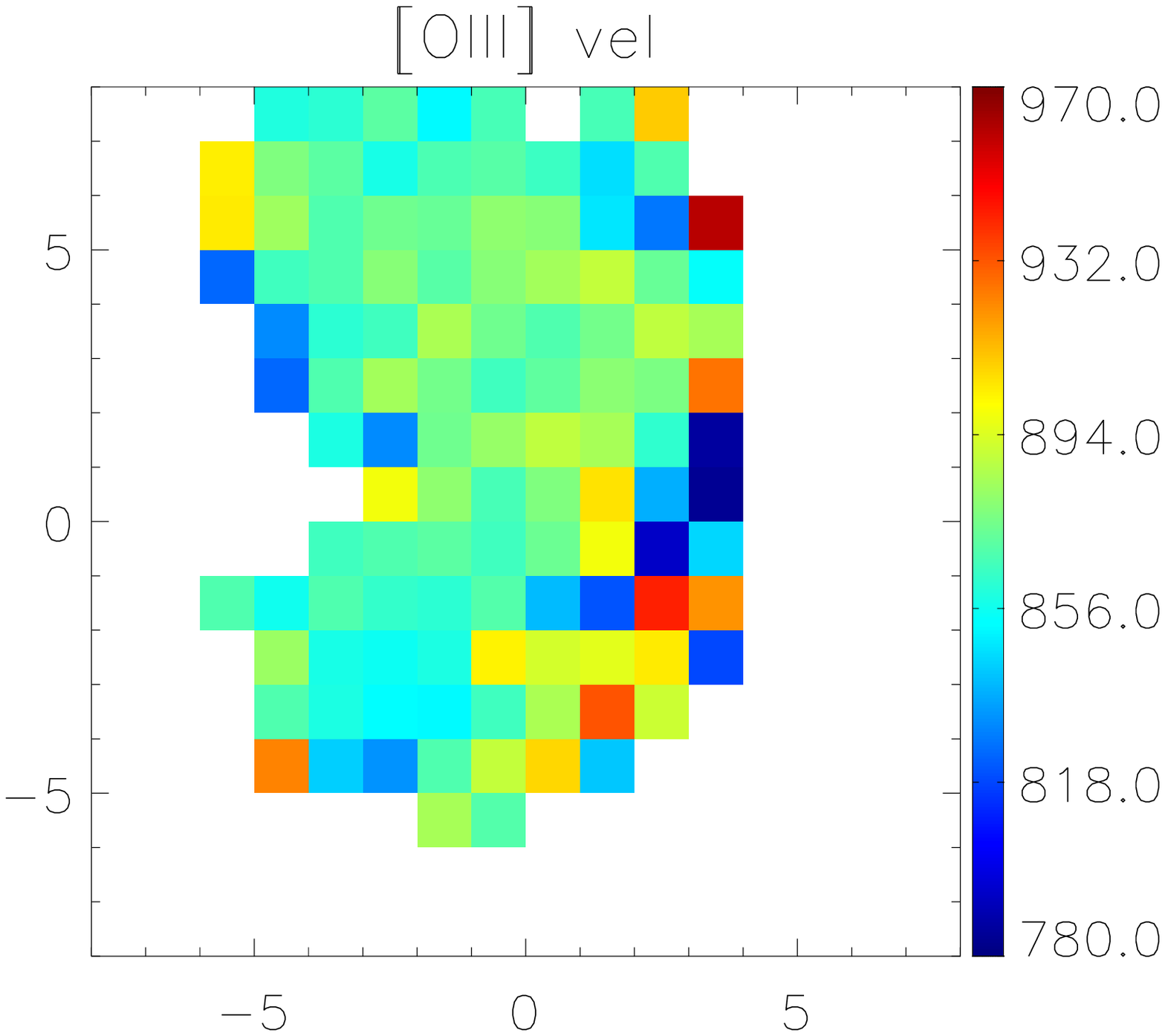}}
}}  
\caption{Same as Fig.~\ref{Figure:mrk407} for Mrk~32. 
The [\ion{O}{i}]~$\lambda6300$ flux map and the 
[\ion{O}{i}]~$\lambda6300$/\Ha\ ionization  ratio map are also included. 
The outline of the identified SF knots (see 
Sect.~\ref{SubSection:IntegratedSpectroscopy}) is shown in the \Ha\ map. 
Spaxels with a marginal detection of the blue WR bump have been marked by 
crosses in the [\ion{O}{iii}] map.}
\label{Figure:mrk32}
\end{figure*}

\begin{figure*}
\mbox{
\centerline{
\hspace*{0.0cm}\subfigure{\includegraphics[width=0.24\textwidth]{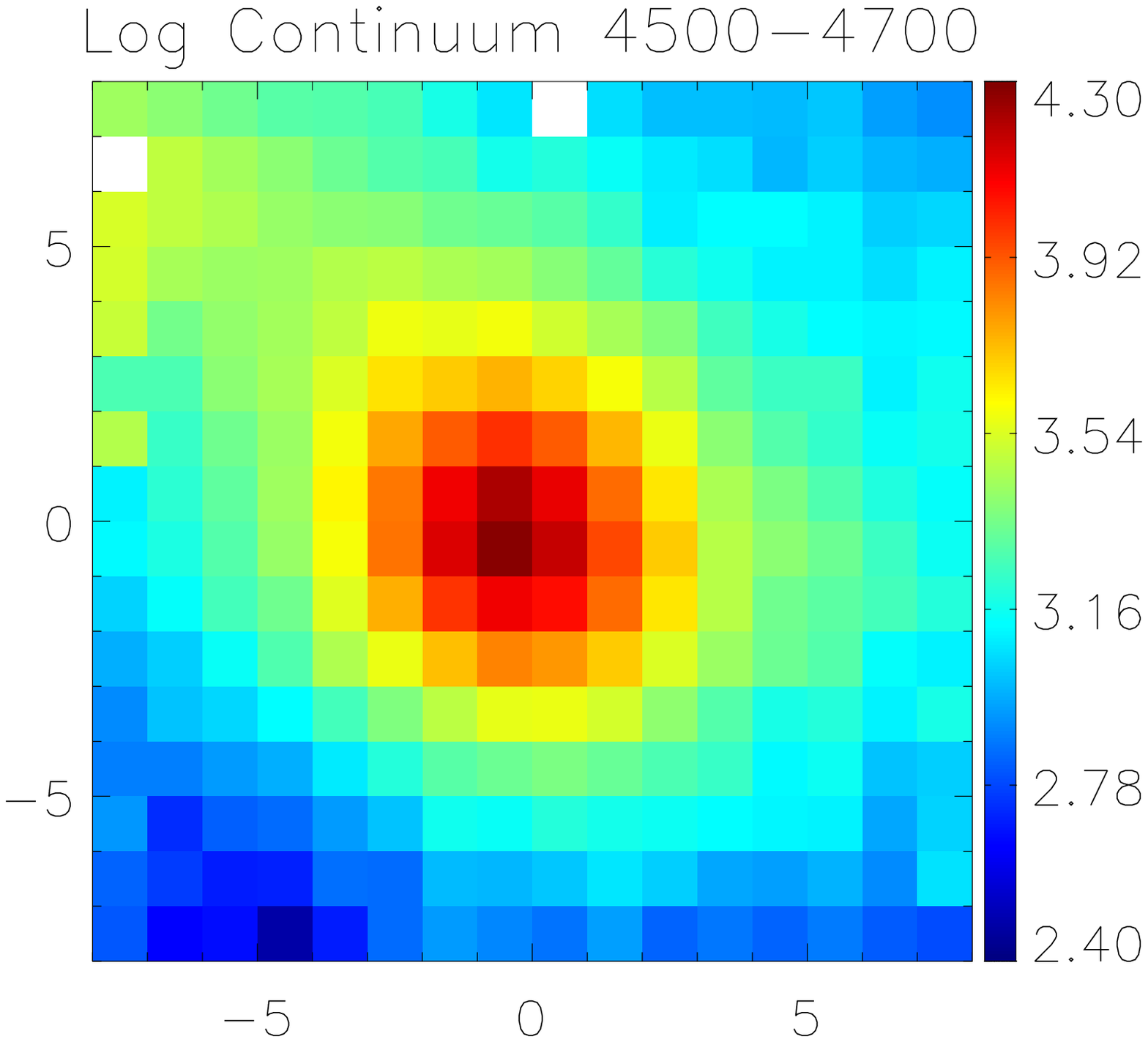}}
\hspace*{0.0cm}\subfigure{\includegraphics[width=0.24\textwidth]{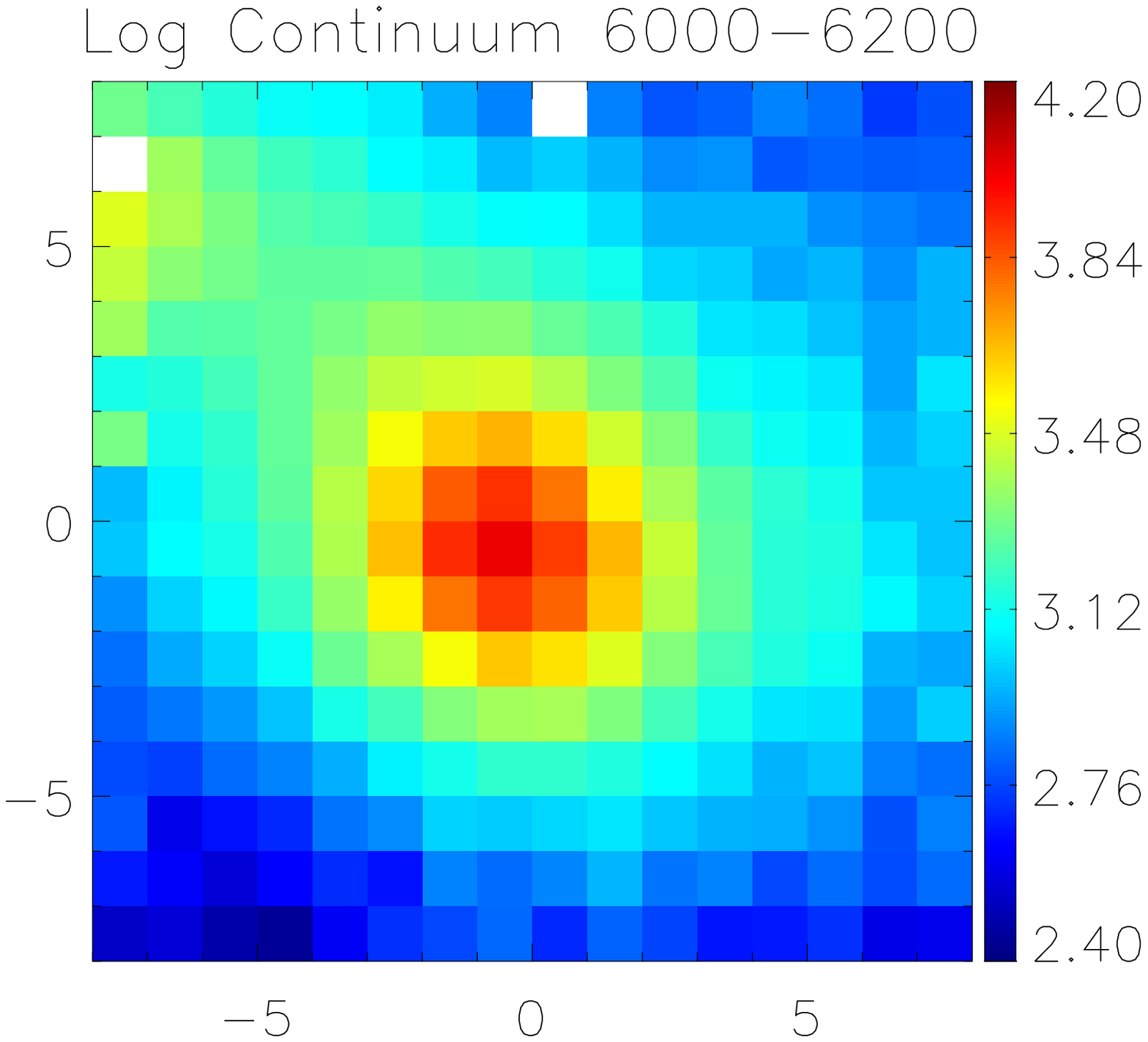}}
\hspace*{0.0cm}\subfigure{\includegraphics[width=0.24\textwidth]{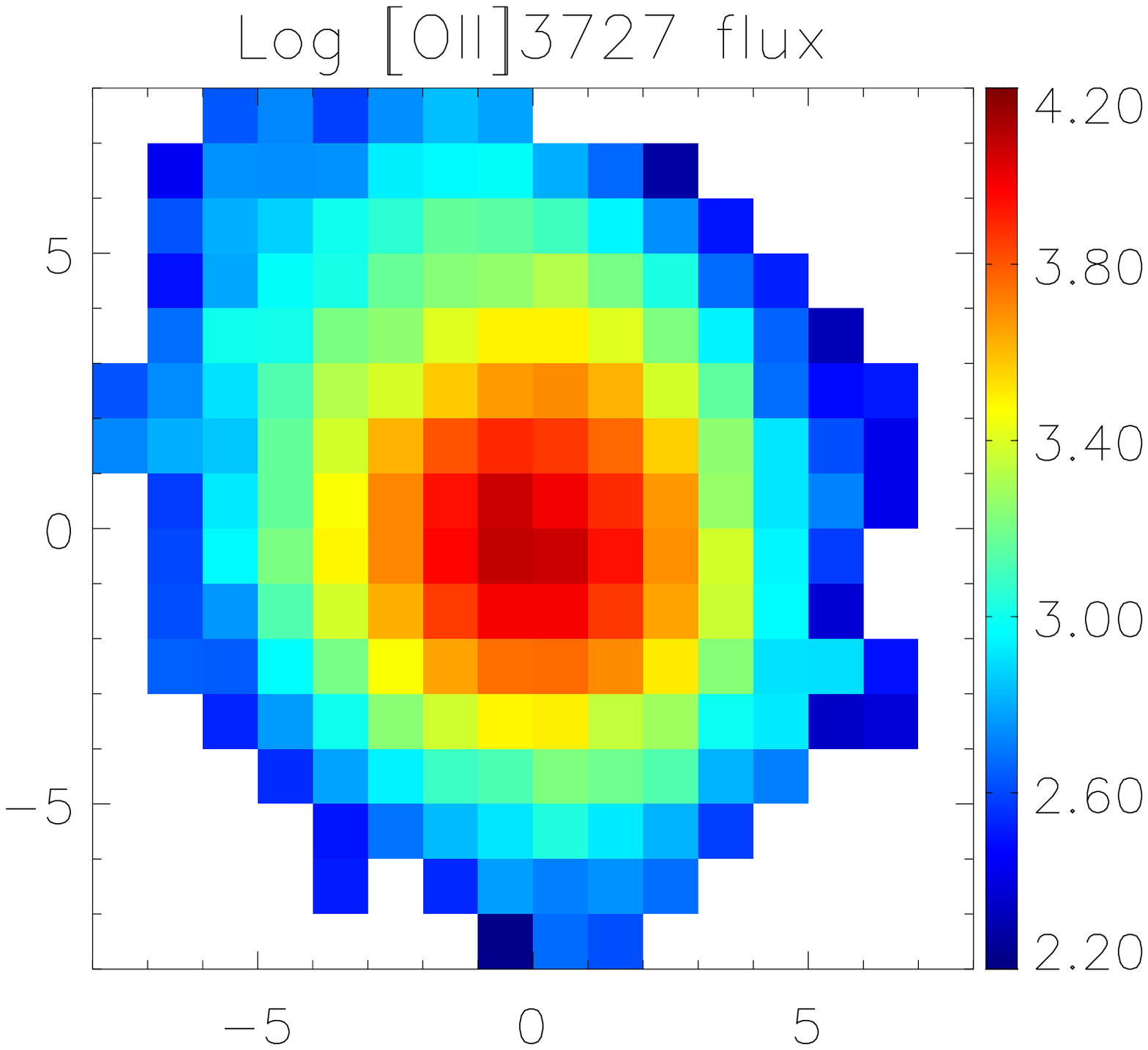}}
\hspace*{0.0cm}\subfigure{\includegraphics[width=0.24\textwidth]{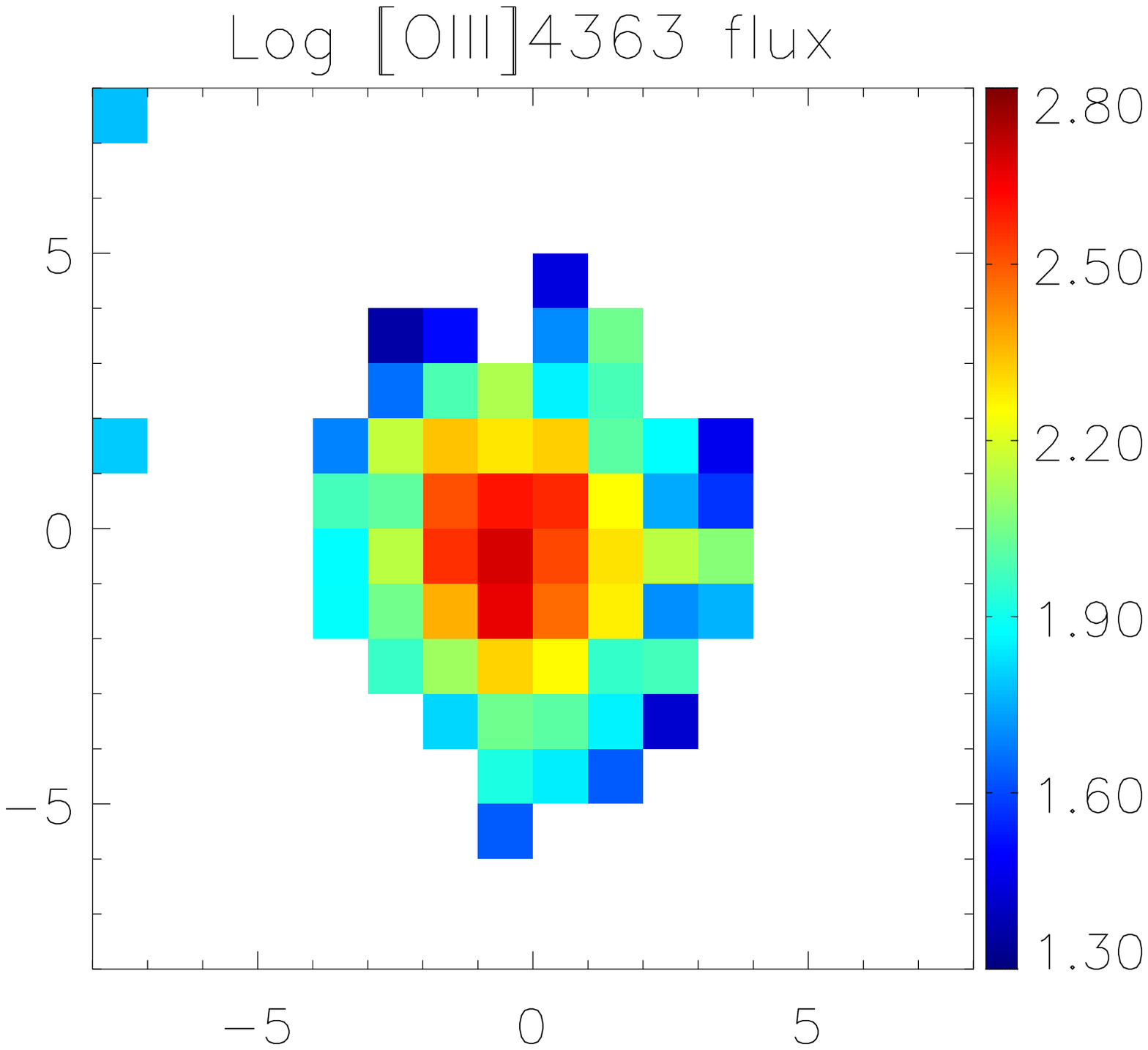}}
}}   
\mbox{
\centerline{
\hspace*{0.0cm}\subfigure{\includegraphics[width=0.24\textwidth]{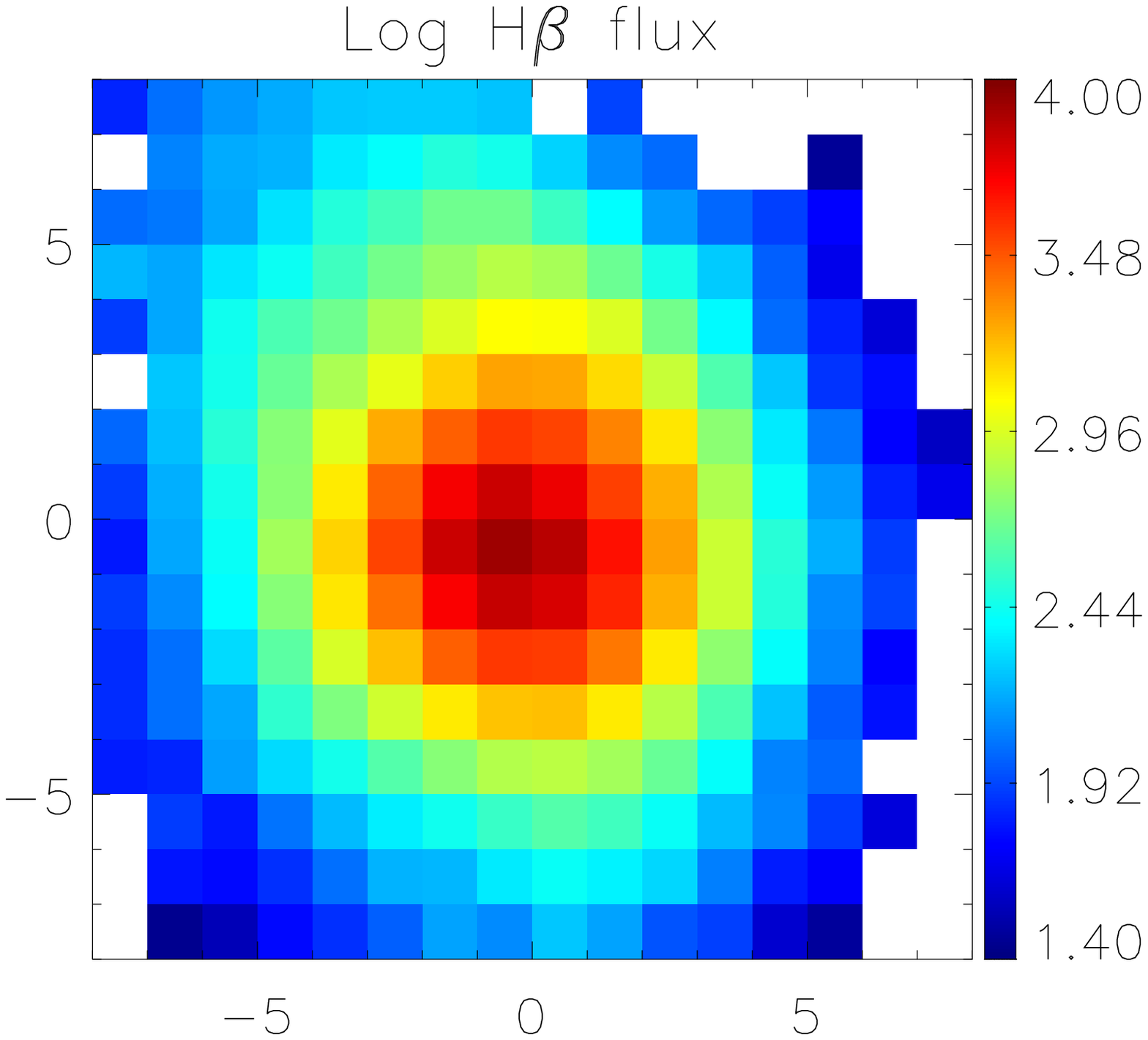}}
\hspace*{0.0cm}\subfigure{\includegraphics[width=0.24\textwidth]{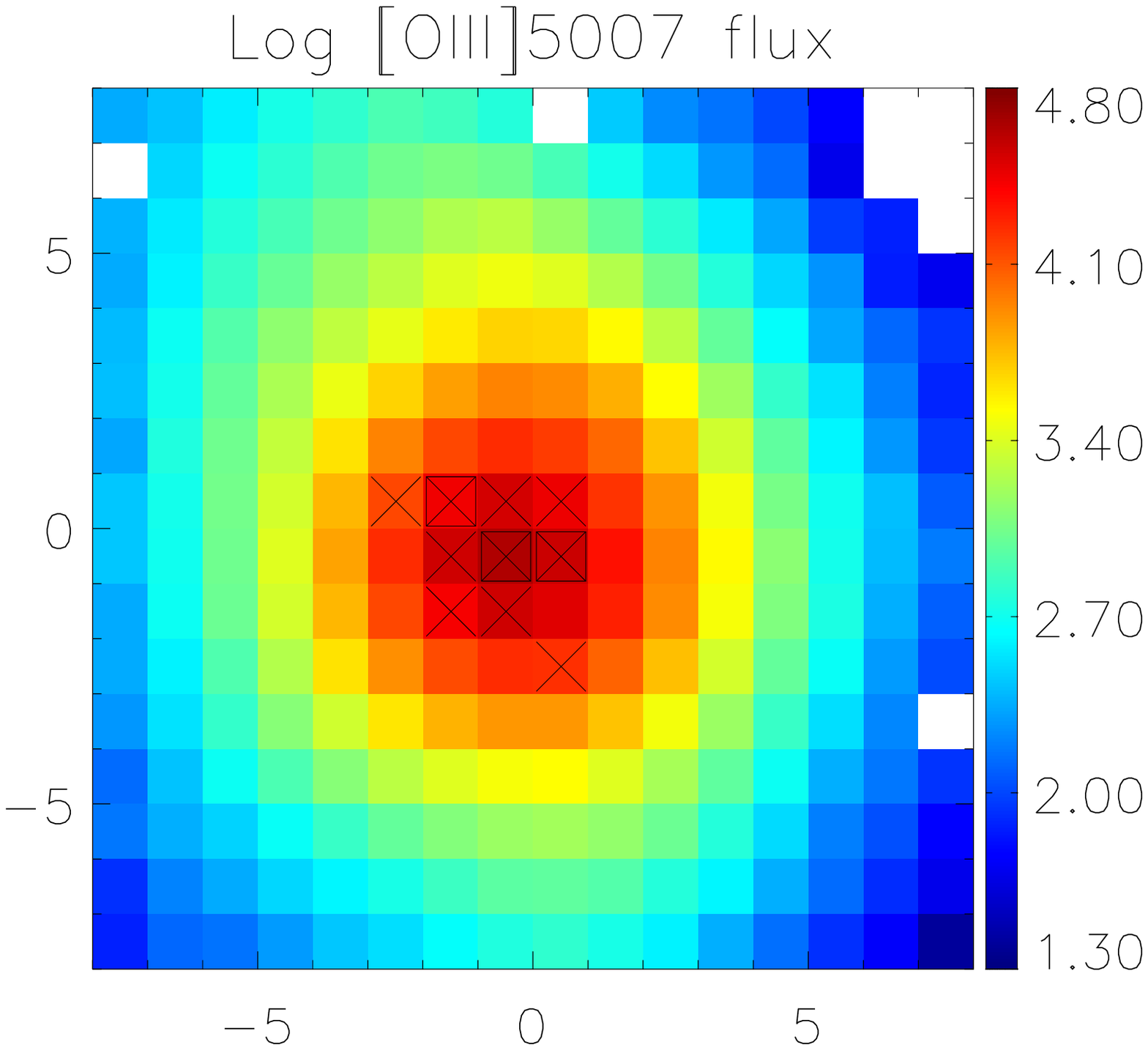}}
\hspace*{0.0cm}\subfigure{\includegraphics[width=0.24\textwidth]{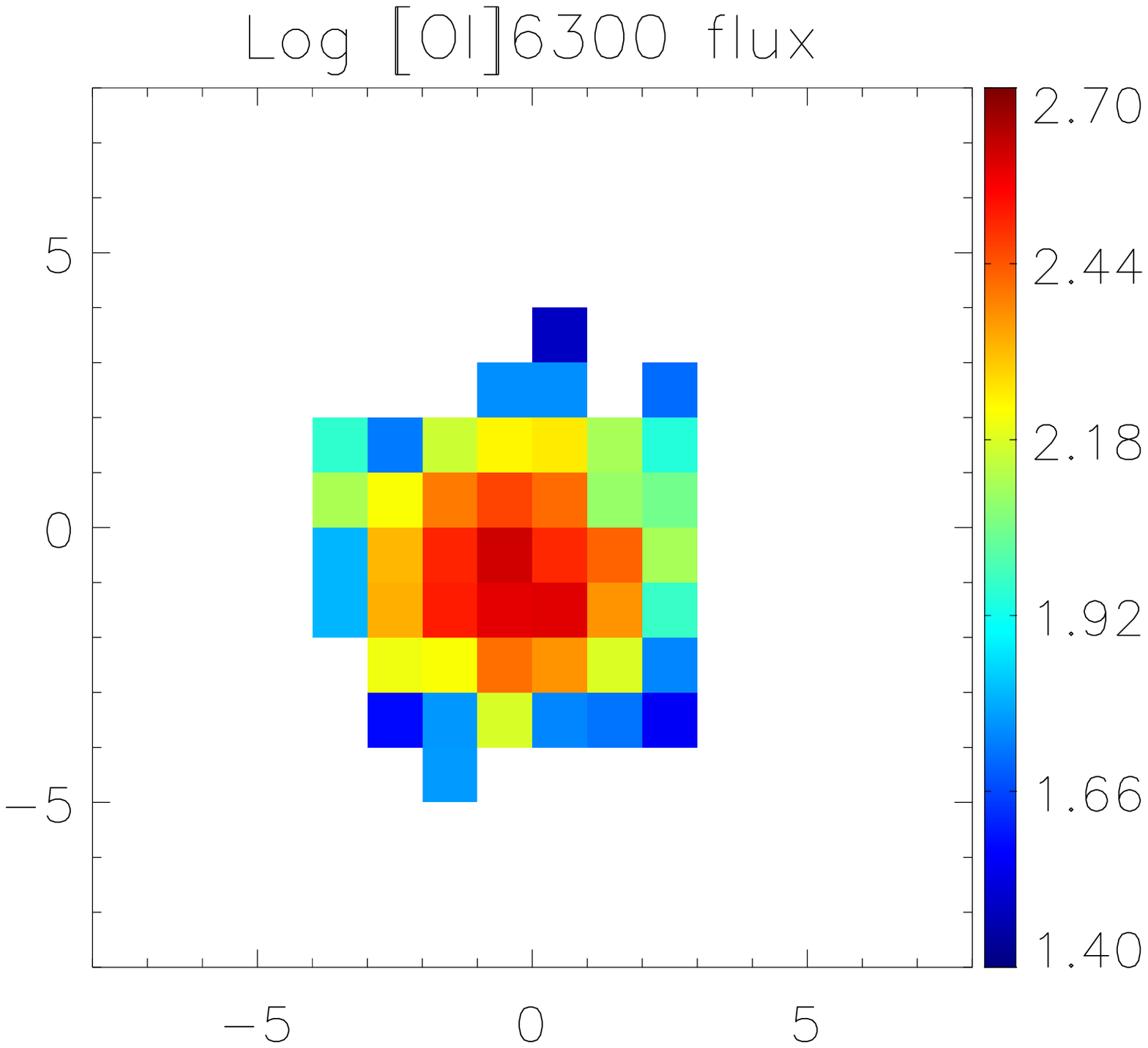}}
\hspace*{0.0cm}\subfigure{\includegraphics[width=0.24\textwidth]{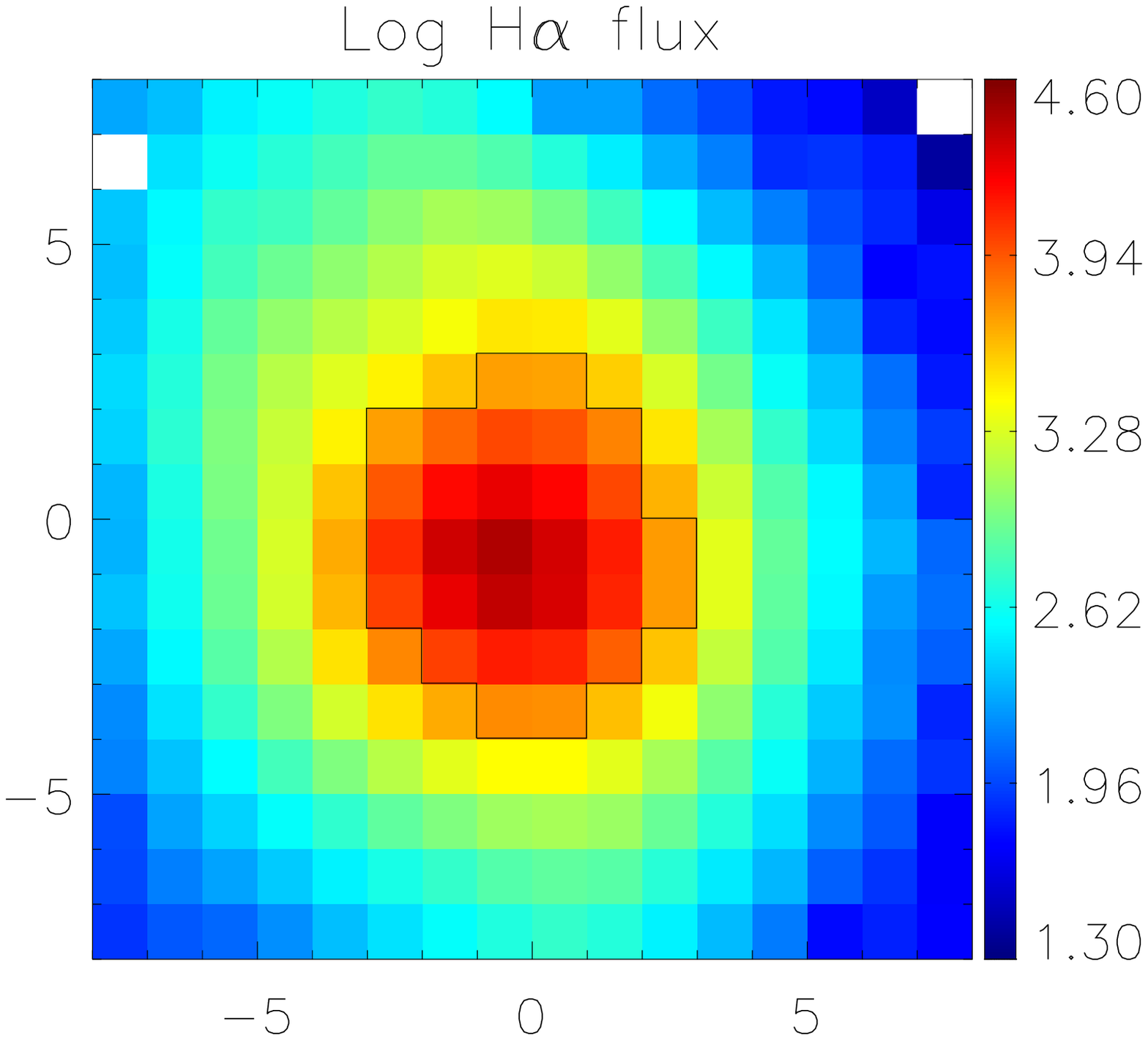}}
}} 
\mbox{
\centerline{
\hspace*{0.0cm}\subfigure{\includegraphics[width=0.24\textwidth]{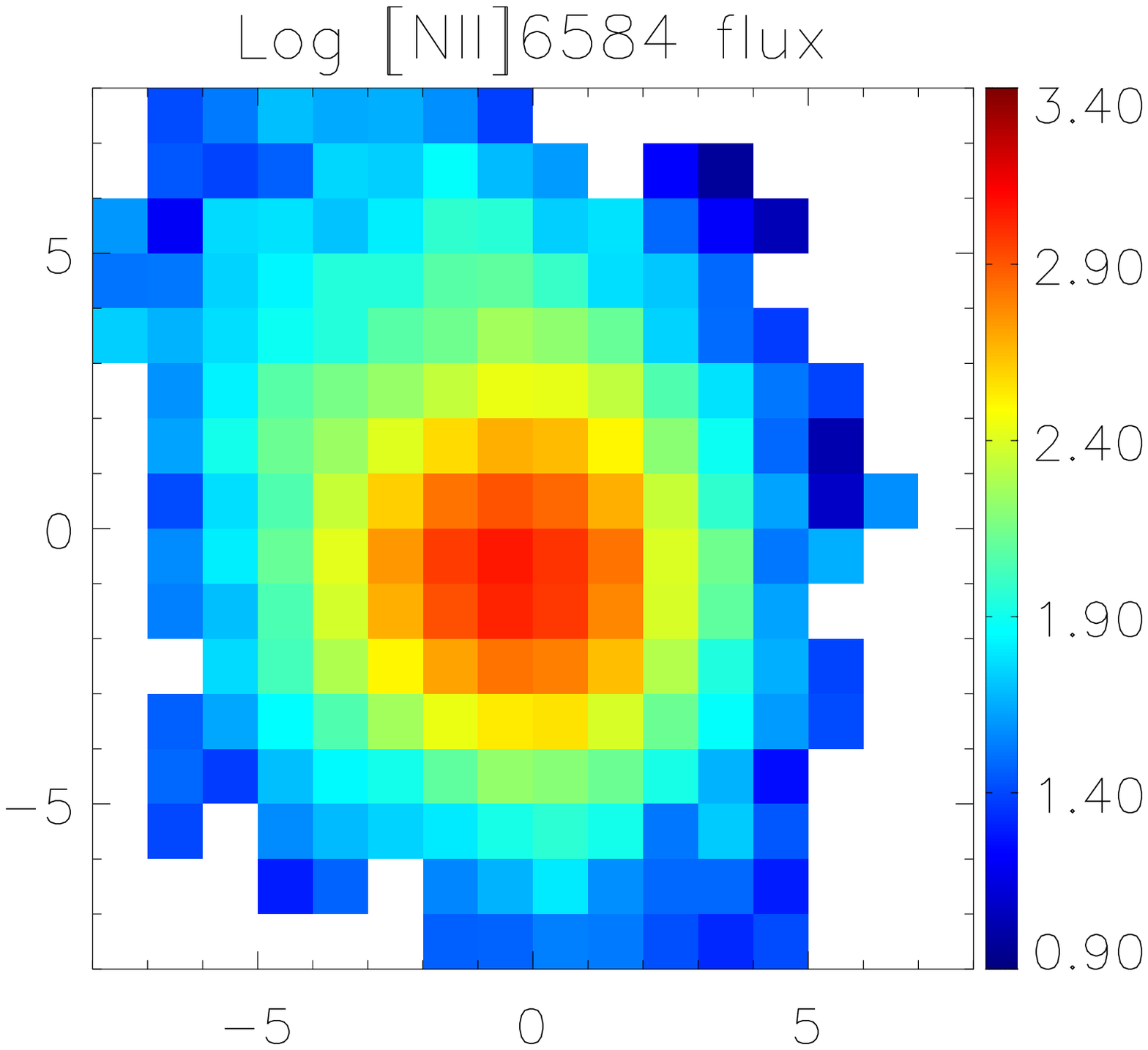}}
\hspace*{0.0cm}\subfigure{\includegraphics[width=0.24\textwidth]{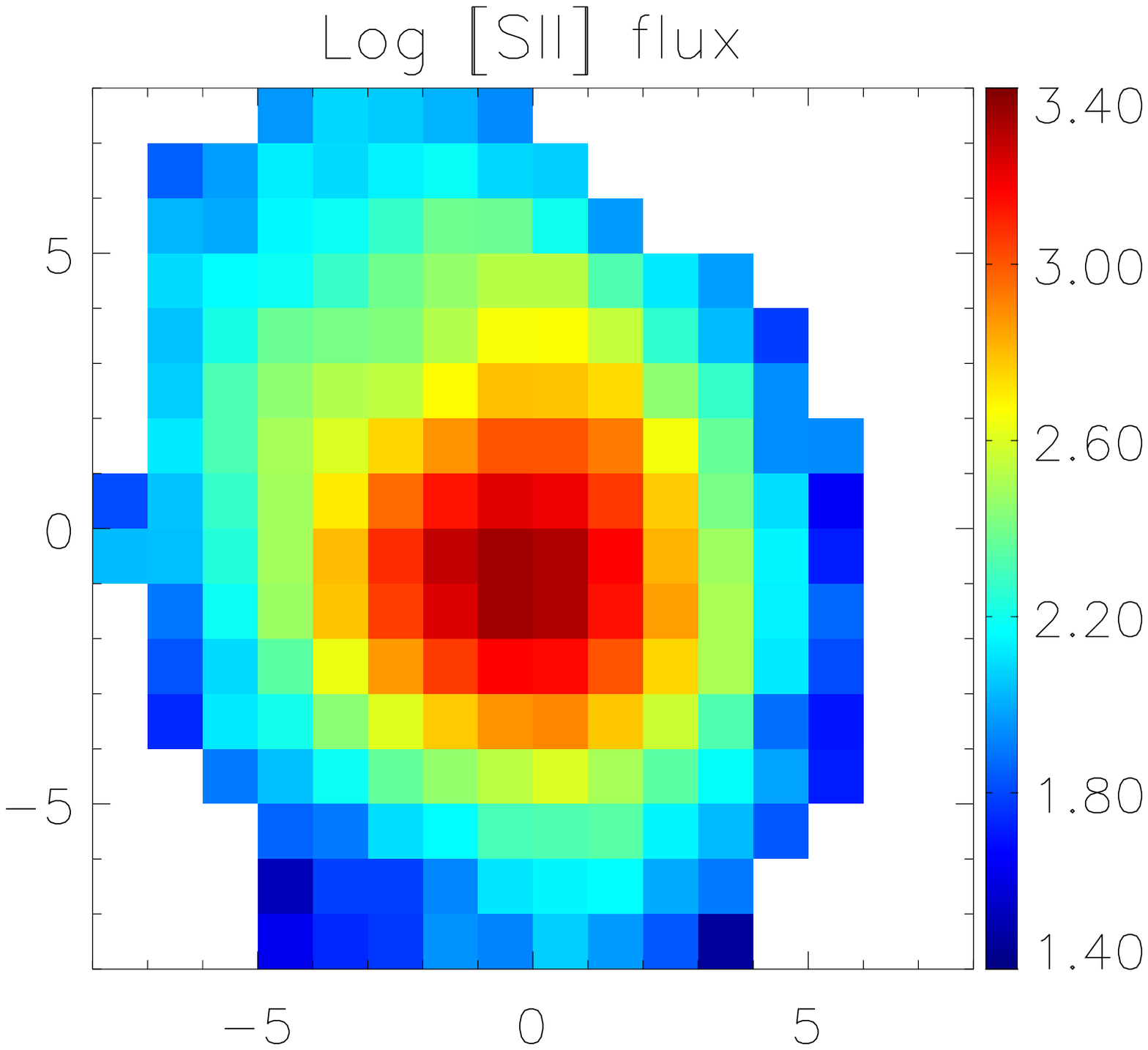}}
\hspace*{0.0cm}\subfigure{\includegraphics[width=0.24\textwidth]{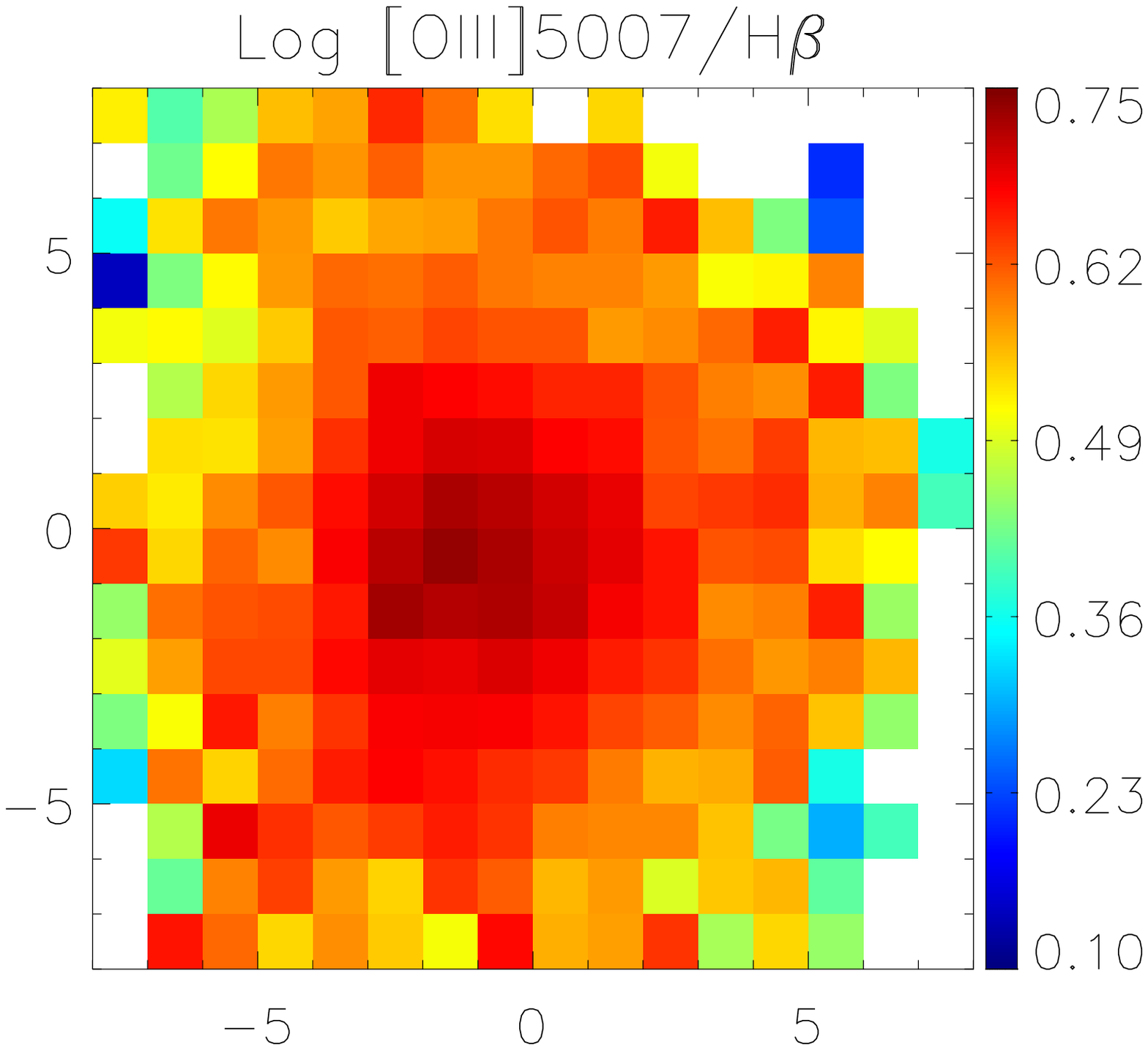}}
\hspace*{0.0cm}\subfigure{\includegraphics[width=0.24\textwidth]{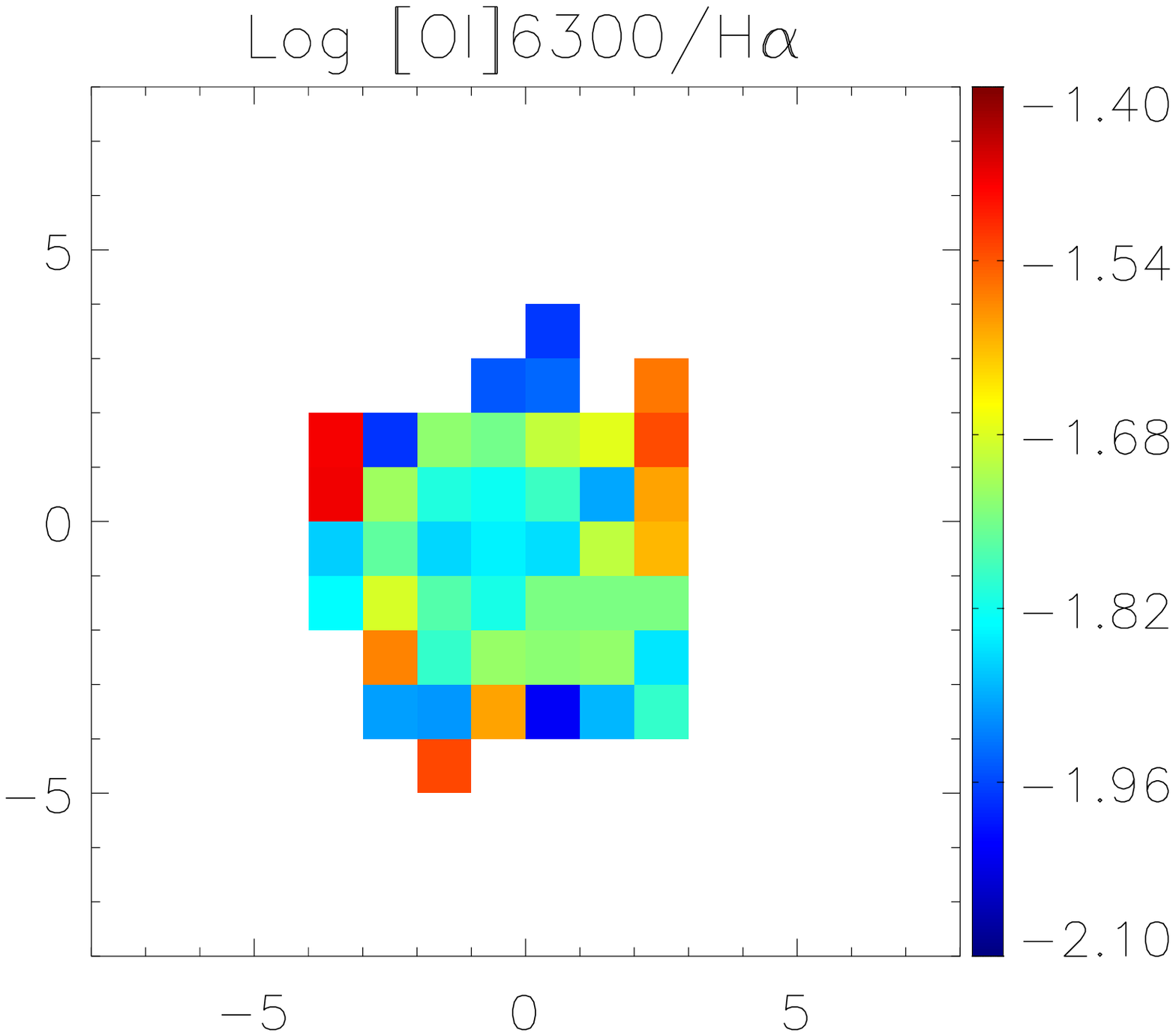}}
}}
\mbox{
\centerline{
\hspace*{0.0cm}\subfigure{\includegraphics[width=0.24\textwidth]{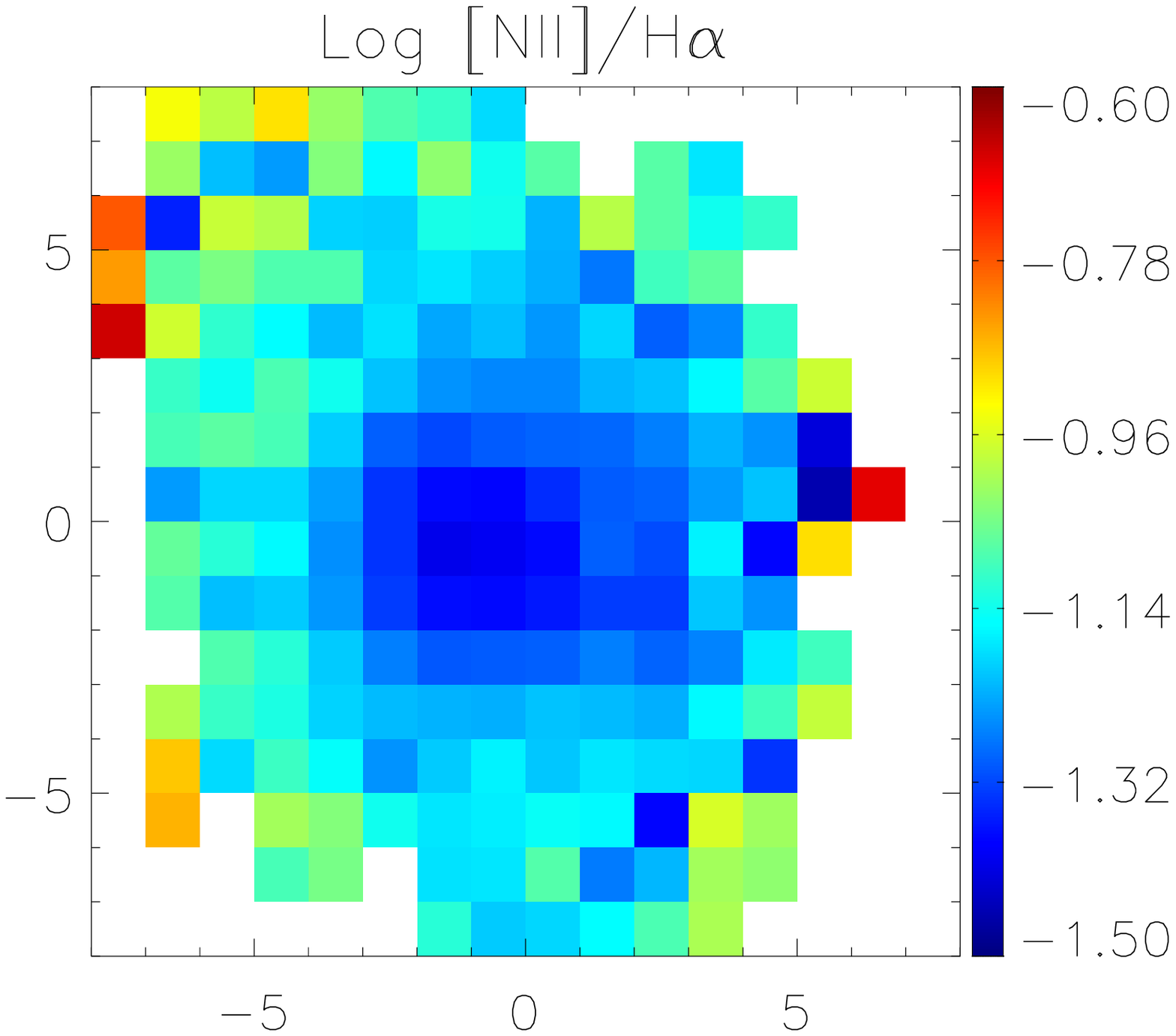}}
\hspace*{0.0cm}\subfigure{\includegraphics[width=0.24\textwidth]{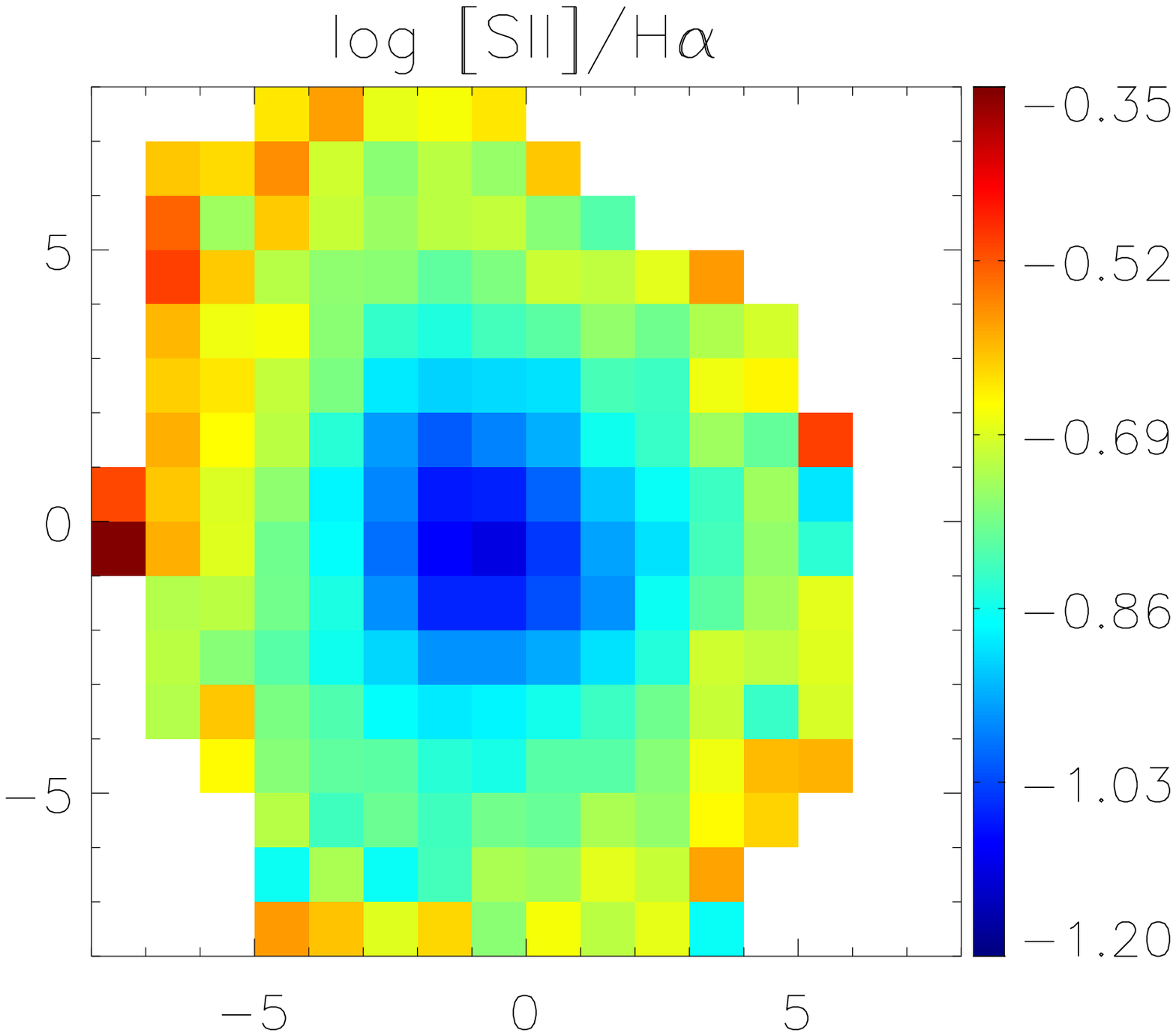}}
\hspace*{0.0cm}\subfigure{\includegraphics[width=0.24\textwidth]{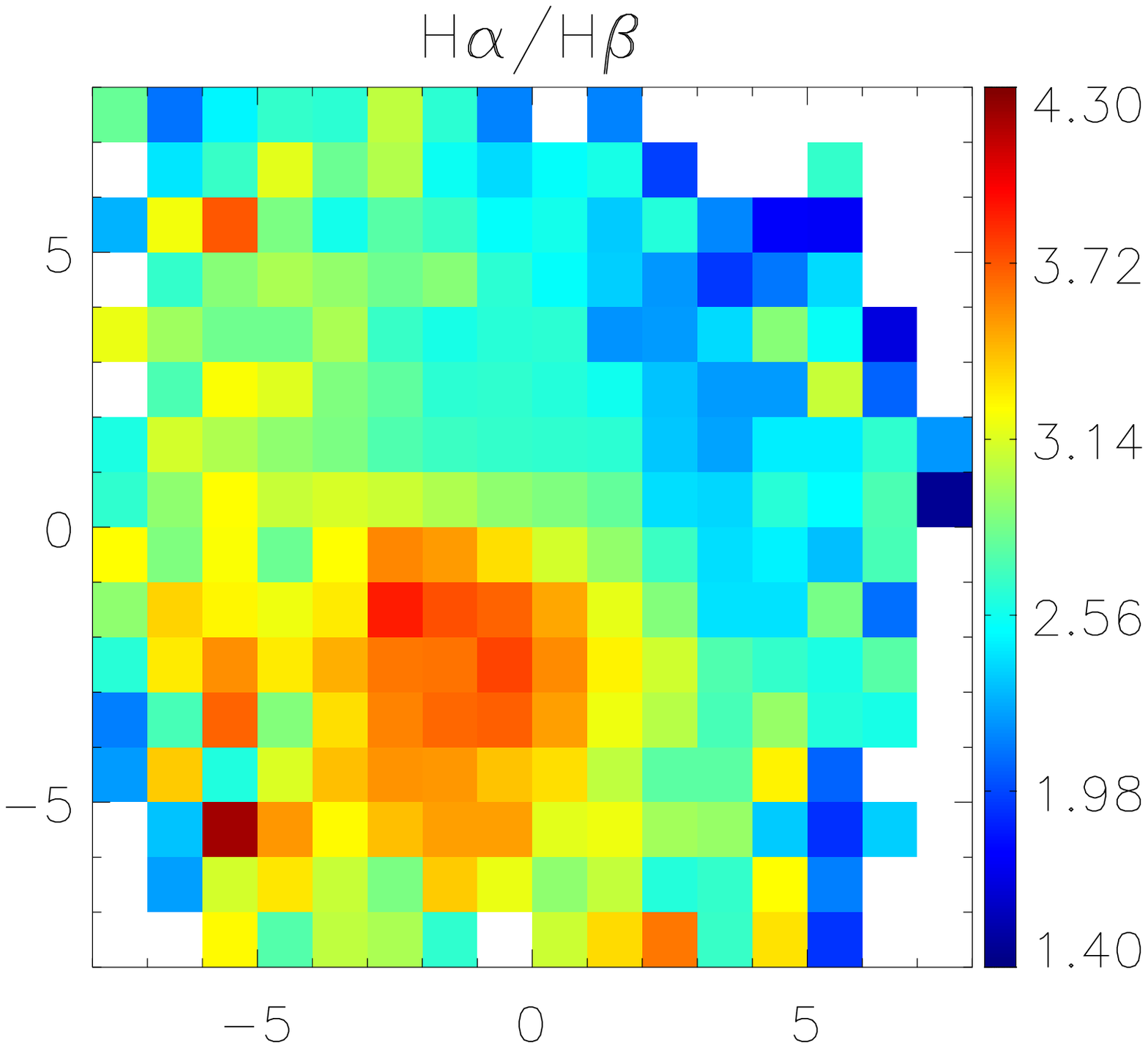}}
\hspace*{0.0cm}\subfigure{\includegraphics[width=0.24\textwidth]{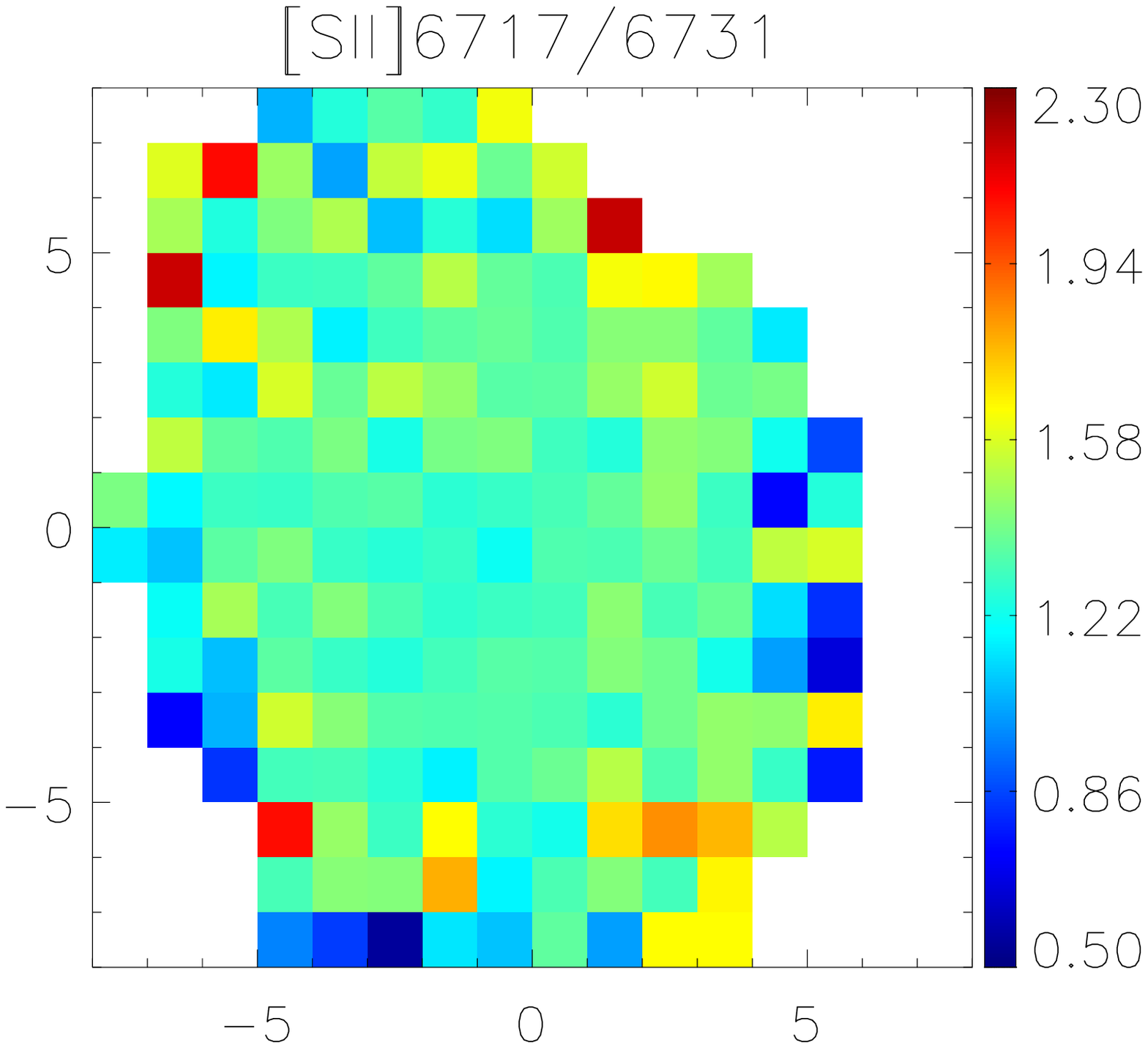}}
}} 
\mbox{
\centerline{
\hspace*{0.0cm}\subfigure{\includegraphics[width=0.24\textwidth]{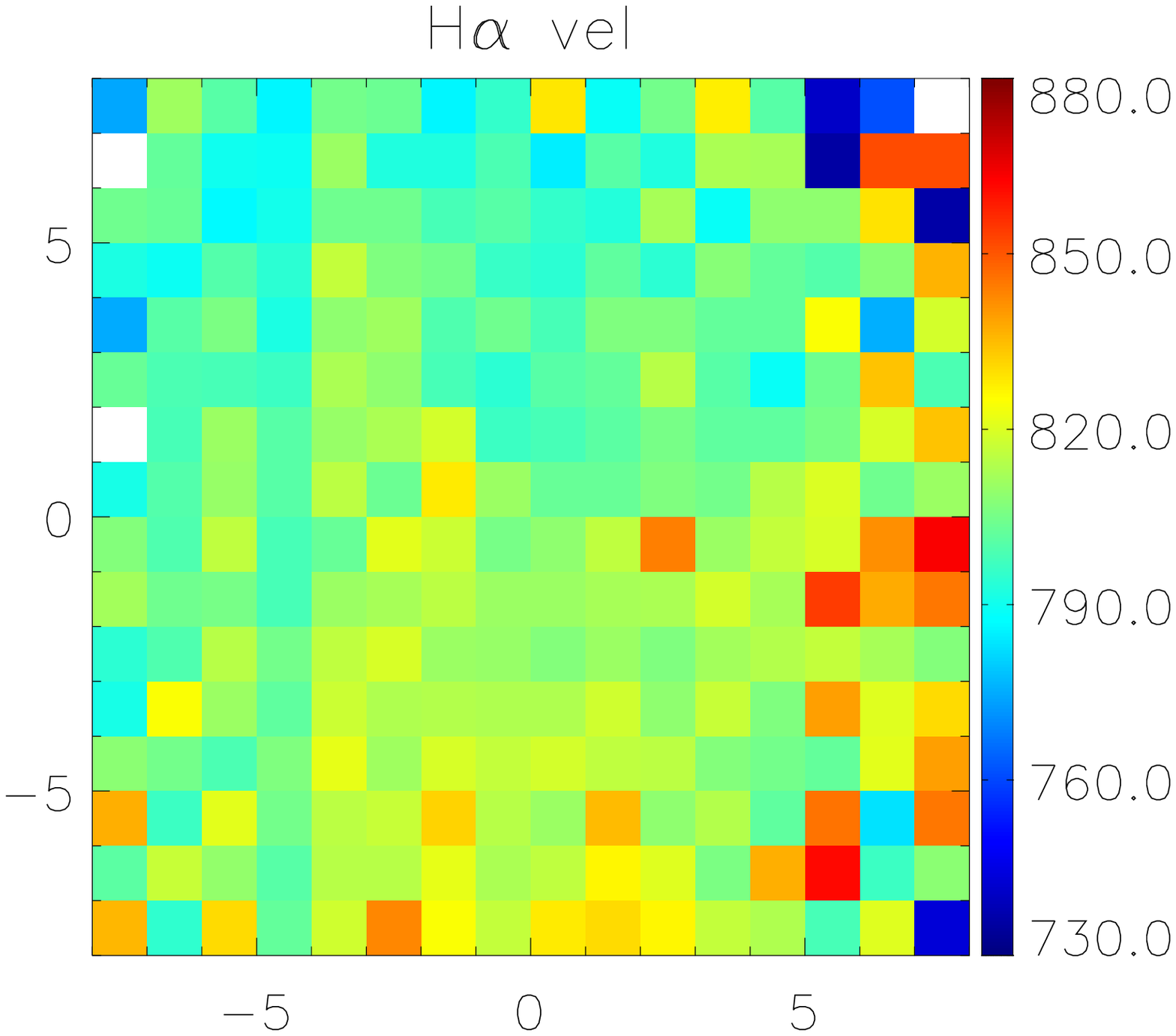}}
\hspace*{0.0cm}\subfigure{\includegraphics[width=0.24\textwidth]{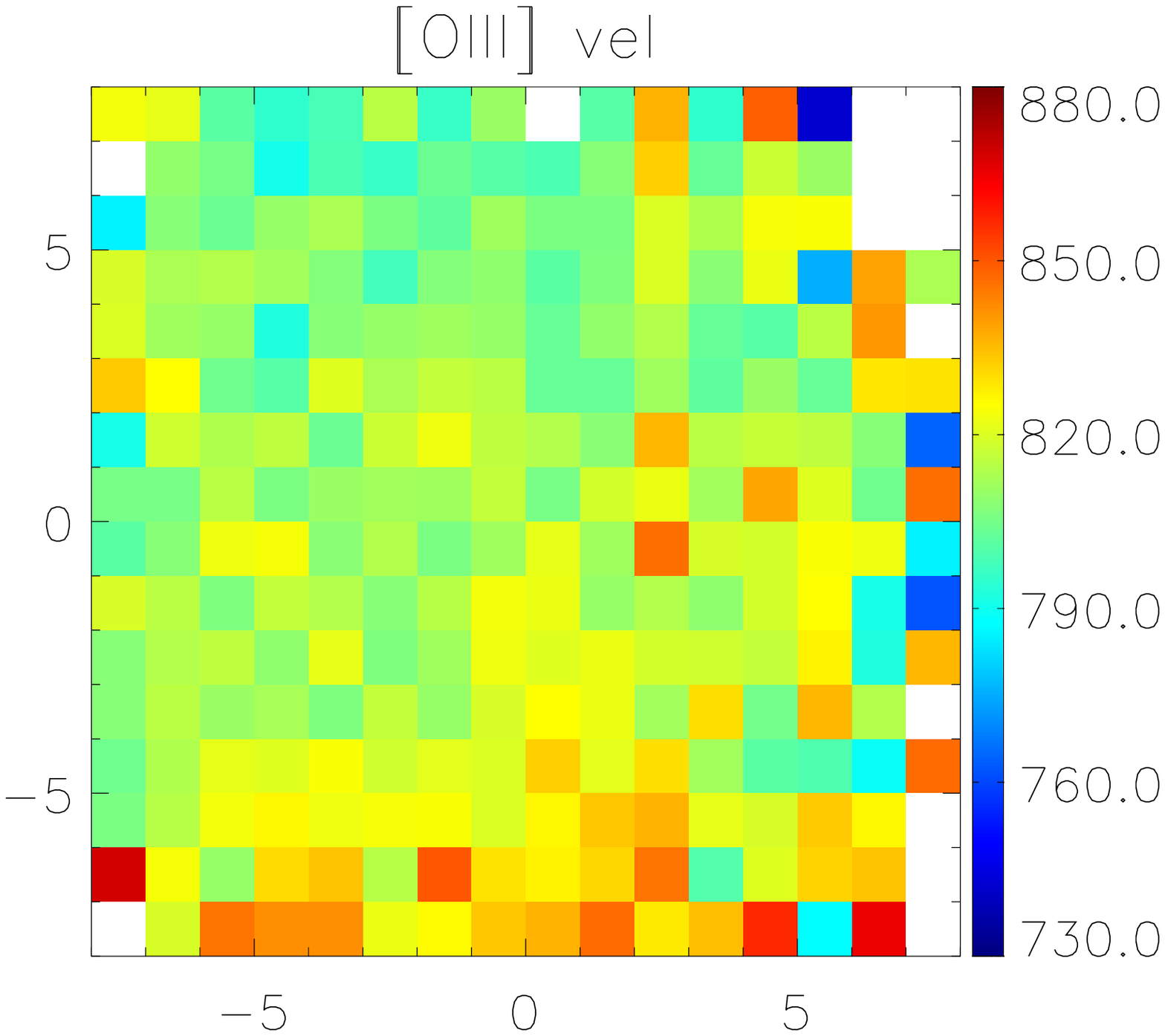}}
}}   
\caption{Same as Fig.~\ref{Figure:mrk407} for Mrk~750. Maps of
[\ion{O}{iii}]~$\lambda4363$, [\ion{O}{i}]~$\lambda6300$, and 
of the [\ion{O}{i}]~$\lambda6300$/\Ha\ ionization ratio are also included.
Spaxels in which the WR feature was 
detected were marked in the [\ion{O}{iii}]~$\lambda5007$ map with crosses 
and squares for the blue and the red bumps respectively.}
\label{Figure:mrk750}
\end{figure*}

\begin{figure*}
\mbox{
\centerline{
\hspace*{0.0cm}\subfigure{\includegraphics[width=0.24\textwidth]{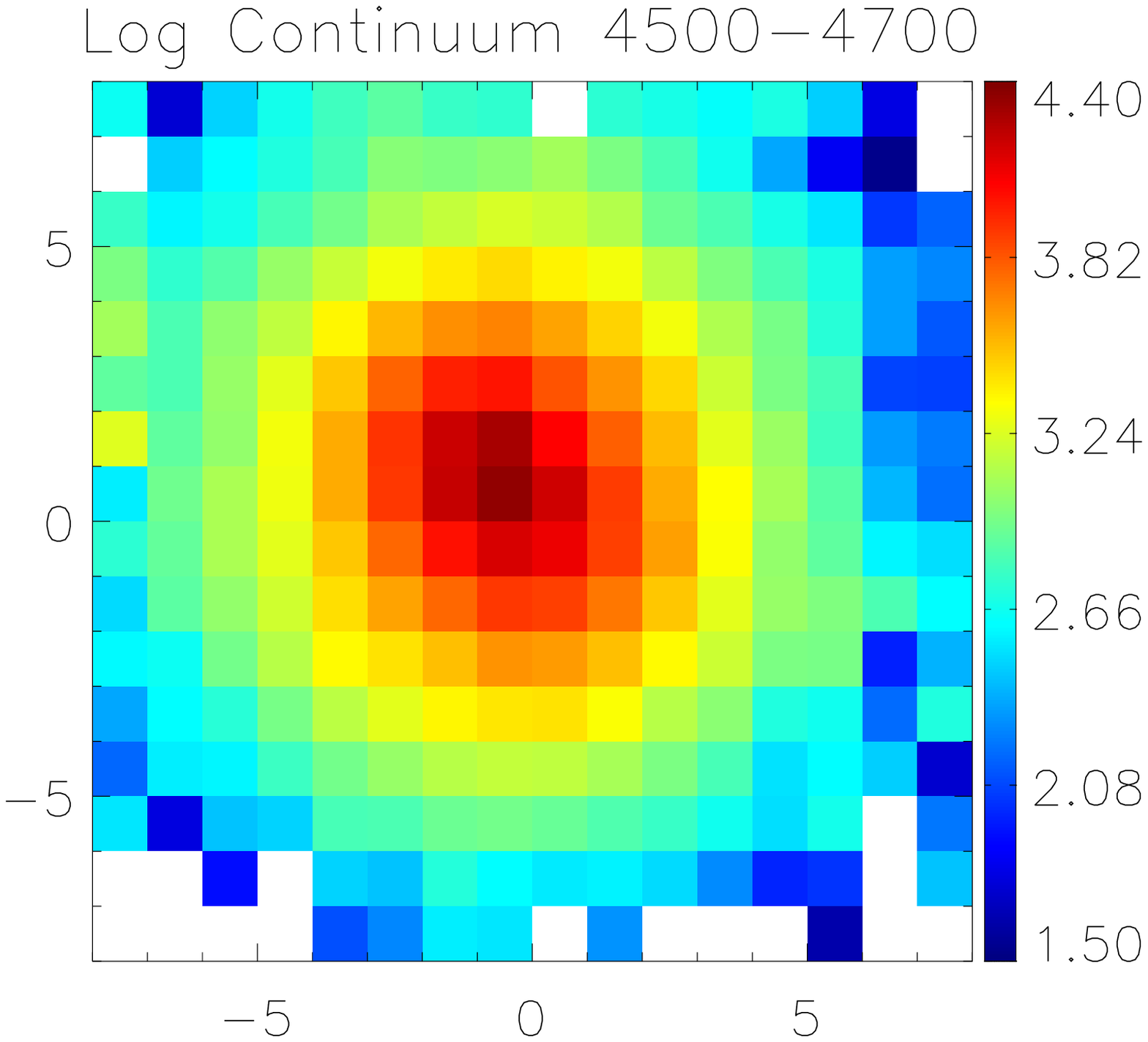}}
\hspace*{0.0cm}\subfigure{\includegraphics[width=0.24\textwidth]{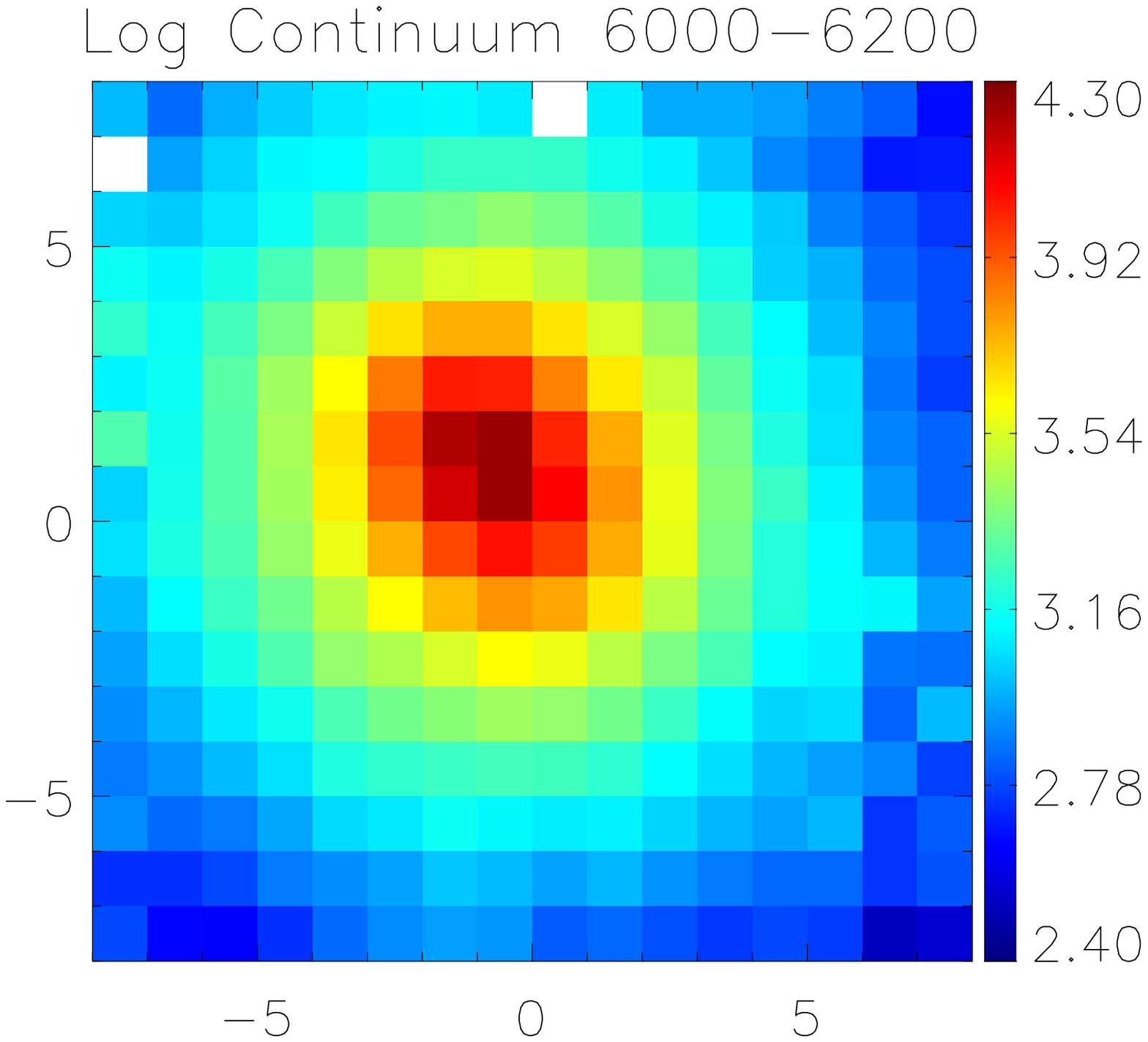}}
\hspace*{0.0cm}\subfigure{\includegraphics[width=0.24\textwidth]{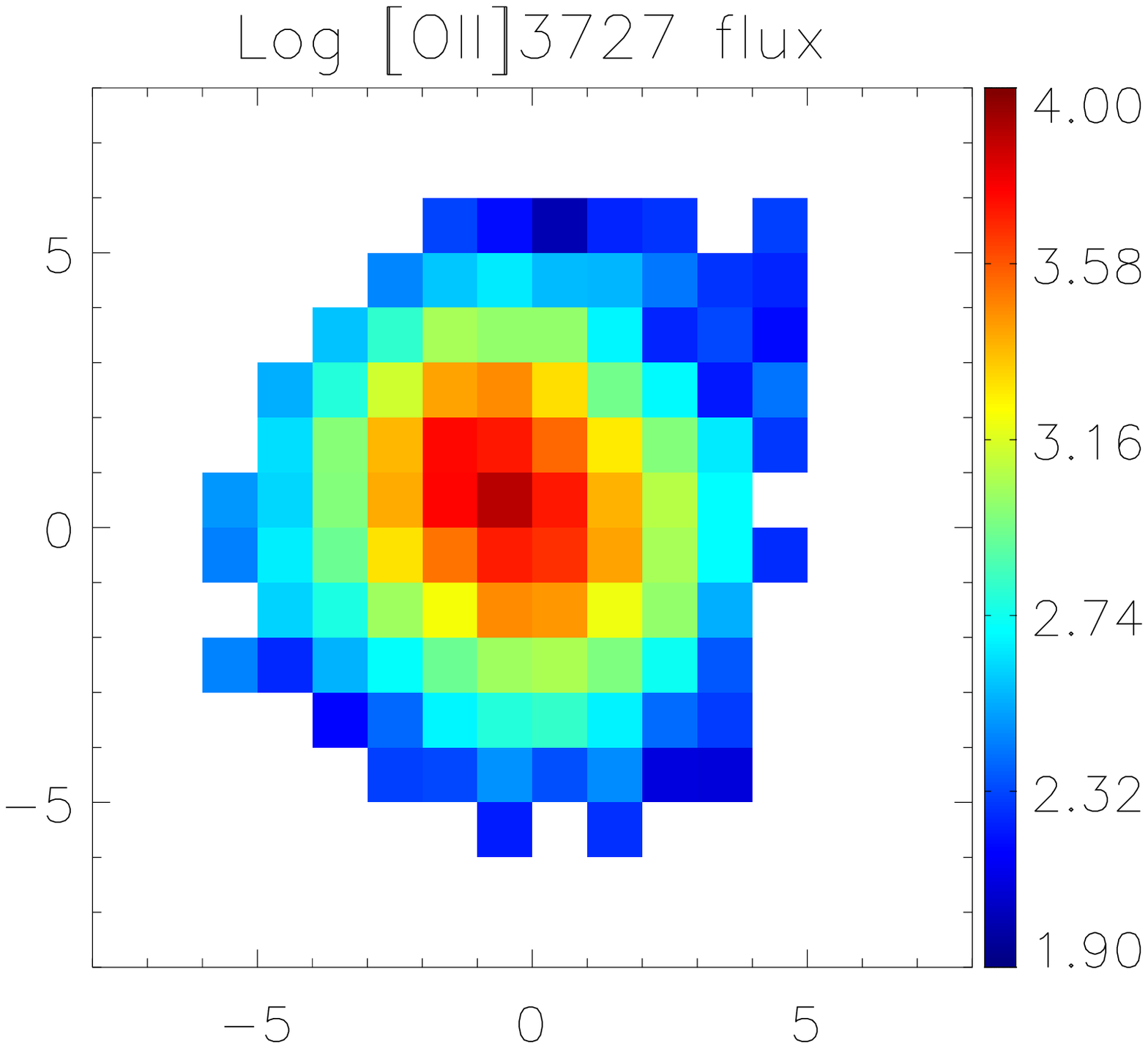}}
\hspace*{0.0cm}\subfigure{\includegraphics[width=0.24\textwidth]{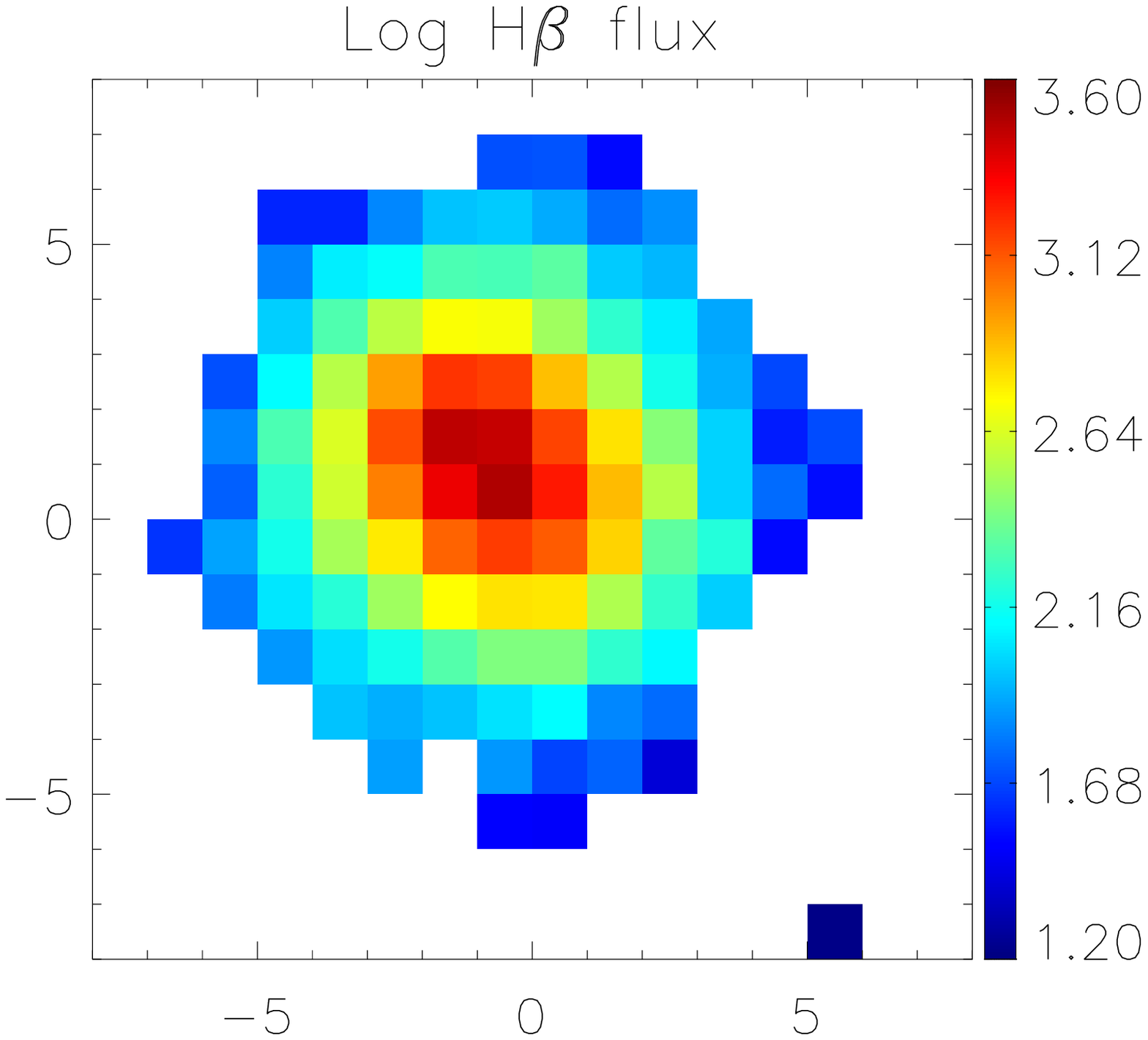}}
}}   
\mbox{
\centerline{
\hspace*{0.0cm}\subfigure{\includegraphics[width=0.24\textwidth]{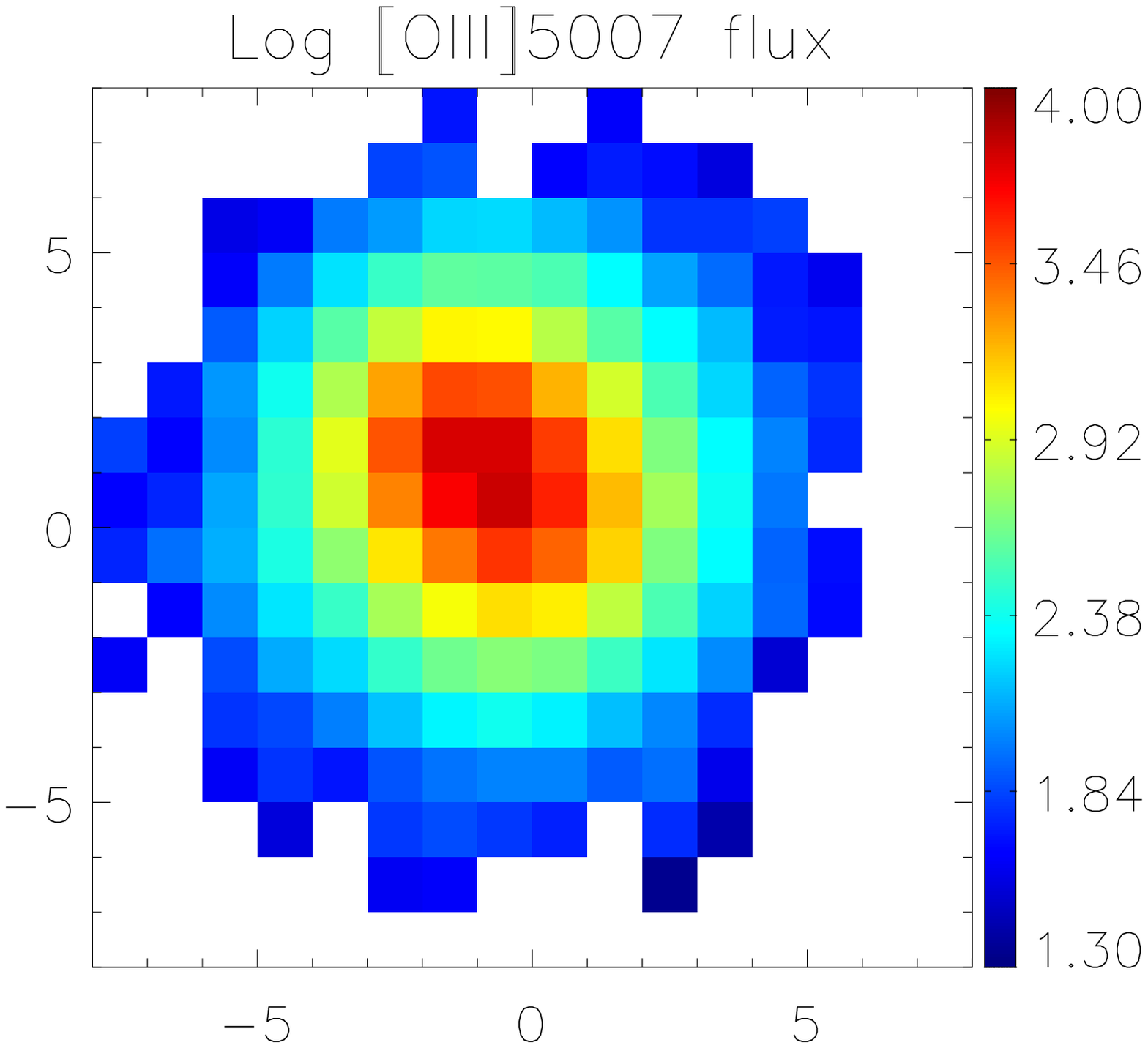}}
\hspace*{0.0cm}\subfigure{\includegraphics[width=0.24\textwidth]{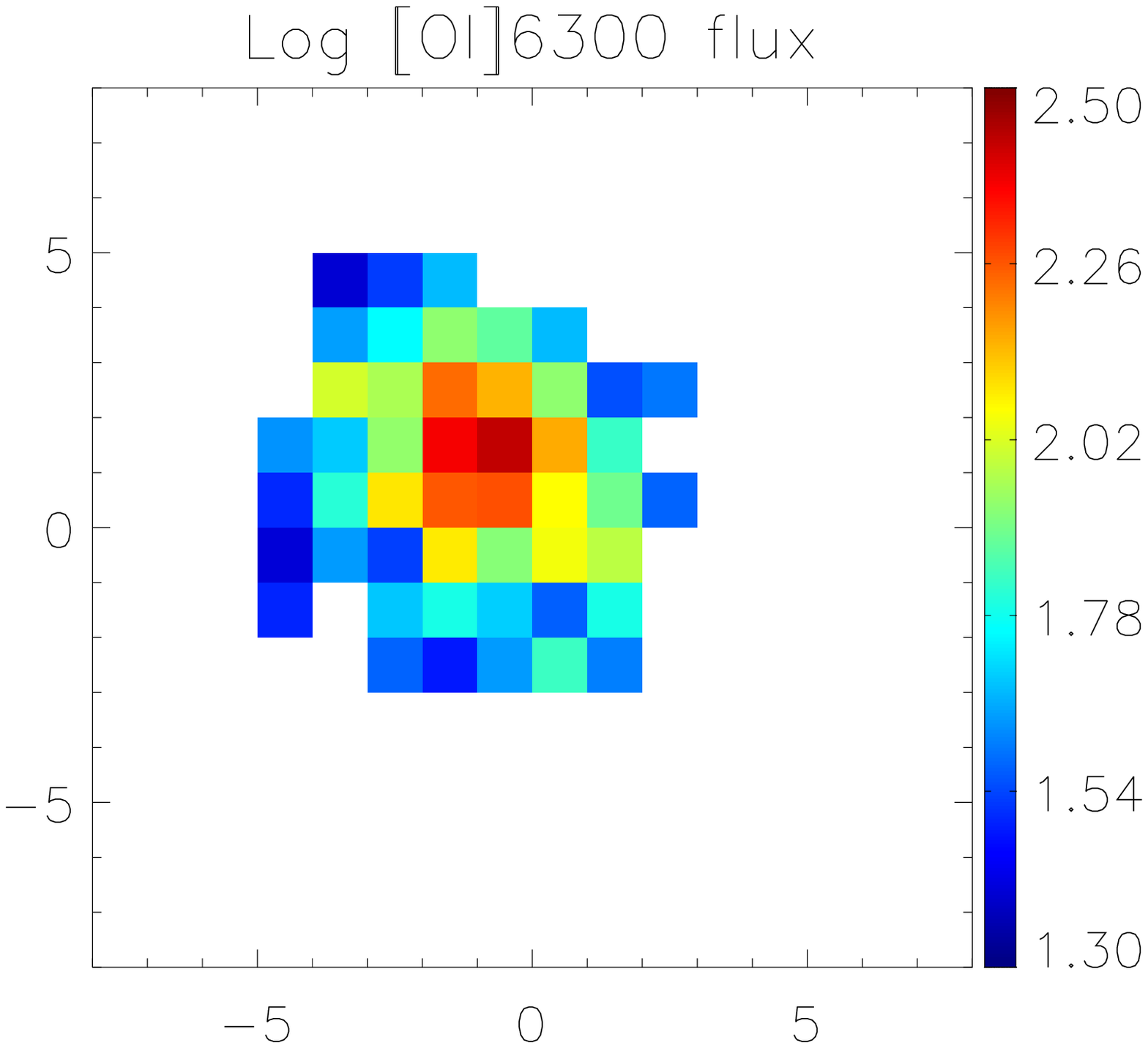}}
\hspace*{0.0cm}\subfigure{\includegraphics[width=0.24\textwidth]{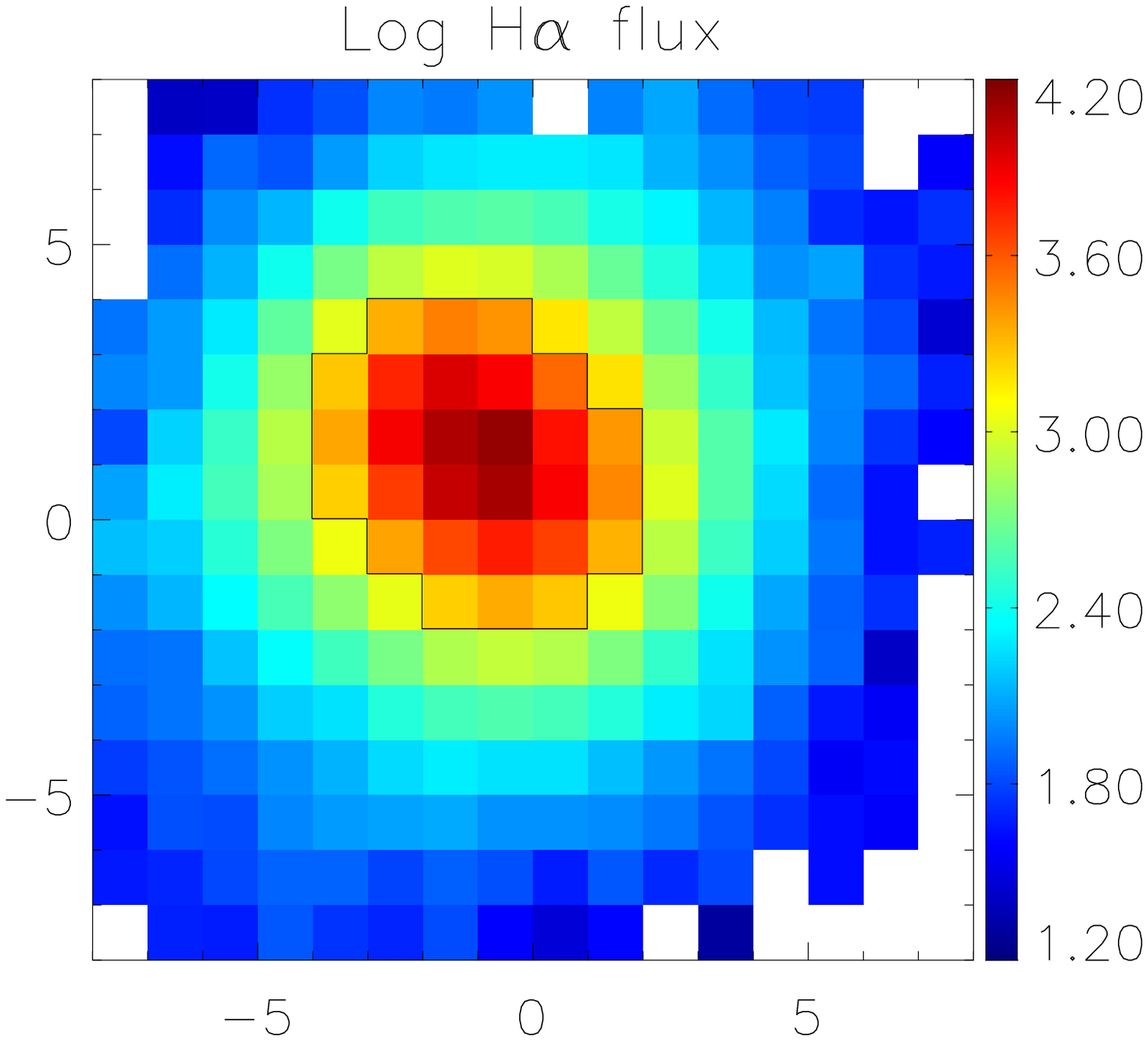}}
\hspace*{0.0cm}\subfigure{\includegraphics[width=0.24\textwidth]{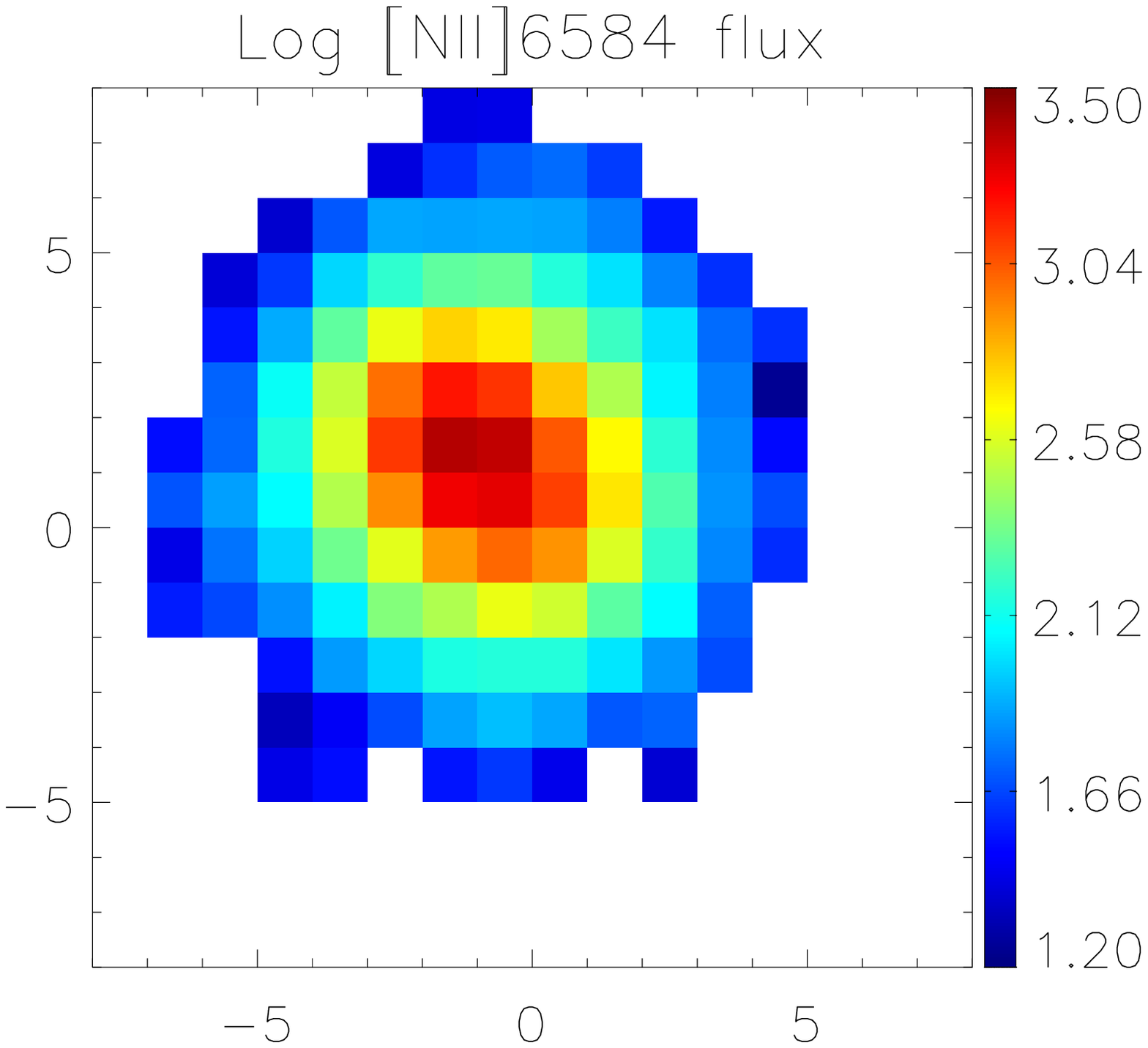}}
}} 
\mbox{
\centerline{
\hspace*{0.0cm}\subfigure{\includegraphics[width=0.24\textwidth]{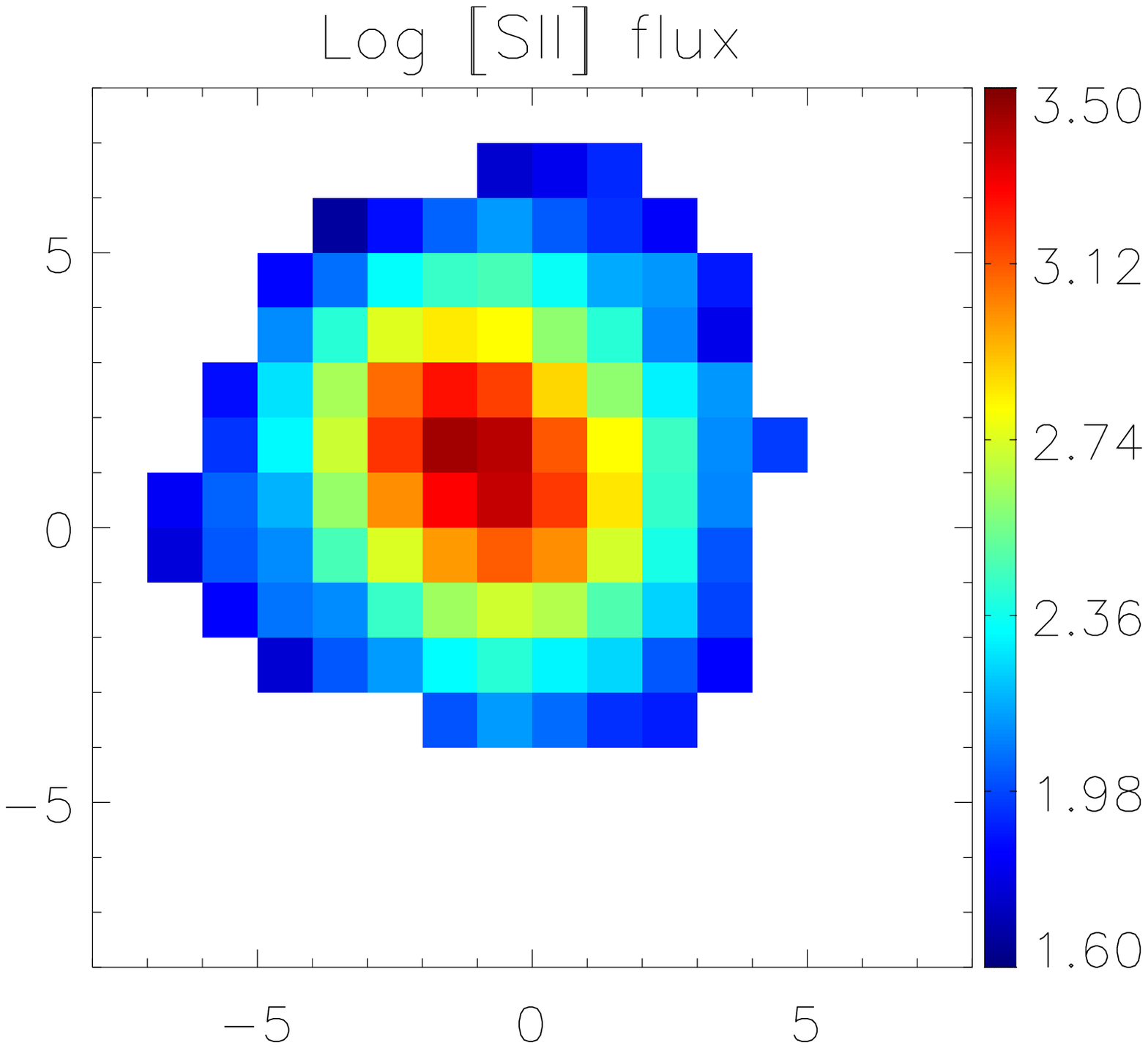}}
\hspace*{0.0cm}\subfigure{\includegraphics[width=0.24\textwidth]{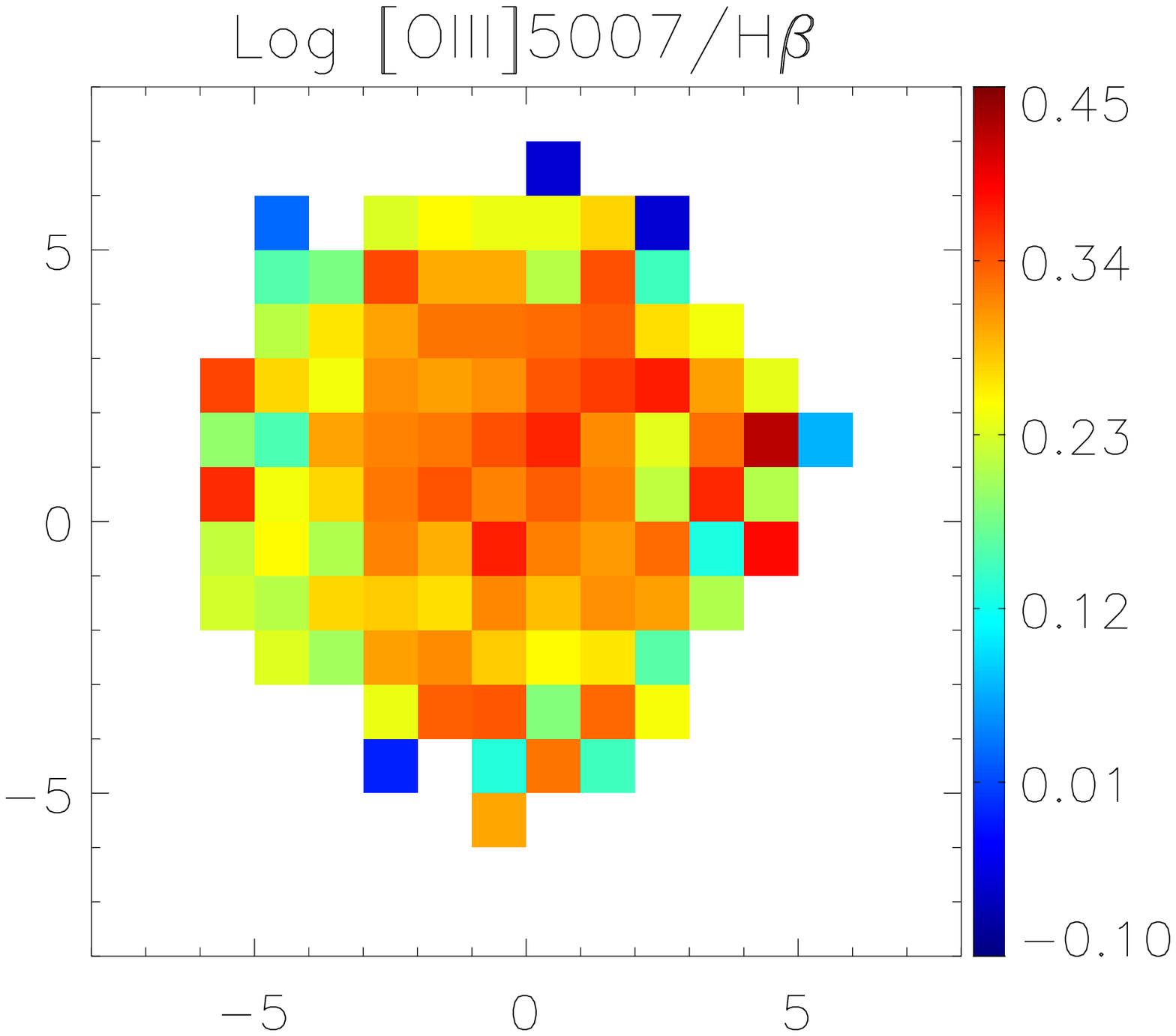}}
\hspace*{0.0cm}\subfigure{\includegraphics[width=0.24\textwidth]{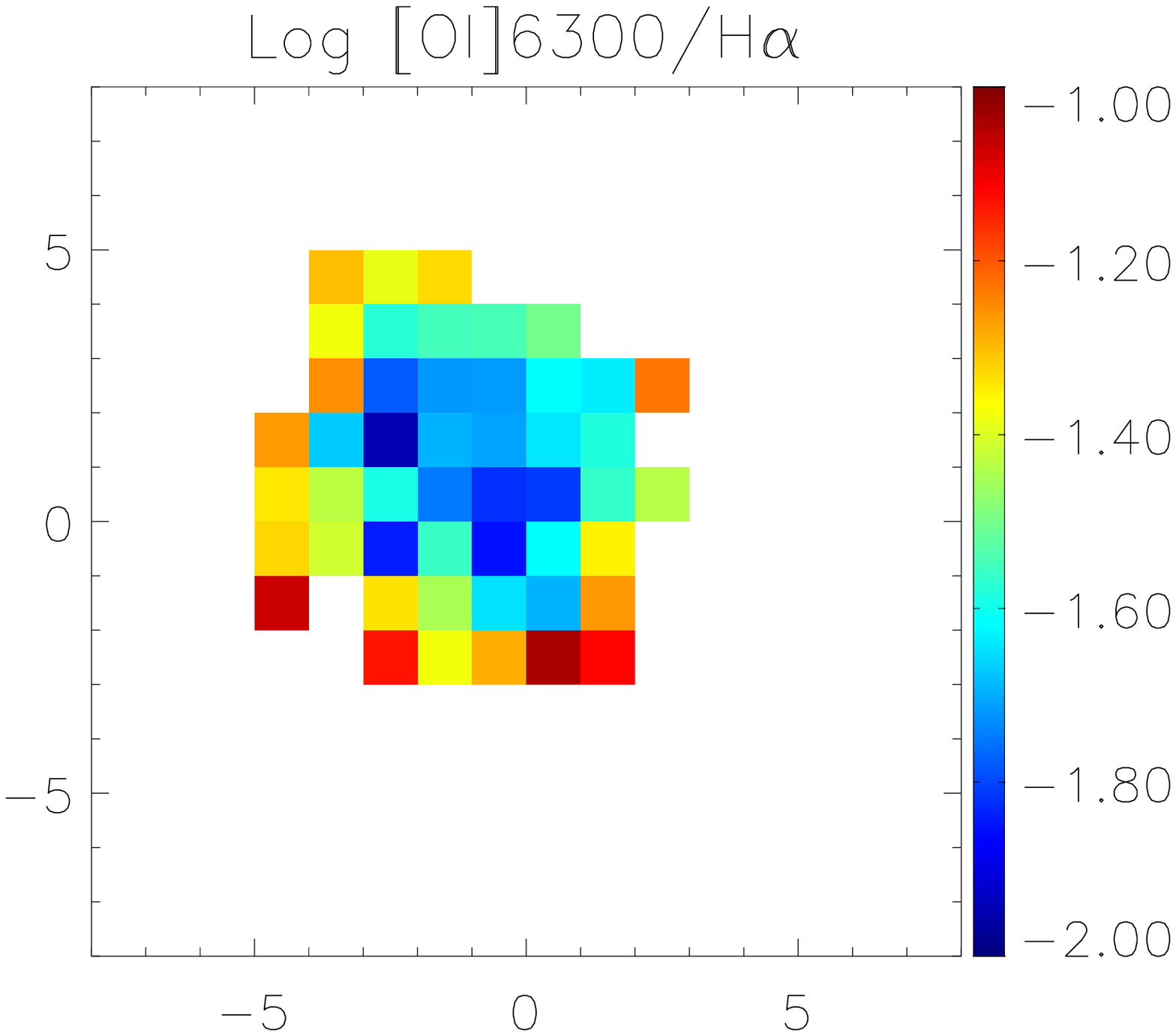}}
\hspace*{0.0cm}\subfigure{\includegraphics[width=0.24\textwidth]{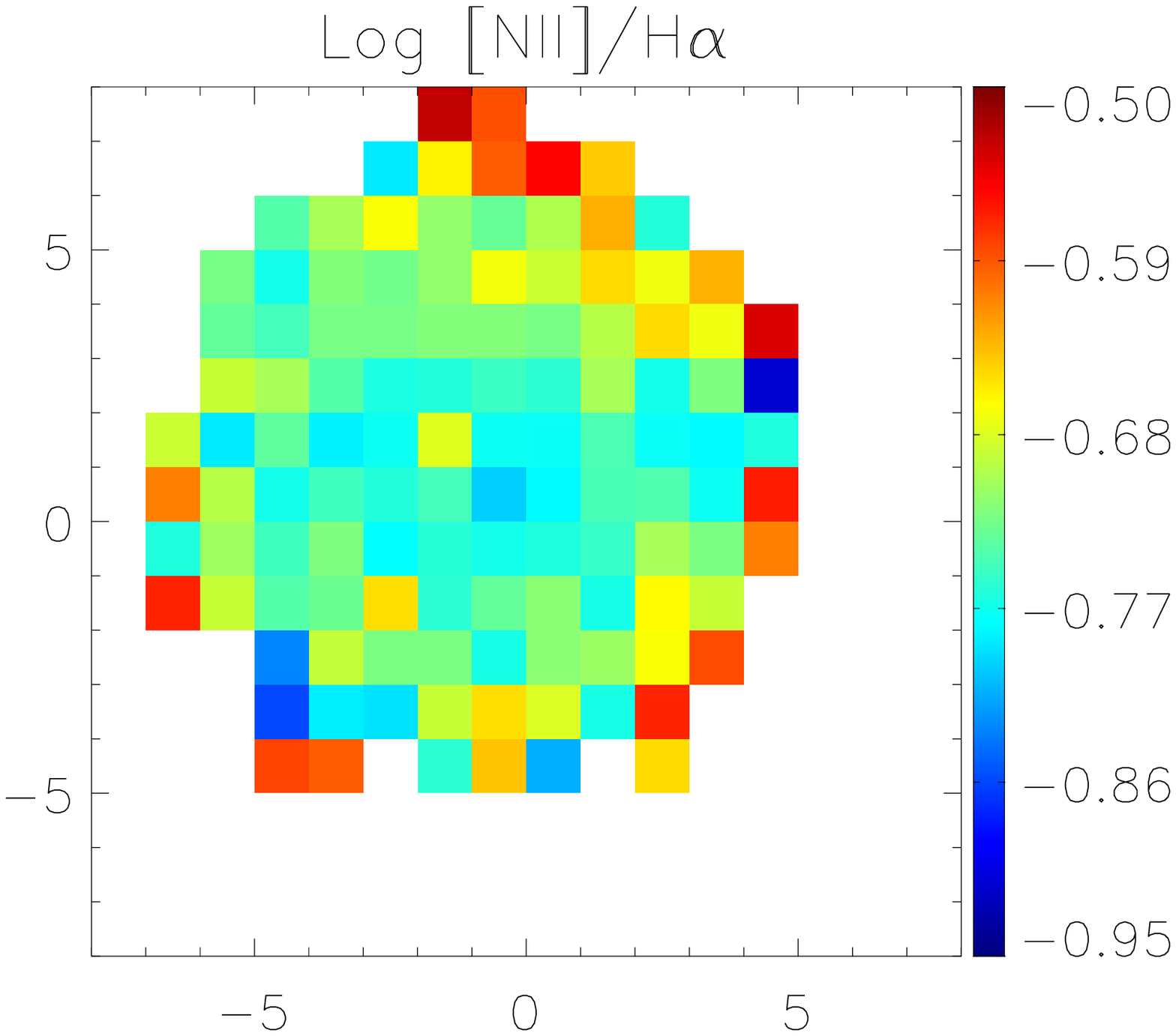}}
}}
\mbox{
\centerline{
\hspace*{0.0cm}\subfigure{\includegraphics[width=0.24\textwidth]{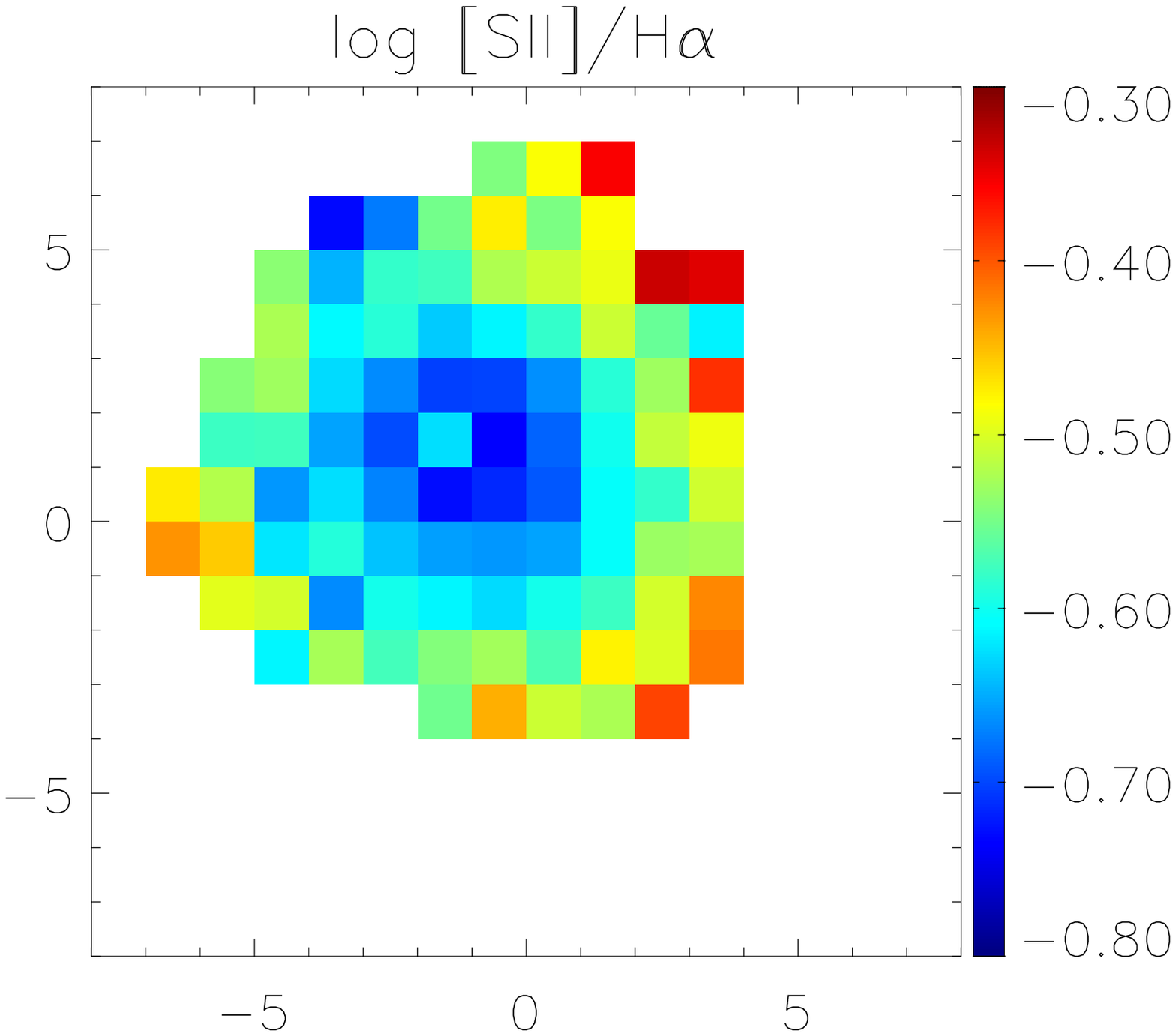}}
\hspace*{0.0cm}\subfigure{\includegraphics[width=0.24\textwidth]{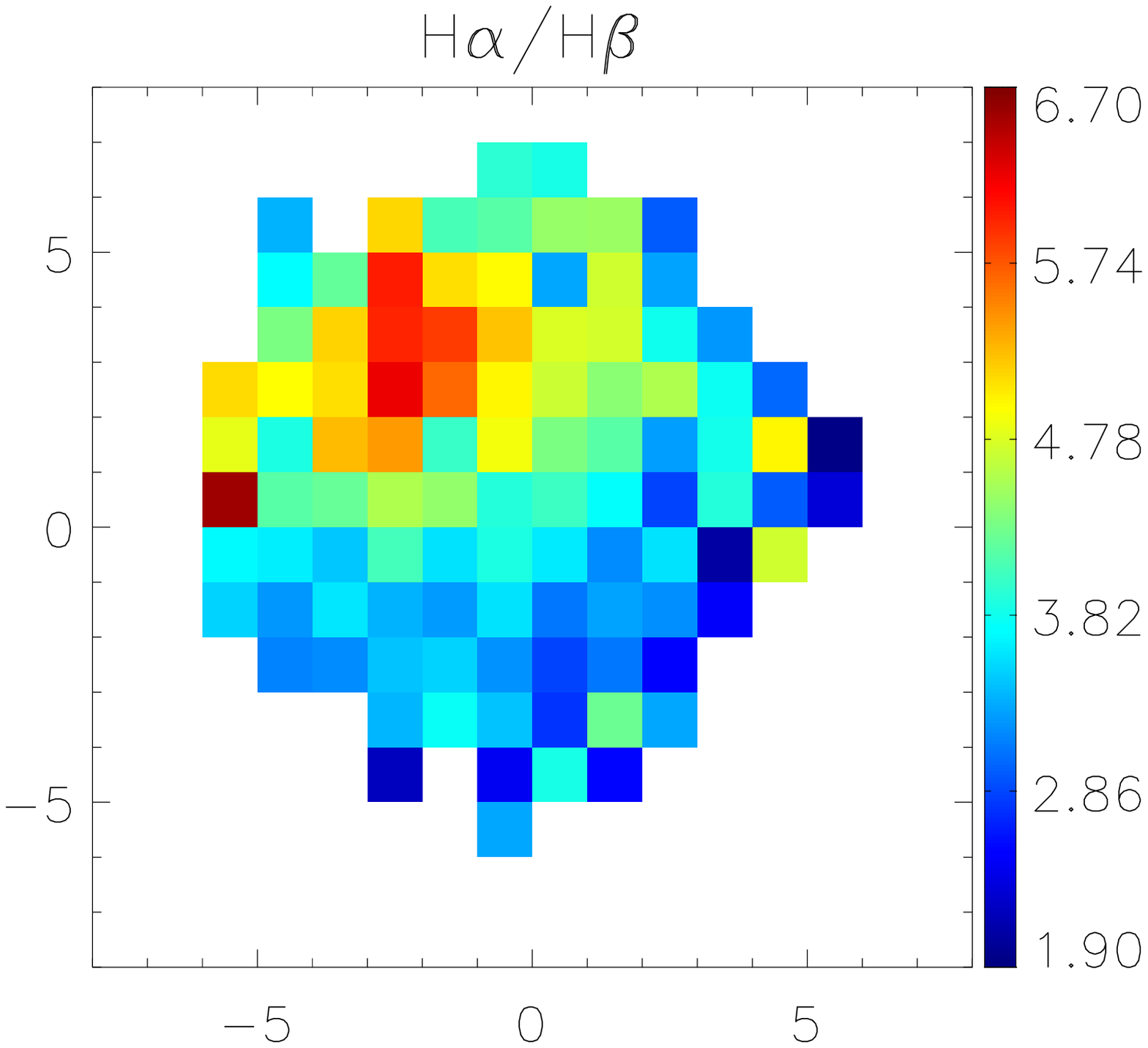}}
\hspace*{0.0cm}\subfigure{\includegraphics[width=0.24\textwidth]{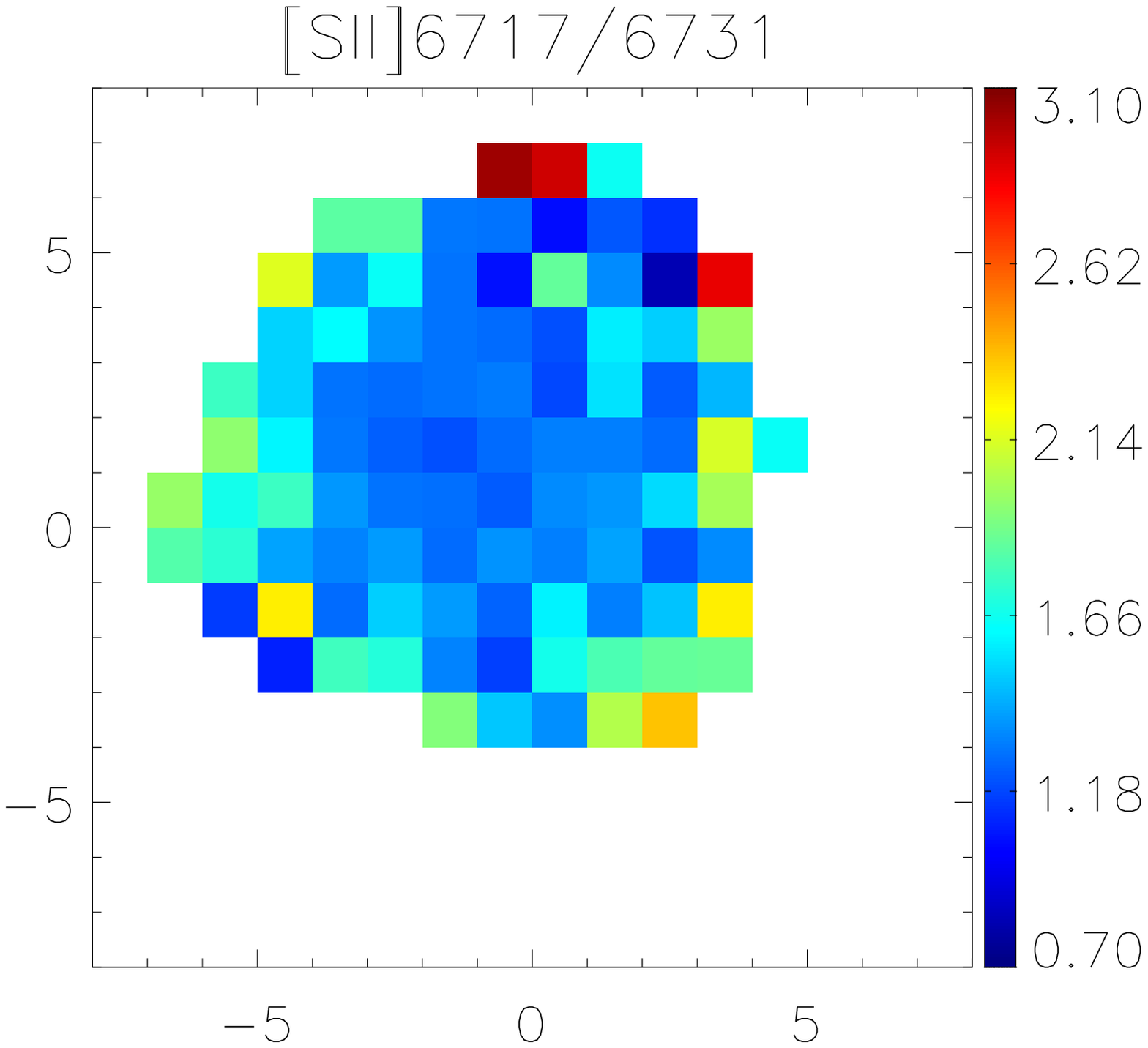}}
}}  
\mbox{
\centerline{
\hspace*{0.0cm}\subfigure{\includegraphics[width=0.24\textwidth]{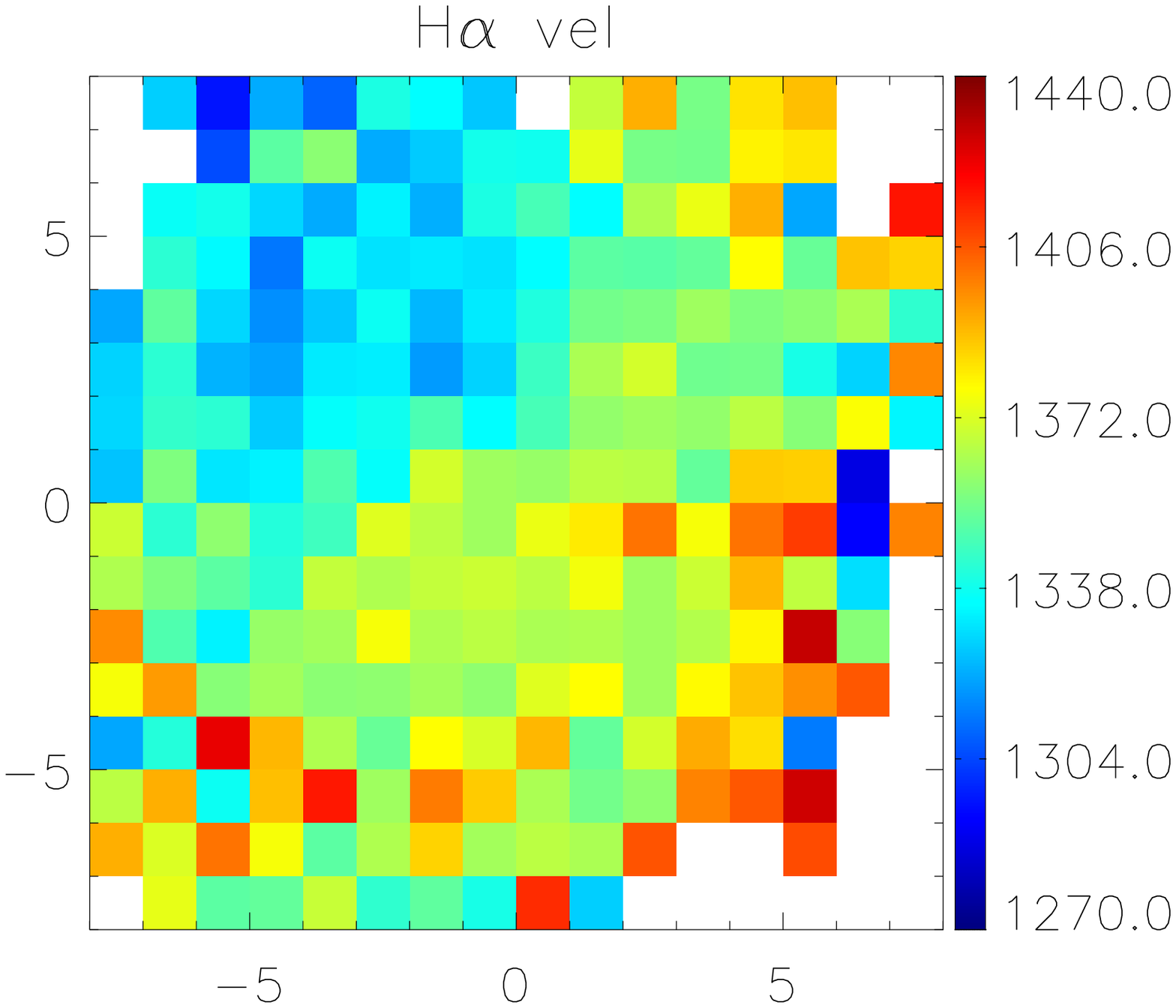}}
\hspace*{0.0cm}\subfigure{\includegraphics[width=0.24\textwidth]{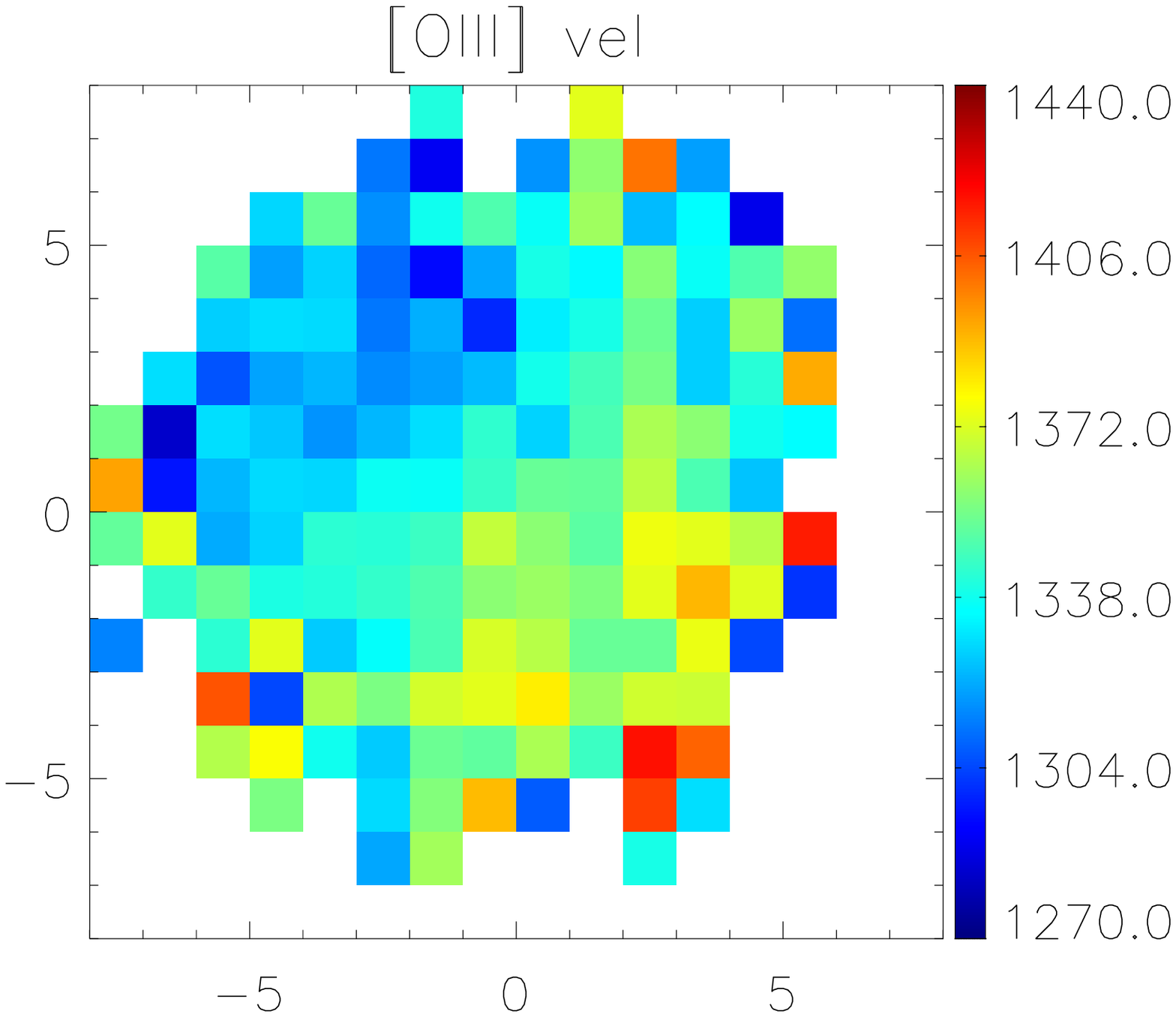}}
}}  
\caption{Same as Fig.~\ref{Figure:mrk407} for Mrk~206. Maps of
[\ion{O}{i}]~$\lambda6300$ and of the [\ion{O}{i}]~$\lambda6300$/\Ha\ 
ionization ratio are also included.}
\label{Figure:mrk206}
\end{figure*}

\begin{figure*}
\mbox{
\centerline{
\hspace*{0.0cm}\subfigure{\includegraphics[width=0.24\textwidth]{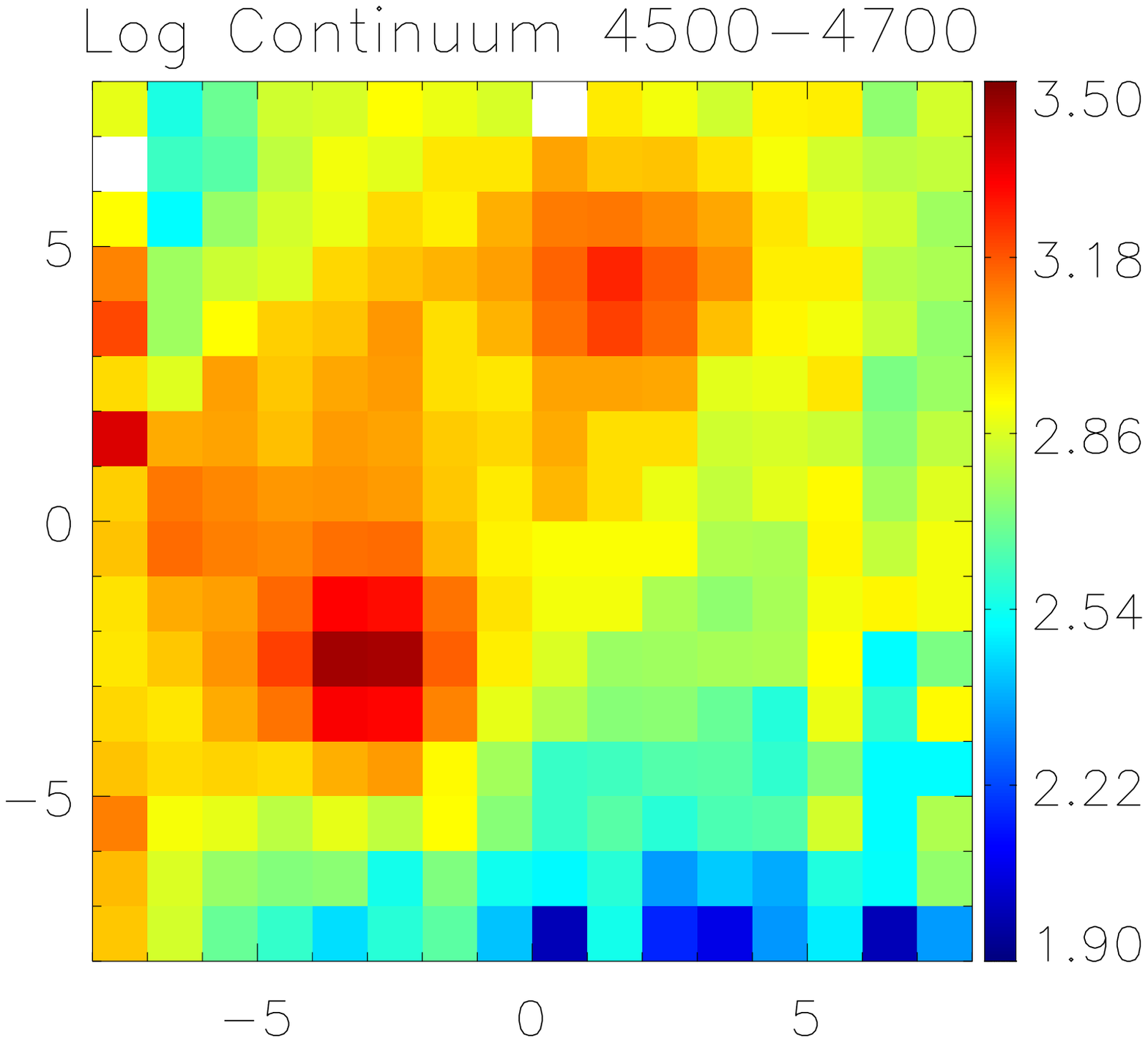}}
\hspace*{0.0cm}\subfigure{\includegraphics[width=0.24\textwidth]{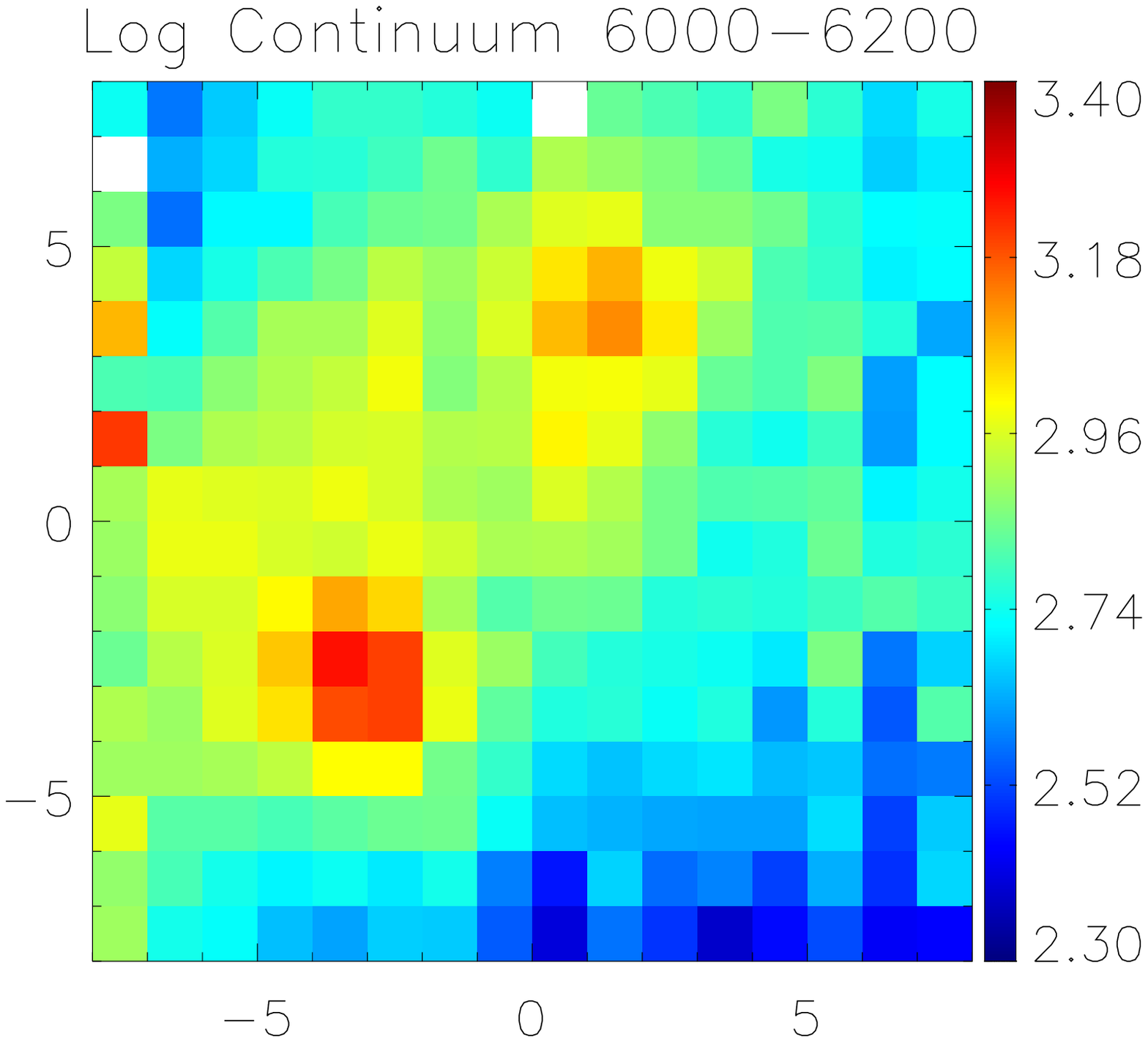}}
\hspace*{0.0cm}\subfigure{\includegraphics[width=0.24\textwidth]{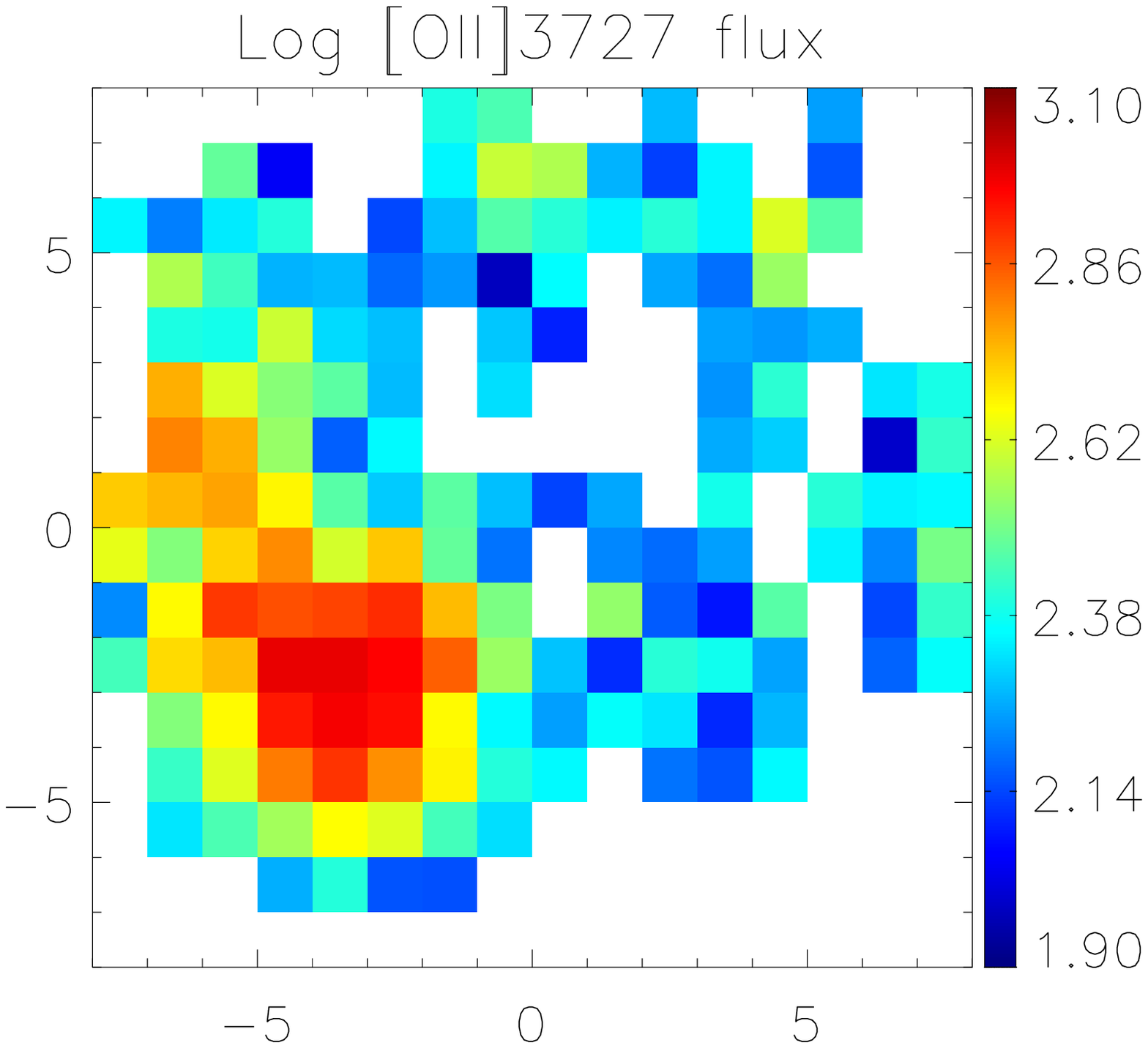}}
\hspace*{0.0cm}\subfigure{\includegraphics[width=0.24\textwidth]{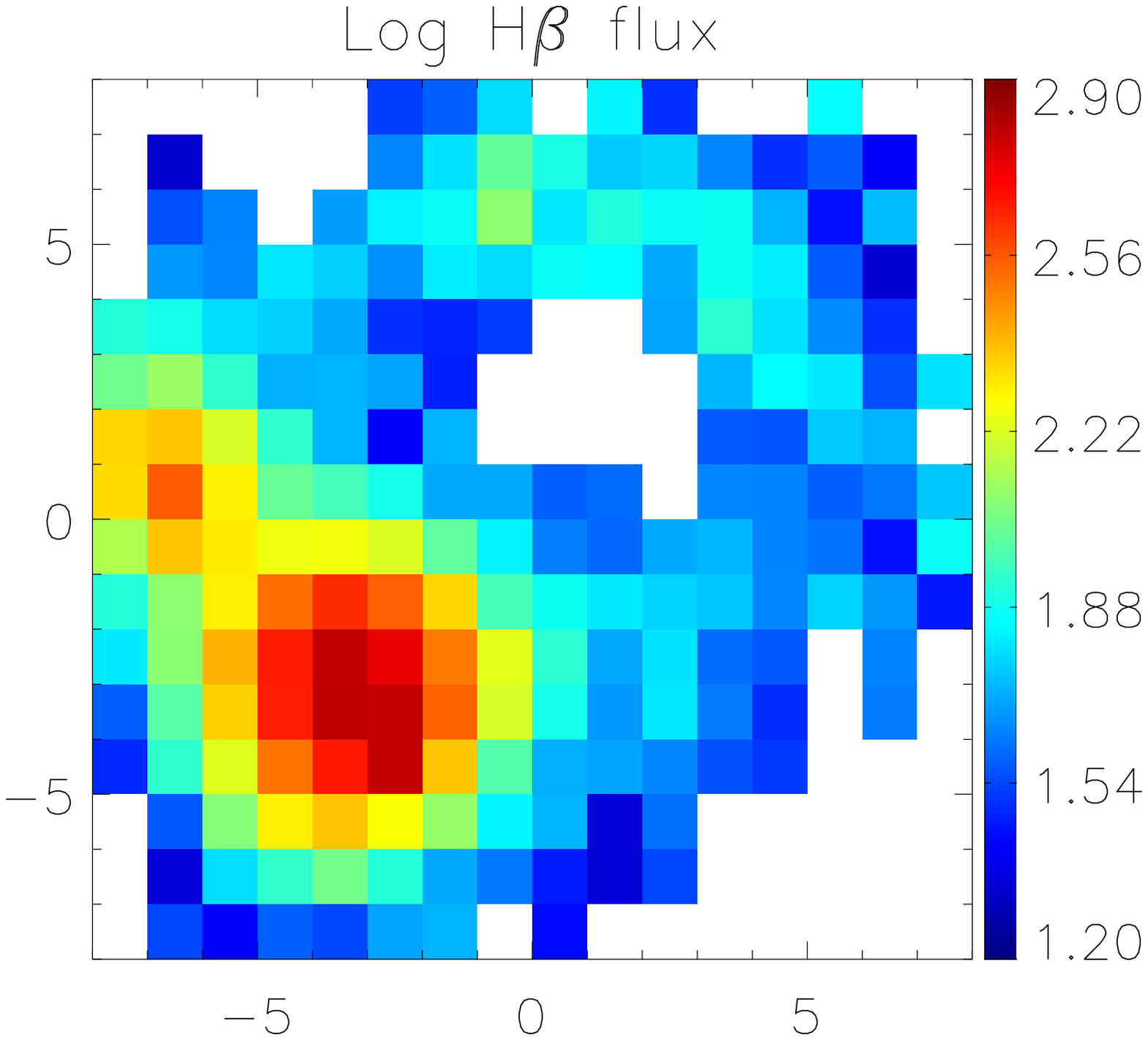}}
}}   
\mbox{
\centerline{
\hspace*{0.0cm}\subfigure{\includegraphics[width=0.24\textwidth]{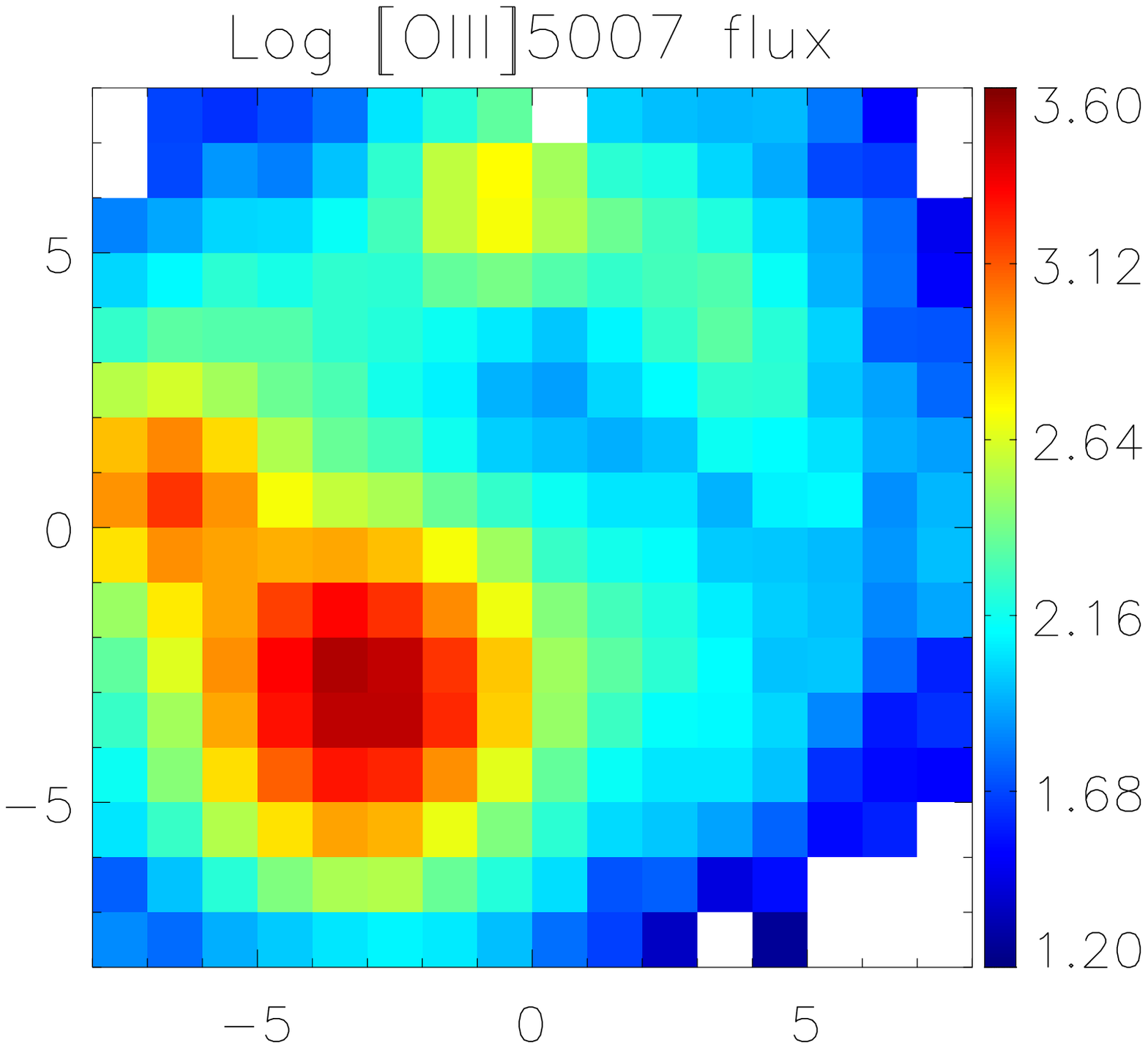}}
\hspace*{0.0cm}\subfigure{\includegraphics[width=0.24\textwidth]{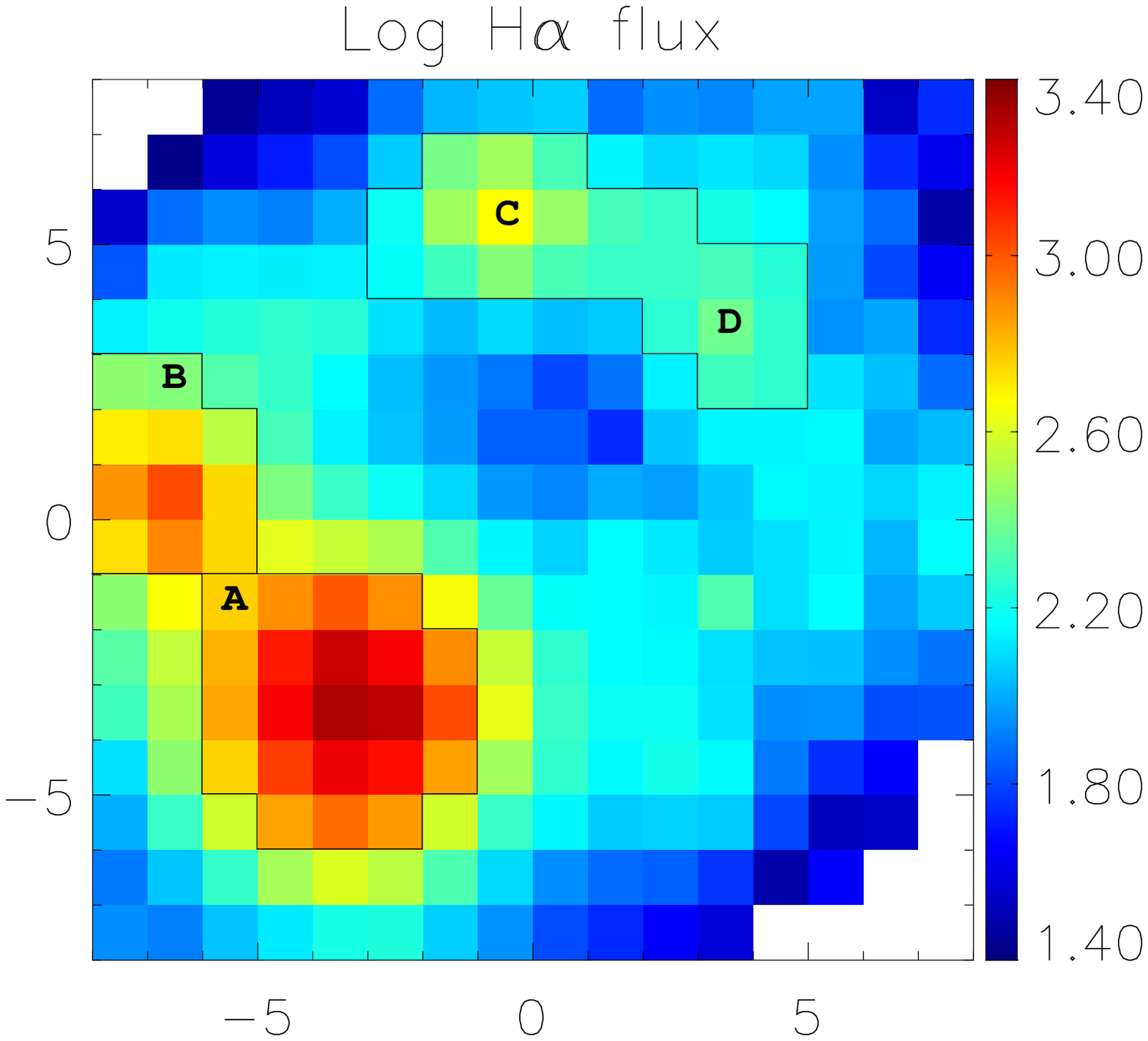}}
\hspace*{0.0cm}\subfigure{\includegraphics[width=0.24\textwidth]{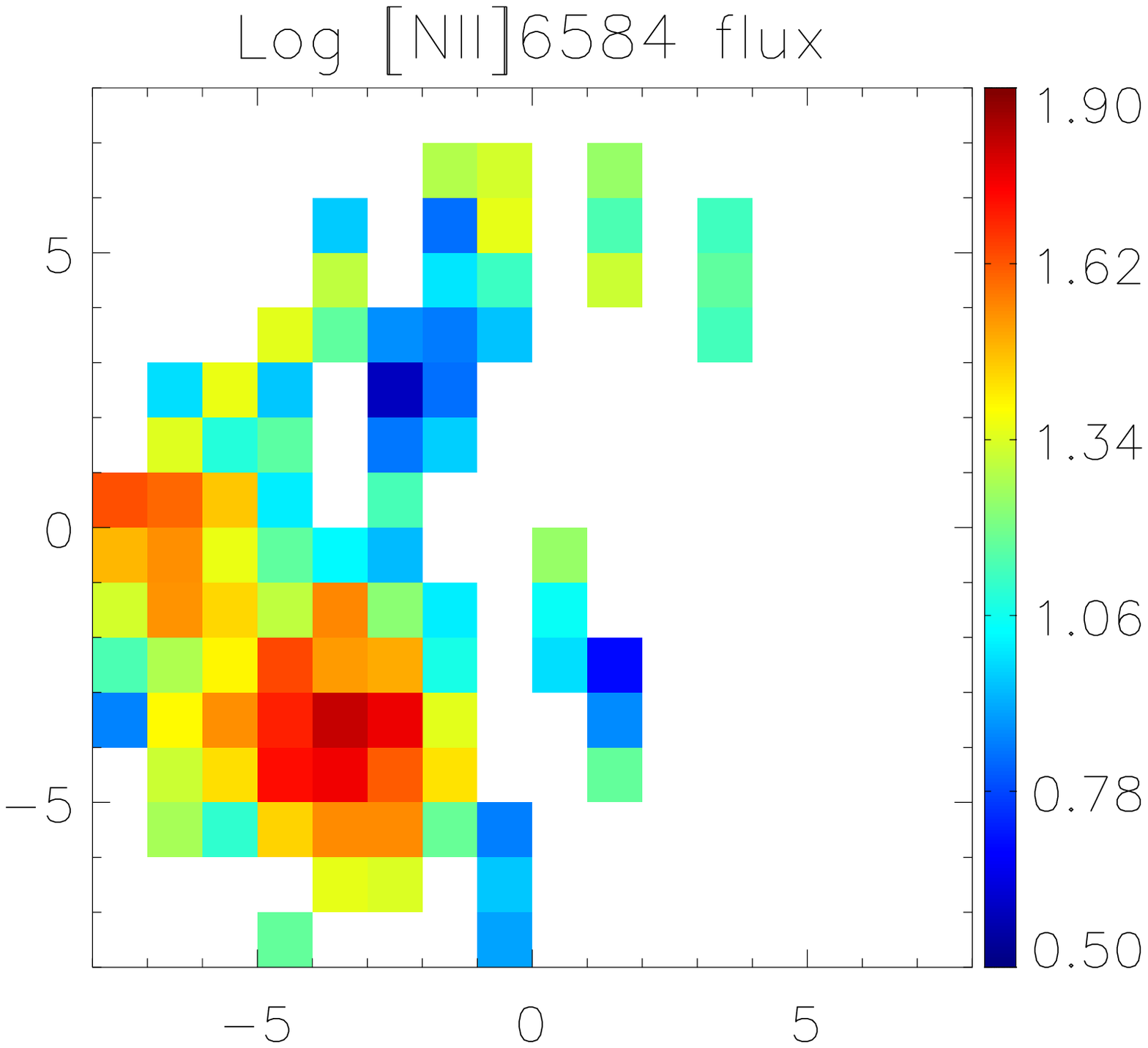}}
\hspace*{0.0cm}\subfigure{\includegraphics[width=0.24\textwidth]{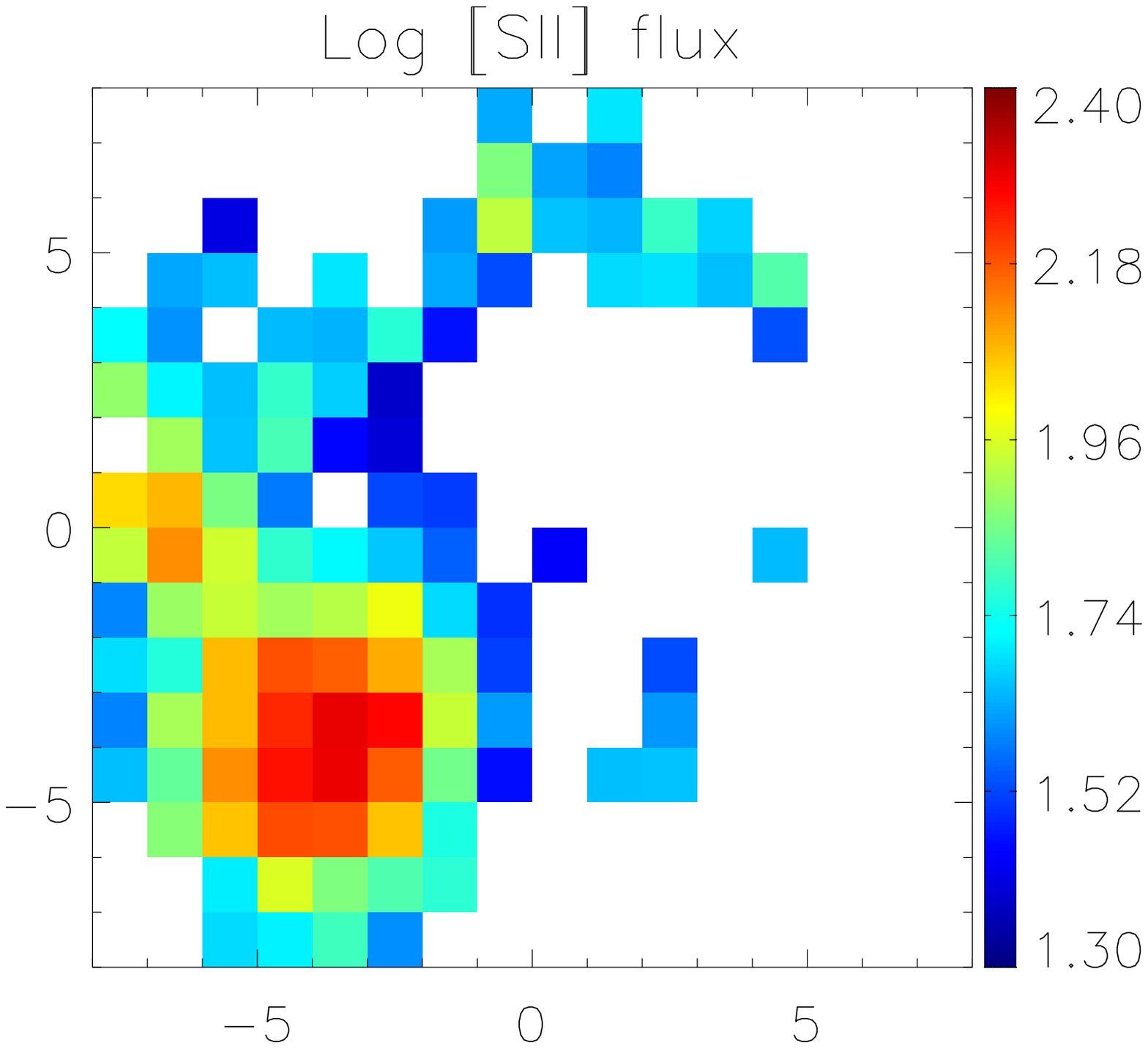}}
}} 
\mbox{
\centerline{
\hspace*{0.0cm}\subfigure{\includegraphics[width=0.24\textwidth]{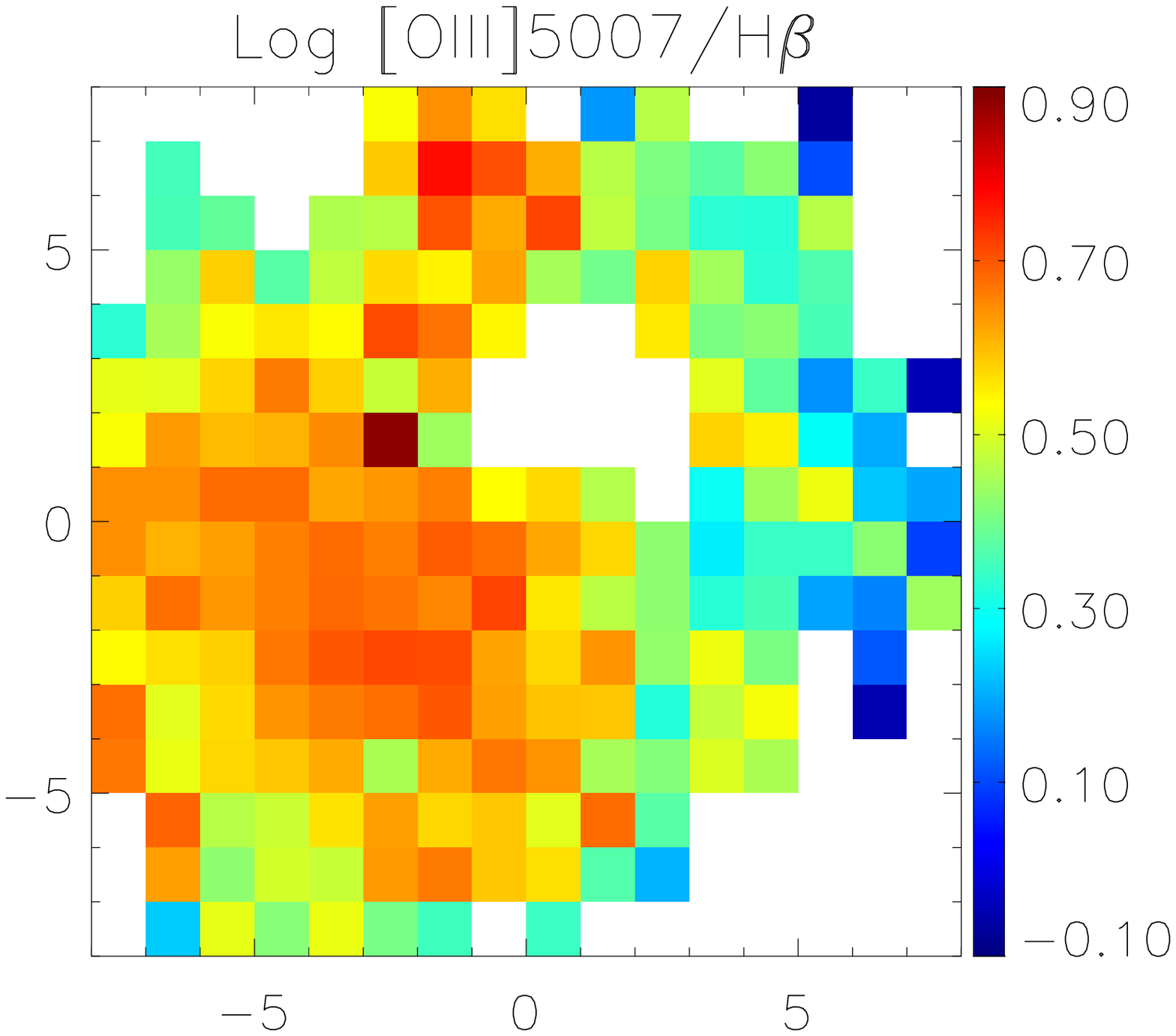}}
\hspace*{0.0cm}\subfigure{\includegraphics[width=0.24\textwidth]{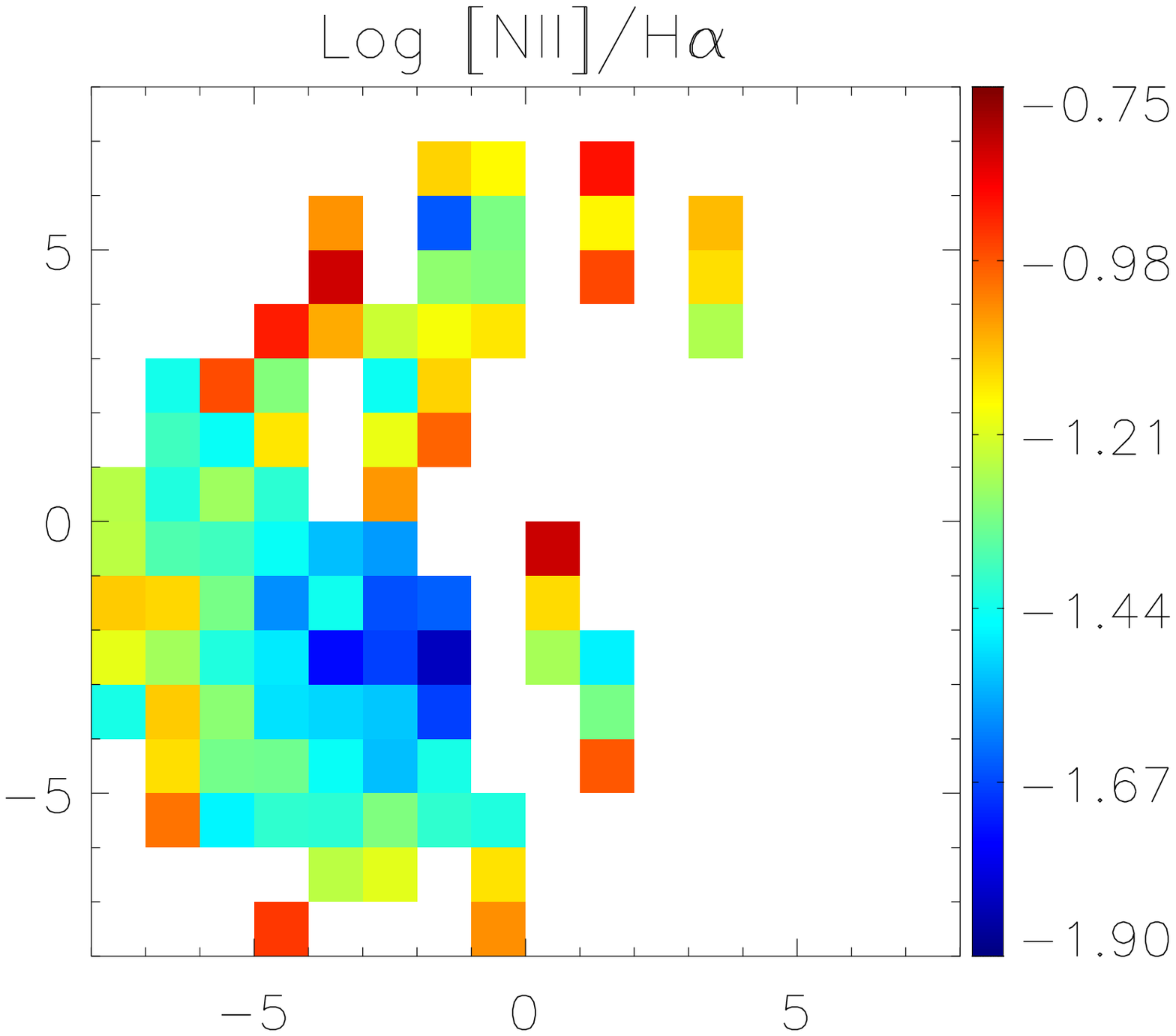}}
\hspace*{0.0cm}\subfigure{\includegraphics[width=0.24\textwidth]{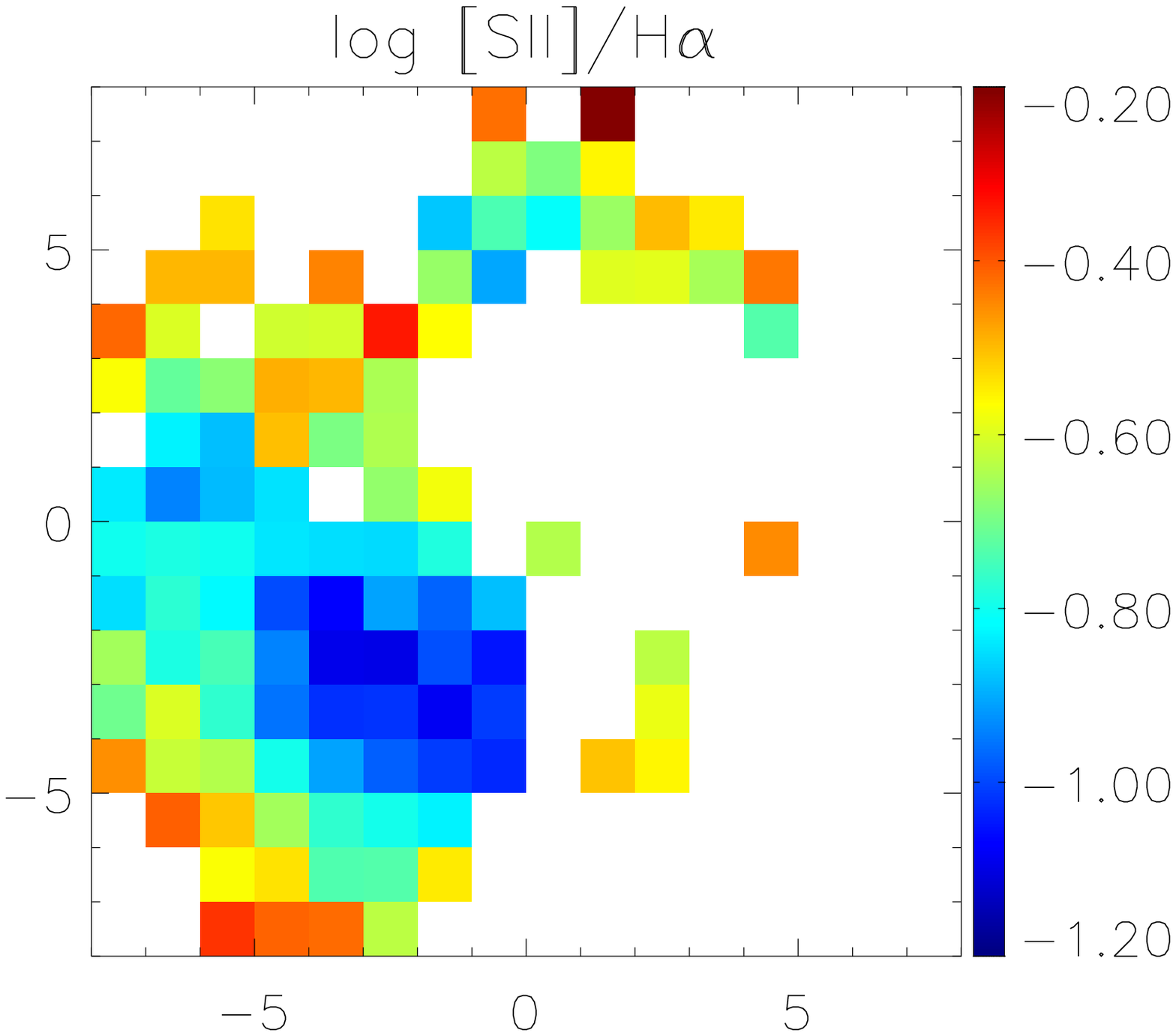}}
\hspace*{0.0cm}\subfigure{\includegraphics[width=0.24\textwidth]{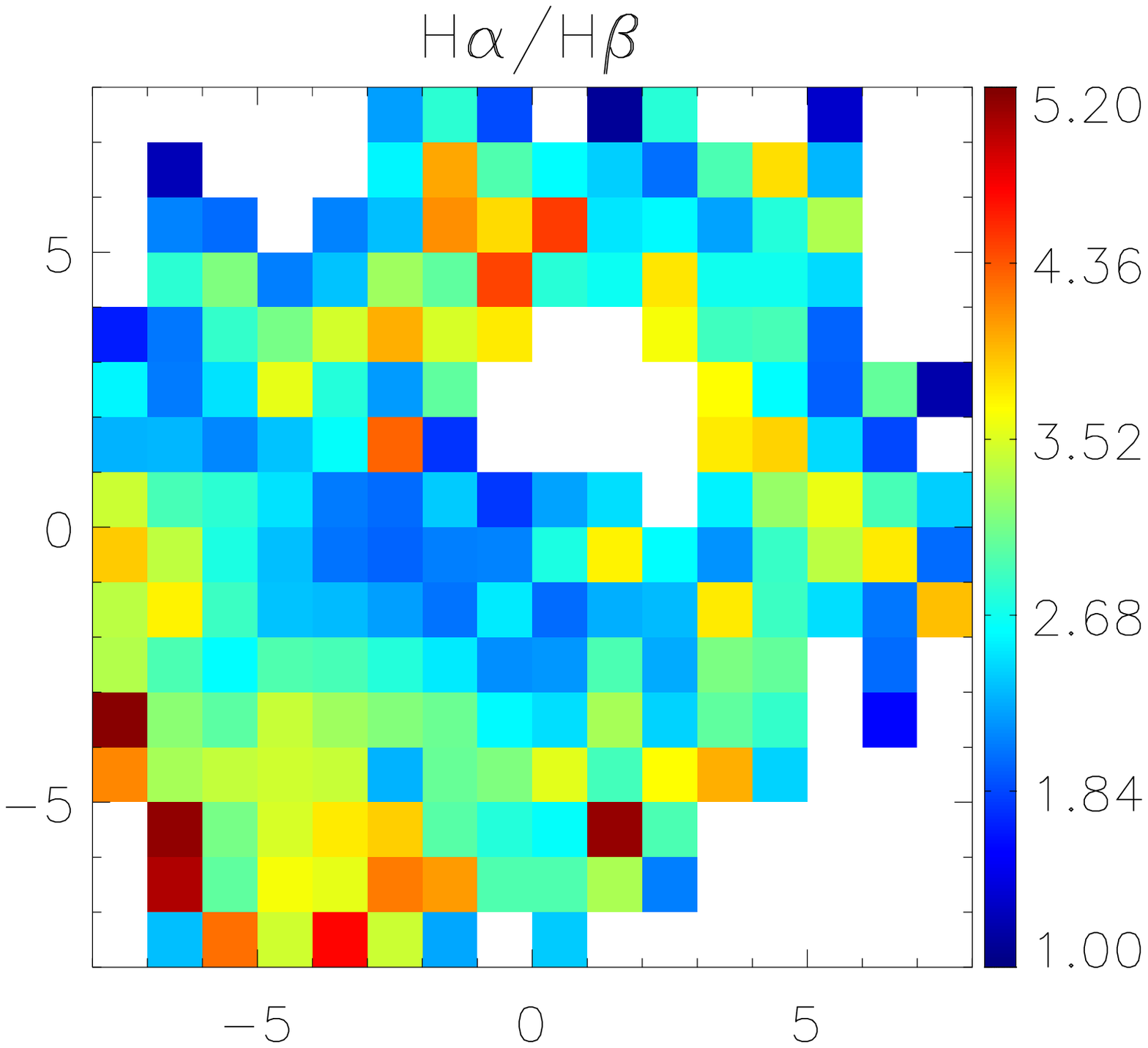}}
}}
\mbox{
\centerline{
\hspace*{0.0cm}\subfigure{\includegraphics[width=0.24\textwidth]{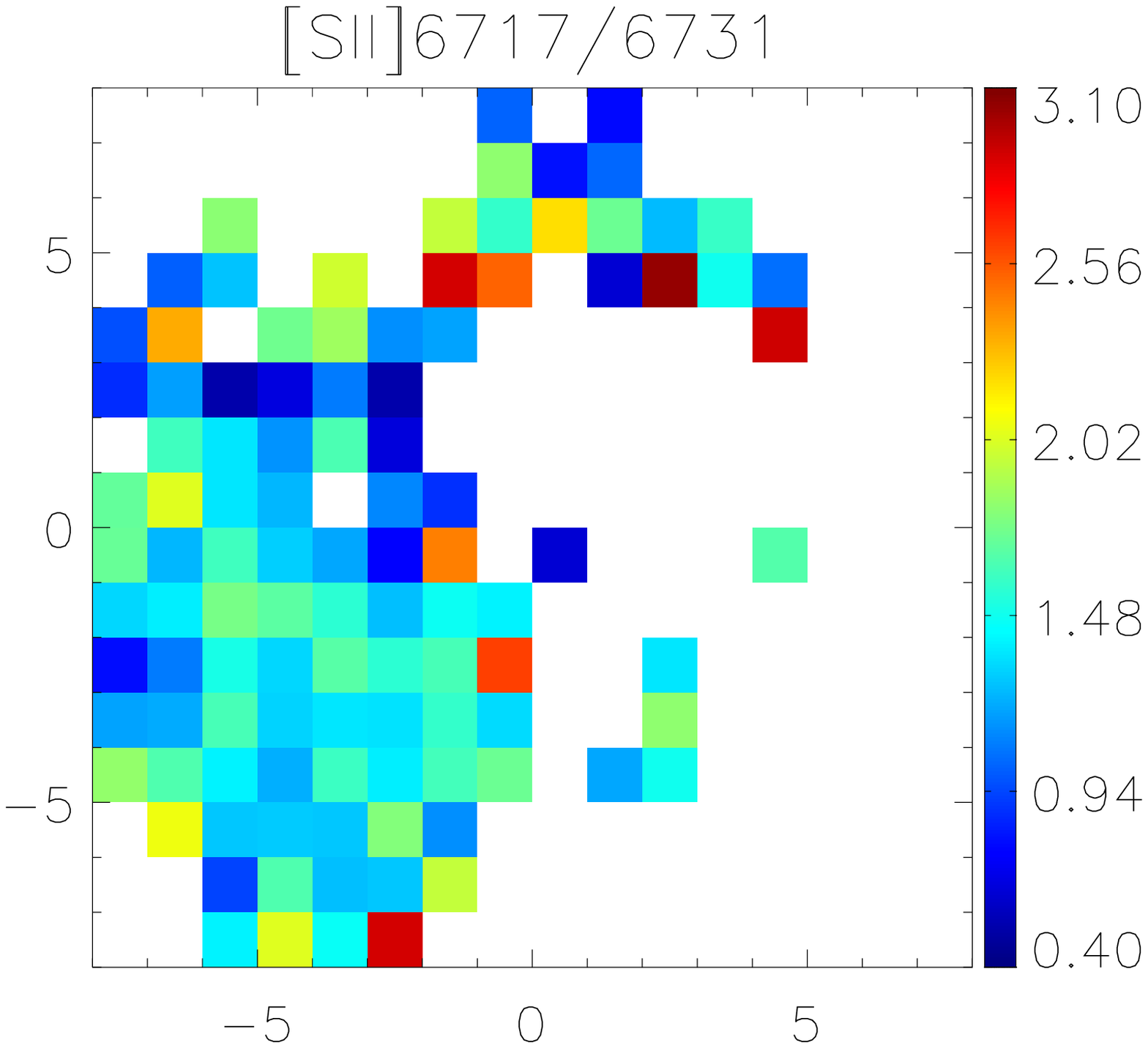}}
\hspace*{0.0cm}\subfigure{\includegraphics[width=0.24\textwidth]{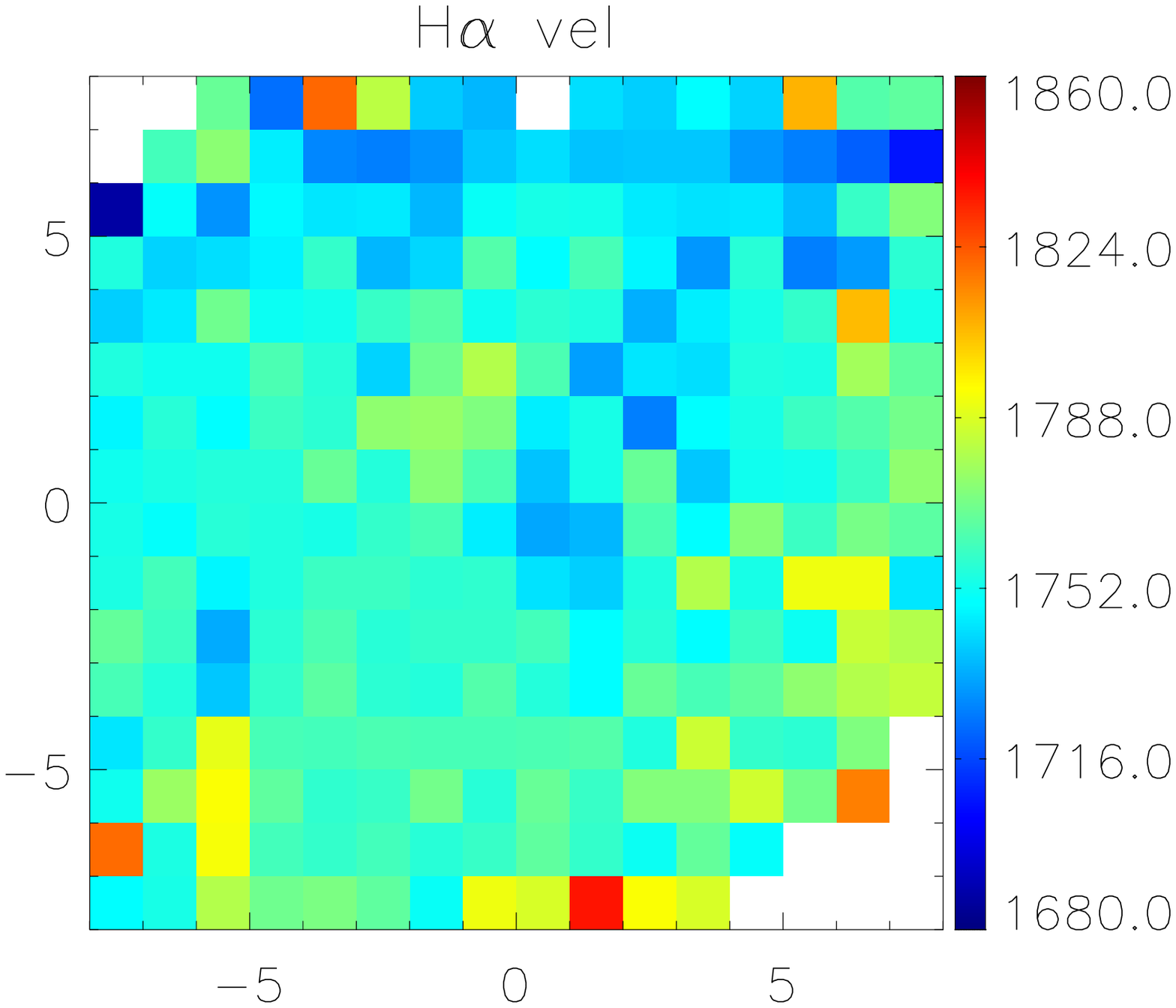}}
\hspace*{0.0cm}\subfigure{\includegraphics[width=0.24\textwidth]{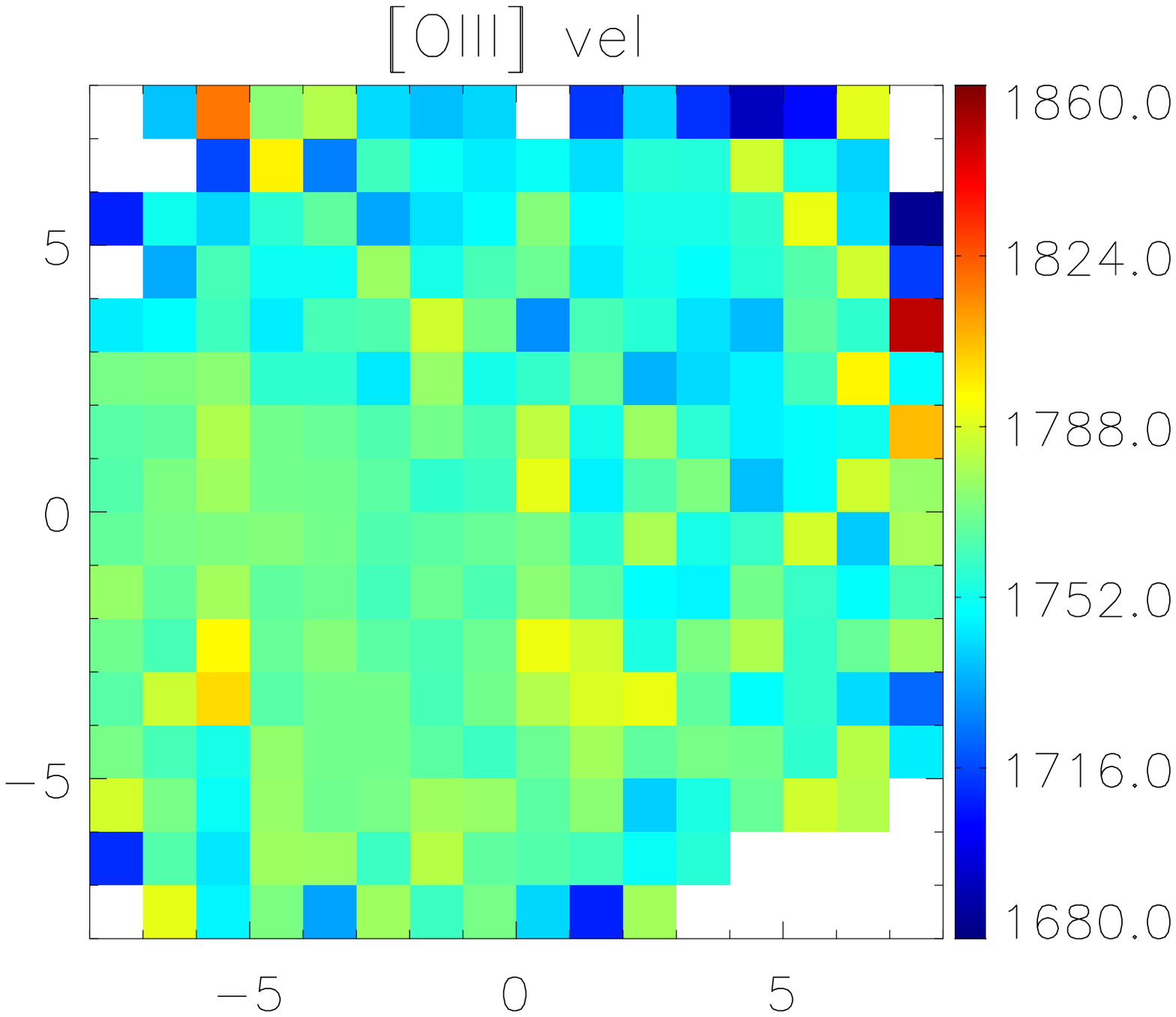}}
}}  
\caption{Same as Fig.~\ref{Figure:mrk407} for Tololo~1434+032. The outline 
of the identified SF knots (see Sect.~\ref{SubSection:IntegratedSpectroscopy})
is shown in the \Ha\ map.}
\label{Figure:tololo1434}
\end{figure*}

\begin{figure*}
\mbox{
\centerline{
\hspace*{0.0cm}\subfigure{\includegraphics[width=0.24\textwidth]{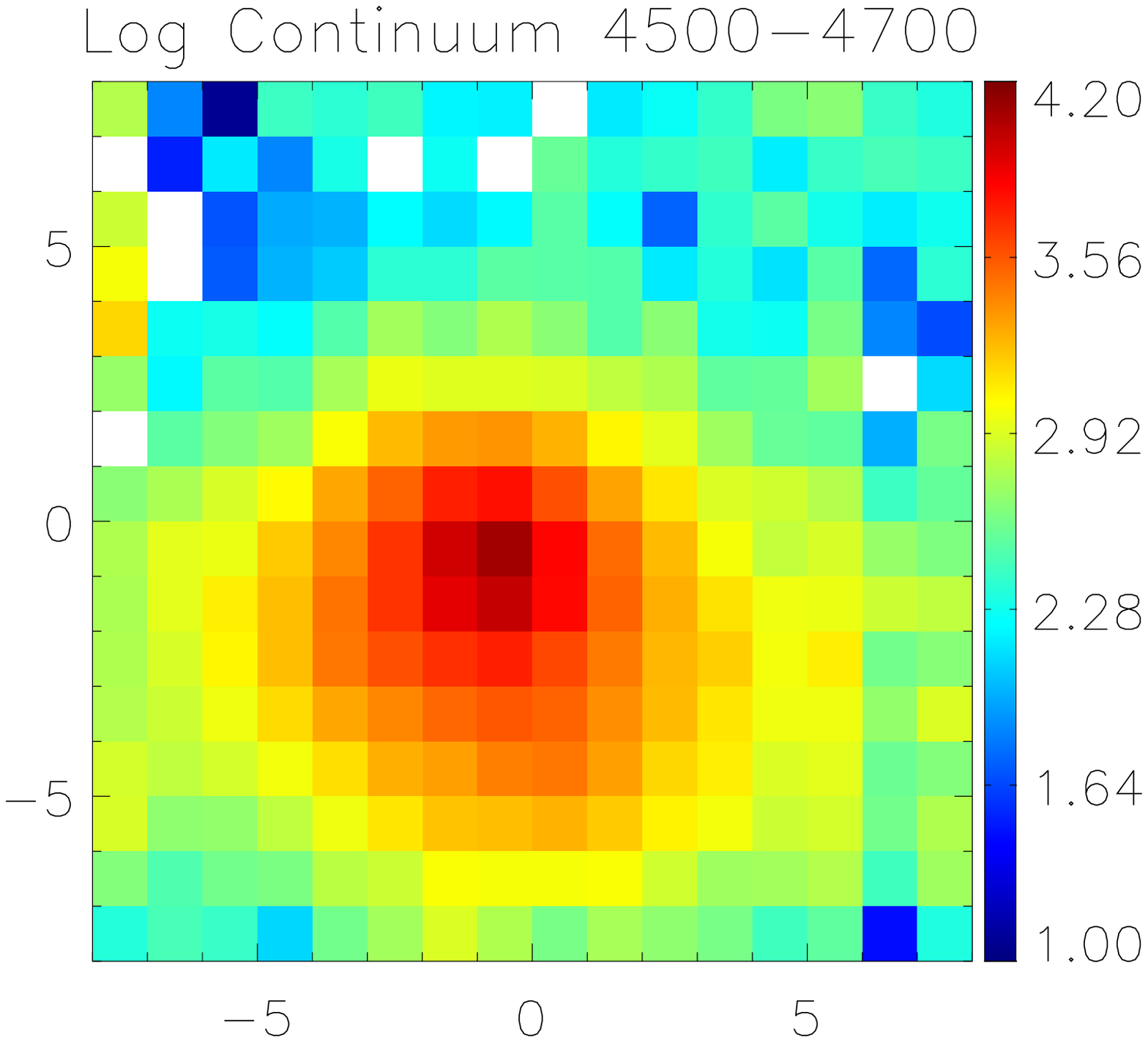}}
\hspace*{0.0cm}\subfigure{\includegraphics[width=0.24\textwidth]{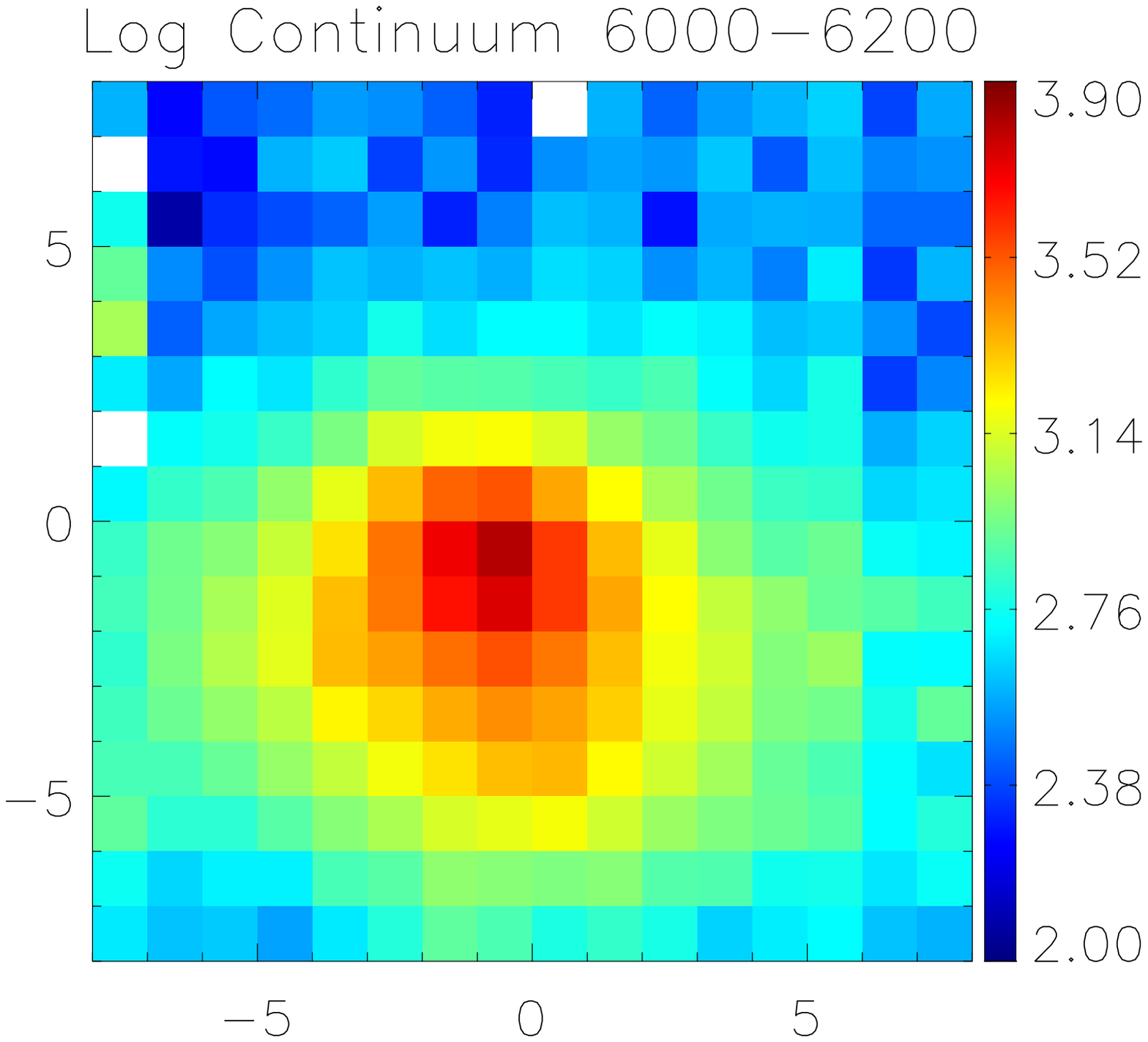}}
\hspace*{0.0cm}\subfigure{\includegraphics[width=0.24\textwidth]{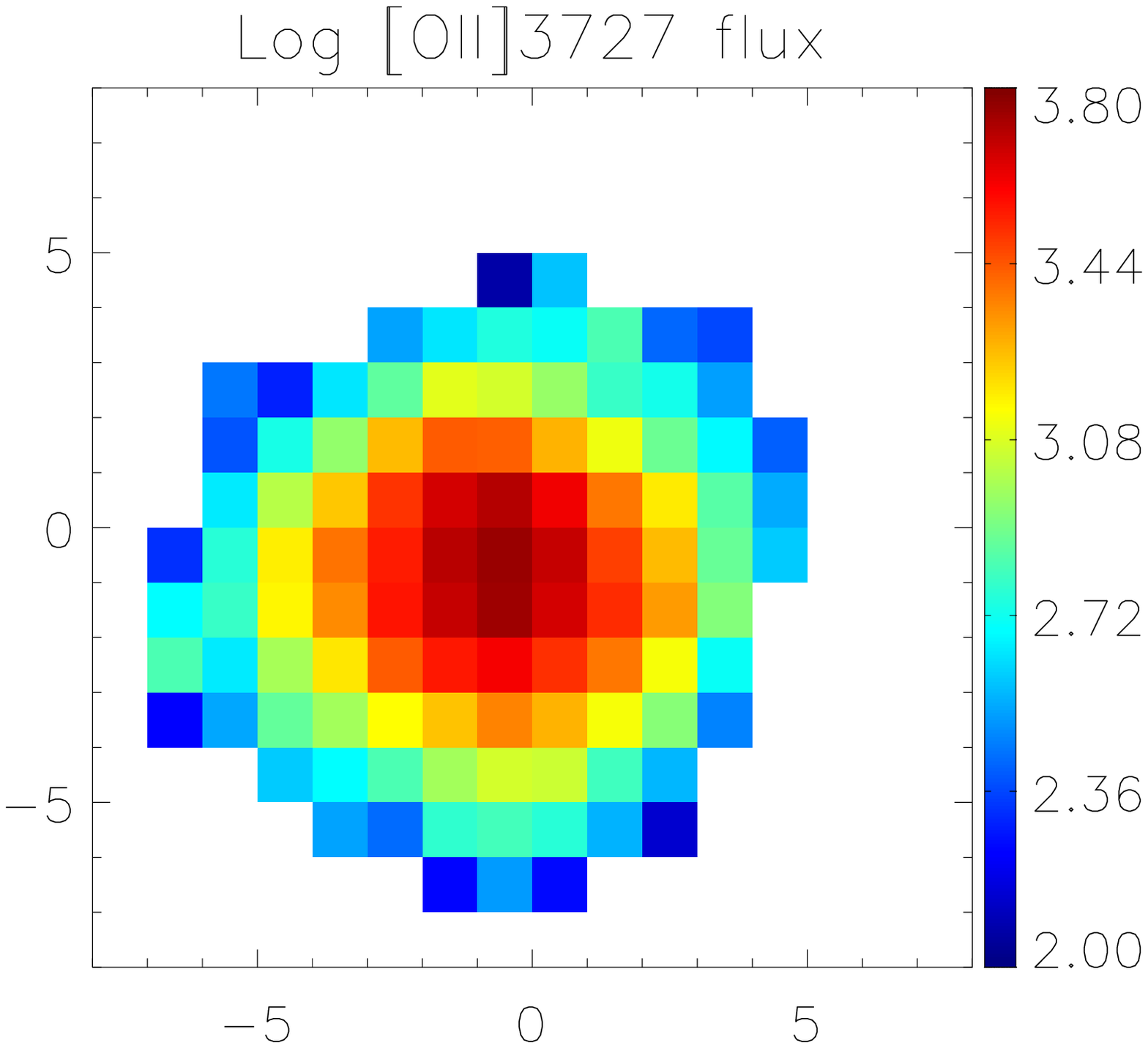}}
\hspace*{0.0cm}\subfigure{\includegraphics[width=0.24\textwidth]{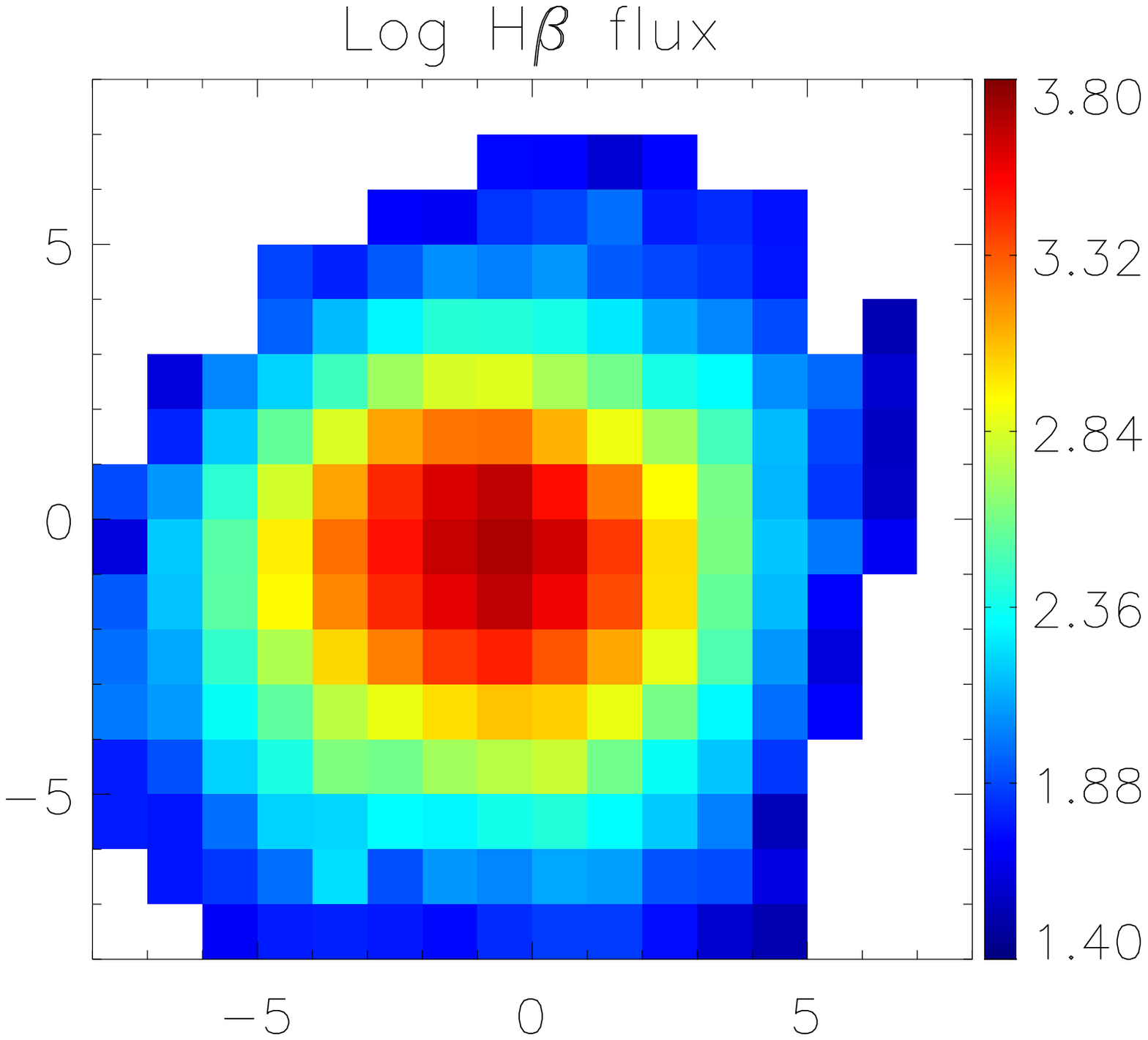}}
}}   
\mbox{
\centerline{
\hspace*{0.0cm}\subfigure{\includegraphics[width=0.24\textwidth]{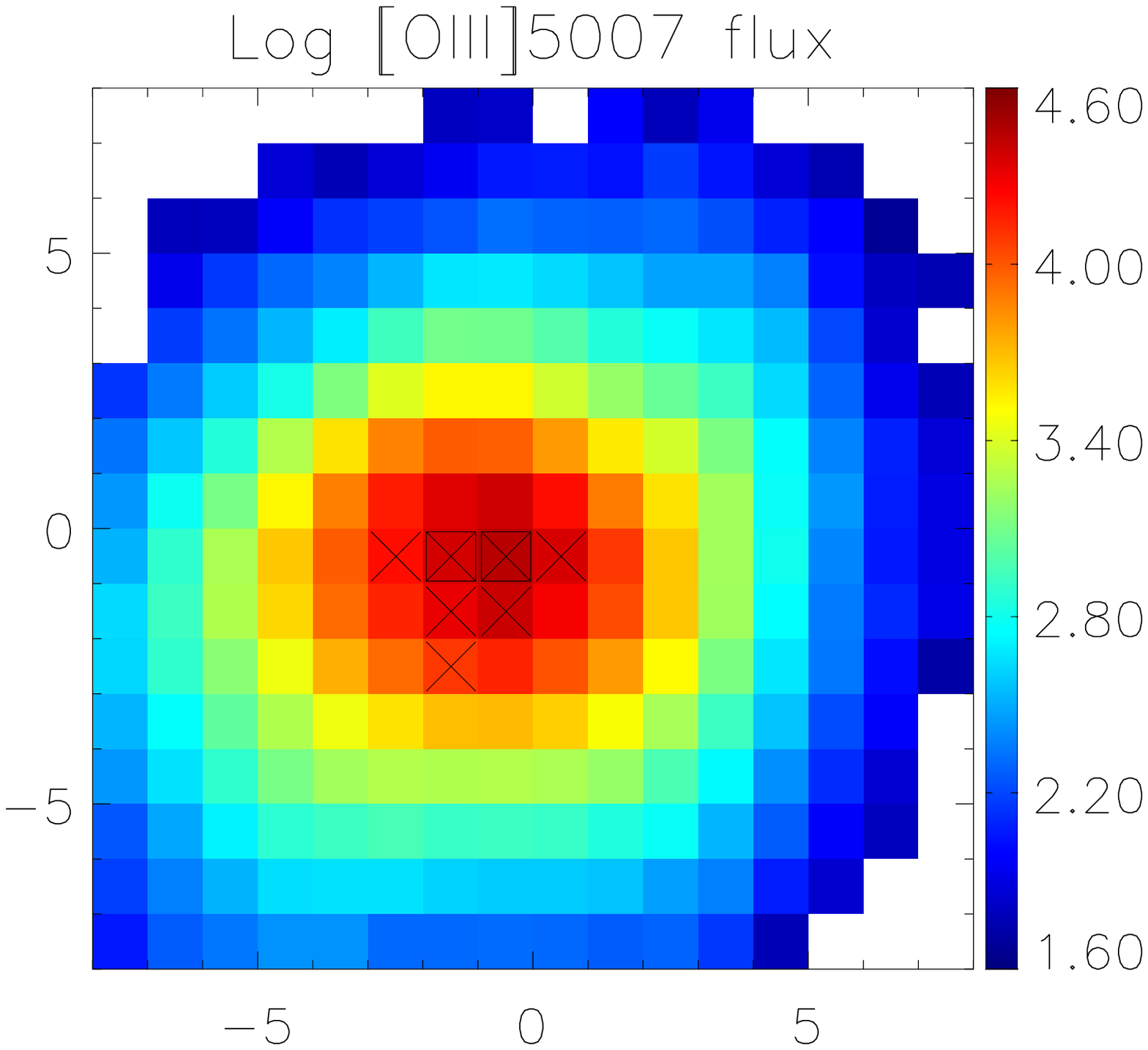}}
\hspace*{0.0cm}\subfigure{\includegraphics[width=0.24\textwidth]{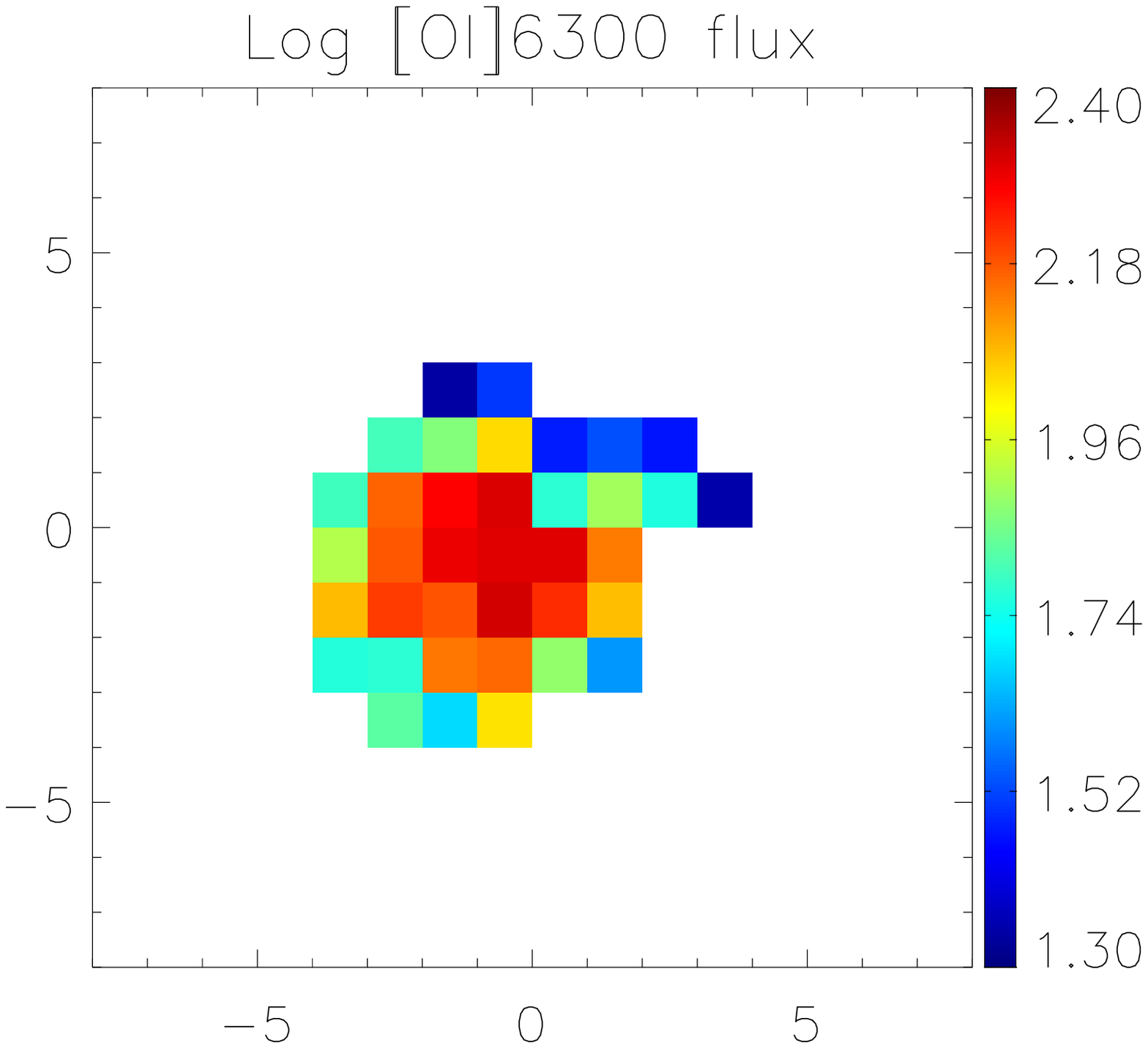}}
\hspace*{0.0cm}\subfigure{\includegraphics[width=0.24\textwidth]{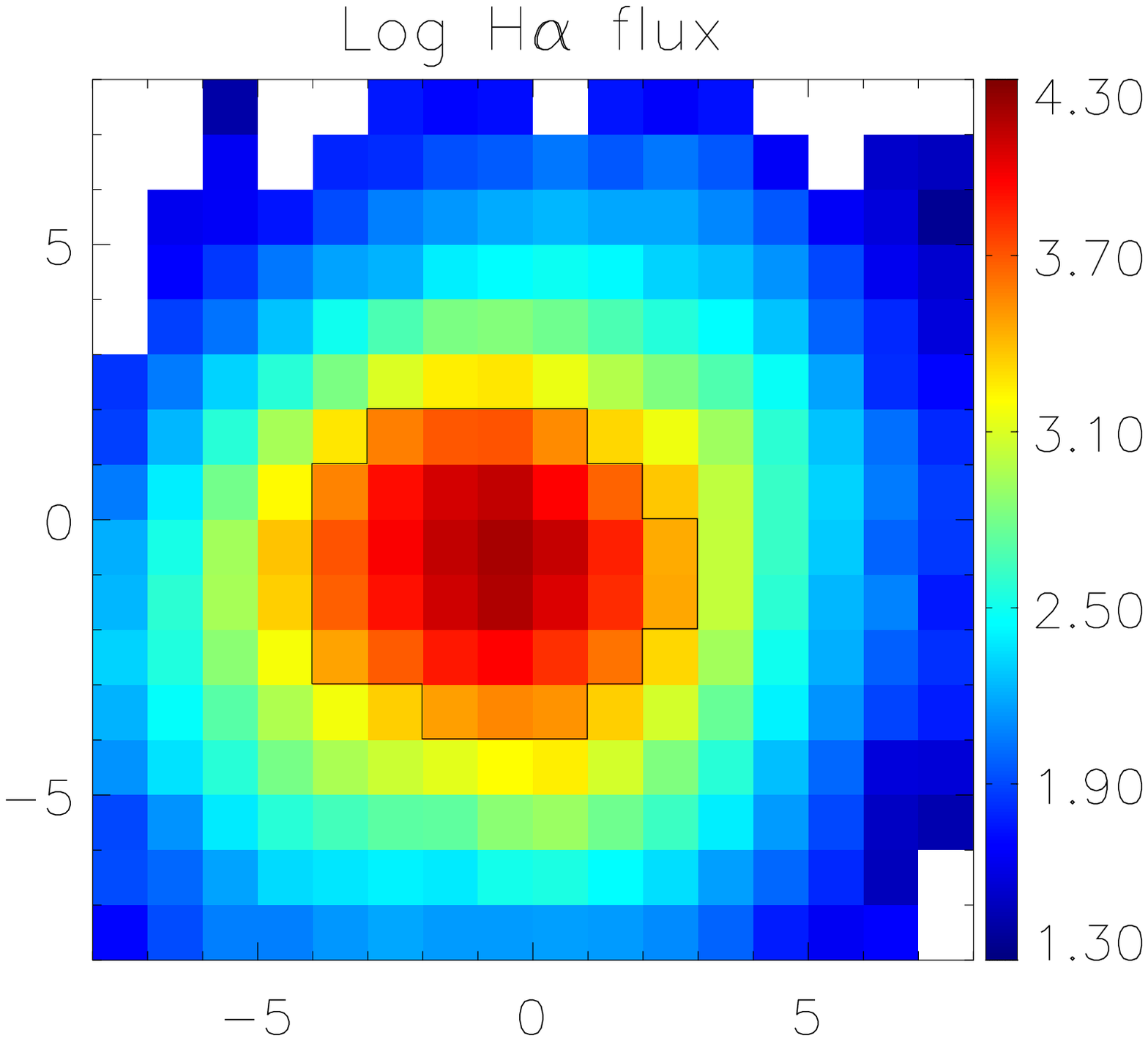}}
\hspace*{0.0cm}\subfigure{\includegraphics[width=0.24\textwidth]{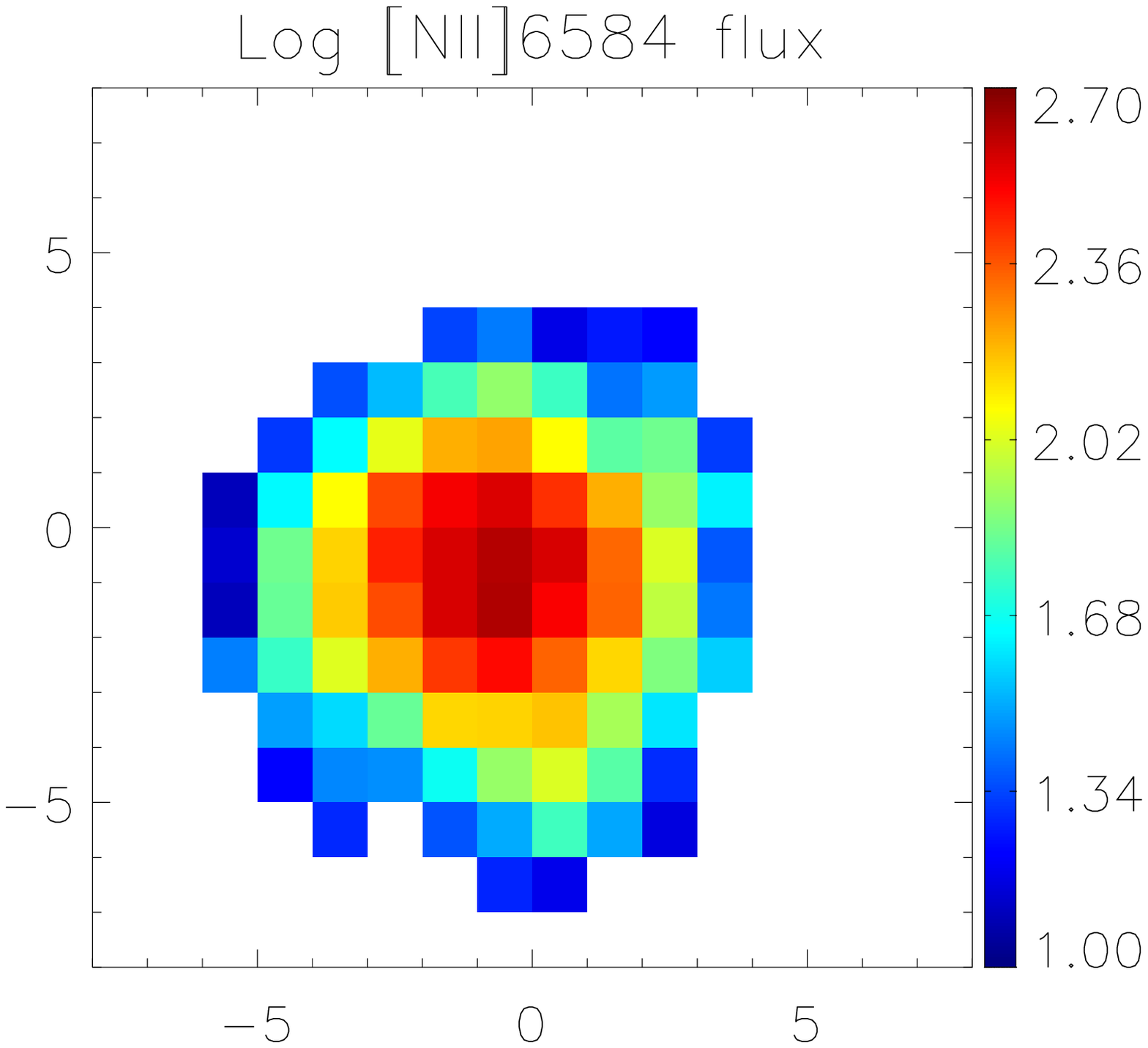}}
}} 
\mbox{
\centerline{
\hspace*{0.0cm}\subfigure{\includegraphics[width=0.24\textwidth]{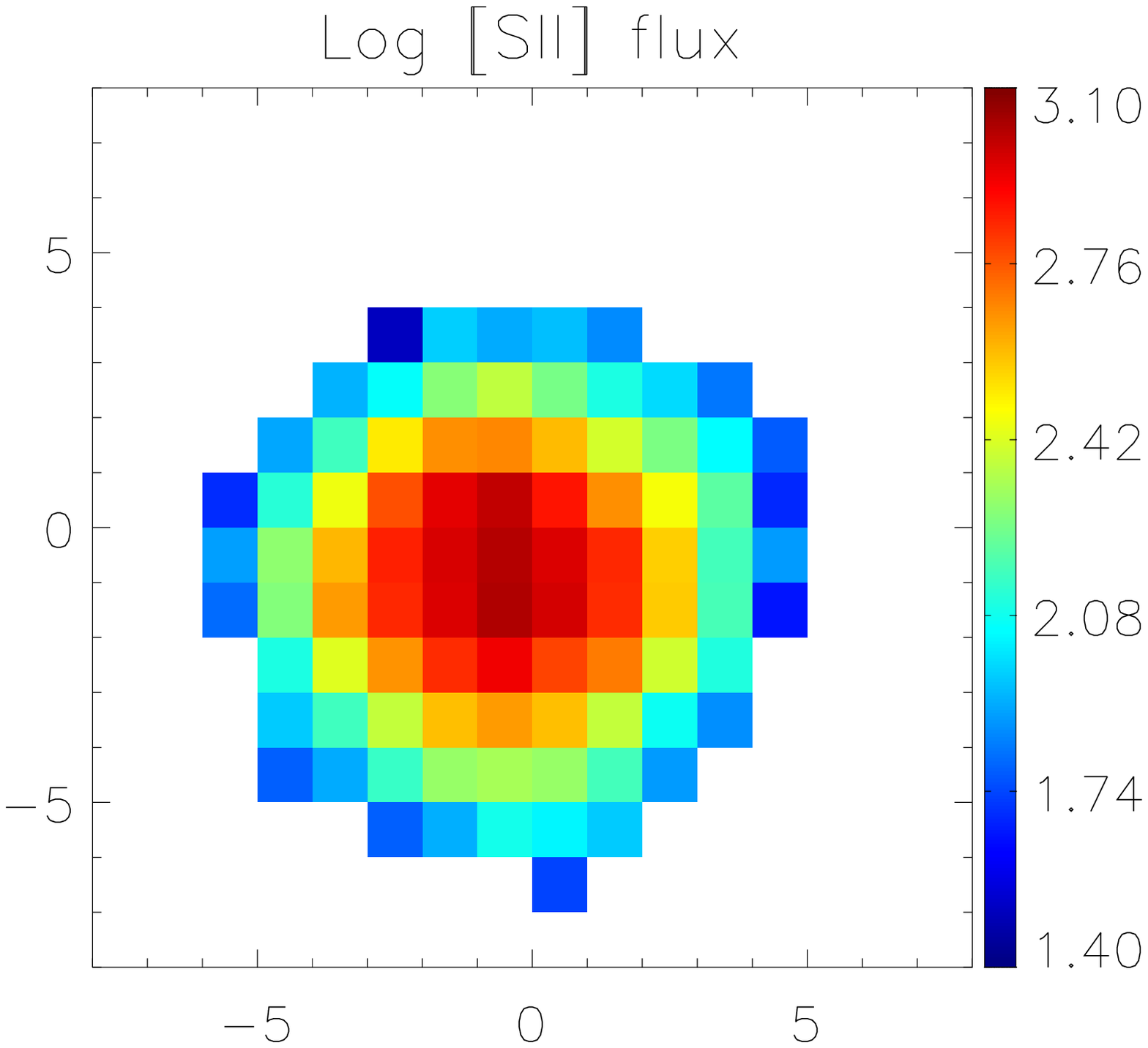}}
\hspace*{0.0cm}\subfigure{\includegraphics[width=0.24\textwidth]{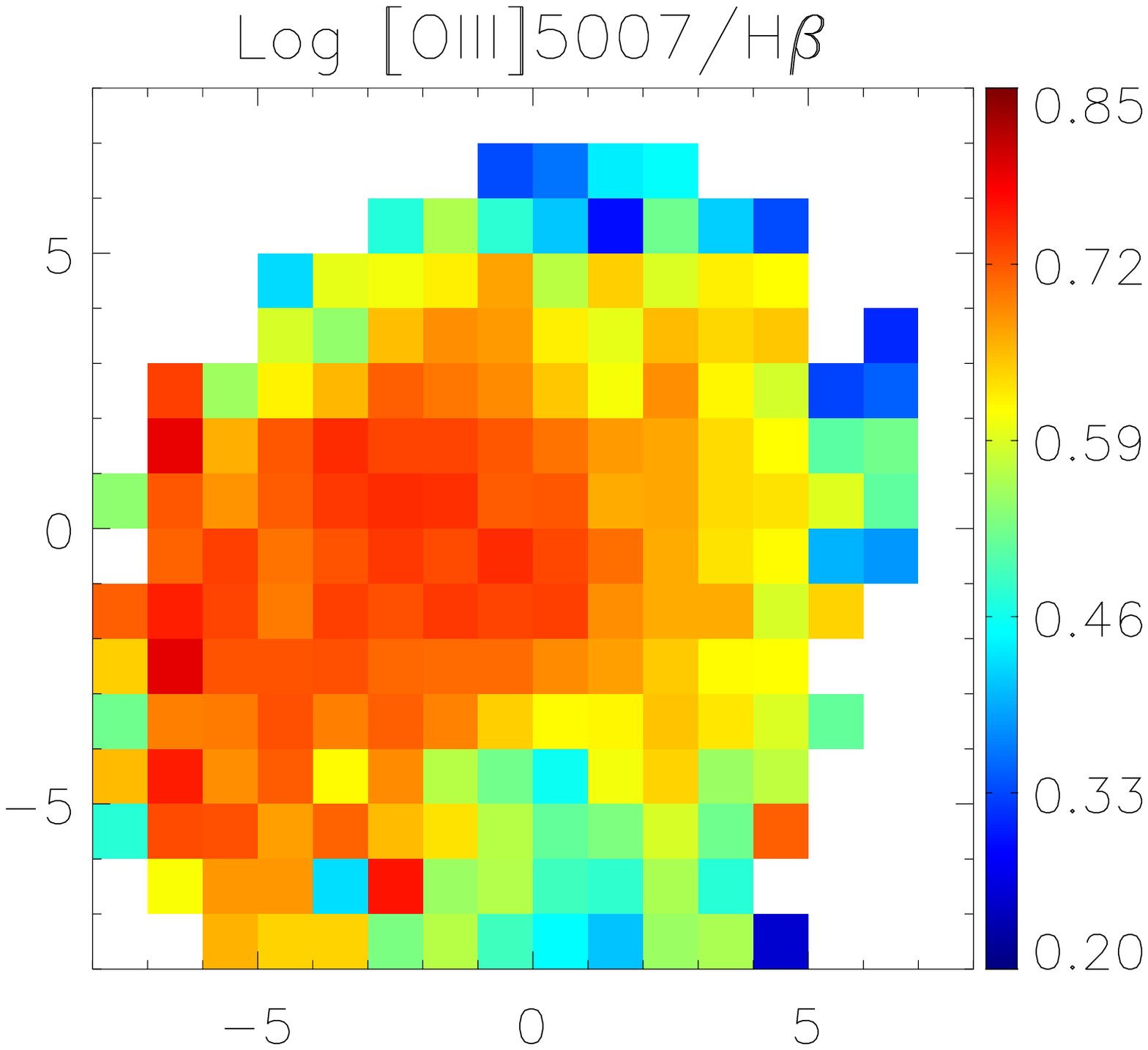}}
\hspace*{0.0cm}\subfigure{\includegraphics[width=0.24\textwidth]{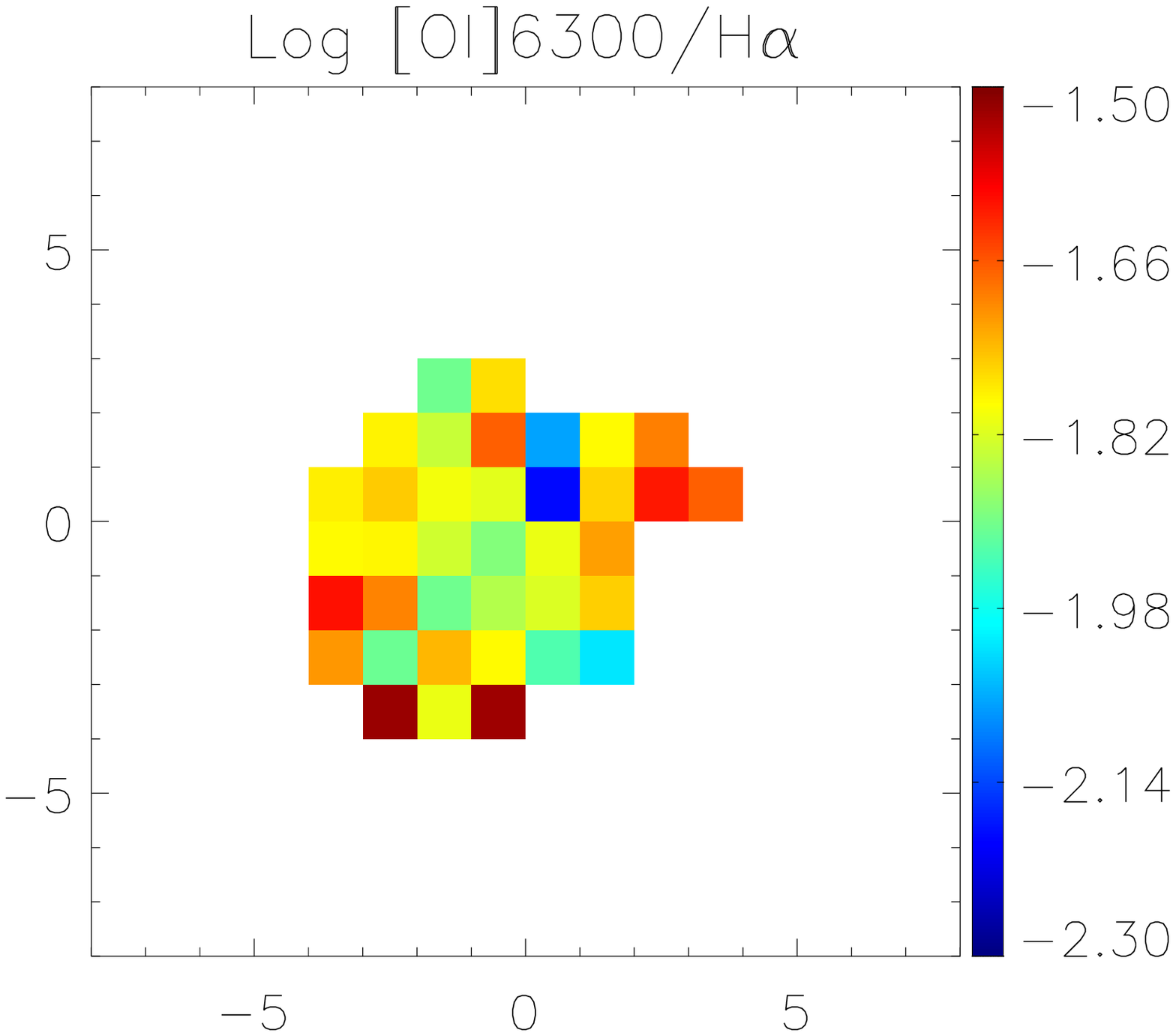}}
\hspace*{0.0cm}\subfigure{\includegraphics[width=0.24\textwidth]{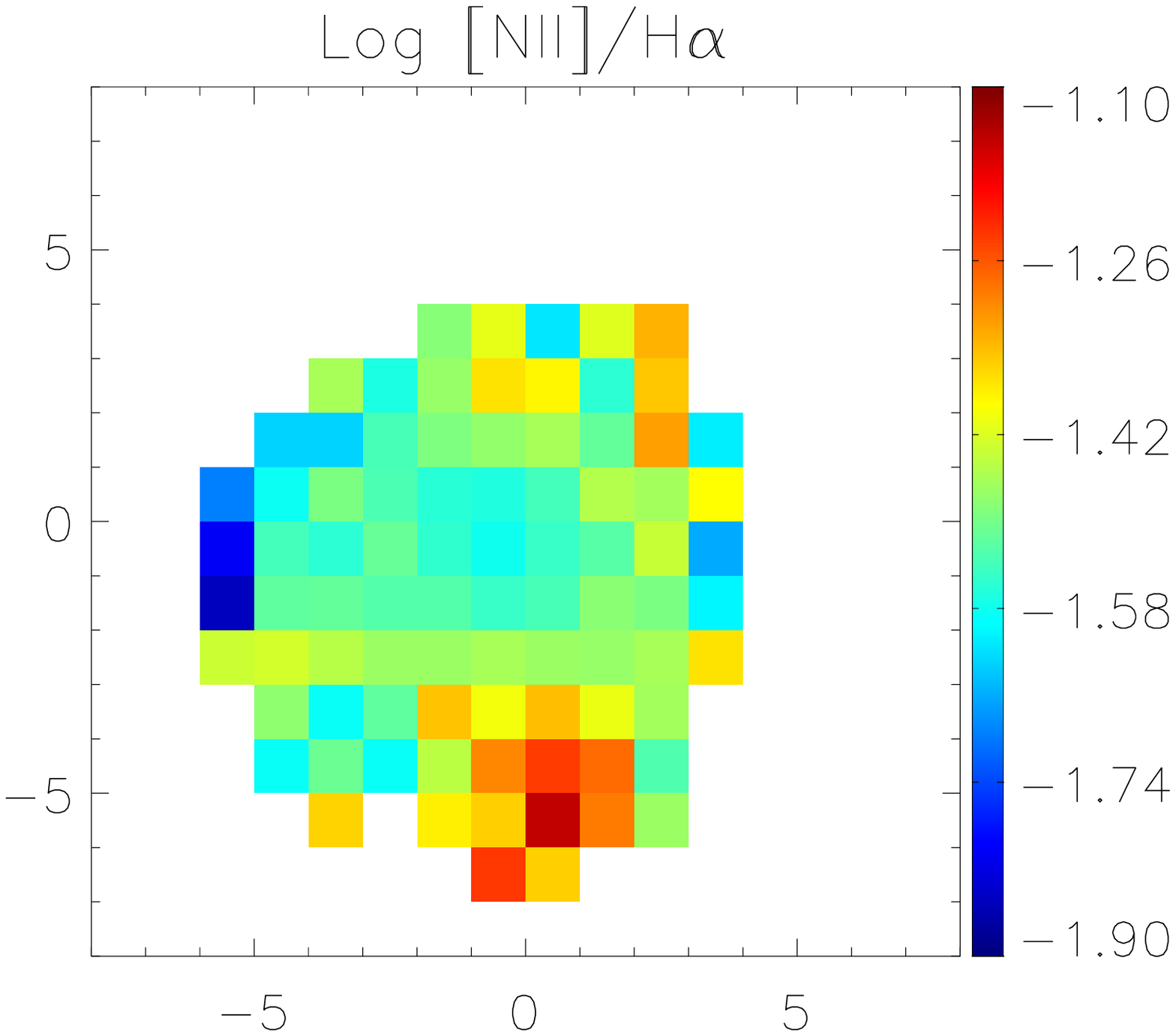}}
}}
\mbox{
\centerline{
\hspace*{0.0cm}\subfigure{\includegraphics[width=0.24\textwidth]{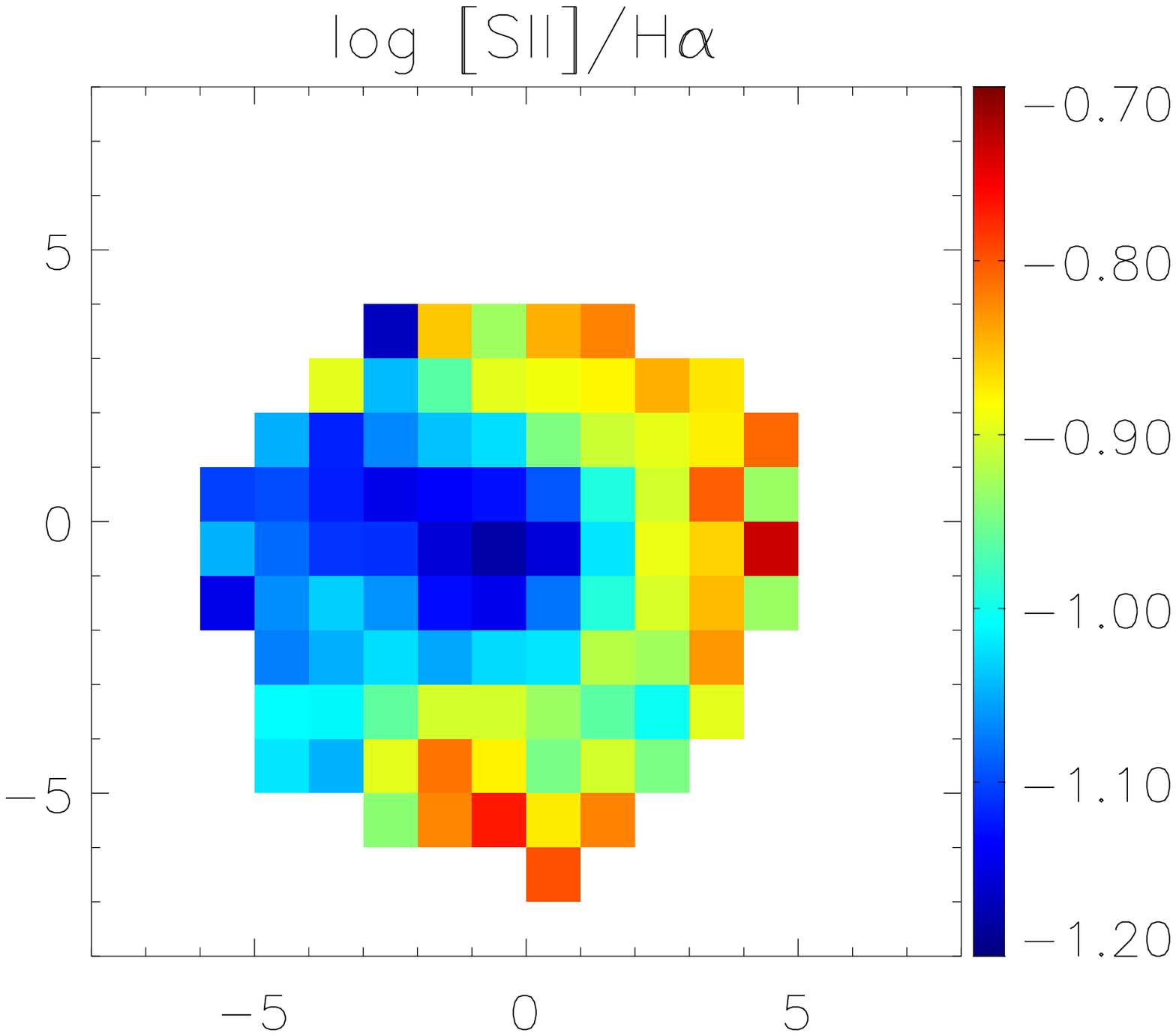}}
\hspace*{0.0cm}\subfigure{\includegraphics[width=0.24\textwidth]{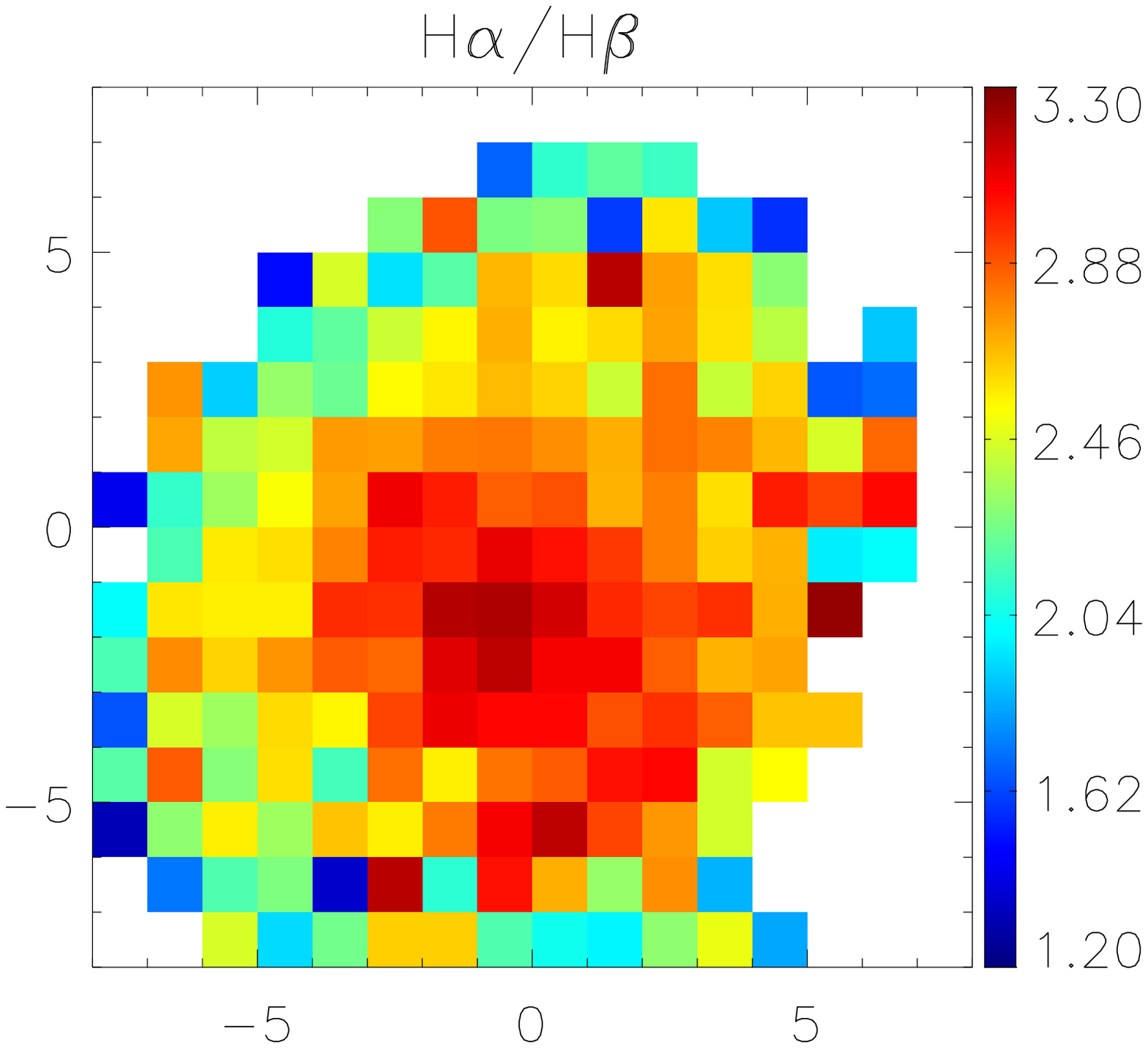}}
\hspace*{0.0cm}\subfigure{\includegraphics[width=0.24\textwidth]{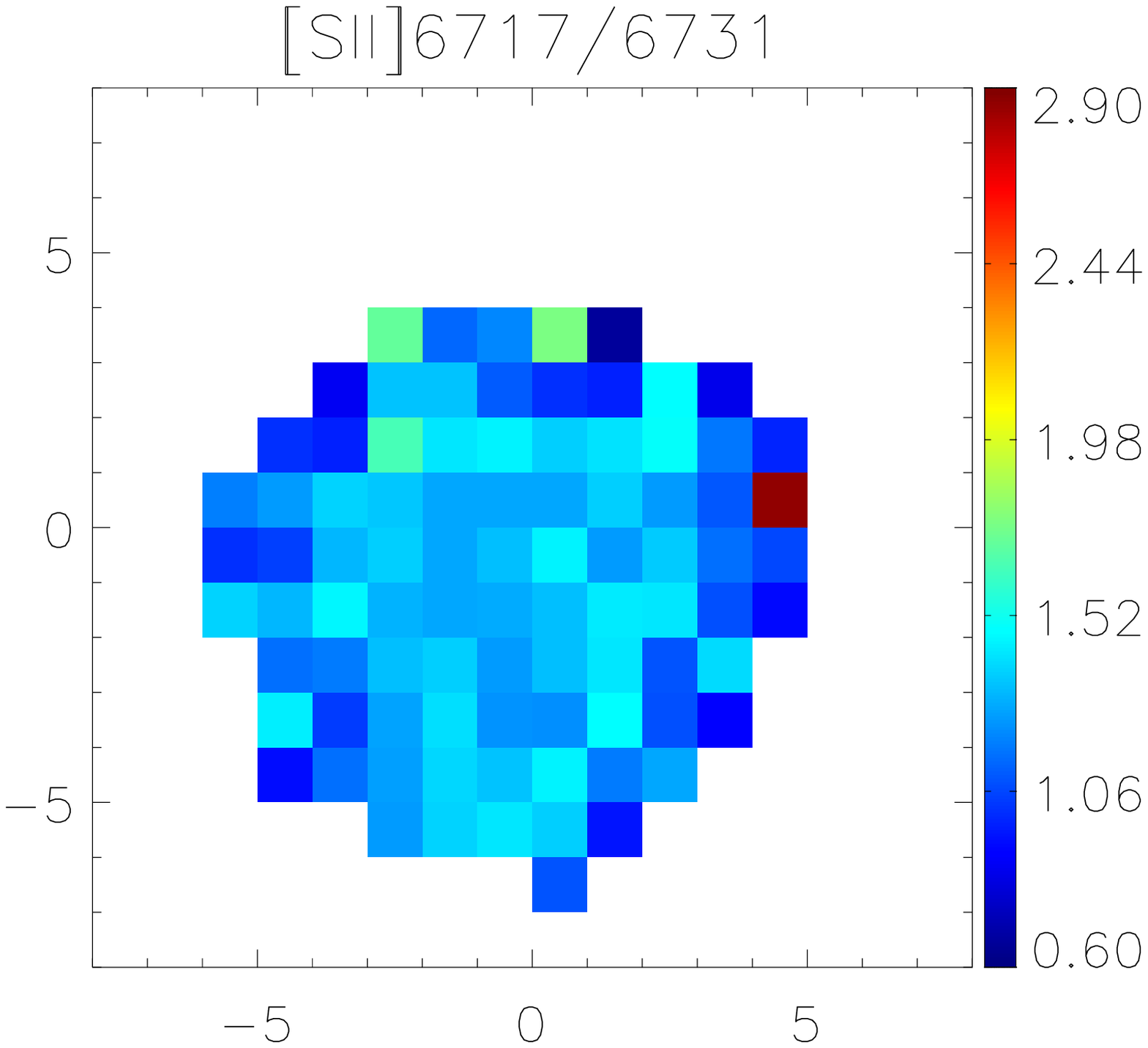}}
}} 
\mbox{
\centerline{
\hspace*{0.0cm}\subfigure{\includegraphics[width=0.24\textwidth]{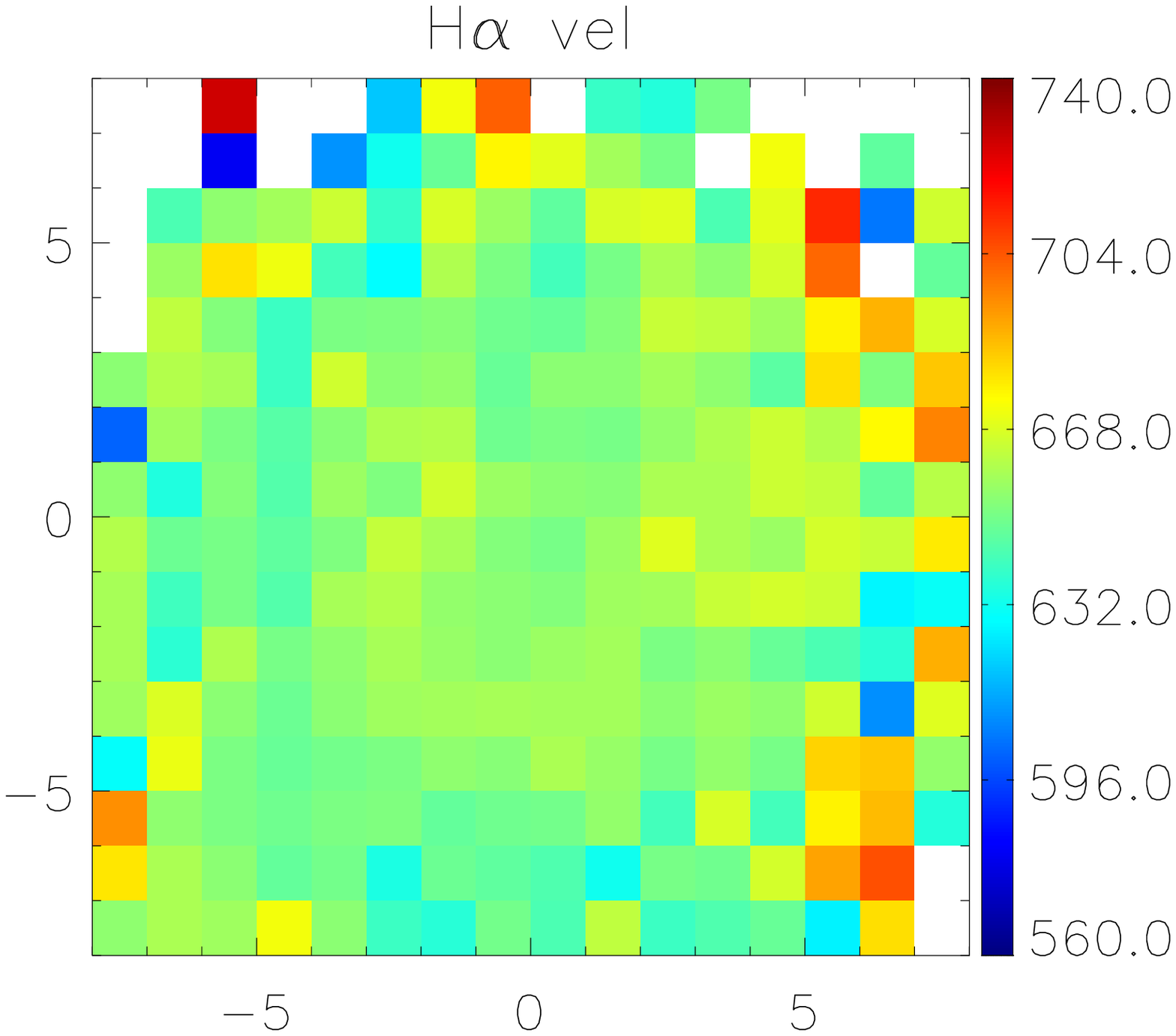}}
\hspace*{0.0cm}\subfigure{\includegraphics[width=0.24\textwidth]{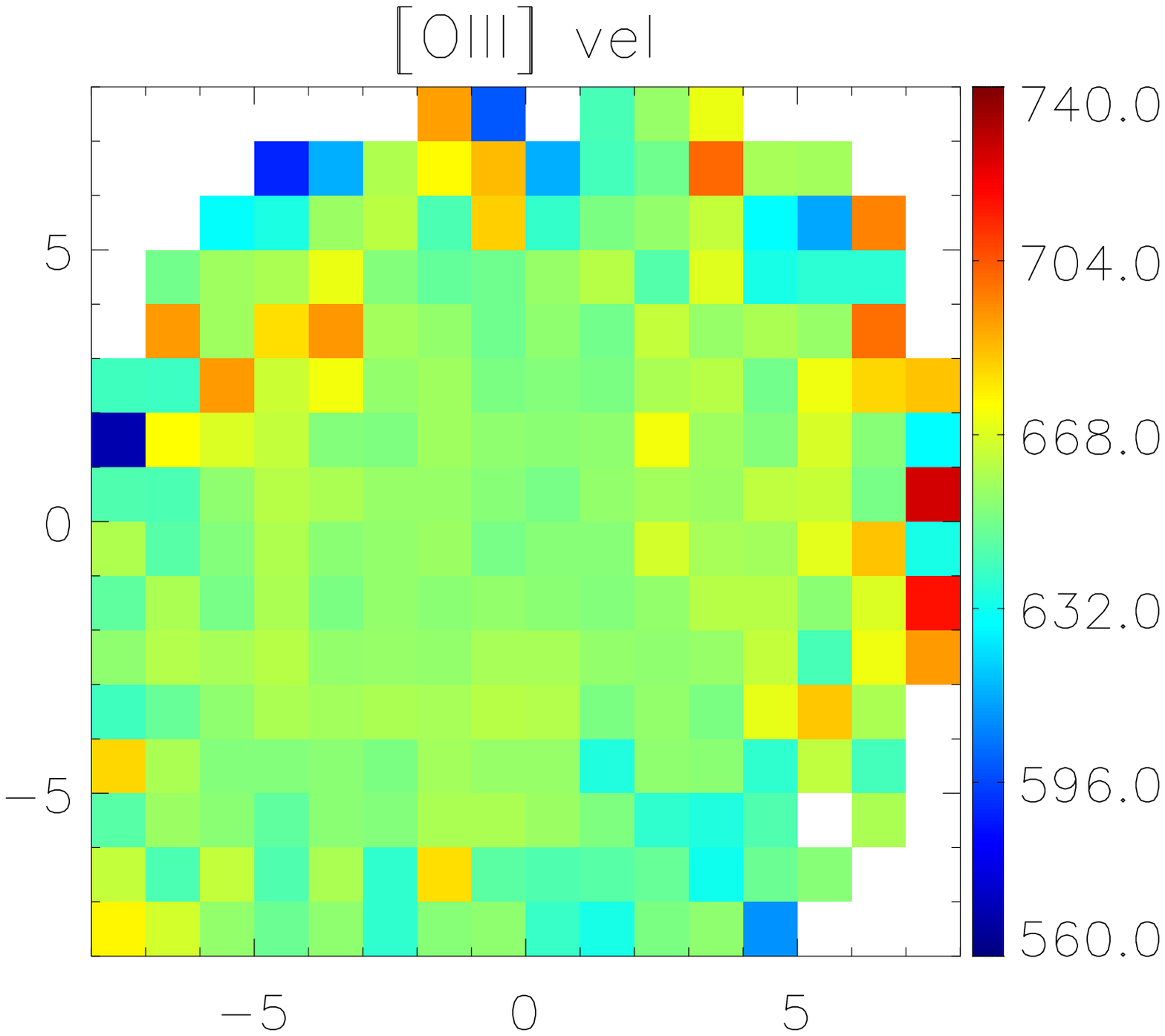}}
}}
 
\caption{Same as Fig.~\ref{Figure:mrk407} for Mrk~475. Maps of
[\ion{O}{i}]~$\lambda6300$ and of the [\ion{O}{i}]~$\lambda6300$/\Ha\ 
ionization ratio are also included. Spaxels in which the WR feature has been 
detected have been marked in the [\ion{O}{iii}]~$\lambda5007$ map with crosses 
and squares for the blue and red bumps respectively.}
\label{Figure:mrk475}
\end{figure*}

\begin{figure*}
\mbox{
\centerline{
\hspace*{0.0cm}\subfigure{\includegraphics[width=0.24\textwidth]{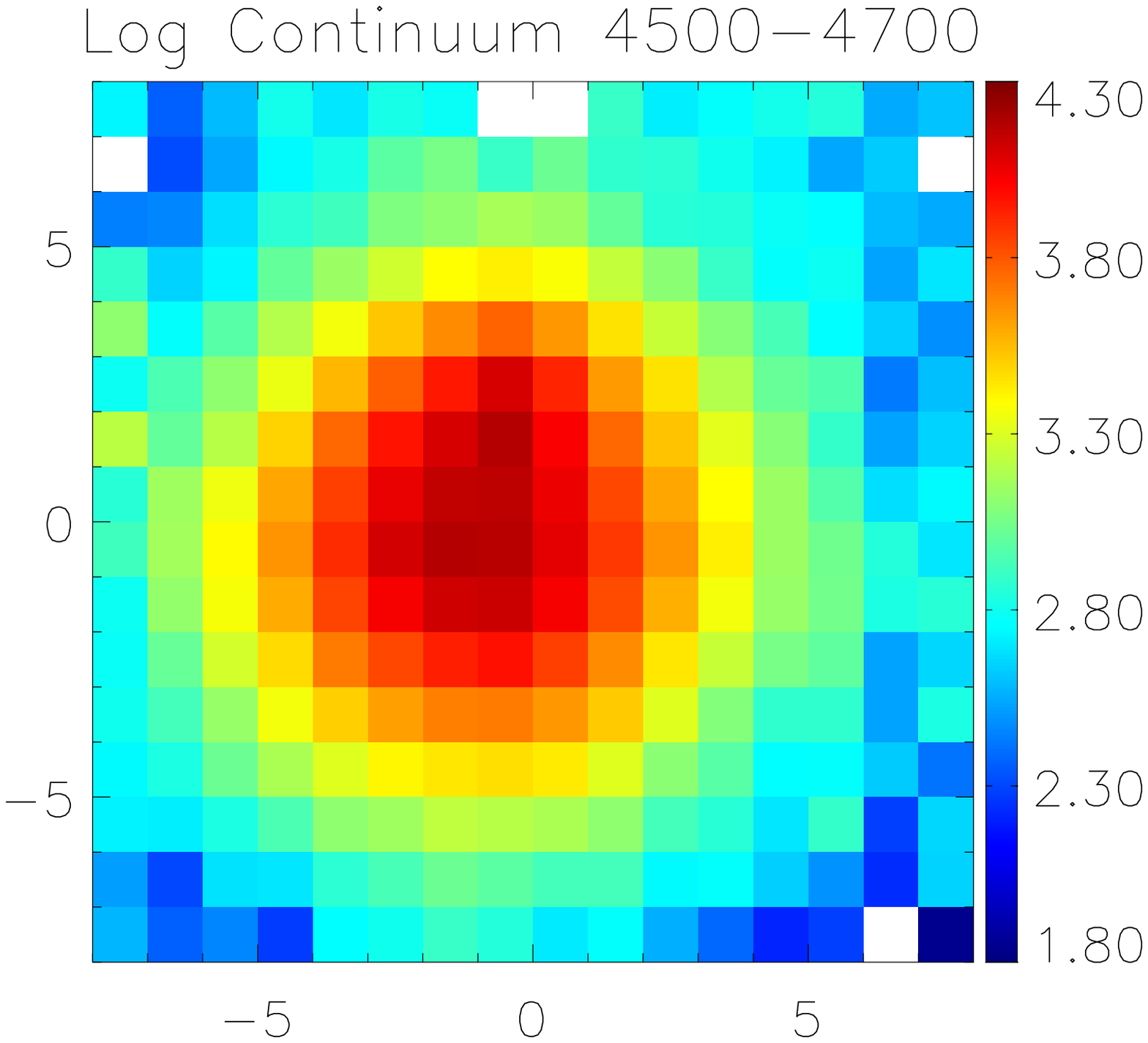}}
\hspace*{0.0cm}\subfigure{\includegraphics[width=0.24\textwidth]{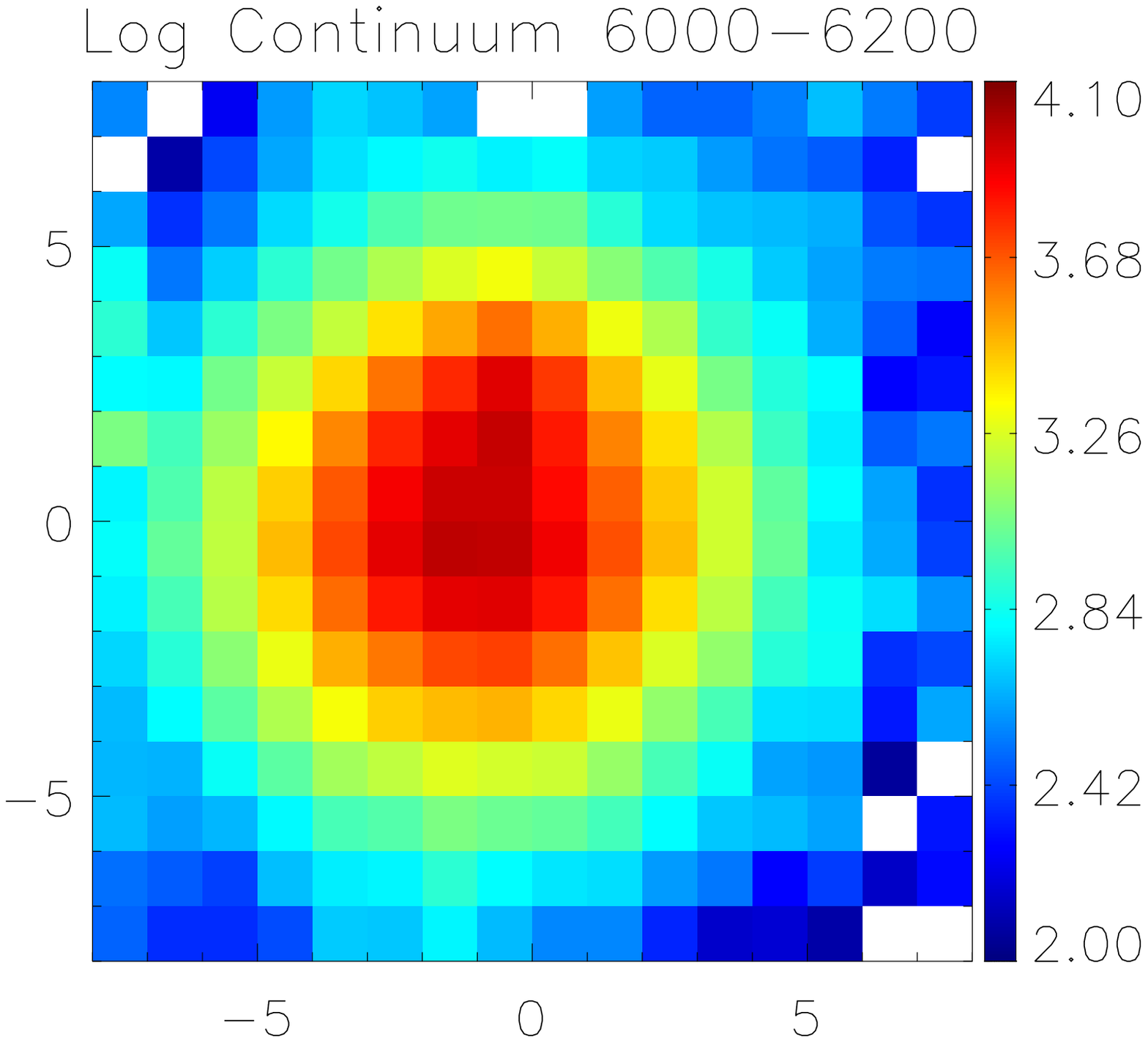}}
\hspace*{0.0cm}\subfigure{\includegraphics[width=0.24\textwidth]{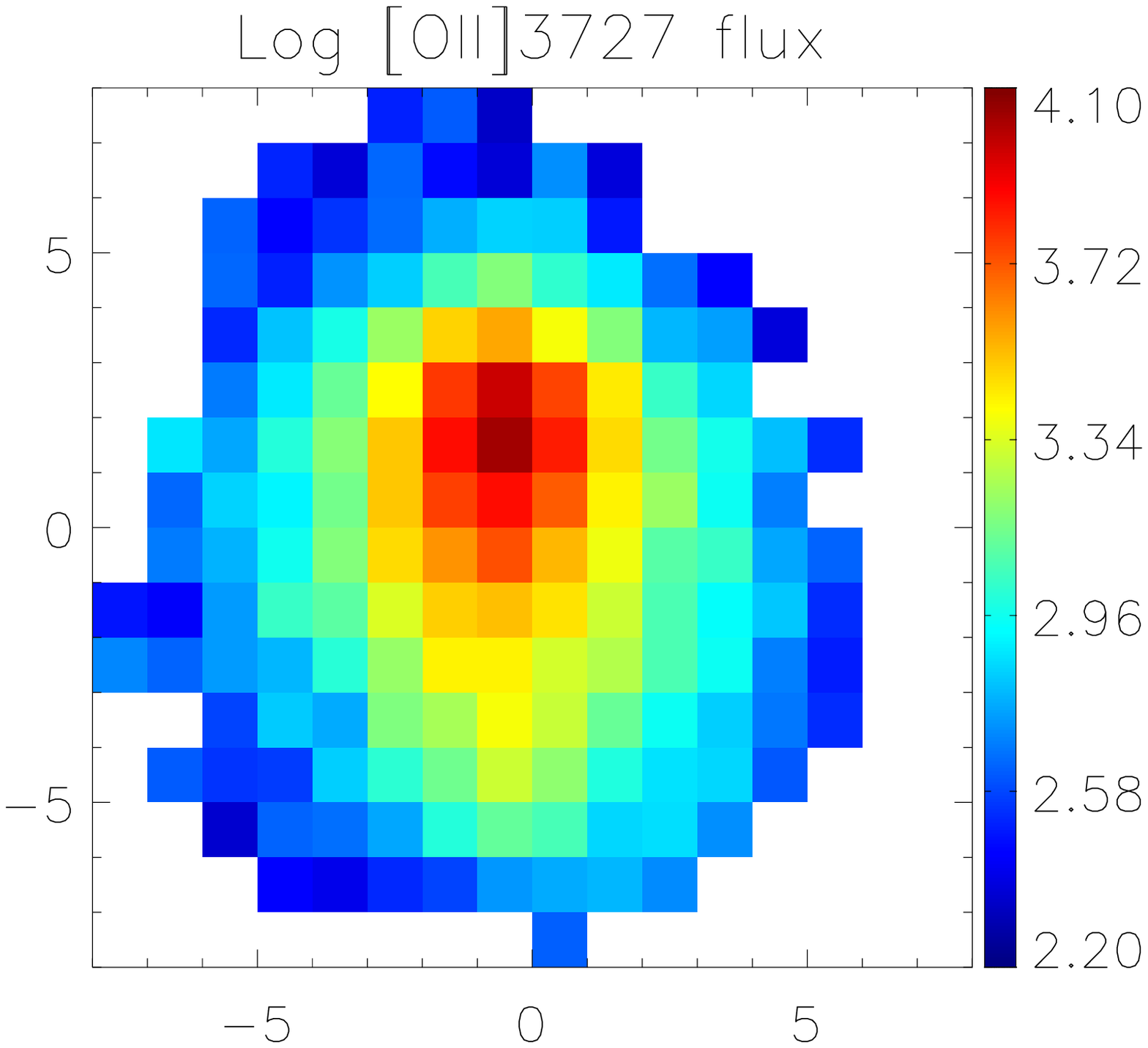}}
\hspace*{0.0cm}\subfigure{\includegraphics[width=0.24\textwidth]{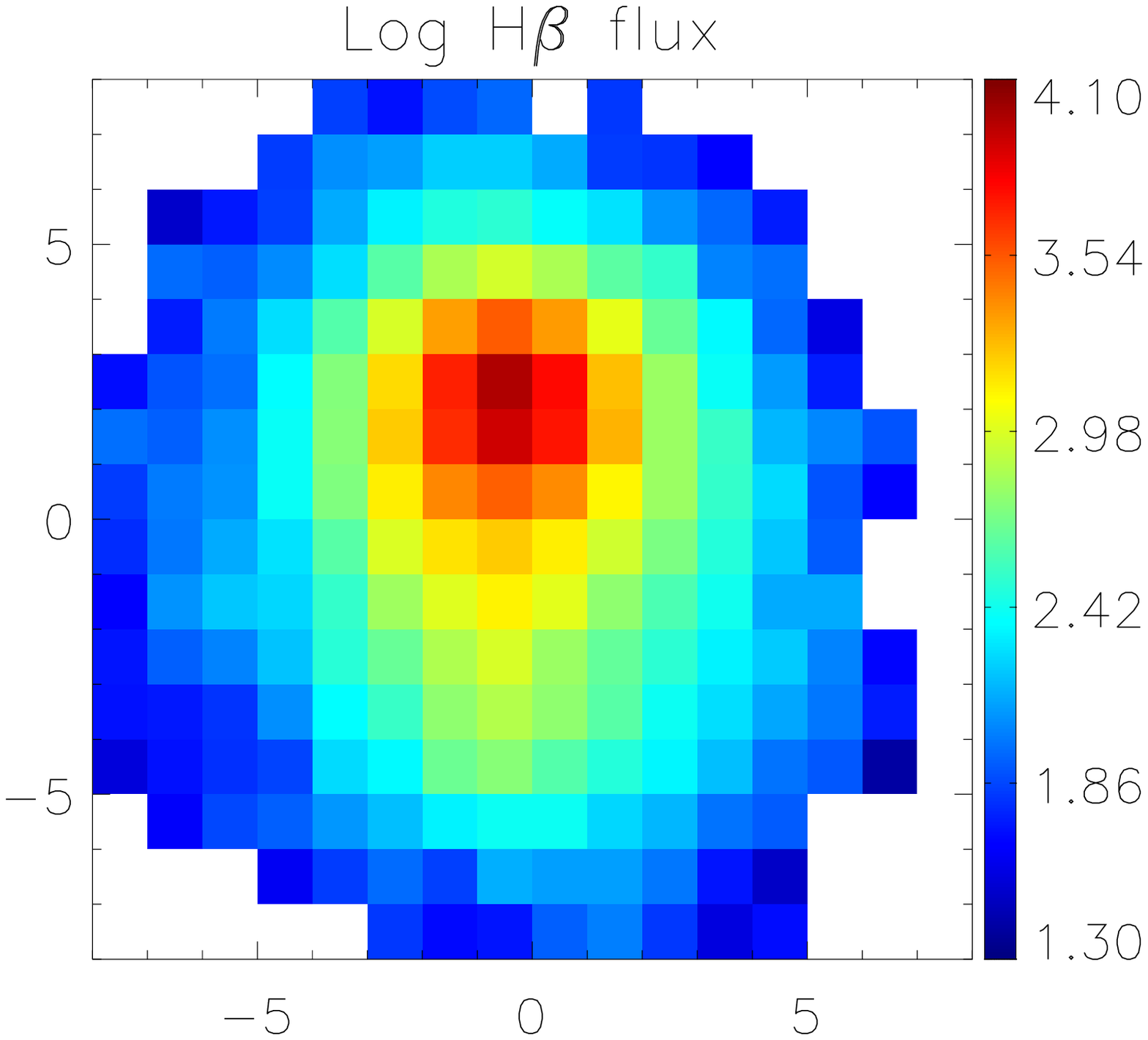}}
}}   
\mbox{
\centerline{
\hspace*{0.0cm}\subfigure{\includegraphics[width=0.24\textwidth]{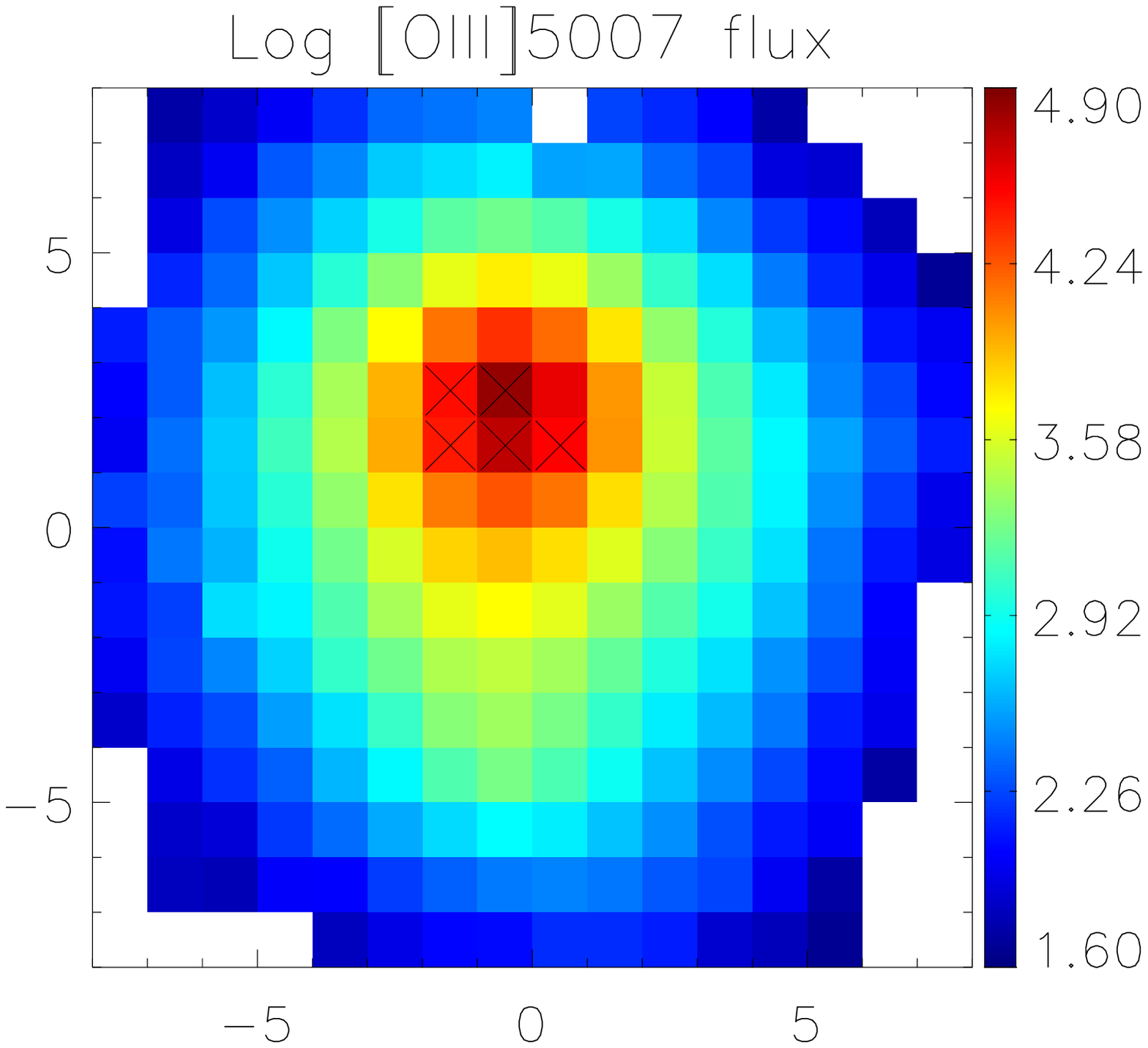}}
\hspace*{0.0cm}\subfigure{\includegraphics[width=0.24\textwidth]{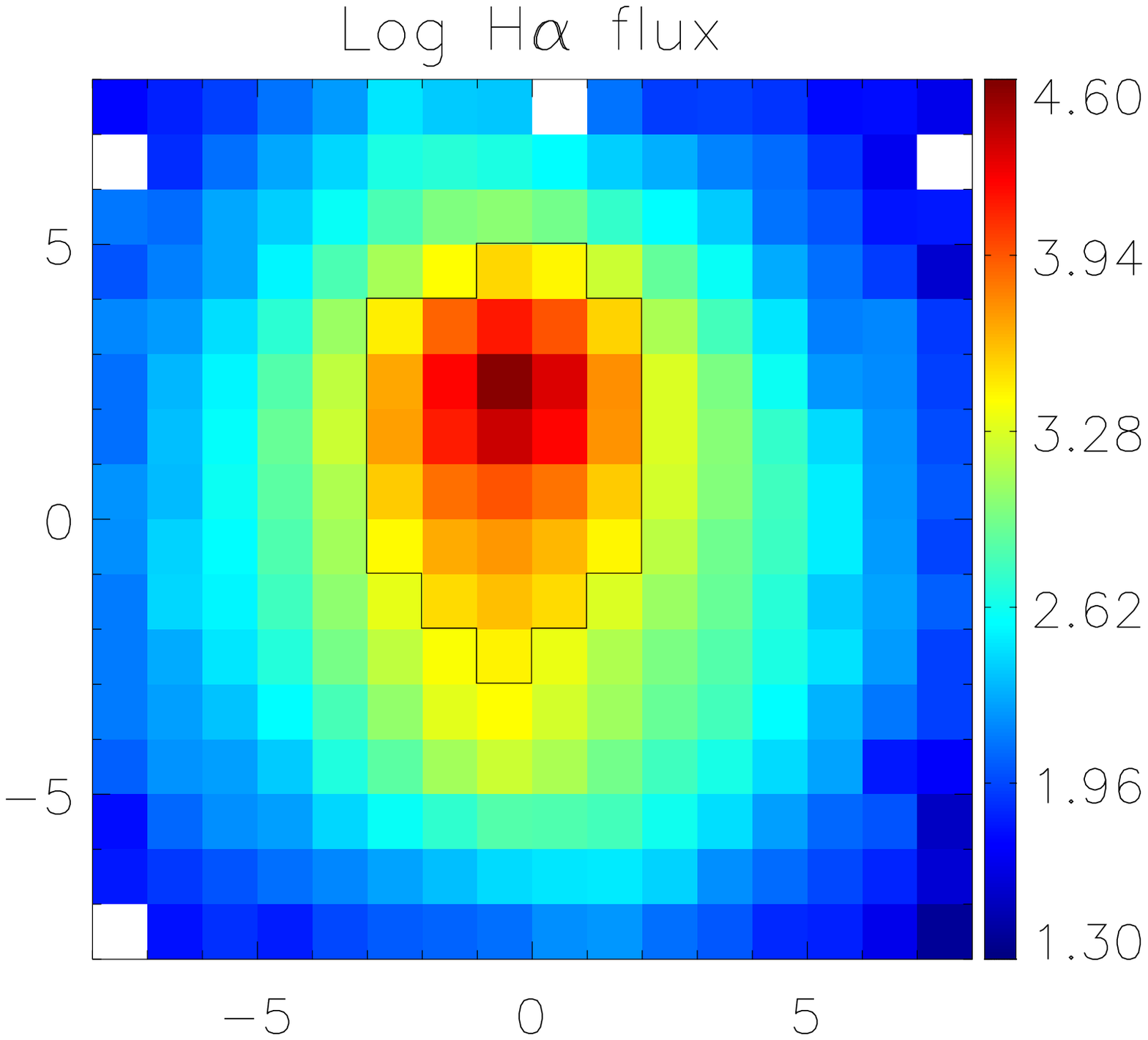}}
\hspace*{0.0cm}\subfigure{\includegraphics[width=0.24\textwidth]{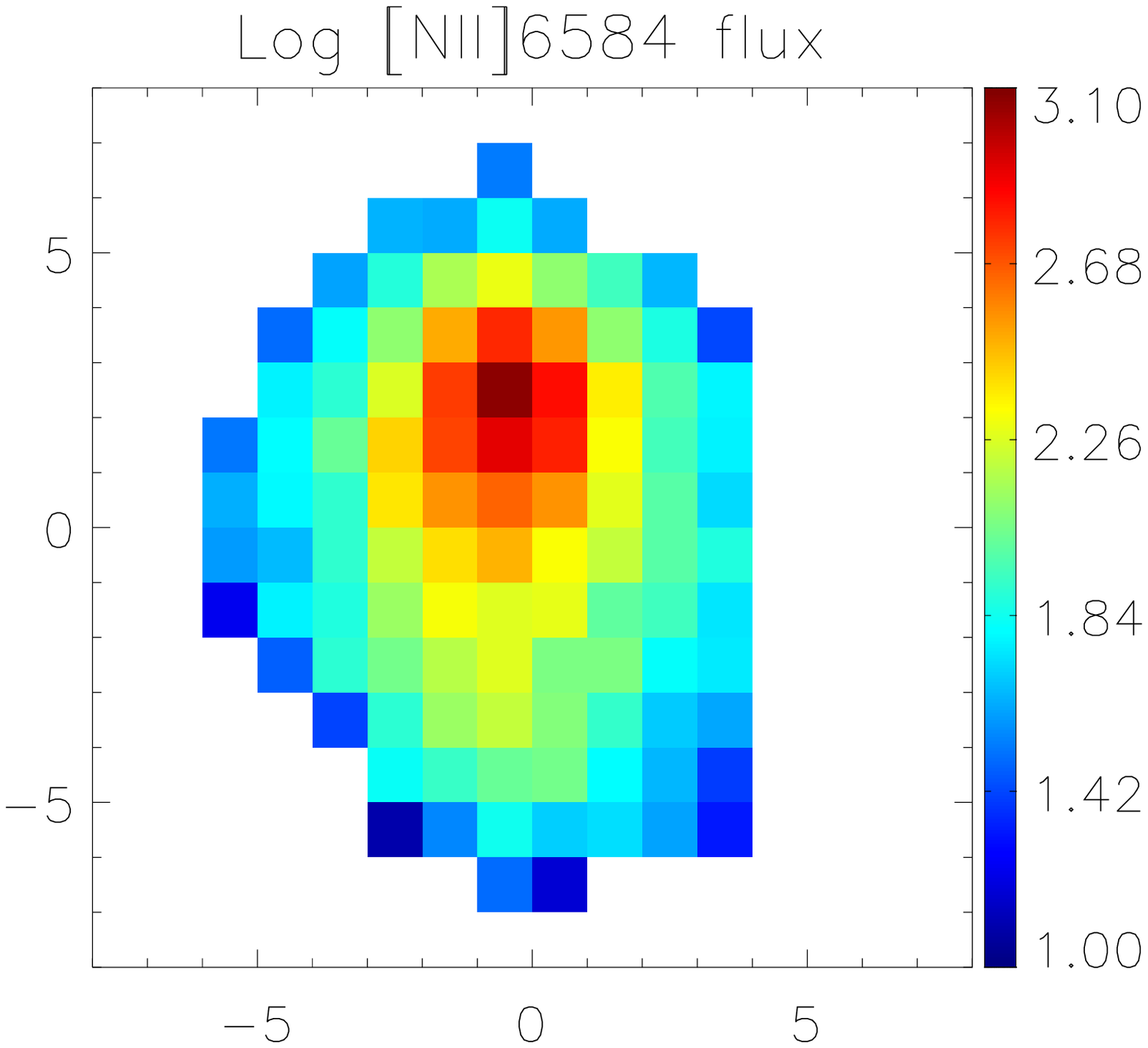}}
\hspace*{0.0cm}\subfigure{\includegraphics[width=0.24\textwidth]{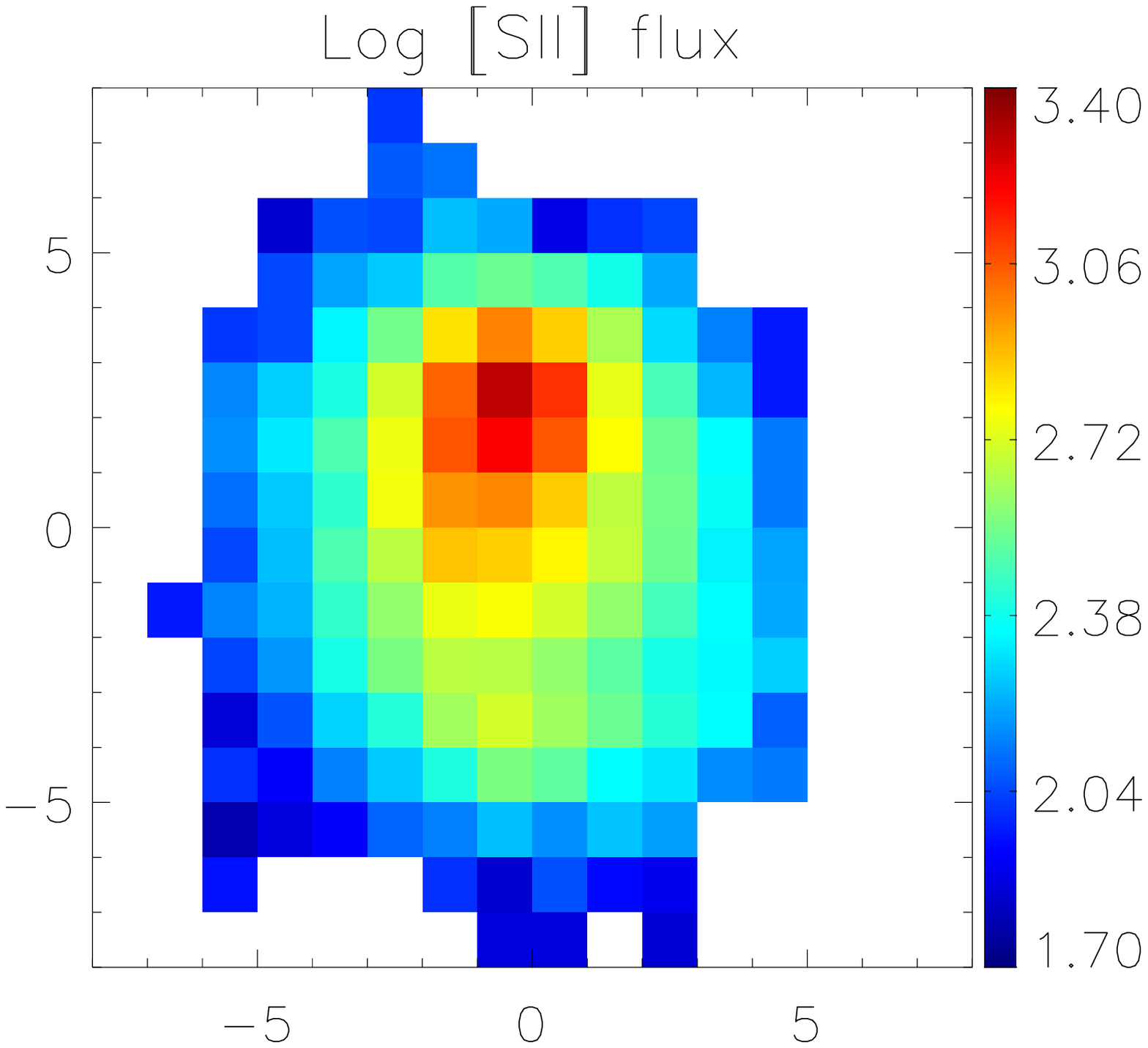}}
}} 
\mbox{
\centerline{
\hspace*{0.0cm}\subfigure{\includegraphics[width=0.24\textwidth]{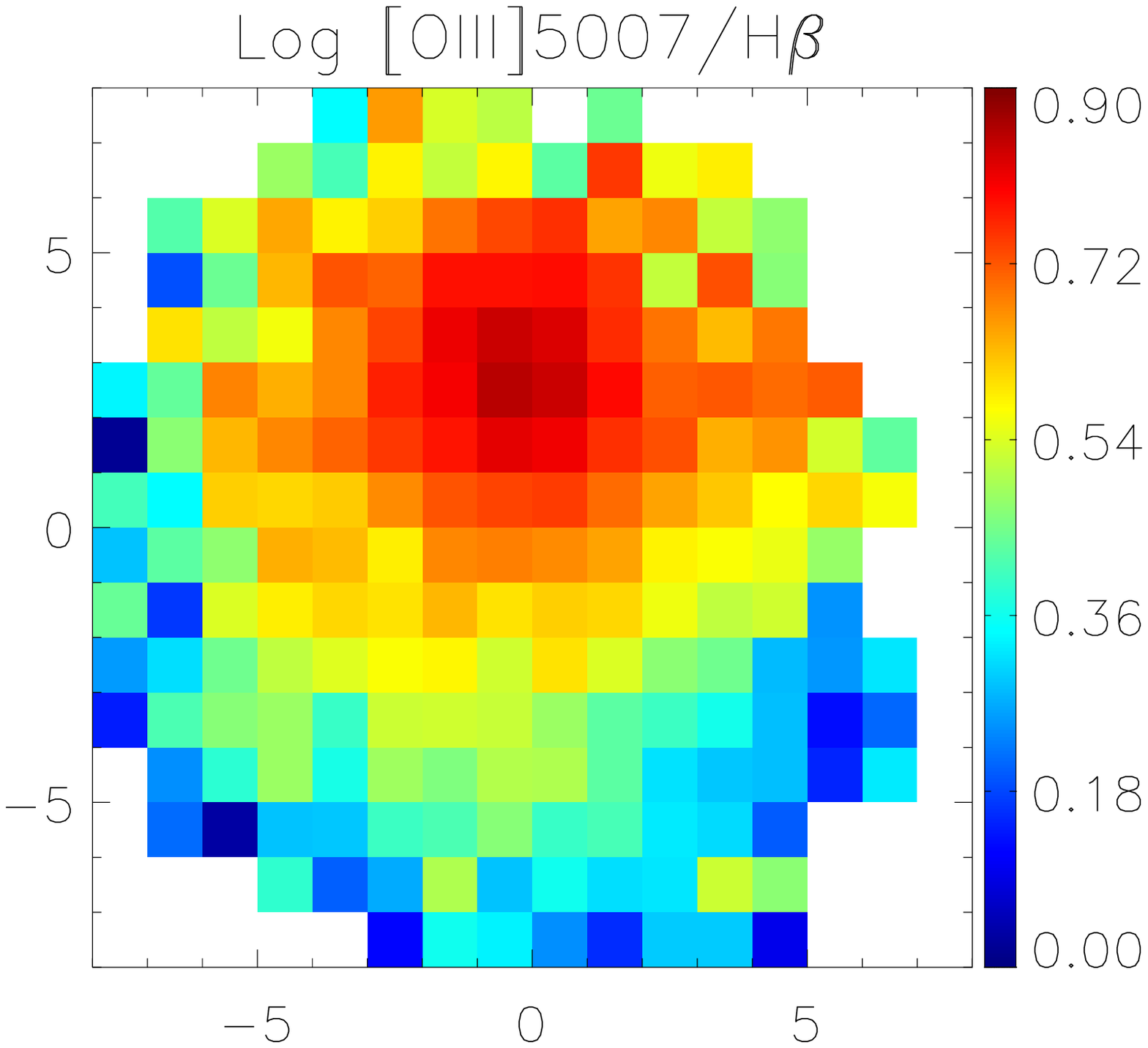}}
\hspace*{0.0cm}\subfigure{\includegraphics[width=0.24\textwidth]{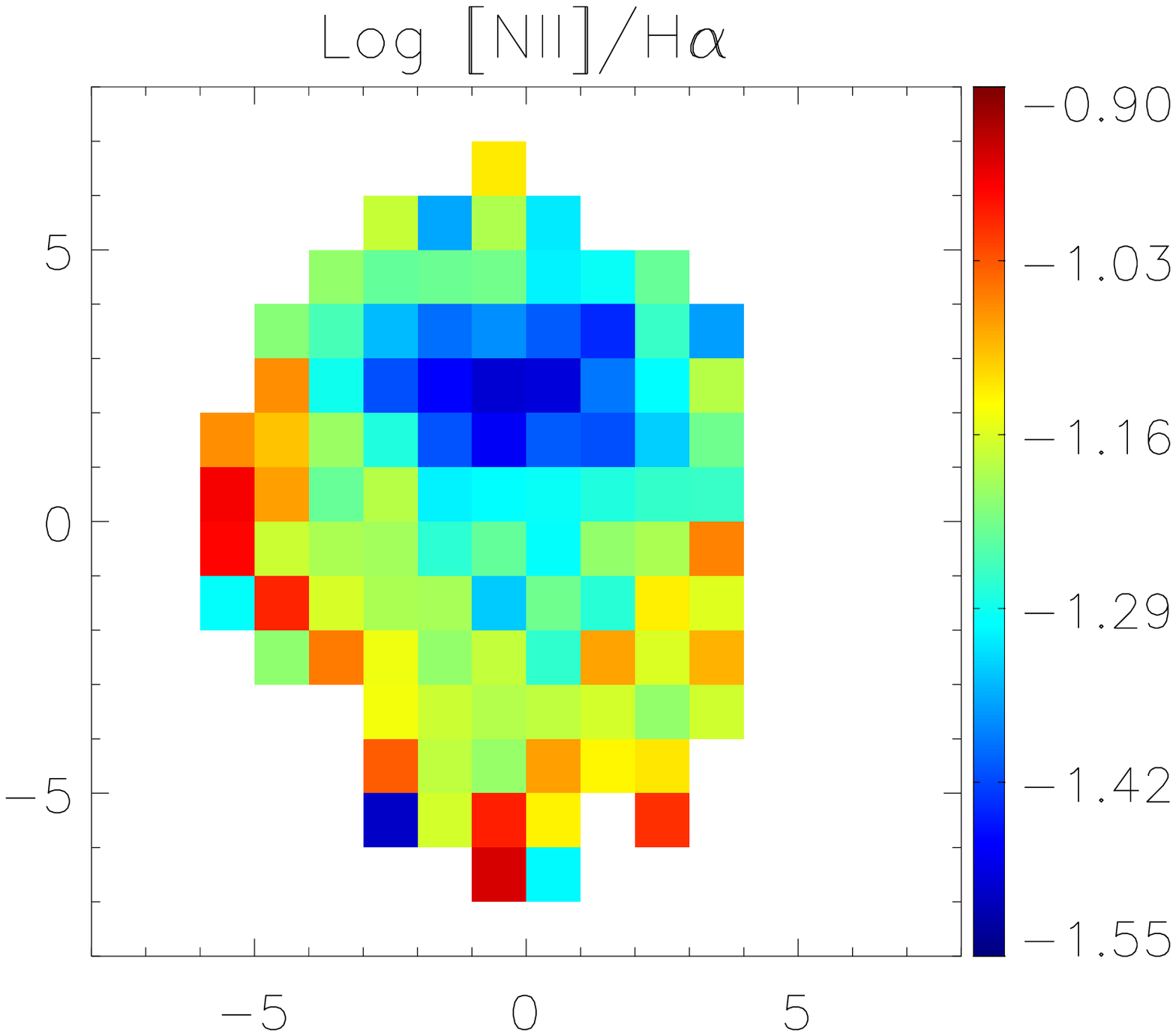}}
\hspace*{0.0cm}\subfigure{\includegraphics[width=0.24\textwidth]{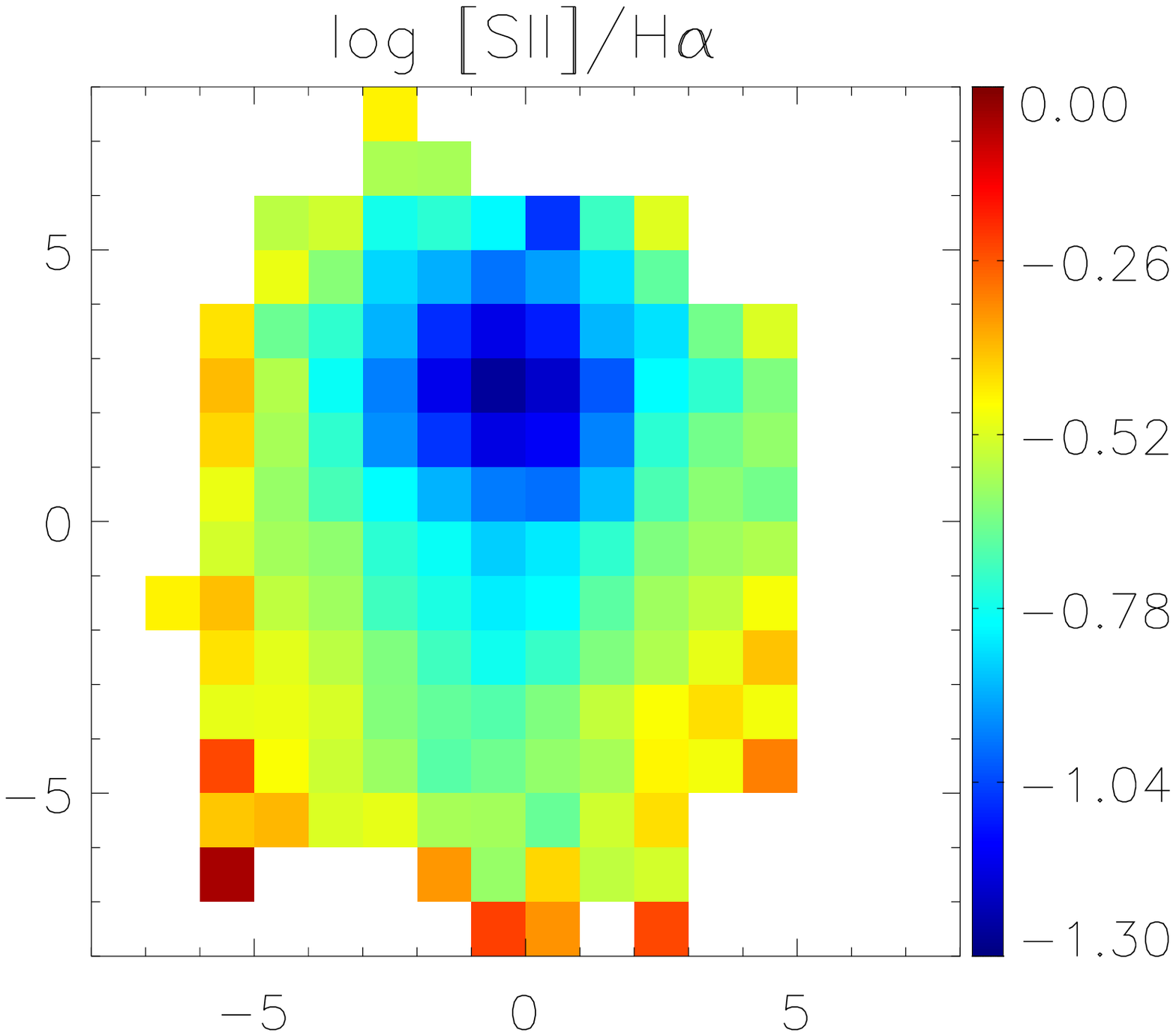}}
\hspace*{0.0cm}\subfigure{\includegraphics[width=0.24\textwidth]{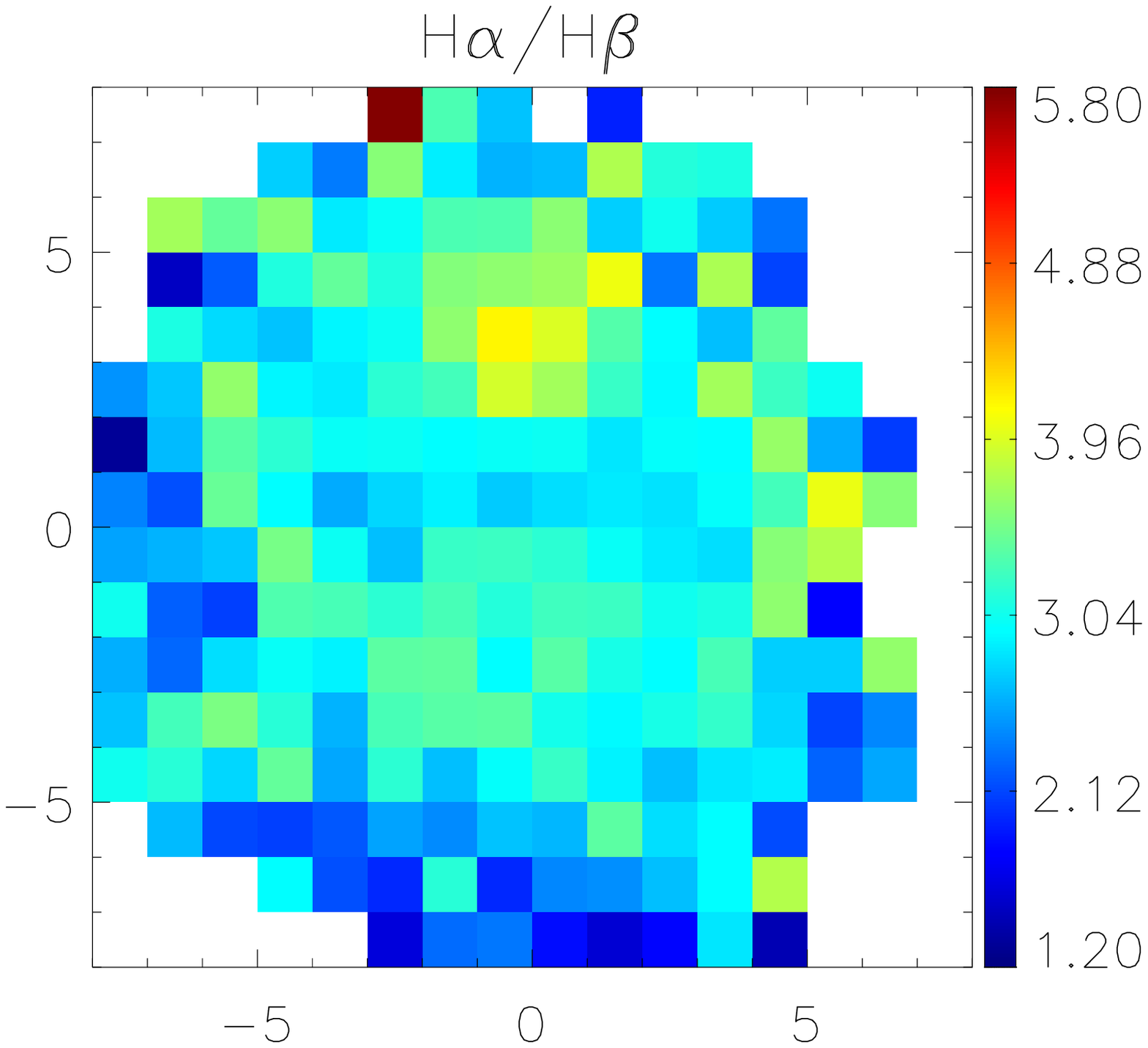}}
}}
\mbox{
\centerline{
\hspace*{0.0cm}\subfigure{\includegraphics[width=0.24\textwidth]{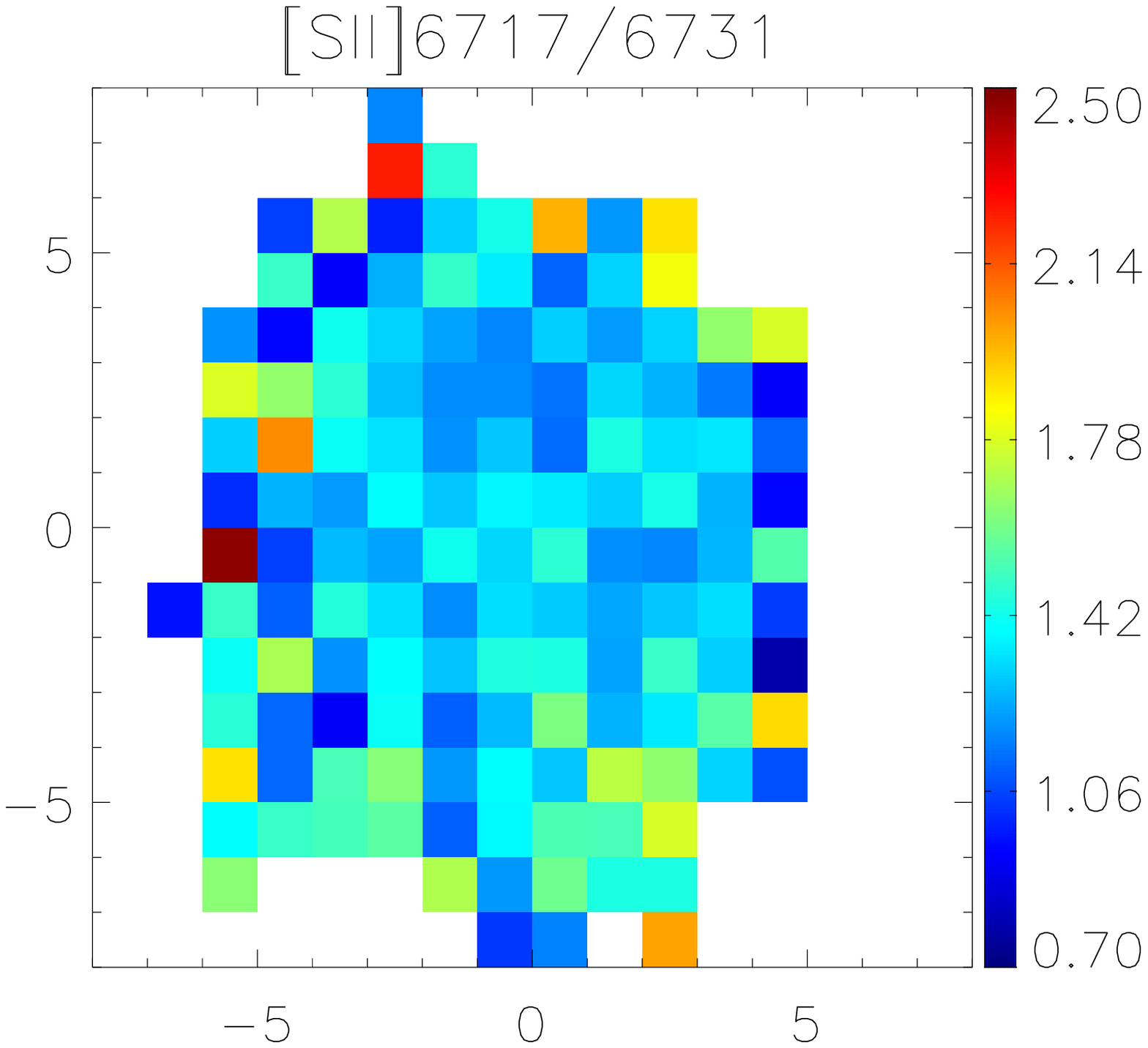}}
\hspace*{0.0cm}\subfigure{\includegraphics[width=0.24\textwidth]{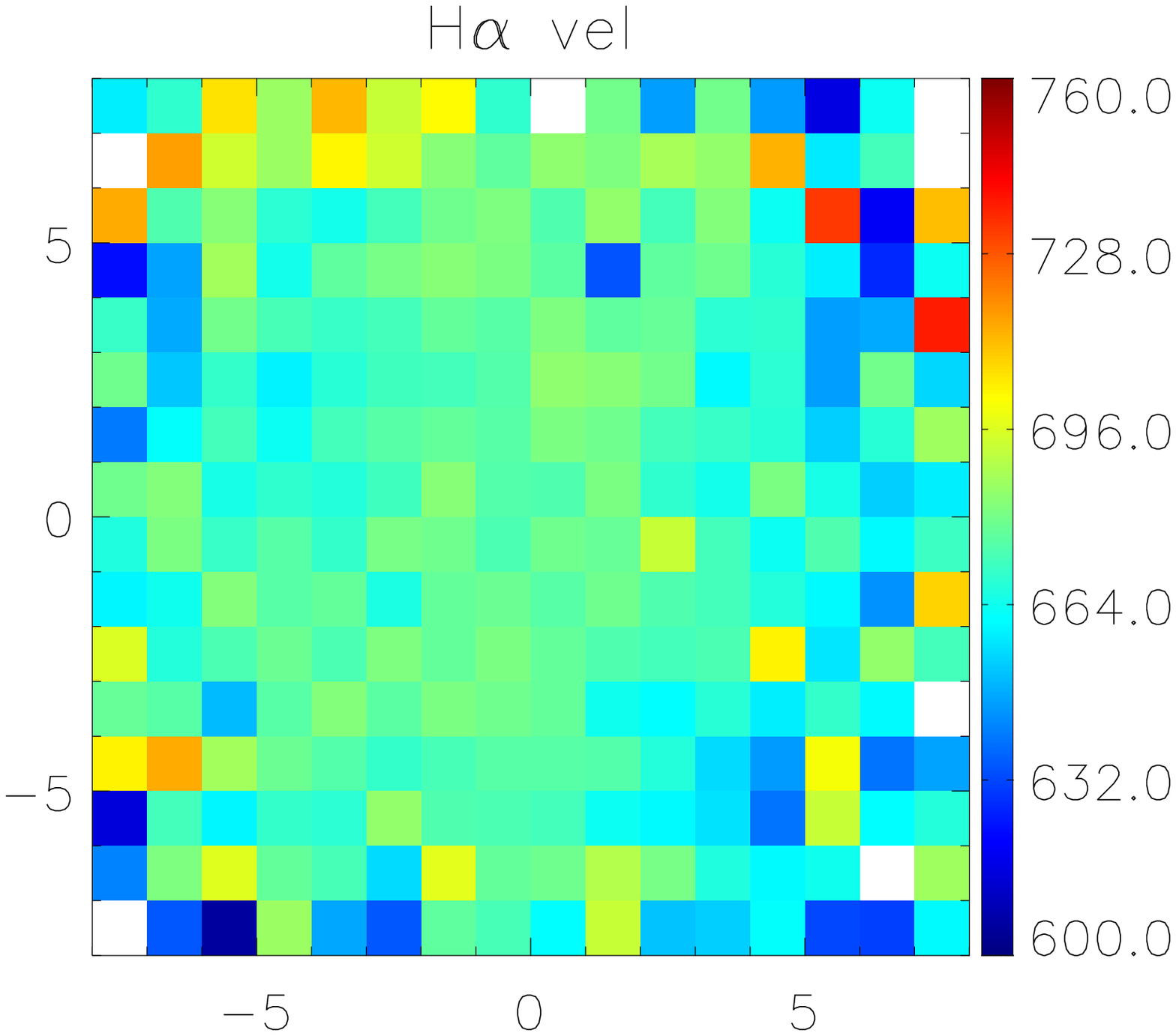}}
\hspace*{0.0cm}\subfigure{\includegraphics[width=0.24\textwidth]{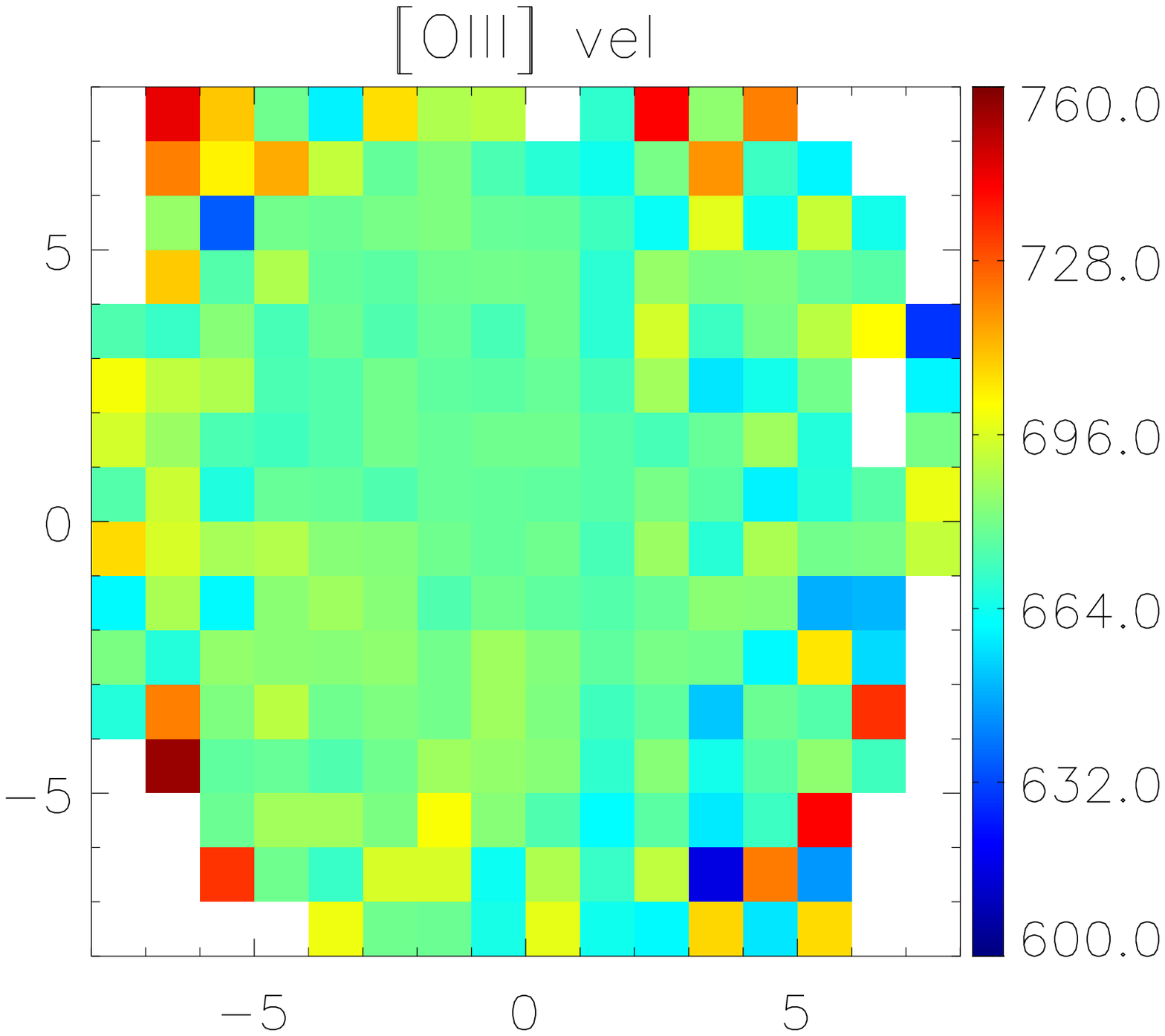}}
}}  
\caption{Same as Fig.~\ref{Figure:mrk407} for I~Zw~123. Spaxels in which the 
WR blue bump was detected were marked in the 
[\ion{O}{iii}]~$\lambda5007$ map.}
\label{Figure:izw123}
\end{figure*}

\begin{figure*}
\mbox{
\centerline{
\hspace*{0.0cm}\subfigure{\includegraphics[width=0.24\textwidth]{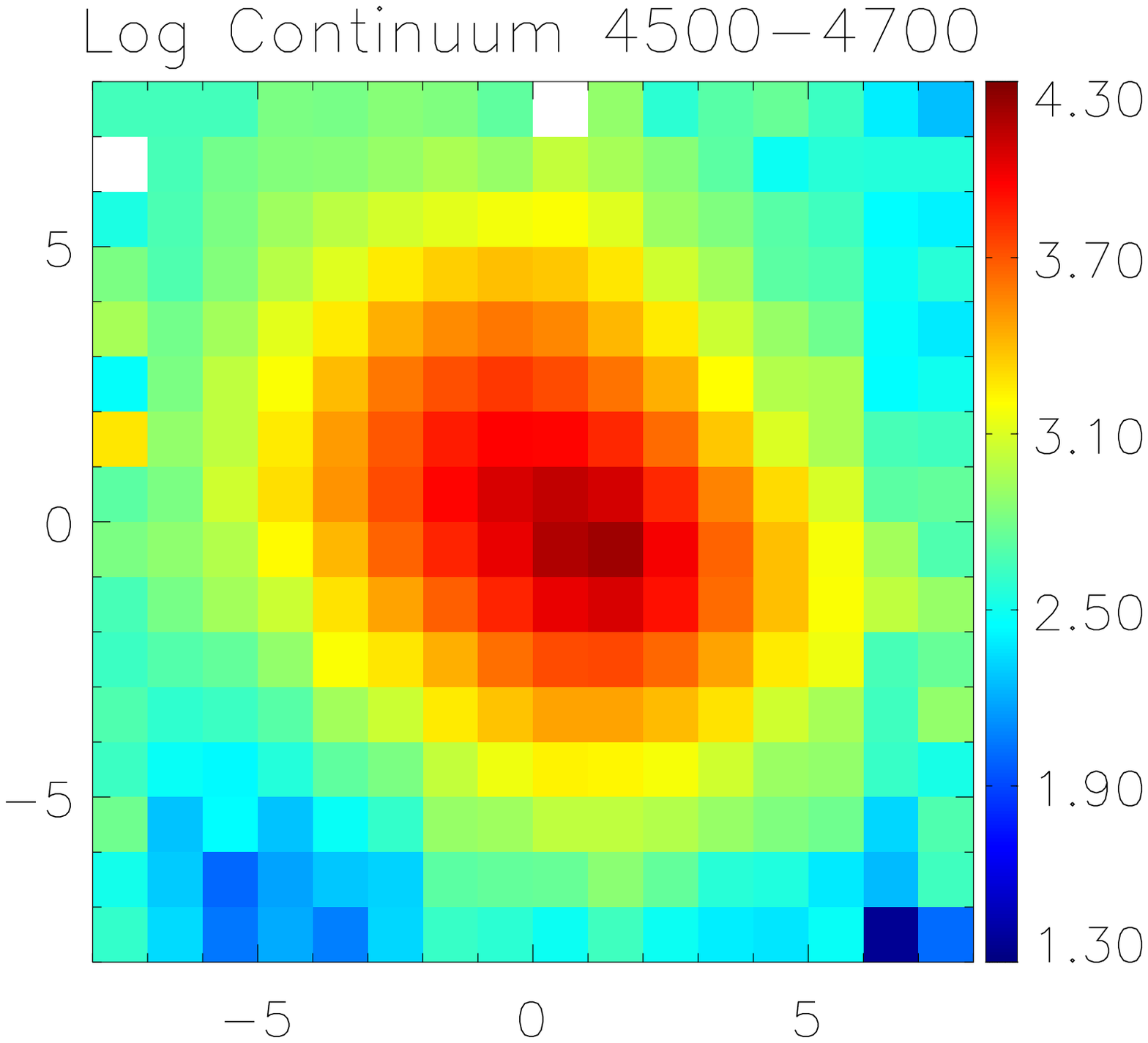}}
\hspace*{0.0cm}\subfigure{\includegraphics[width=0.24\textwidth]{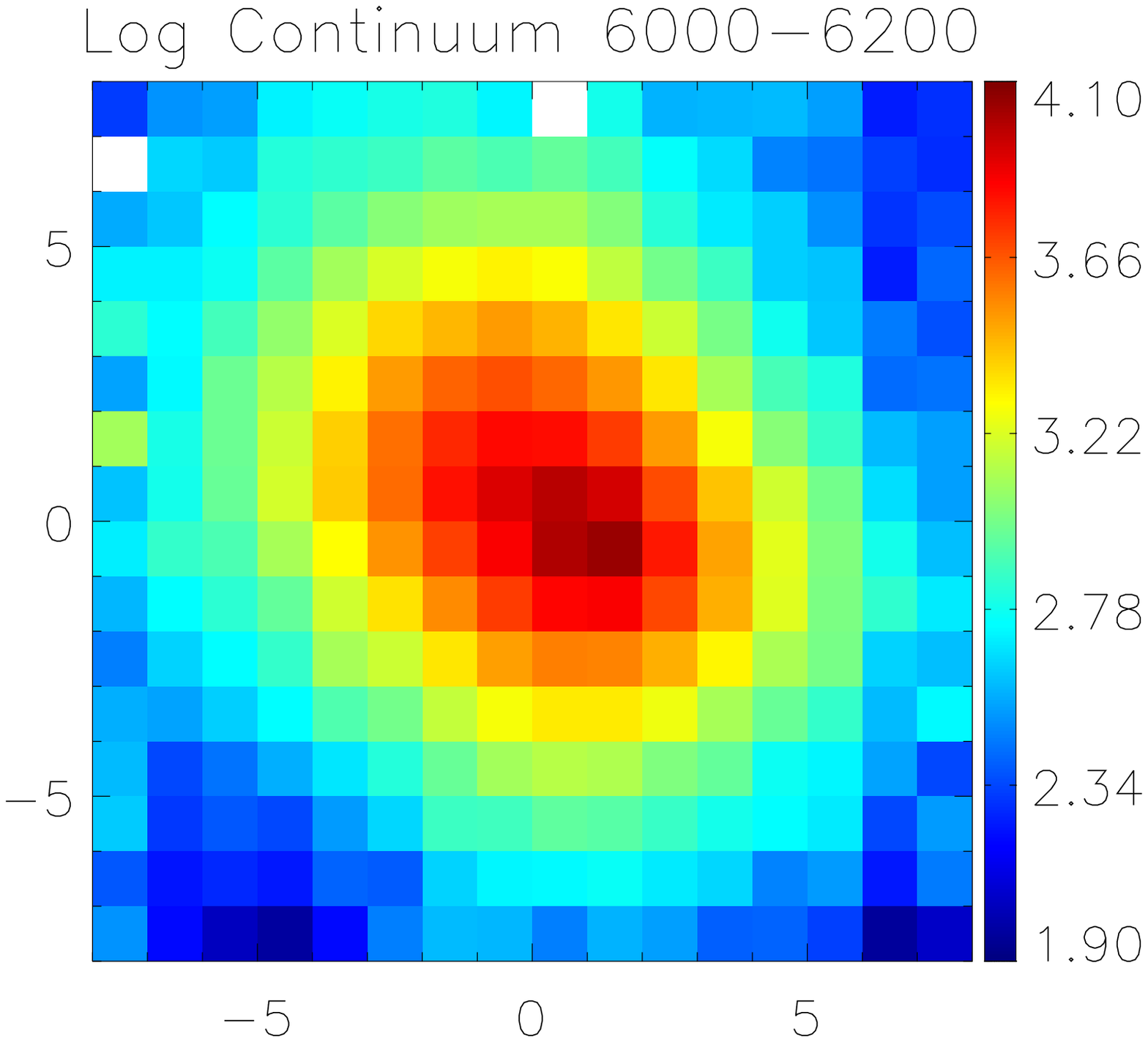}}
\hspace*{0.0cm}\subfigure{\includegraphics[width=0.24\textwidth]{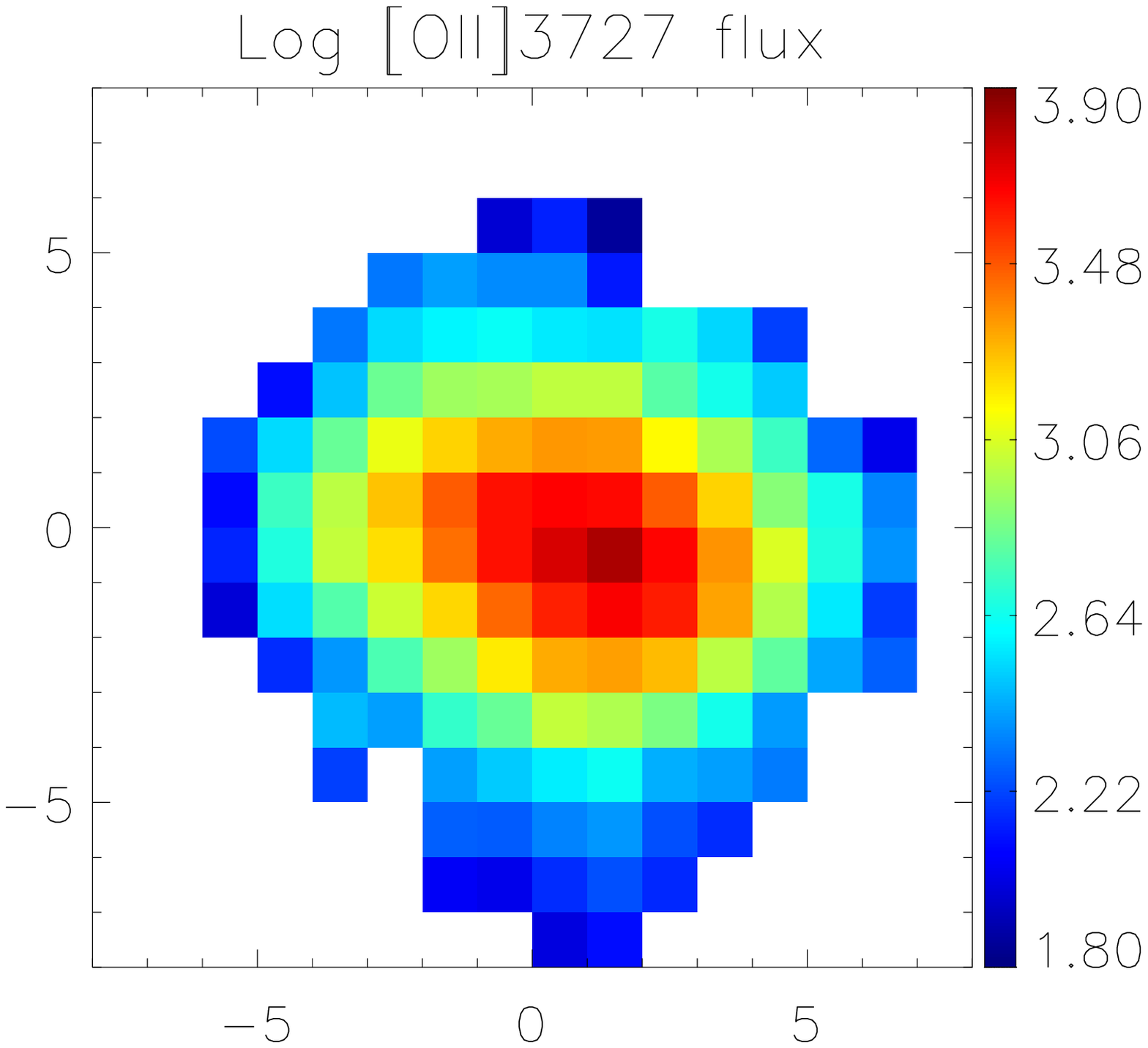}}
\hspace*{0.0cm}\subfigure{\includegraphics[width=0.24\textwidth]{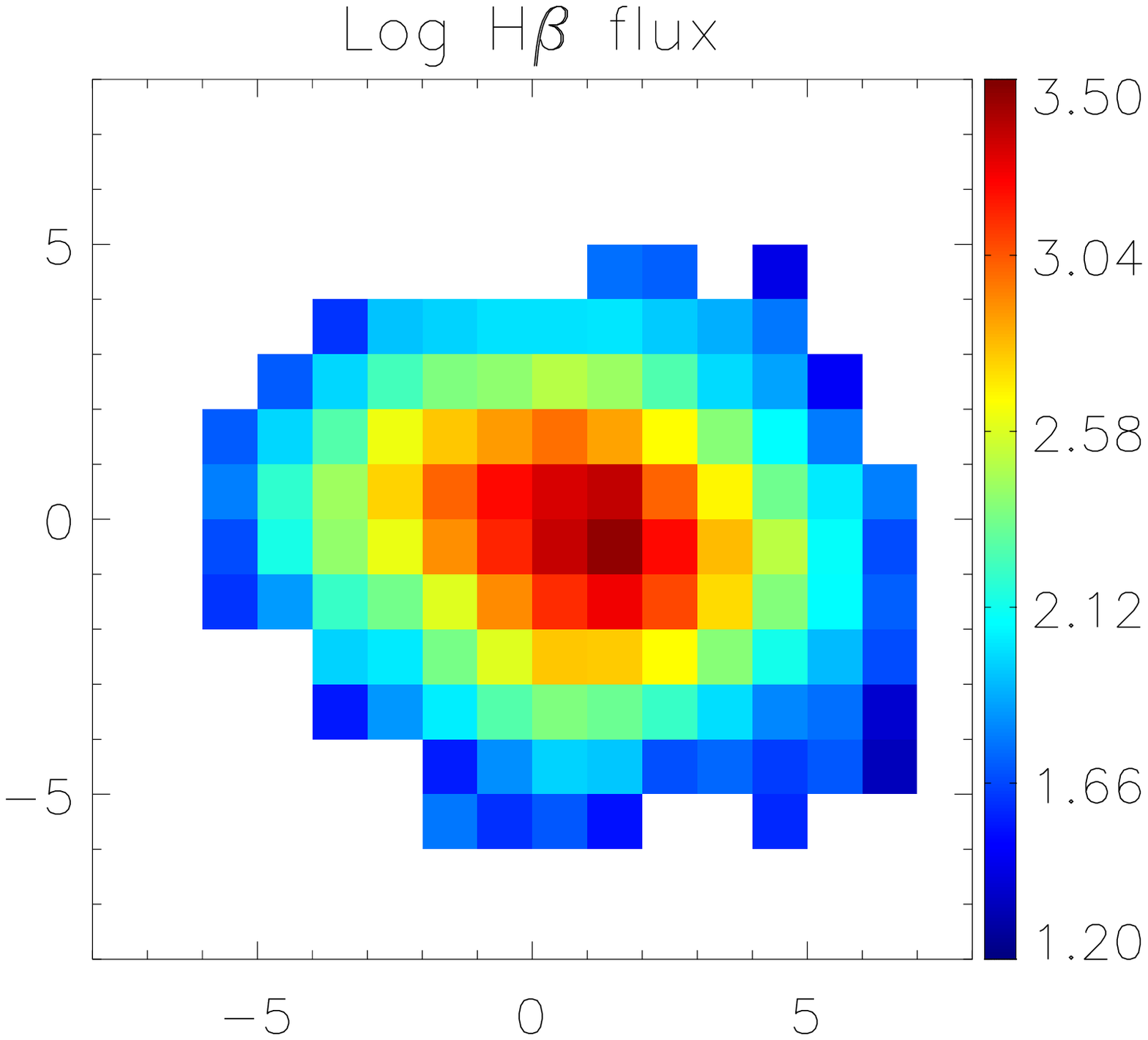}}
}}   
\mbox{
\centerline{
\hspace*{0.0cm}\subfigure{\includegraphics[width=0.24\textwidth]{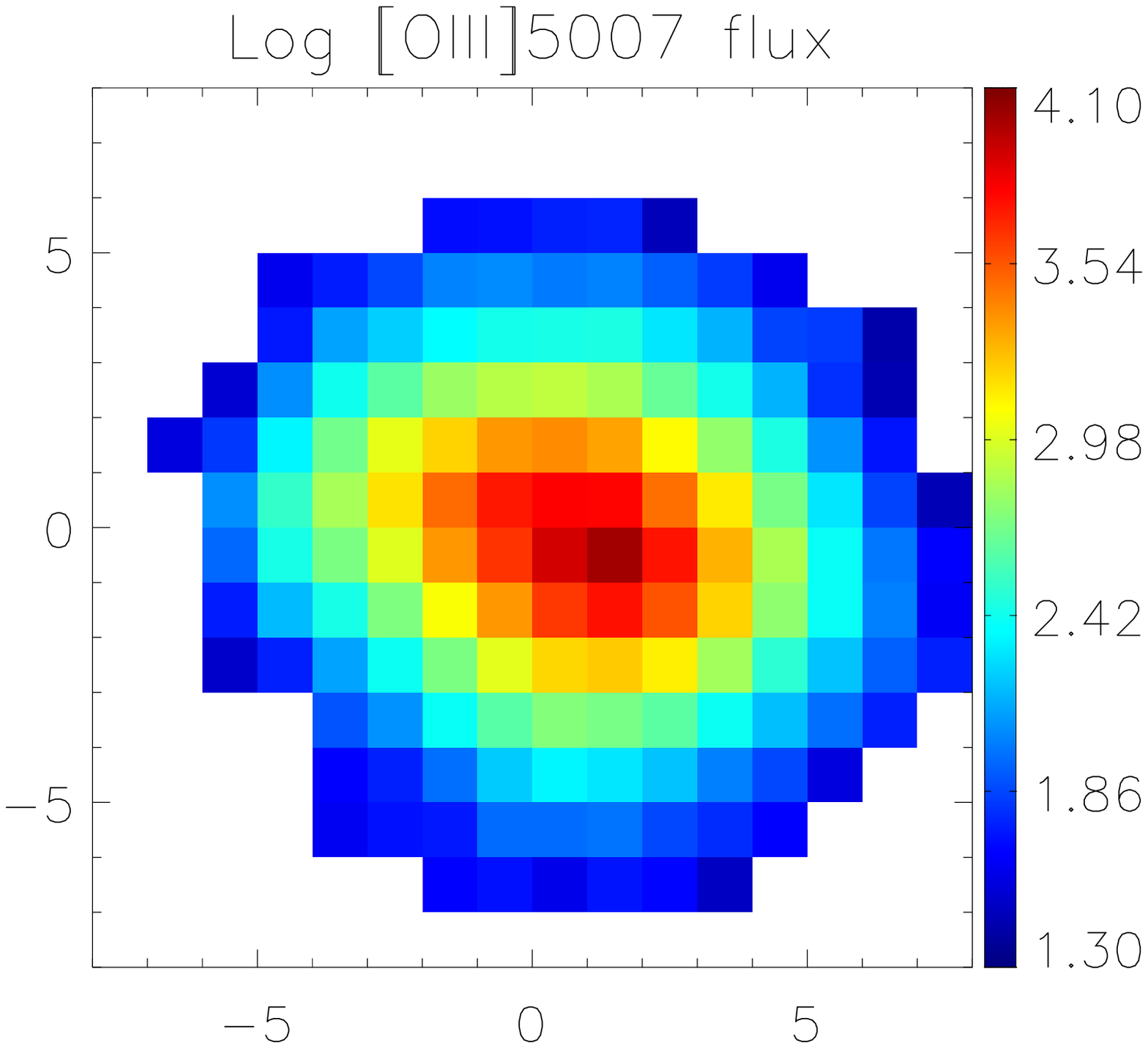}}
\hspace*{0.0cm}\subfigure{\includegraphics[width=0.24\textwidth]{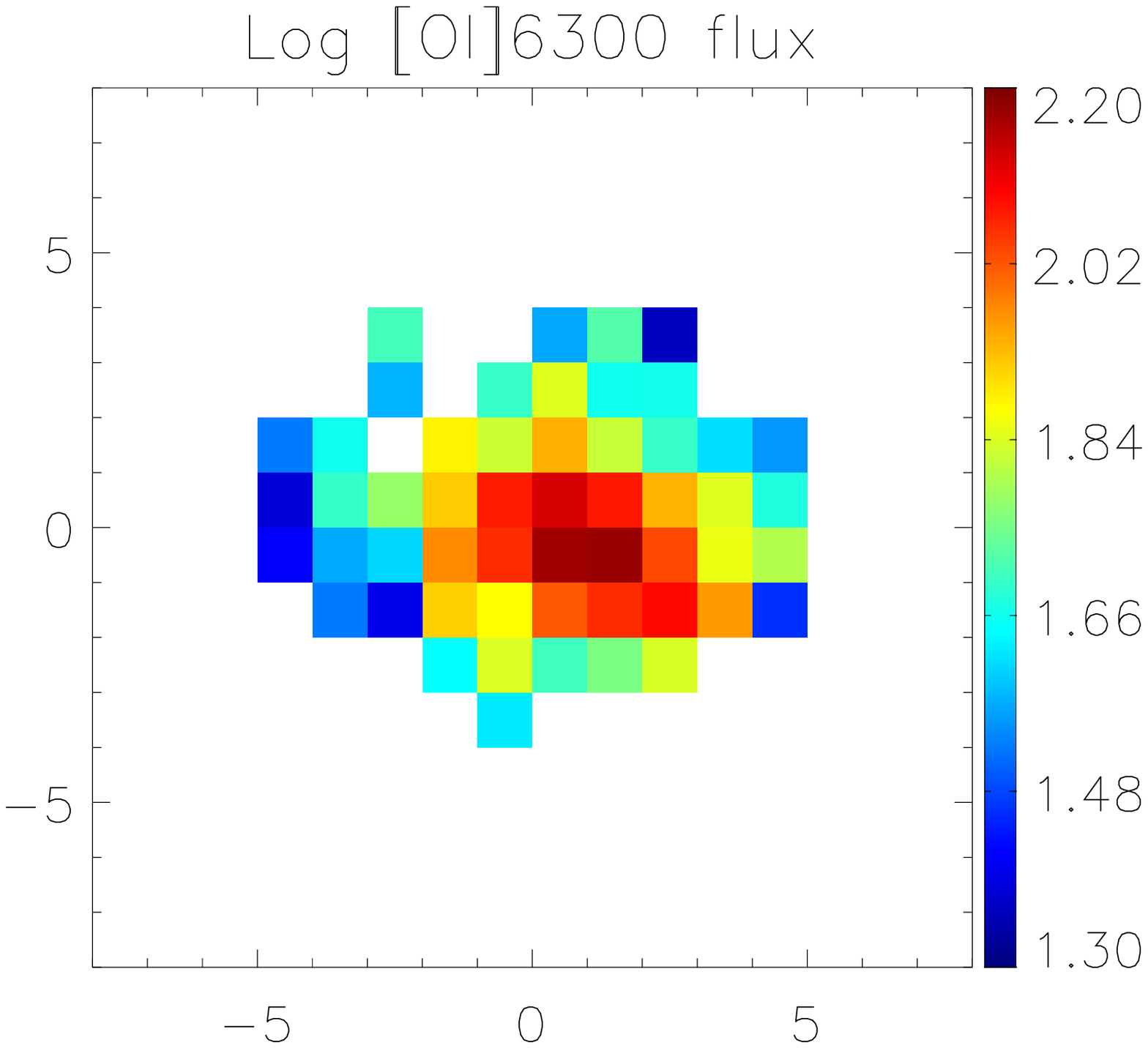}}
\hspace*{0.0cm}\subfigure{\includegraphics[width=0.24\textwidth]{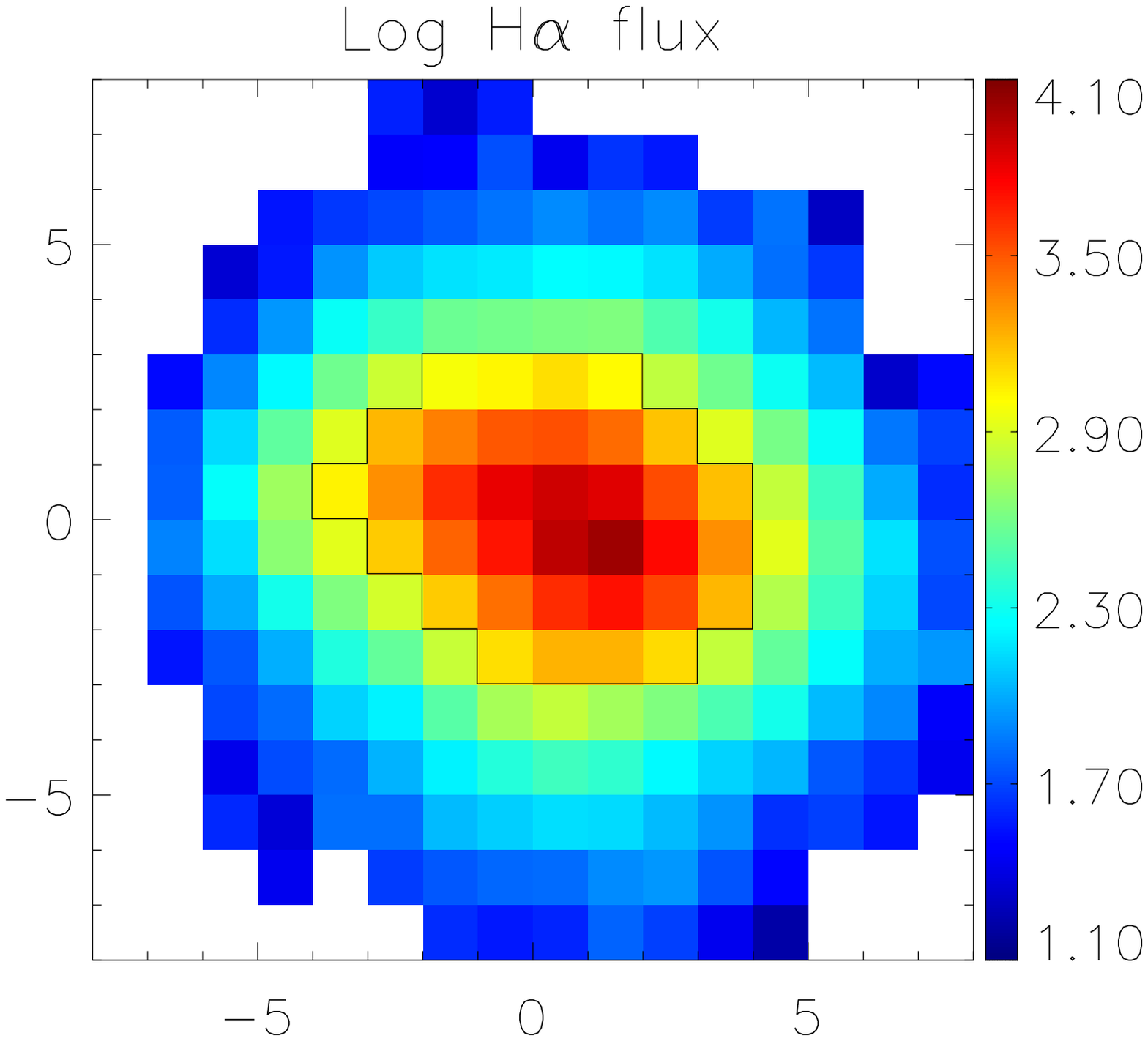}}
\hspace*{0.0cm}\subfigure{\includegraphics[width=0.24\textwidth]{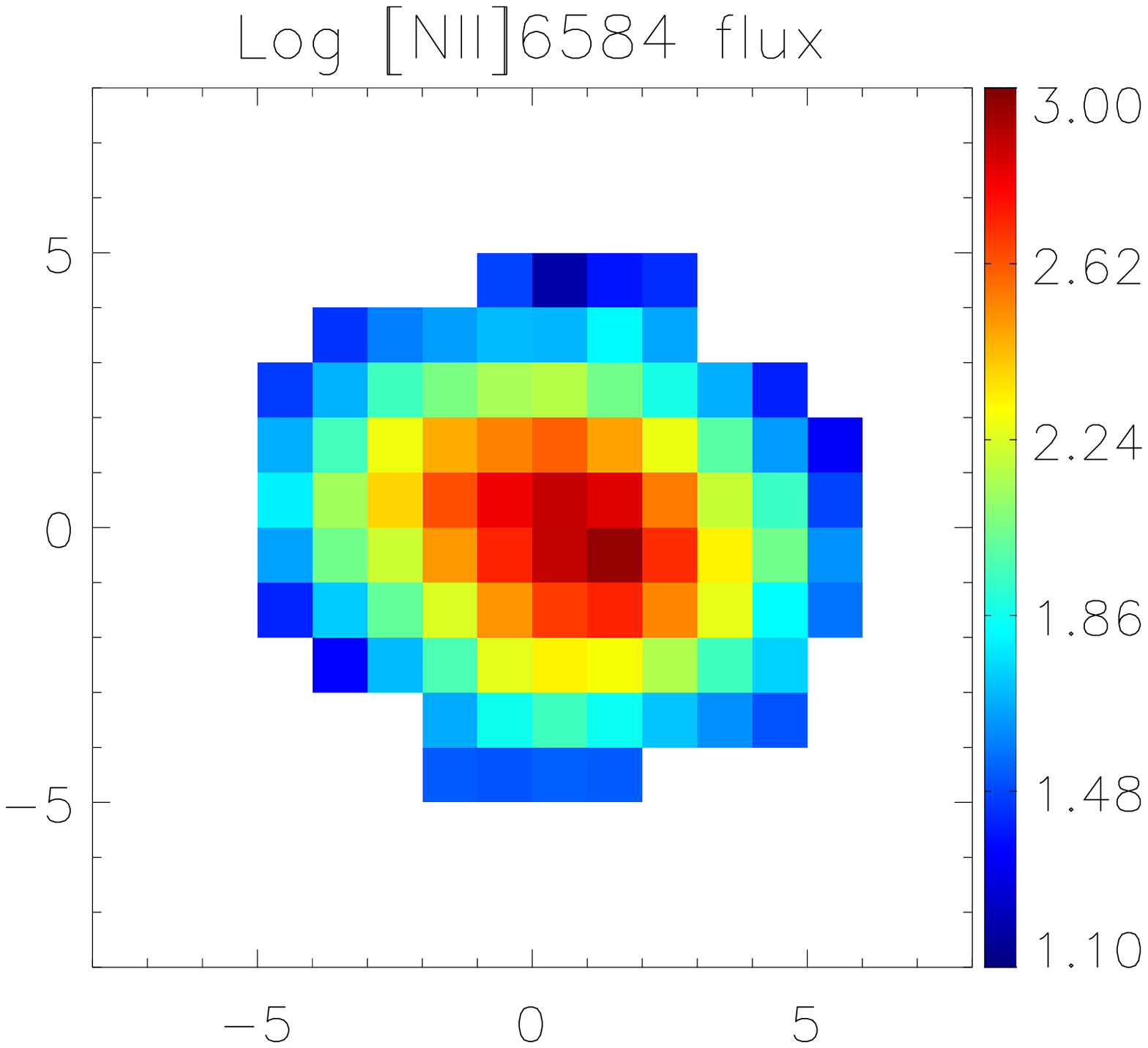}}
}} 
\mbox{
\centerline{
\hspace*{0.0cm}\subfigure{\includegraphics[width=0.24\textwidth]{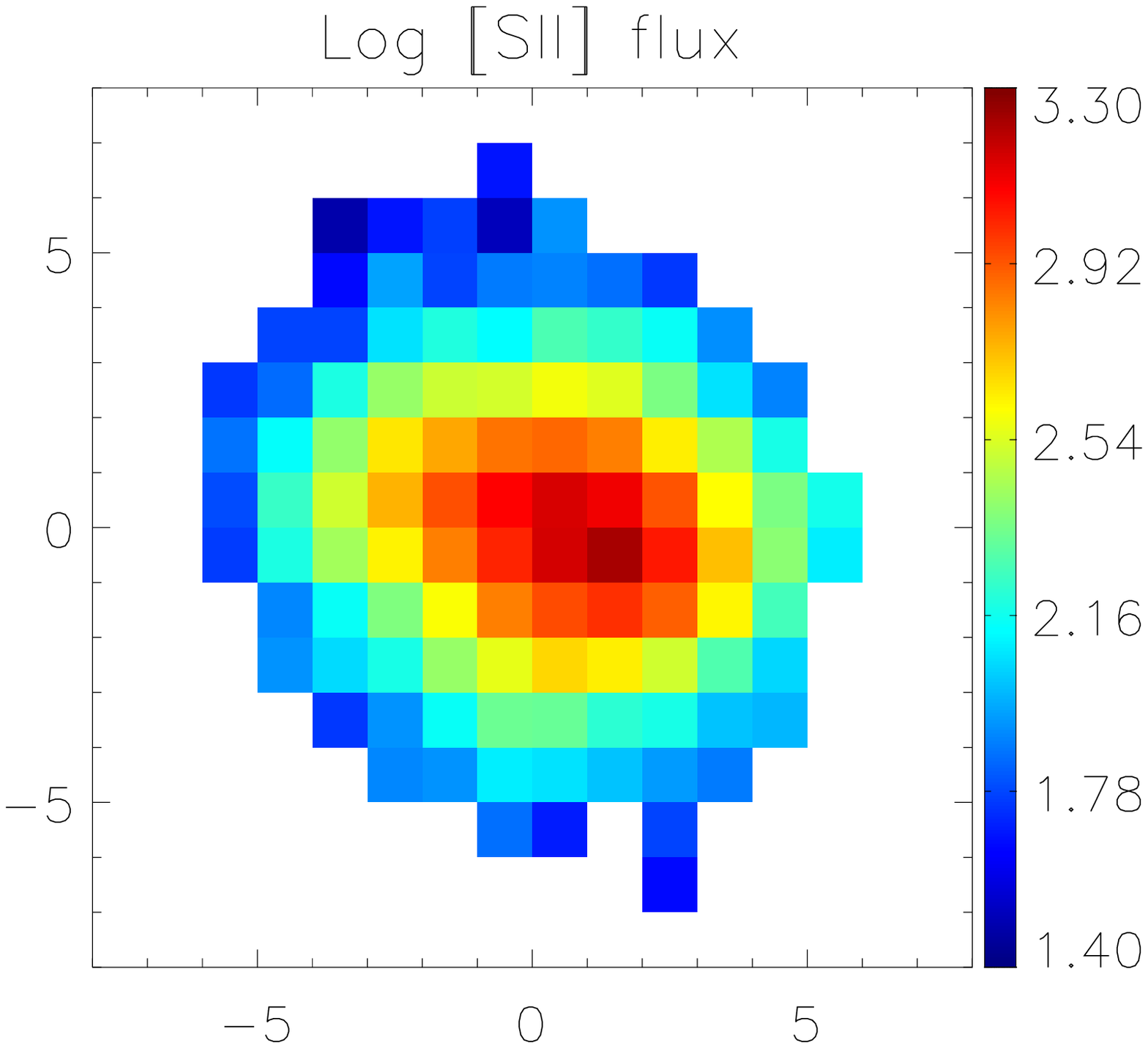}}
\hspace*{0.0cm}\subfigure{\includegraphics[width=0.24\textwidth]{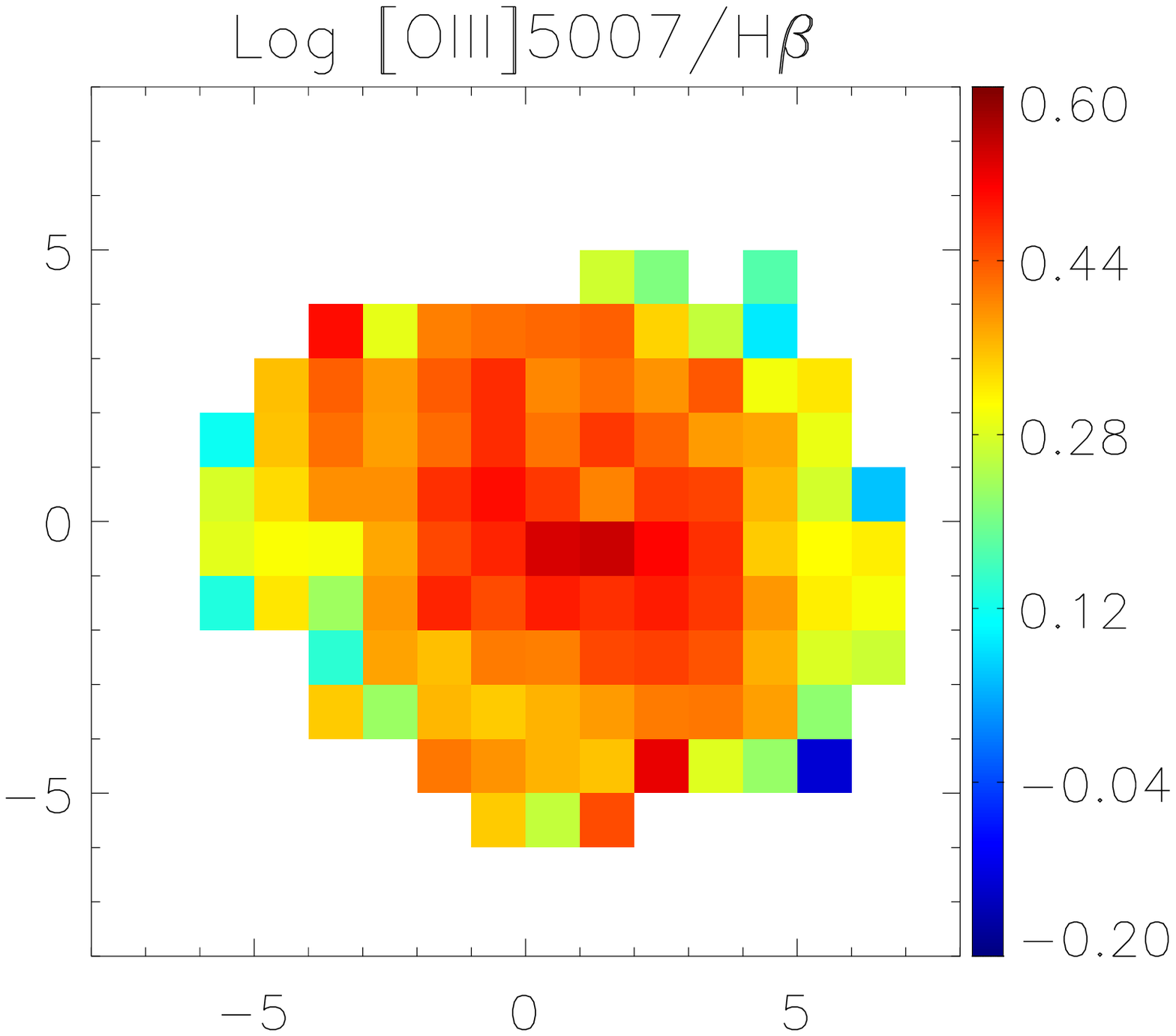}}
\hspace*{0.0cm}\subfigure{\includegraphics[width=0.24\textwidth]{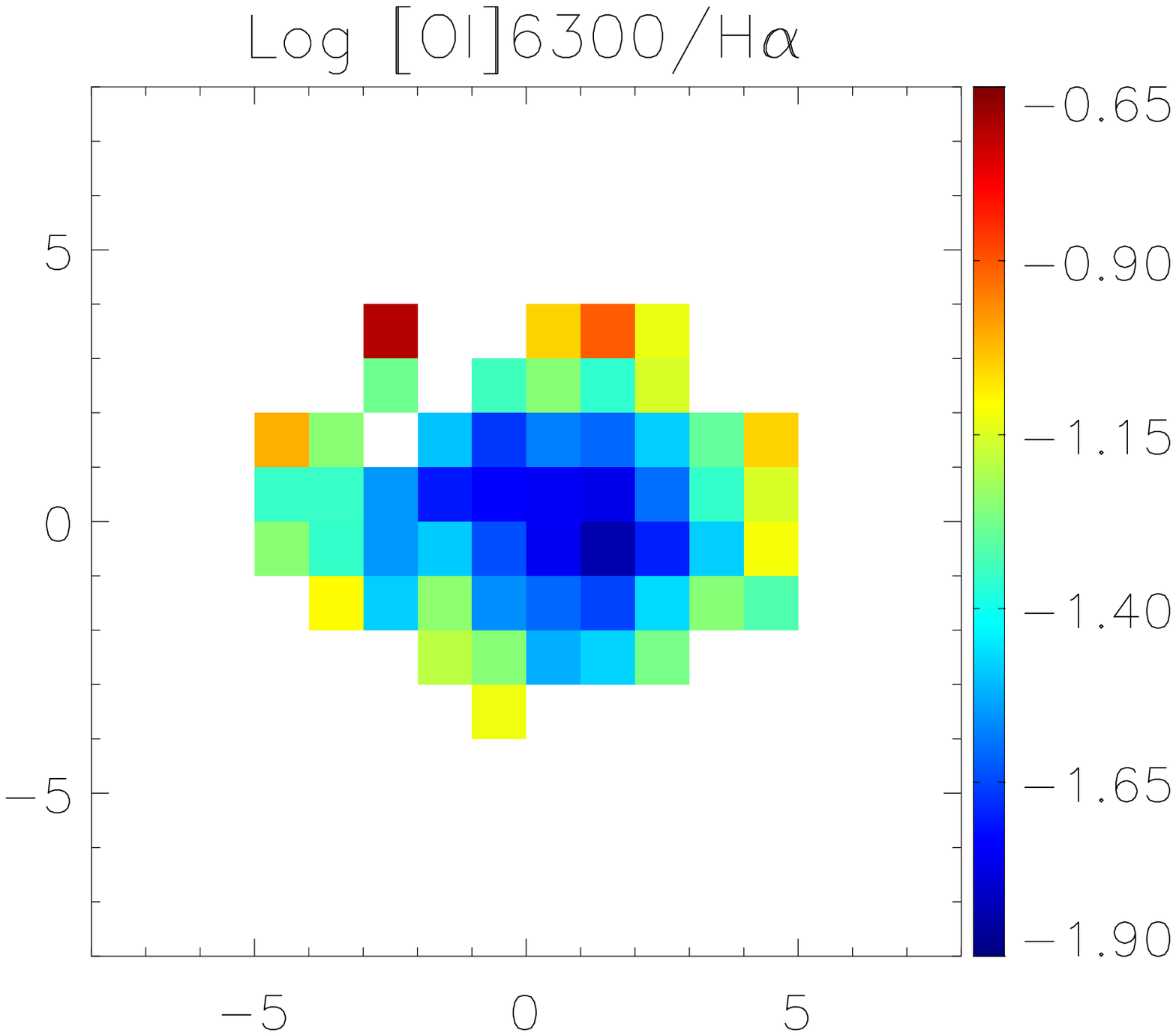}}
\hspace*{0.0cm}\subfigure{\includegraphics[width=0.24\textwidth]{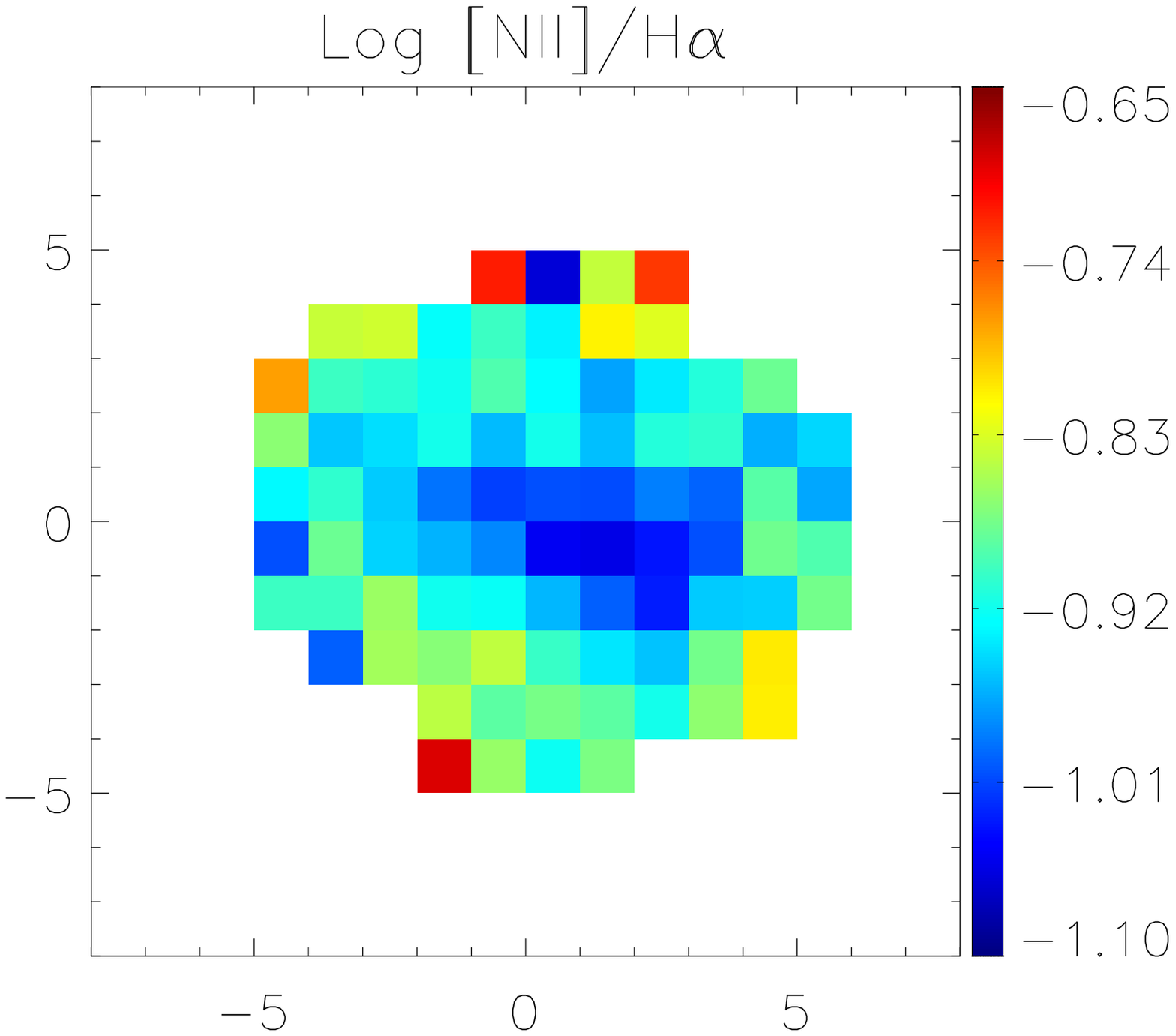}}
}}
\mbox{
\centerline{
\hspace*{0.0cm}\subfigure{\includegraphics[width=0.24\textwidth]{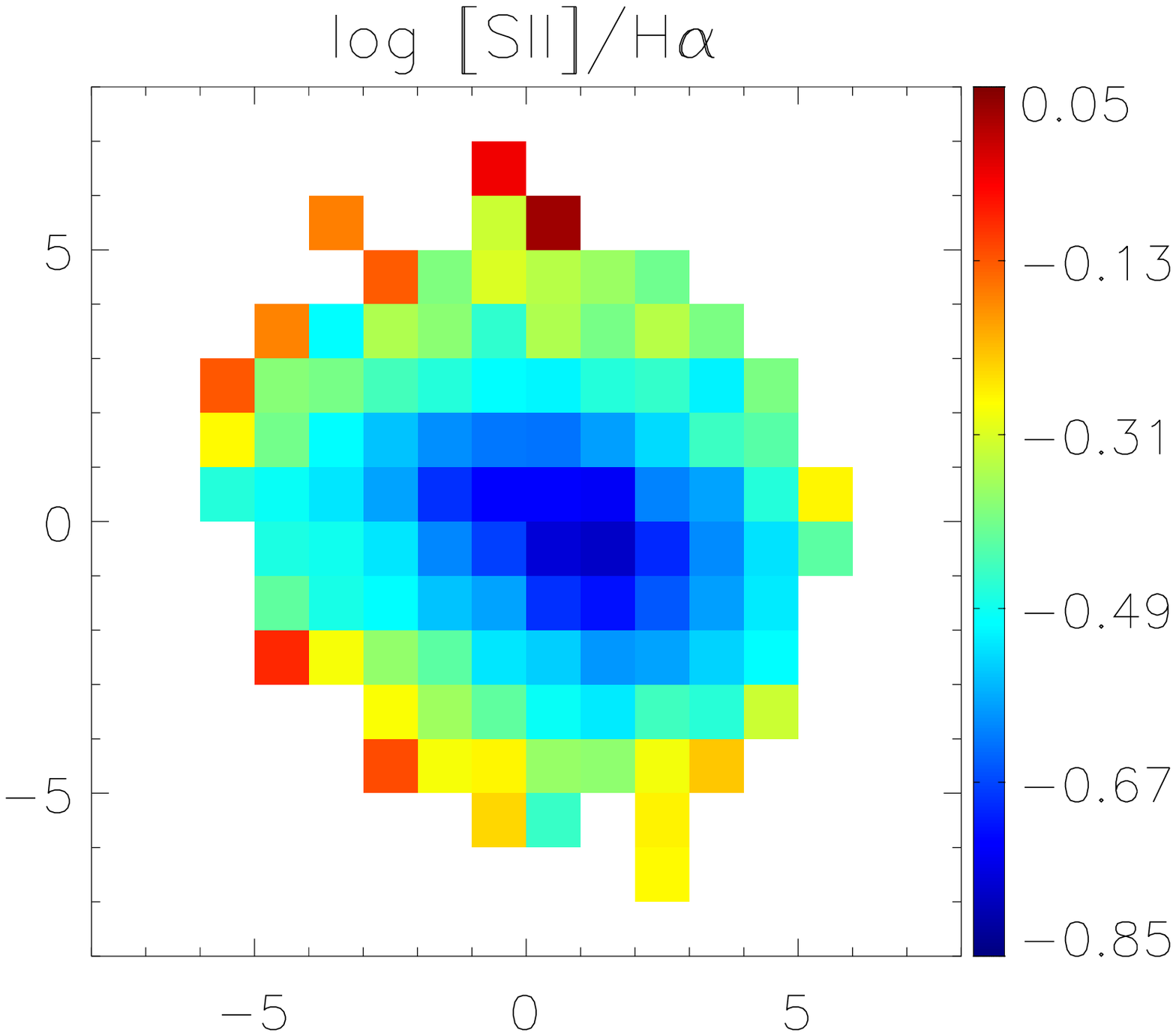}}
\hspace*{0.0cm}\subfigure{\includegraphics[width=0.24\textwidth]{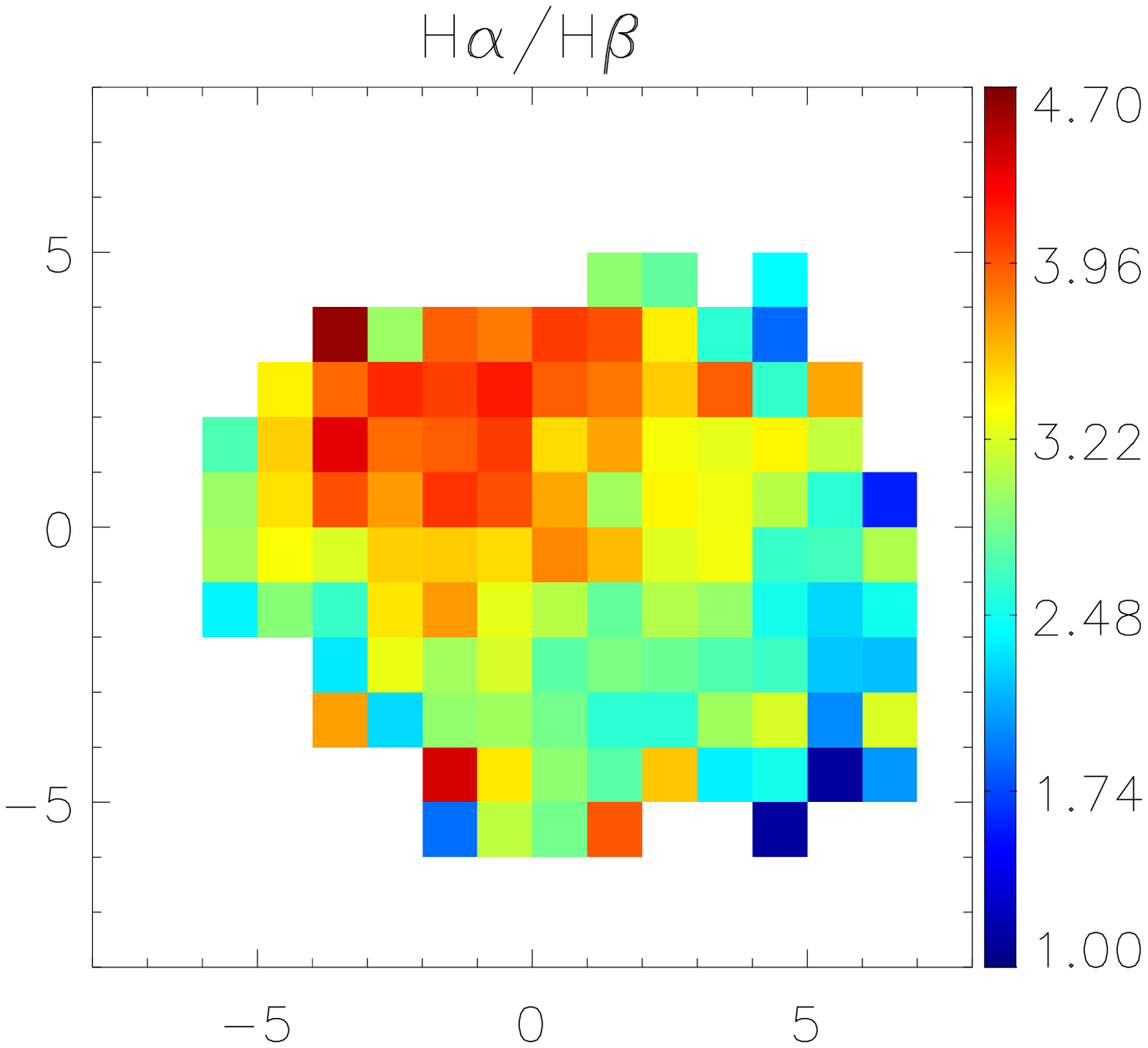}}
\hspace*{0.0cm}\subfigure{\includegraphics[width=0.24\textwidth]{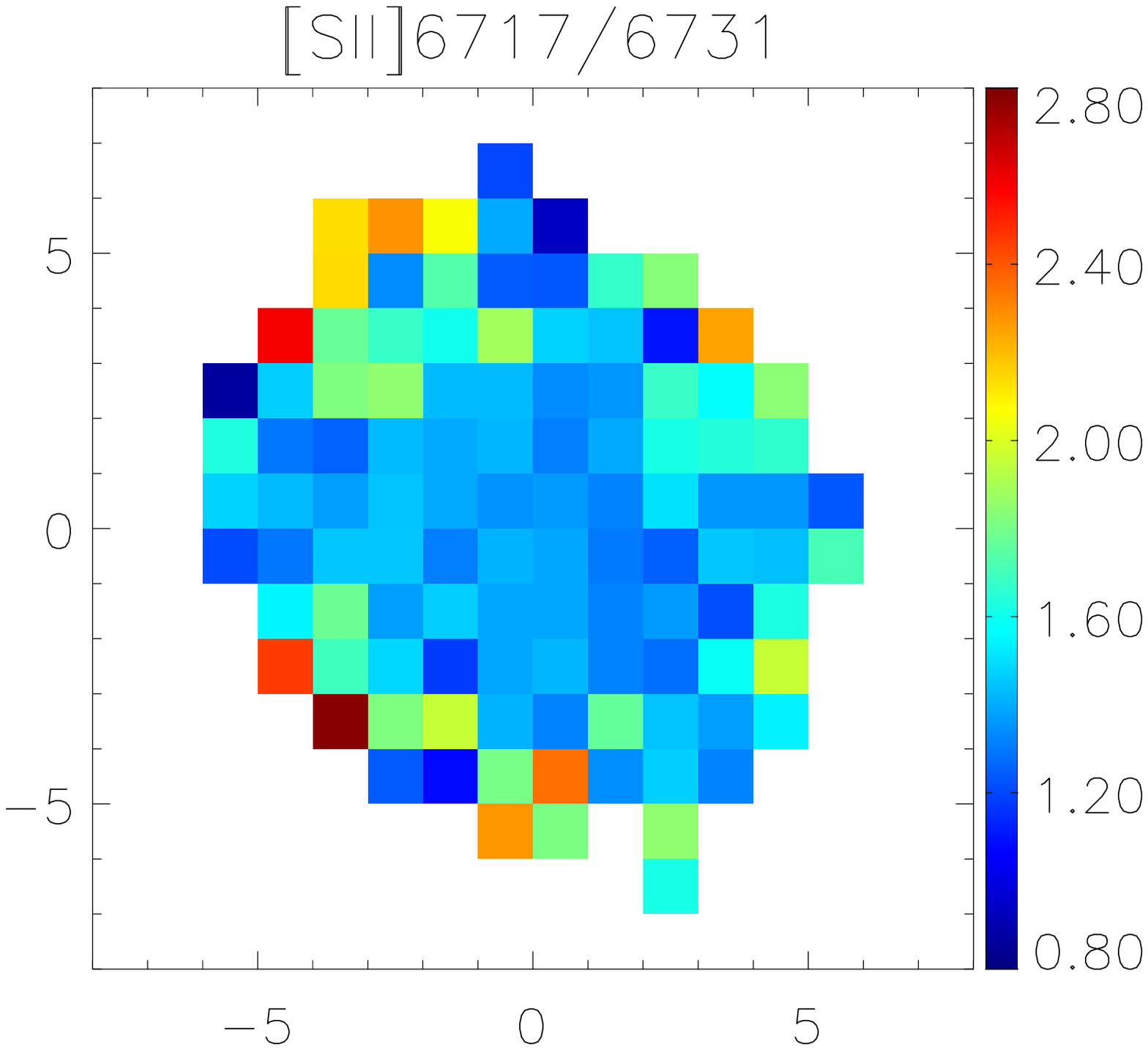}}
}}
\mbox{
\centerline{
\hspace*{0.0cm}\subfigure{\includegraphics[width=0.24\textwidth]{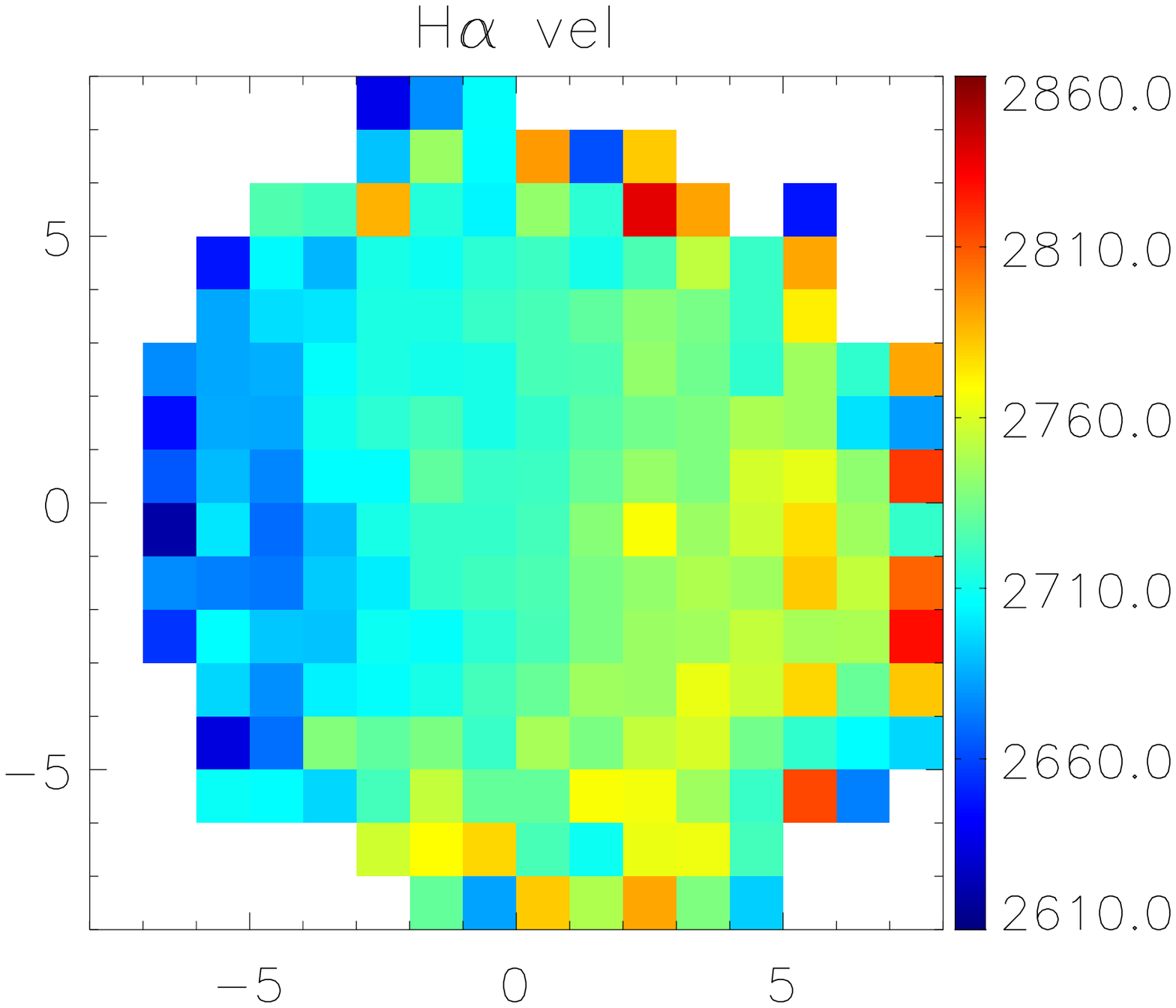}}
\hspace*{0.0cm}\subfigure{\includegraphics[width=0.24\textwidth]{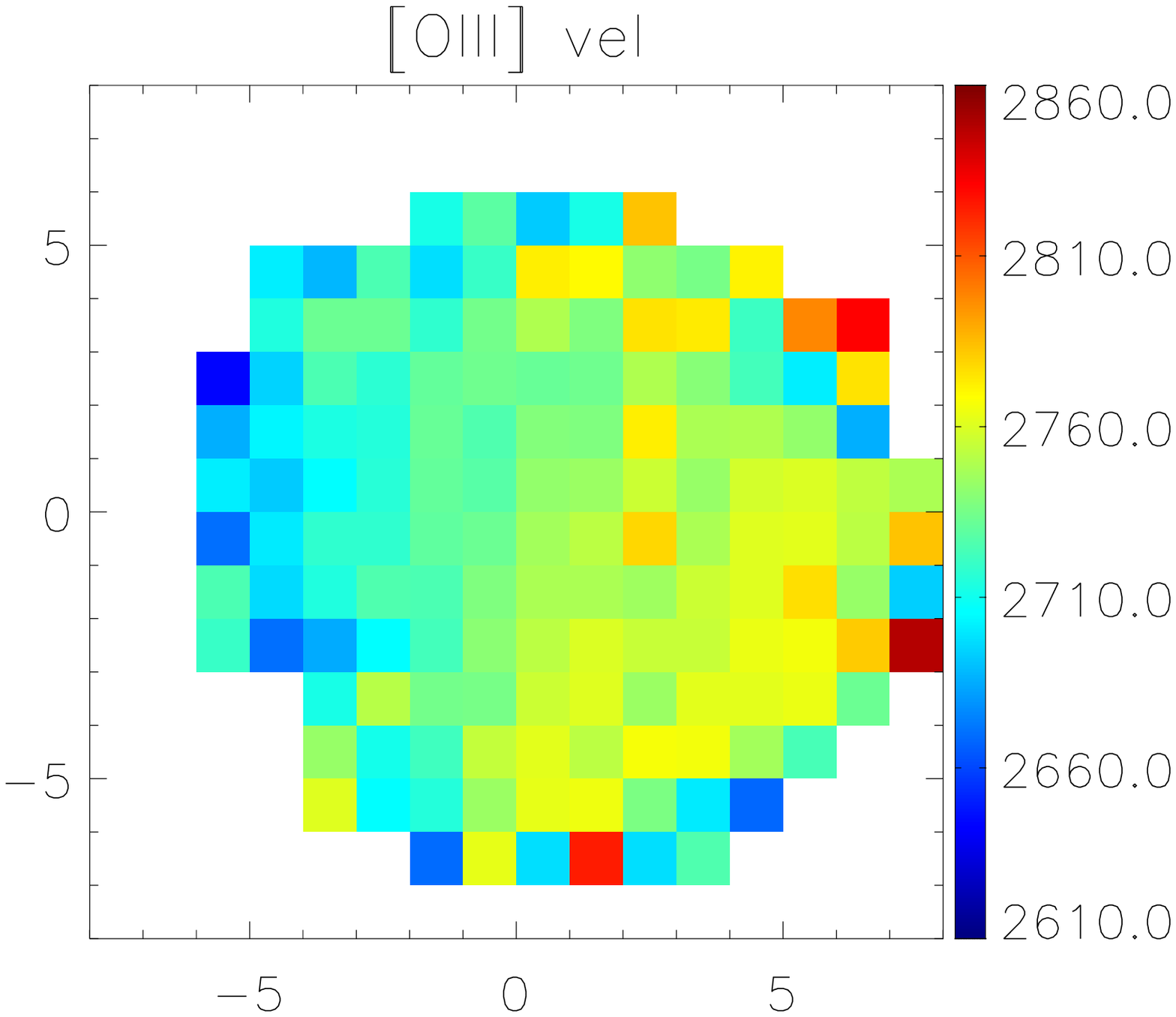}}
}}  
\caption{Same as Fig.~\ref{Figure:mrk407} for I~Zw~159. Maps of
[\ion{O}{i}]~$\lambda6300$ and of the [\ion{O}{i}]~$\lambda6300$/\Ha\ 
ionization ratio are also included.}
\label{Figure:izw159}
\end{figure*}

To study the properties of the stellar component, we built continuum maps
within selected wavelength intervals free from emission lines (``pure
continua''). Figs.~\ref{Figure:mrk407}--\ref{Figure:izw159} show the
``blue'' and ``red'' continuum maps for the sample galaxies, obtained by
integrating the spectrum in the regions 4500--4700 and 6000--6200 \AA\
respectively. 

All objects except two show an overall regular morphology in the continuum,
with a well defined central peak and roughly circular isophotes. The
exceptions are Mrk~750, whose outer isophotes are elongated in the northeast
direction, and Tololo~1434+032, which displays a very irregular and clumpy
morphology.

\subsubsection{Emission line maps}
\label{SubSubSection:EmissionLineMaps}

In order to study the ionized gas morphology of the sample galaxies we built
continuum-subtracted emission line flux maps for the most prominent emission
lines: [\ion{O}{ii}]~$\lambda3727$, \Hb, [\ion{O}{iii}]~$\lambda5007$,
\Ha, [\ion{N}{ii}]~$\lambda6584$ and
[\ion{S}{ii}]~$\lambda\lambda6717,\;6731$).

All the galaxies show a similar morphology in the different emission
lines, as expected in objects ionized by stars. Six of the eight galaxies,
namely Mrk~407, Mrk~750, Mrk~206, Mrk~475, I~Zw~123 and I~Zw~159, appear very
compact in emission lines and have a single central starburst. In Mrk~32 the
starburst is resolved into three smaller SF regions, aligned along a
north-south axis, while in Tololo~1434+032 the current star formation activity
spreads all over the mapped region.

\subsection{Line ratio maps}
\label{SubSection:LineRatioMaps}

\subsubsection{Ionization sources}
\label{SubSubSection:IonizationSources}

To investigate the excitation mechanisms acting in the galaxies we computed
the line ratio maps for [\ion{O}{iii}]~$\lambda$5007/\Hb,
[\ion{N}{ii}]~$\lambda$6584/\Ha, [\ion{S}{ii}]~$\lambda\lambda6717,6731$/\Ha\
and, in those cases in which the [\ion{O}{i}]~$\lambda$6300 line has a
reasonably good signal-to-noise ratio, also for
[\ion{O}{i}]~$\lambda$6300/\Ha. 

High excitation values correspond to high values of the [\ion{O}{iii}]/\Hb\ 
ratios and to low values of [\ion{N}{ii}]/\Ha\ and [\ion{S}{ii}]/\Ha. High
values for [\ion{O}{iii}]/\Hb\ are expected when the ionization is produced
predominantly by UV photons, especially when the ionization parameter is high.
On the other hand, low excitation values can be associated with an ionizing
mechanism different from photoionization \citep{Veilleux1987}.

Excitation maps for the galaxy sample are displayed  in
Figs.~\ref{Figure:mrk407}--\ref{Figure:izw159}. All objects show
the same pattern in the different ratio maps: they trace the regions of
star formation, with [\ion{O}{iii}]/\Hb\ ([\ion{N}{ii}]/\Ha,
[\ion{S}{ii}]/\Ha) peaking (having a minimum) in the SF knots, and
decreasing (increasing) with the distance to the center of the region. This
is the expected behavior in regions ionized by UV photons coming from
massive stars: the metal-to-hydrogen line ratios change as a function of the
ionization parameter ($U$), and therefore increasing the distance from the
ionization source decreases the value of $U$, lowering the [\ion{O}{iii}]/\Hb\
ratio and increasing the [\ion{S}{ii}]/\Ha\ ratio \citep{Domgorgen1994}.

\subsubsection{Extinction maps}
\label{SubSubSection:ExtinctionMaps}

Interstellar extinction can be probed by comparing the observed ratios of
hydrogen recombination lines with their theoretical values
\citep{Osterbrock2006}. In the optical domain the extinction is derived from
the ratio of the different Balmer line series to \Hb. In our case, although
the \Hd\ and \Hg\ emission lines are usually visible, their weakness and that 
they are superposed on a strong underlying stellar absorption line
make it impossible to obtain reliable measurements of their flux \textnormal{for
individual spaxels}. Therefore the extinction maps have been derived from
the \Ha/\Hb\ ratio. 

For $T=10\,000$ K and electron densities $\sim100$~cm$^{-3}$, the theoretical 
\Ha/\Hb\ ratio should be close to 2.86. Because extinction is stronger at \Hb\
that at \Ha\ wavelengths, its effect is to increase the observed ratio. 

\textnormal{In computing the extinction map, we corrected the \Hb\ line for 
underlying stellar absorption by fitting a Gaussian profile to its absorption 
wings. For the \Ha\ line, the absence of visible absorption wings makes 
this decomposition impossible. 
To account for the \Ha\ absorption component, several approaches could be 
adopted in principle. The most popular one is to set the equivalent width of
the \Ha\ absorption, $W(\Ha)_\mathrm{abs}$, to the same value as found for
\Hb.
A second more conservative approach is to set it to some fixed value (for 
instance 2 \AA) or simply to assume that the absorption in \Ha\ is negligible.} 

\textnormal{While the first strategy might be appropriate when dealing with with 
integrated spectra, the relatively high uncertainties on the measurements of 
$W(\Hb)_\mathrm{abs}$ makes it unfeasible for a spaxel-to-spaxel correction, 
thus we adopted here the more conservative approach 
(setting $W(\Ha)_\mathrm{abs}=0$).
Under this assumption, the computed extinction is actually a lower limit to 
its actual value.}

\Ha/\Hb\ line ratio maps of the sample galaxies are displayed in
Figs.~\ref{Figure:mrk407}--\ref{Figure:izw159}. Six out of the eight objects
have significant interstellar extinction values ---~up to $\EBV=0.8$ (herein
we use the relation $\EBV=0.69\,\CHbeta$)~--- and a patchy dust
distribution.  


\subsubsection{Electron density}
\label{SubSubSection:ElectronDensity}

We also produced maps of the
[\ion{S}{ii}]~$\lambda6717$/[\ion{S}{ii}]~$\lambda6731$ ratio, an electron 
density diagnostic sensitive ratio in the range 100--10\,000 cm$^{-3}$. The 
density maps of the sample galaxies are displayed in
Figs.~\ref{Figure:mrk407}--\ref{Figure:izw159}.


\subsection{Kinematics of the ionized gas}
\label{SubSection:GasKinematics}

We studied the kinematics of the ionized gas using the
[\ion{O}{iii}]~$\lambda5007$ and \Ha\ emission lines. We fitted the peak
wavelength of the above emission lines with a Gaussian to obtain the radial
velocity of the ionized gas at each spaxel. Due to the low spectral resolution
of our data (about 6.8 \AA\ FWHM) no reliable velocity dispersion
measurements could be obtained.

The velocity fields are shown in
Figs.~\ref{Figure:mrk407}--\ref{Figure:izw159}; in these maps red colors
represent redshifts and blue colors blueshifts. The mean uncertainty
in the velocity data, estimated from the scatter of velocity maps built on the
5577 \AA\ skyline (which in our spectra has intensities comparable with the
\Ha\ and [\ion{O}{iii}]~$\lambda5007$ lines in the galaxies brightest parts),
is about 10--15 km s$^{-1}$ in the central regions and increases outwards
with decreasing line emission intensity.

\subsection{Integrated spectroscopy}
\label{SubSection:IntegratedSpectroscopy}

In this section we present results of the integrated spectroscopy. For each
galaxy we extracted a one-dimensional spectrum of the SF regions in our
maps and the integrated spectrum inside the whole mapped FOV. Six galaxies
have just one nuclear SF region; only in Mrk~32 and Tololo~1434+032 we
identified two or more SF regions.

Because of the heterogeneity of the sample in terms of apparent luminosity,
distance, morphology, brightness of the SF regions, and signal-to-noise
ratio of the observed spectra, there is no unique, clear-cut criterion for
delineating the area of the SF regions.
Thus we followed a more pragmatic approach, based on the specific morphology
of the SF knots: we integrated within a boundary that follows the shape of the
SF knot with a minimum area of $\sim 20$ square arcsec (that is an equivalent
radius of about twice the seeing FHWM).
For galaxies with multiple SF knots, the above criterion was relaxed; see the
discussion on each individual object (Sect.~\ref{Section:Discussion}).
The SF knots are outlined and labeled in the \Ha\ maps in 
Figs.~\ref{Figure:mrk407}--\ref{Figure:izw159}. 

As for the integrated spectrum of a galaxy within the PMAS FOV, in order to
not degrade its signal-to-noise ratio  unnecessarily by including outer
spaxels with no gaseous emission, we only summed those spaxels with
measured \Ha\ emission (that is those shown in the \Ha\ flux maps).

\begin{figure*}
\mbox{
\centerline{
\hspace*{0.0cm}\subfigure{\includegraphics[width=0.48\textwidth]{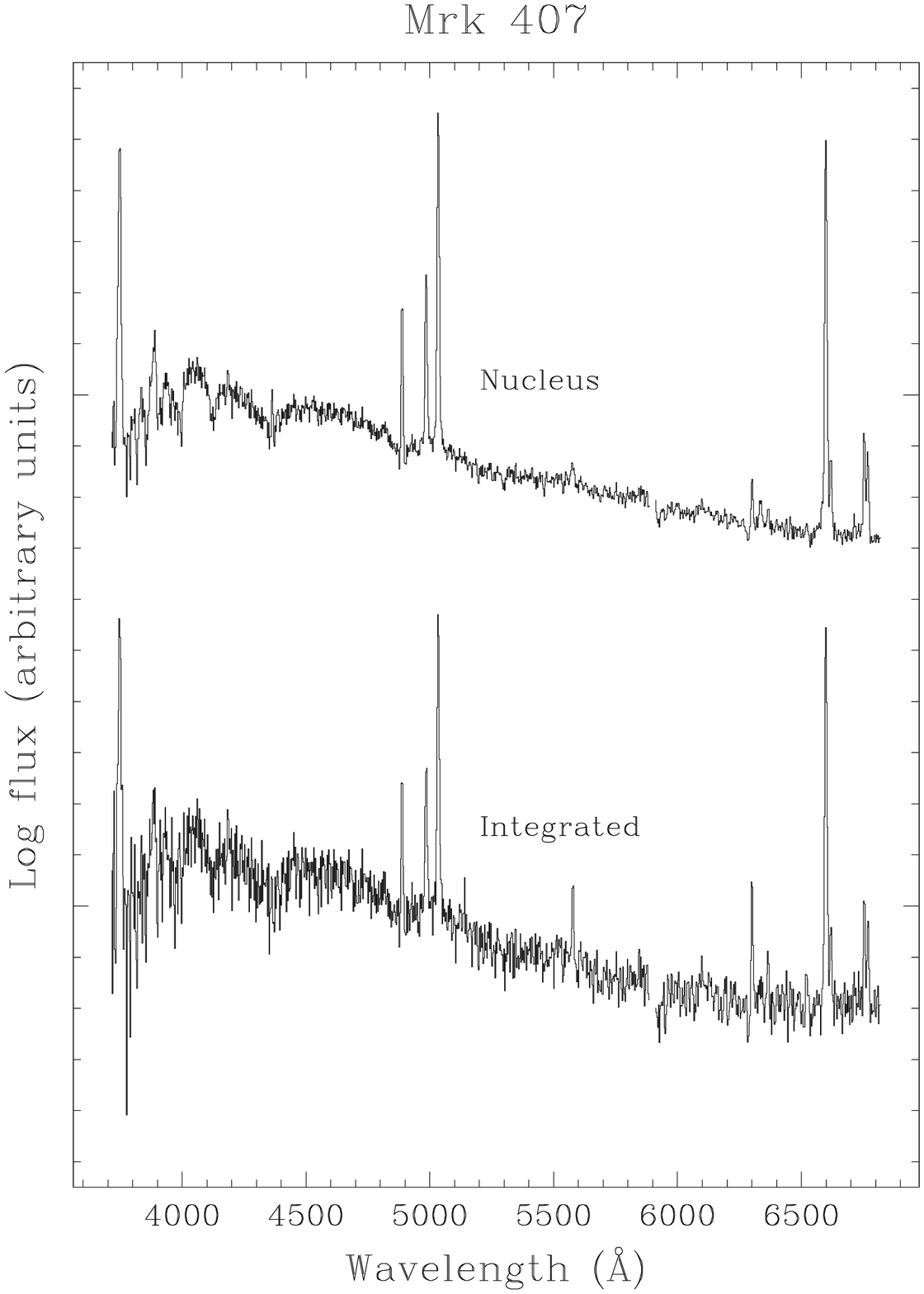}}
\hspace*{0.0cm}\subfigure{\includegraphics[width=0.48\textwidth]{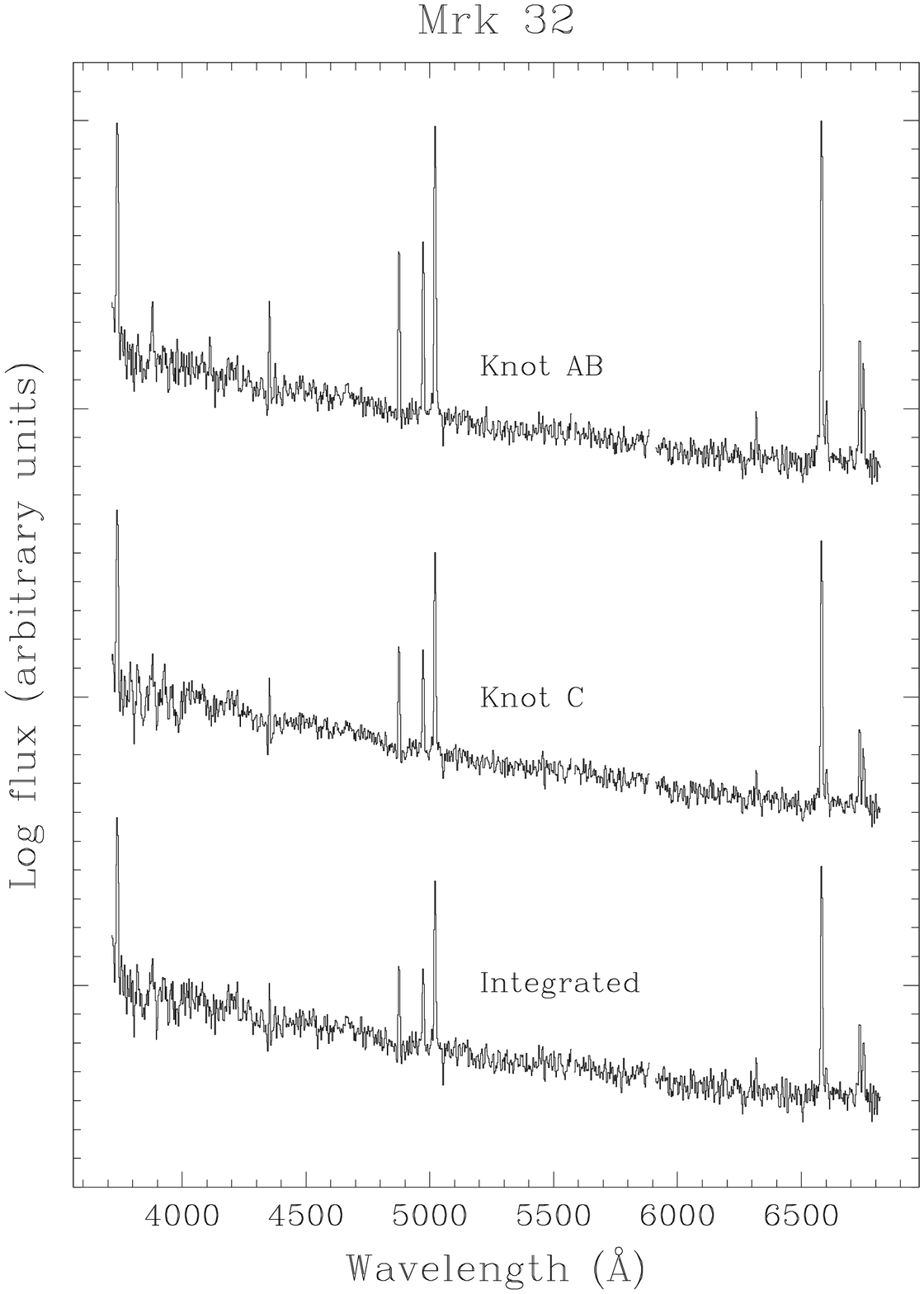}}
}}  
\mbox{
\centerline{
\hspace*{0.0cm}\subfigure{\includegraphics[width=0.48\textwidth]{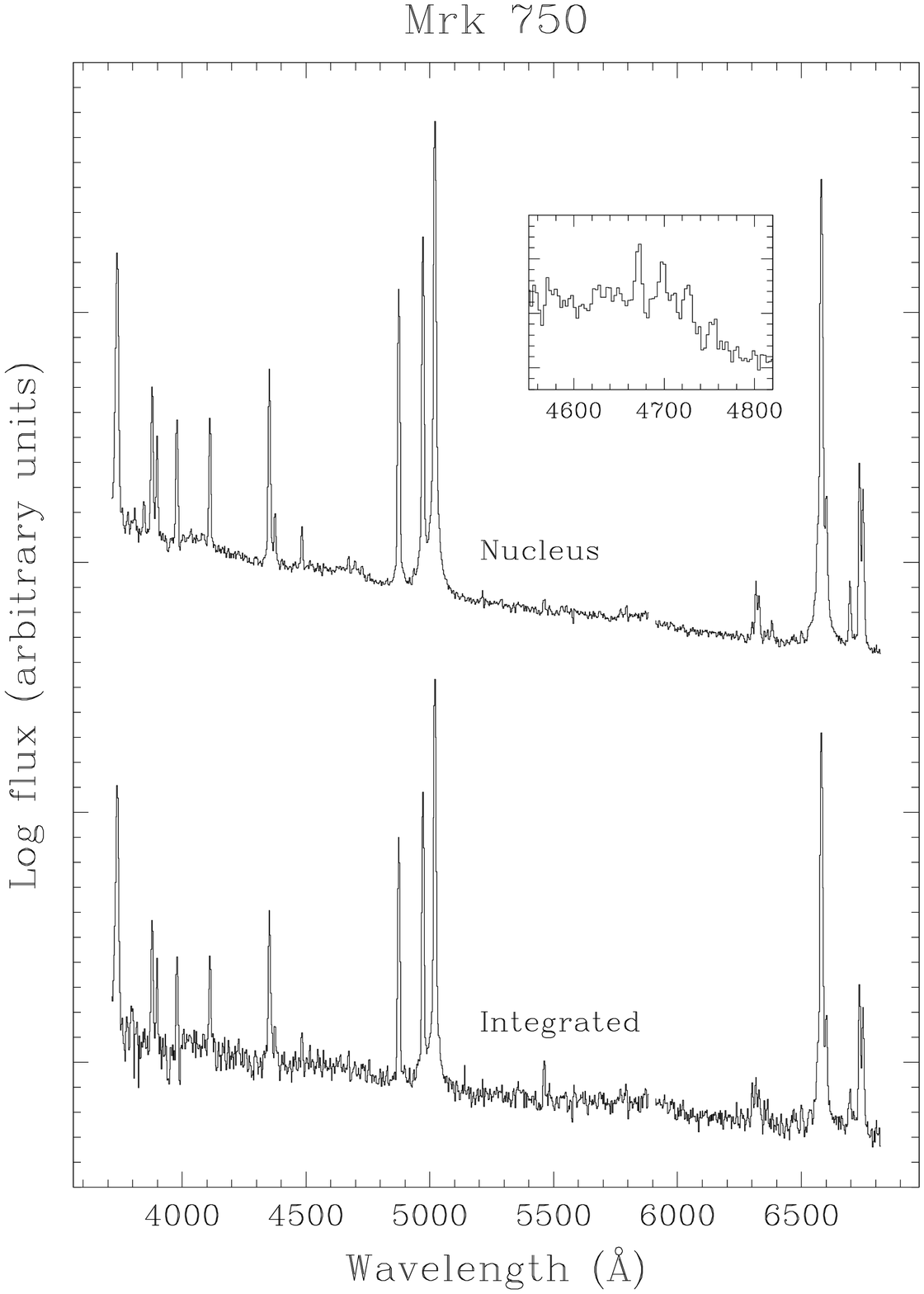}}
\hspace*{0.0cm}\subfigure{\includegraphics[width=0.48\textwidth]{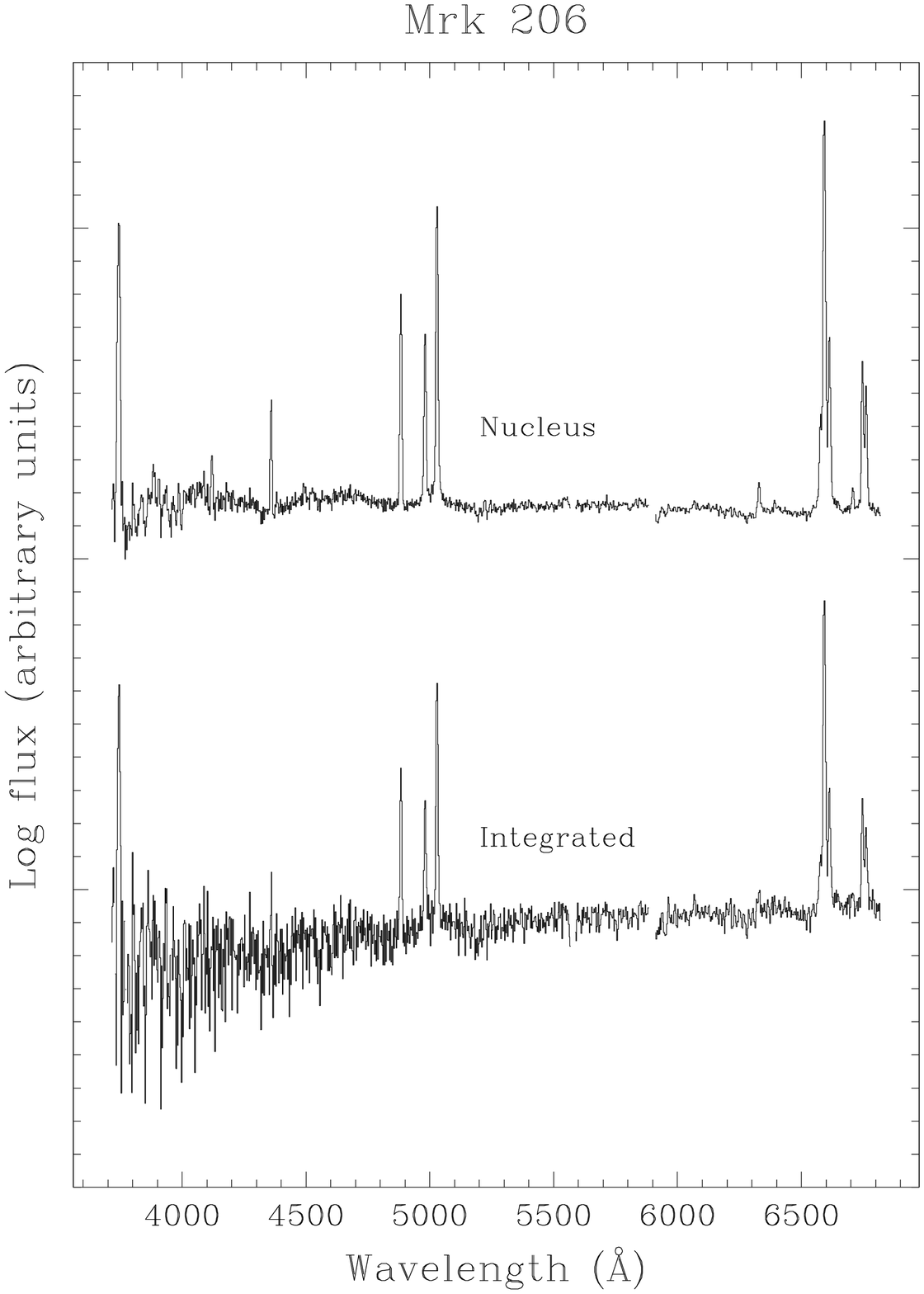}}
}}
\caption{Nuclear and integrated spectra for Mrk~407, Mrk~750 and Mrk~206; 
for Mrk~32, spectra of the identified SF regions and the integrated spectrum.
The inset in the Mrk~750 figure shows in detail the blue WR bump 
region in the nuclear spectrum. Spectra are shown in logarithmic scale and 
are offset for clarity.
The interval between large tickmarks is 1 dex (0.05 dex in the inset).
}
\label{Figure:spectra}
\end{figure*}

\begin{figure*}
\mbox{
\centerline{
\hspace*{0.0cm}\subfigure{\includegraphics[width=0.48\textwidth]{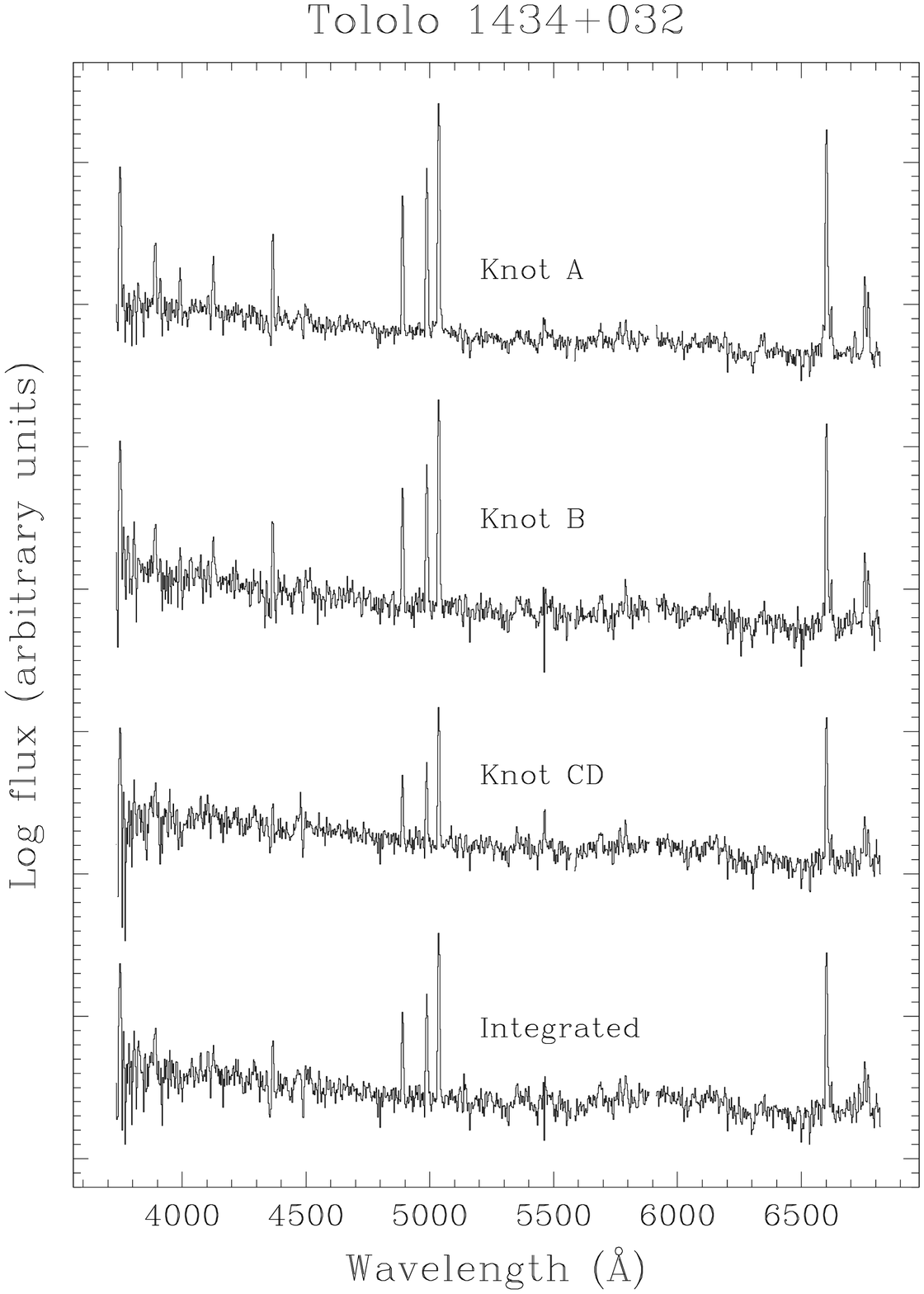}}
\hspace*{0.0cm}\subfigure{\includegraphics[width=0.48\textwidth]{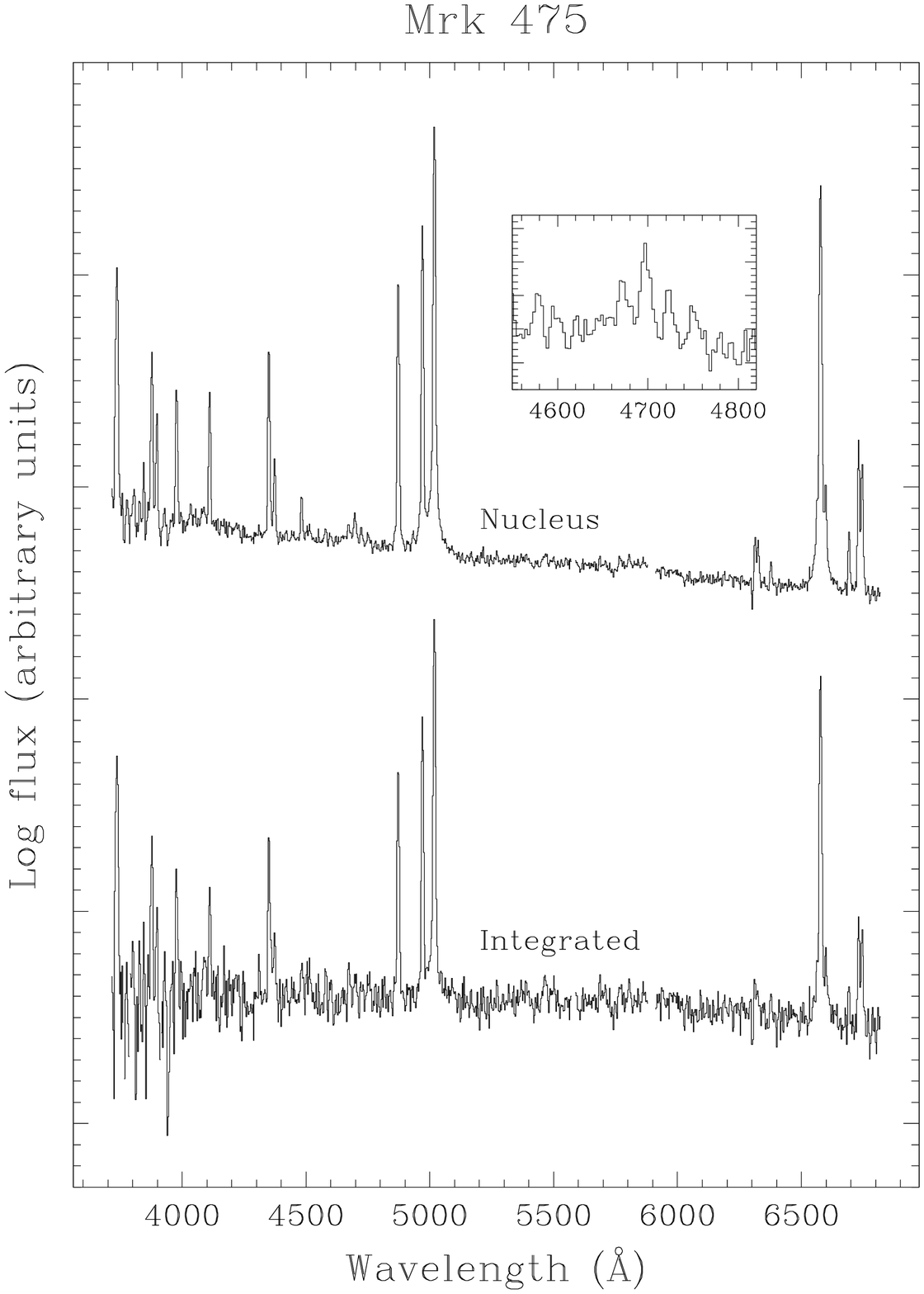}}
}}
\mbox{
\centerline{
\hspace*{0.0cm}\subfigure{\includegraphics[width=0.48\textwidth]{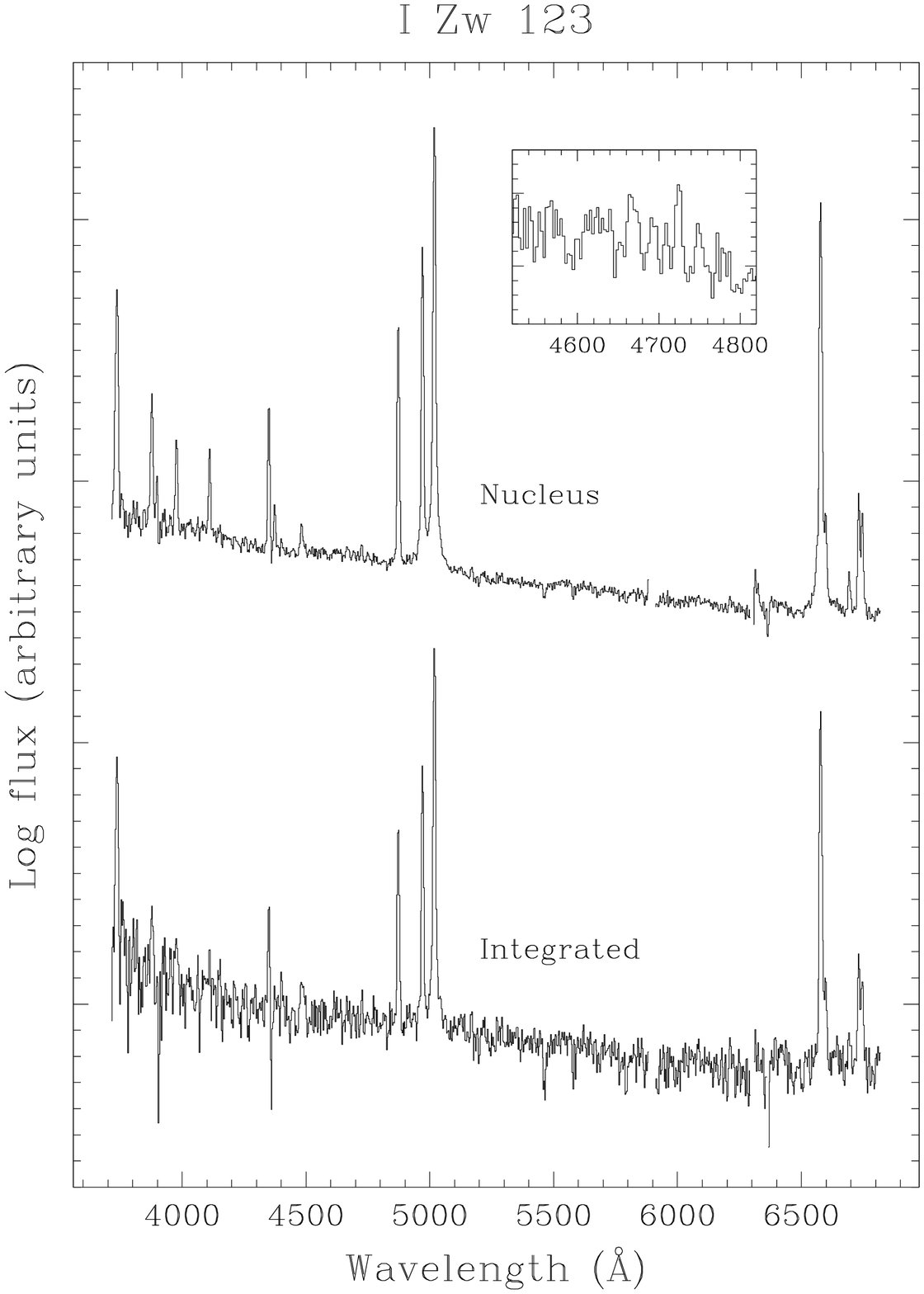}}
\hspace*{0.0cm}\subfigure{\includegraphics[width=0.48\textwidth]{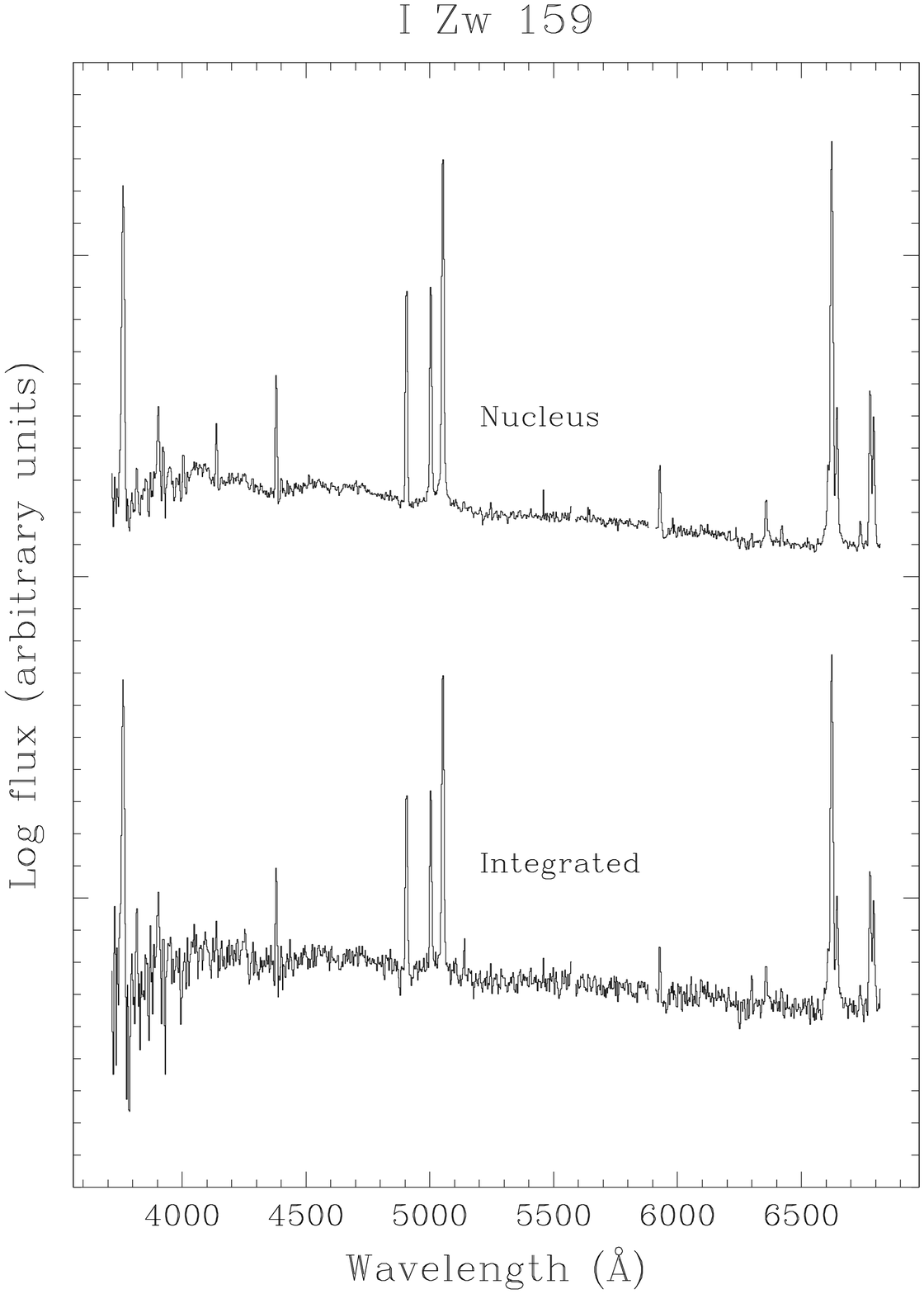}}
}} 
\caption{Nuclear spectra and integrated spectra for Mrk~475, I~Zw~123 and 
I~Zw~159;  for Tololo~1434+032, spectra of the identified SF regions and the 
integrated spectrum.
The insets in the Mrk~475 and I~Zw~123 figures show in detail the blue WR  
bump region in the nuclear spectrum.}
\label{Figure:spectra2}
\end{figure*}

Figs.~\ref{Figure:spectra} and \ref{Figure:spectra2} display the spectra of
the sample galaxies. In general all spectra are dominated by bright,
narrow emission lines, indicating an important contribution of ionizing stars,
but with significant differences among the objects: several galaxies display a
very high continuum with strong absorption features, whereas some of them show
a flat spectrum, characteristic of an OB population. A more detailed
description of the spectral characteristics of the individual objects is
provided in Sect.~\ref{Section:Discussion}.

\subsubsection{Line fluxes and reddening correction}
\label{SubSubSection:fluxes}

\textnormal{The higher S/N ratio of the integrated spectra allows us on one hand
a more accurate measurement of the Balmer line fluxes, and on the
other a more careful and reliable determination of the extinction coefficient.}

For each spectrum we measured fluxes and equivalent widths of the
emission lines using the Gaussian profile fitting option in the IRAF task
\emph{splot}. In order to obtain reliable values of Balmer fluxes in emission
we must take into account the underlying stellar absorption
\citep{McCall1985,Diaz1988}. To do that, we followed two different approaches,
depending on the characteristics of the spectra. 

When the absorption wings around the Balmer lines were not visible, we assumed
that the equivalent width in absorption is the same for all the lines. We
first adopted an initial estimate for the absorption equivalent width,
EW$_\mathrm{abs}$, corrected the measured fluxes, and computed the extinction
coefficient \CHbeta\ through a least-square fit to the Balmer decrement. We
then varied the value of EW$_\mathrm{abs}$, until we found the one that
provided the best match (e.g. the minimum scatter in the fit) between the
corrected and the theoretical line ratios. A more detailed description of this
method can be found in \cite{Izotov1994} and \cite{Cairos2007}.

We adopted the ``case B'' Balmer recombination decrement for
$T_\mathrm{e}=10\,000$ K and $N_\mathrm{e}=10^{4}$ cm$^{-3}$ 
\citep{Brocklehurst1971} and the \cite{Cardelli1989} reddening curve.

In those cases where absorption wings around the Balmer lines were visible, we
simultaneously fitted an absorption and an emission component.
We then applied the same method as before, varying the equivalent width in
absorption only in those lines in which it could not be fitted.

In several cases the values of the \Hd\ and \Hg\ fluxes are doubtful due to
their intrinsic weakness and the large uncertainties in the correction for the
underlying stellar absorption. In these cases we computed \CHbeta\ directly 
from the \Ha/\Hb\ ratio: (i) if the absorption in \Hb\ was fitted, we set
$\mathrm{EW}(\Ha)_{abs}=\mathrm{EW}(\Hb)_{abs}$; (ii) if not, 
following \cite{McCall1985} we set the equivalent width of both lines in 
absorption to 2 \AA.

Reddening-corrected intensity ratios and equivalent widths for the different
spatial regions are listed in 
Tables~\ref{Table:Linesflux1}--\ref{Table:Linesflux4}. 


\subsubsection{Physical parameters and abundances} 
\label{SubSubSection:PhysicalParameters}

Physical properties and ionic abundances were derived from the
reddening-corrected emission line fluxes, following the 5-level atom 
\emph{fivel} program in the IRAF \emph{nebular} package
\citep{deRobertis1987,ShawDufour1995}. 

Electron densities were measured from the emission line ratio 
[\ion{S}{ii}]$\lambda$6717/$\lambda$6731; electron temperatures
($T_\mathrm{e}$[\ion{O}{iii}]) were derived from the [\ion{O}{iii}]~$\lambda
$4363/($\lambda$4959+$\lambda$5007) ratio, in those spectra where
[\ion{O}{iii}]~$\lambda4363$ was measured with high enough S/N. In these
cases, $T_\mathrm{e}$[\ion{O}{ii}] was calculated from the relation between
$T_\mathrm{e}$[\ion{O}{ii}] and $T_\mathrm{e}$[\ion{O}{iii}] provided in
\cite{Pilyugin2006}. 

We then adopted $T_\mathrm{e}$[\ion{O}{ii}] for the calculation of 
N$^{+}$, O$^{+}$ and S$^{+}$ abundances, and $T_\mathrm{e}$[\ion{O}{iii}] 
for the calculation of O$^{2+}$ and Ne$^{2+}$ abundances.  
We used [\ion{Ne}{iii}]~$\lambda3869$ to
derive the neon abundance, [\ion{O}{ii}]~$\lambda3727$ and
[\ion{O}{iii}]~$\lambda\lambda4959,5007$ for the oxygen abundance, 
[\ion{N}{ii}]~$\lambda\lambda6548,6584$ for nitrogen abundance and 
[\ion{S}{ii}]~$\lambda\lambda6717,6731$ for the sulfur abundance. The total
neon and  nitrogen abundances were obtained as Ne/O = Ne$^{++}$/O$^{++}$ and
N/O = N$^{+}$/O$^{+}$, respectively, and the total oxygen abundance was
calculated as O/H = (O$^{+}$/H$^{+}$ + O$^{++}$/H$^{+}$) . 

To obtain oxygen abundances in those knots in which
[\ion{O}{iii}]~$\lambda4363$ could not be measured, we applied the commonly
used strong-line method from \cite{PettiniPagel2004}.  

Physical properties and chemical abundances are listed in
Tables~\ref{Table:Parameters} and \ref{Table:Parameters2}, where the most
common ratios for diagnostic are also included. The final quoted uncertainties
were derived by error propagation taking into account the errors in flux
measurements.

\section{Discussion}
\label{Section:Discussion}

\subsection{Mrk~407}
\label{SubSection:Mrk407}

The galaxy Mrk~407 is included in the \cite{Petrosian2007} Atlas of Markarian
galaxies, where it appears classified as an S0. $J$, $H$, and $K$ NIR surface
brightness photometry was published in \cite{Cairos2003}, but the quality of
these data was insufficient for assessing whether or not this galaxy has an
older stellar low surface brightness component underlying the SF regions. The
only  spectroscopic data available in the literature are its redshift and the
equivalent widths and flux ratios of the strongest lines
\citep{Ugryumov1998}. 

The PMAS FOV covers an area of $2.1 \times 2.1$ kpc with a spatial sampling of
130 pc per spaxel. Two-dimensional maps of Mrk~407 are shown in
Fig.~\ref{Figure:mrk407}.

Emission line and continuum maps show an overall regular morphology, with the
intensity distribution peaking roughly at the center of the outer
elliptical isophotes. In the emission line maps the isophotes are elongated 
along the northwest-southeast axis, whereas in the continuum frames they
are of a roughly circular morphology, although in the outer regions they
appear slightly distorted and elongated along the northwest-southeast
direction ---~as previously seen in the NIR contour maps published in
\cite{Cairos2003}.

The excitation maps trace the SF knots, as expected in regions photoionized
by stars, with values typical of \ion{H}{ii} regions \citep{Veilleux1987}. The
[\ion{O}{iii}]/\Hb\ map also peaks southeast of the central SF region, at the
same spatial location as the extinction maximum. 

The galaxy has an inhomogeneous extinction pattern, with a dust lane crossing
it southwest--northeast, where \Ha/\Hb\ reaches values
of up to 5.5 ---~which translates into an interstellar reddening \EBV\ of up
to 0.6. In the rest of the galaxy, the values of the \Ha/\Hb\ ratio are closer 
to the theoretical value of 2.86.

Both the nuclear and the integrated spectra of Mrk~407 display strong emission
lines atop a blue continuum, with several absorption features (high order
Balmer lines in absorption and pronounced absorption wings in \Hd, \Hg\ and
\Hb; see Fig.~\ref{Figure:spectra}); the absorption features are indicative
of a substantial contribution from older stars. As we could not detect
[\ion{O}{iii}]~$\lambda4346$, the oxygen abundance was derived from the
empirical calibrations. We found a value of $0.4 Z_{\sun}$, which places
Mrk~407 in the high metallicity BCD group.

While the \Ha\ and [\ion{O}{iii}] velocity maps are quite noisy, they both 
seem to indicate an overall rotation around an axis roughly oriented on the
northwest-southeast direction, with a velocity amplitude of about 30--40
km s$^{-1}$.

\subsection{Mrk~32 (=~UGCA~211, SBS~1023+565)}  
\label{SubSection:Mrk32}

With $M_{B}=-14.99$, Mrk~32 is one of the faintest galaxies in our sample. In
the \cite{Petrosian2007} Atlas it is classified as an Im/BCD. Broad-band
imaging and optical spectroscopy were published in \cite{Hunter2006} and
\cite{Hunter1999}, respectively.

The PMAS FOV covers an area of $1.27\times1.27$ kpc with a sampling of 80 pc
per spaxel. Two-dimensional maps of Mrk~32 are shown in
Fig.~\ref{Figure:mrk32}.

In the continuum the galaxy shows an overall regular behavior, with elliptical
isophotes, much elongated in the north-south direction. Although the intensity
distribution clearly increases towards the central regions, there is not a
clear peak, but a somehow diffuse maximum, whose position seems to be
displaced to the west when moving towards red wavelengths.  All the
emission-line maps display a similar pattern, with the SF knots roughly
aligned in the south-north direction, and all, except [\ion{O}{ii}], peak at
the northeast knot (A, as labeled in Fig.~\ref{Figure:mrk32}); none of the
three SF knots seen in emission-lines coincides with the central intensity
peak in the continuum maps. 

The excitation maps trace the three SF regions; the maximum in
[\ion{O}{iii}]/\Hb\ (the minimum in the other maps) is located in the
brightest knot A. Line ratio maps display values typical of \ion{H}{ii}
regions, aside from the southern regions whose higher values for 
[\ion{O}{i}]/\Ha\ and [\ion{S}{ii}]/\Ha\ are more characteristic of
LINERS/Seyfert galaxies (log[\ion{O}{i}]/\Ha\ $\geq -1.3$,
log[\ion{S}{ii}]/\Ha\ $\geq -0.4$; \citealp{Veilleux1987}); this suggests that
another mechanism, most probably shocks, is contributing to the gas
excitation.

The \Ha/\Hb\ ratio map displays a noisy pattern, with most of the spaxels
showing values close to 2.86; however, several spaxels located in the outer
regions and close to knot C have higher values ---~consistent with the
higher reddening values derived for the integrated spectrum of knot C. 

We produced the integrated spectrum of knot AB (because of their closeness,
knots A and B were lumped together although they seem to be two separate
SF regions) and C (whose small size is dictated by the need of keeping it
separated from AB), and of the integrated galaxy spectrum. All three
spectra are dominated by young stars, with strong emission lines atop a blue
continuum; in knot C, and in the integrated spectrum, absorption wings around
Balmer lines are detected. We also have a marginal detection of the Wolf-Rayet
(WR) bump at $\lambda4650$ \AA\ in the two spaxels marked in
Fig.~\ref{Figure:mrk32}. Unfortunately, the S/N ratio of the WR bump is too
low for a reliable measurement of its flux.

We found a low interstellar extinction, $\EBV\leq0.12$, considerably lower
than the $\EBV=0.48$ value reported by \cite{Hunter1999}.

\textnormal{For knot AB (the only one where we could measure the
[\ion{O}{iii}]~$\lambda4363$ line), we derived an oxygen abundance of
$12+\log(\mathrm{O/H})=7.56$  (about $1/13Z\odot$), which places Mrk~32 among
the extremely metal-deficient BCDs ---~galaxies having 
$12+\log(\mathrm{O/H})\leq7.6$ \citep{Papaderos2008,Kunth2000}.}

\textnormal{This value is considerably lower (by 0.6 dex) than that derived
with the empirical calibrations. This is in line with previous works that
found that empirical calibrations yield oxygen abundances systematically
higher than the $T_\mathrm{e}$-based abundances (see for instance
\citealp{Shi2005}, where discrepancies on the order of 0.45 dex have been
found). However, this considerable disagreement may also arise from the
relatively large measurement uncertainties for the faint
[\ion{O}{iii}]~$\lambda4363$ emission line (according to
\cite{Kobulnicky1999}, Gaussian fits to emission lines with very low S/N
ratios are systematically biased towards higher values). Deep spectroscopic
observations are required to derive accurate  $T_{\mathrm e}$-based
abundances.}

The velocity field appears essentially flat, with perhaps a hint of a small
increase to the southwest.

\subsection{Mrk~750}
\label{SubSection:Mrk750}

The galaxy Mrk~750 is an extremely faint ($M_{B}=-12.82$) BCD, classified as
BCD/Im in  \cite{Petrosian2007}. It belongs to the sample of low metallicity
galaxies \citep{Izotov1999,Izotov2007}, and it is also a well-known Wolf-Rayet
galaxy: \cite{KunthJoubert1985} first pointed out the broad
\ion{He}{ii}~$\lambda4686$ \AA\ emission, and \cite{Conti1991} reported the
detection of \ion{N}{iii}~$\lambda4640$ \AA; these results are consistent with
those found in \cite{Izotov1998}; \cite{Guseva2000} also noticed 
\ion{C}{iv}~$\lambda5808$ \AA.  Broad-band optical and \Ha\ observations
were published by \cite{Mendez2000}. From  high spatial resolution
\ion{H}{i} synthesis observations, \cite{VanZee2001} found that the neutral
gas extends to approximately twice the optical diameter of the galaxy,
peaking in the central region of star formation.  

Our IFU data cover a region of $400\times400$ pc, with a sampling of 25 pc
per spaxel. Two-dimensional maps of Mrk~750 are shown in
Fig.~\ref{Figure:mrk750}.

The galaxy has a similar morphology both in the emission line and in the
continuum maps, with roughly circular isophotes and a single central peak,
whose position is the same in all maps (emission lines and continuum).
However, while in the emission line maps the isophotes are circular at all
intensity levels, in the continuum maps the outer ones appear elongated to the
north-east, indicating that the central, bigger knot of Mrk~750 is connecting
with a continuum source. This peculiar morphology was interpreted as an
interaction sign in \cite{Mendez2000}.

The excitation maps all display the same pattern and all trace the central SF
knot. The ratio values in the four maps are consistent with ionization by hot
stars.

In the extinction map we clearly distinguish a dust patch dominating  the
southeast part of the galaxy, with \Ha/\Hb\ values up to 3.8 --- \EBV\ up to
0.27. 

The integrated and nuclear spectra are both blue, OB young stars dominated
spectra with very prominent emission lines; neither Balmer absorption wings
nor absorption lines are visible. In agreement with previous results, we 
detected the blue Wolf Rayet bump around 4640\AA\ (see
Fig.~\ref{Figure:spectra}); this feature is prominent in ten spaxels, located
all in the nuclear region of the galaxy. We also have a marginal detection of
\ion{C}{iv}~$\lambda5808$ \AA\ (the red Wolf-Rayet bump) in three central
fibers.  The position of the spaxels with the Wolf-Rayet signature is shown
in the [\ion{O}{iii}] flux map with crosses and squares for the blue and the 
red bumps respectively (see Fig.~\ref{Figure:mrk750}). 

\textnormal{We computed the fluxes and equivalent width of the blue WR bump
in the nuclear spectrum by a plain integration of the signal in the 
corresponding spectral region after fitting and subtracting the underlying 
continuum. 
Because the WR bumps are a blend of both genuine WR features and nebular lines, 
our measurement is actually an upper limit (the S/N ratio and the resolution 
of the spectrum do not allow us to fit and subtract separately the nebular
lines).  The extinction-corrected flux is $F(\mathrm{WR_{blue}}) \sim 100
\times 10^{-16}$ ergs cm$^{-2}$ s$^{-1}$, and the equivalent width is
$\mathrm{EW(WR_{blue})} \sim 10$.}


The oxygen abundance derived for the nuclear spectra using the $T_\mathrm{e}$
based method, $12+\log(\mathrm{O/H})=8.12\pm0.03$, excellently agrees
with the value $12+\log(\mathrm{O/H})=8.11\pm0.02$ reported by
\cite{Izotov1998}. There is also good agreement with the values we derived 
with empirical calibrations.

Both the \Ha\ and [\ion{O}{iii}] velocity maps, while somewhat noisy, display
a rotation pattern around an axis roughly oriented east-west, with an
amplitude of about 20 km s$^{-1}$; this result qualitatively agrees
with the \ion{H}{i} kinematics published by \cite{VanZee2001}.

\subsection{Mrk~206 (=UGCA~280, SHOC~371)} 
\label{SubSection:Mrk206}

The galaxy Mrk~206 is classified as a BCD in \cite{Petrosian2007}. Intensities
and equivalent widths of hydrogen and oxygen emission lines as well as oxygen
abundances are published in \cite{Kniazev2004}.

Our IFU data cover a region of $1.9\times1.9$ kpc with a spatial sampling of
120 pc per spaxel. Fig.~\ref{Figure:mrk206} displays the two-dimensional
maps of Mrk~206.

The galaxy shows a regular morphology both in emission line and continuum
maps with a central SF region and roughly circular isophotes. 

All the excitation maps trace this central SF region, and in all of them the
ratio values are consistent with ionization by young stars.  

The extinction pattern on the other hand is highly inhomogeneous, with a
substantial amount of dust located on the northeast region of the galaxy,
with \Ha/\Hb\ ratios up to 6 ---~\EBV\ up to 0.7.

The galaxy displays a flat, typical \ion{H}{ii} region spectrum, with no
evident absorption features. With a metallicity of about $0.6Z_{\sun}$ it is
the galaxy with the highest metallicity in our sample. However, as we fail to
detect [\ion{O}{iii}]~$\lambda$4363, only empirical abundances could be
obtained. \cite{Kniazev2004} published a $T_\mathrm{e}$  based method
abundance of $12+\log(\mathrm{O/H})=8.04$, $\approx 0.4$ dex lower than that
derived here.

The velocity field shows a clear overall rotation along a southeast-northwest
axis and interestingly seems much like the extinction pattern seen in the
\Ha/\Hb\ map. The northeast side is approaching, the southwest side is
receding; the velocity amplitude inside the mapped region is about 50--60 km
s$^{-1}$.

\subsection{Tololo~1434+032 (=~SHOC~474)}
\label{SubSection:Tol1434}

Tololo~1434+032 is a low-metallicity BCD \citep{Izotov2007,Kniazev2004}. $B$
and $R$ broad-band photometry was published in \cite{Doublier1997} and
\cite{GildePaz2003}, and \Ha\ imaging in  \cite{GildePaz2003}. 

The PMAS FOV covers a region of $2.3\times2.3$ kpc, with a spatial sampling
of 140 pc per spaxel. Two-dimensional maps of Tololo~1434+032 are shown in 
Fig.~\ref{Figure:tololo1434}.

The galaxy shows a clumpy morphology both in emission lines and in the
continuum. In the continuum maps we resolve two major emission peaks. The
strongest is located in the southeast, while the other is displaced
about 9 arcsec (1.3 kpc) to the northwest. All emission lines maps have the
same structure:  four SF knots are distributed in a roughly circular pattern,
with the peak of emission located in knot A, whose position coincides with the
continuum peak (knots are labeled in the \Ha\ map in
Fig.~\ref{Figure:tololo1434}). Northeast of A there is a smaller knot, B,
while two fainter knots, C and D, are seen in the north side: neither
coincides spatially with the secondary continuum peak. (Because of their
small size and luminosity, C and D were lumped together to obtain their
integrated spectrum.)

The excitation maps trace the regions of star formation and display in the
whole field of view values typical of \ion{H}{ii} regions. 

The galaxy shows an homogeneous extinction pattern with values close to the
theoretical value of 2.86 across the whole FOV.

The spectra of the resolved SF knots are very similar and also resemble well
the integrated spectrum: they are all flat, young star-dominated spectra
with no evidence of absorption features. The oxygen abundance we found
for the brightest knot, using the $T_\mathrm{e}$ method, 
$12+\log(\mathrm{O/H})=8.05\pm0.07$, compares well with the value derived in 
\cite{Kniazev2004}, $12+\log(\mathrm{O/H})=7.97\pm0.04$. The values derived 
through empirical calibrations are also very similar.

Both the \Ha\ and the [\ion{O}{iii}] velocity maps seem to marginally indicate
a low amplitude rotation ($\lesssim 20$ km s$^{-1}$) around a
northeast-southwest axis.

\subsection{Mrk~475}
\label{SubSection:Mrk475}

The galaxy Mrk~475 is a low luminosity object, included in Izotov's sample of
metal poor galaxies \citep{Izotov1994,Izotov2007}, and is classified as a BCD
in  \cite{Petrosian2007}. $B$ and $R$ broad-band surface brightness photometry
and \Ha\ imaging were published by \cite{GildePaz2003}. It is a WR
galaxy, where nebular and broad \ion{He}{ii}~$\lambda4686$ lines and
\ion{C}{iv}~$\lambda5808$ were detected
\citep{Conti1991,Schaerer1999,Guseva2000}. 

Our IFU data cover a region of $920\times920$ pc, with a spatial sampling 
of 58 pc per spaxel. Two-dimensional maps of Mrk~475 are displayed in 
Fig.~\ref{Figure:mrk475}.

This is a compact, regular object, with a single central SF knot. Emission
lines and continuum maps display all the same morphology.

The ionization maps show all the same complex pattern: while in the
western galaxy regions the excitation ratio decreases with the distance from
the central SF region, in the eastern part it displays a constant value. The
four ratio maps show values typical of excitation by hot stars.

The extinction map is inhomogeneous and has a peak in the southwest;
extinction values are moderate, with \Ha/\Hb\ peaking around 3.

The galaxy displays a blue spectrum, with prominent emission lines and no 
visible absorption features. We detected the blue WR bump in eight spaxels
(marked with crosses in Fig.~\ref{Figure:mrk475}); in two of them the red WR
bump is also visible (squares in Fig.~\ref{Figure:mrk475}).  
\textnormal{We measured the flux and equivalent width of both WR bumps in the
nuclear spectrum in the same way we did for Mrk~750, finding fluxes of $\sim
80 \times 10^{-16}$ and $\sim 20 \times 10^{-16}$, and equivalent widths of 15
and 4 \AA\ for the blue and the red bump respectively.}

The oxygen abundance we found for the nuclear region, using the $T_\mathrm{e}$
method, $12+\log(\mathrm{O/H})=7.92\pm0.02$, agrees well with the value of
$7.97\pm0.04$ reported by \cite{Izotov1994}. Values derived using the
empirical calibrations are slightly higher, but still in good agreement
(difference $\leq0.15$ dex).

Both the \Ha\ and the [\ion{O}{iii}] kinematical maps show a flat
velocity field.

\subsection{I~Zw~123 (=~UGCA~410, Mrk~487)}  
\label{SubSection:Izw123}

The galaxy I~Zw~123 is classified as a BCD in \cite{Petrosian2007} and belongs
also to the metal-poor BCD class \citep{Izotov1997,Izotov1999}. I~Zw~123 is a
relatively well-studied object: optical surface photometry was published in
several papers 
\citep{Cairos2001II,Cairos2001I,GildePaz2003,Caon2005,Amorin2007}, NIR
photometry in \cite{Cairos2003}. A thorough spectroscopic study, including an
analysis of its stellar content in terms of population synthesis models, was
published in the series of papers by \cite{Kong2002spectra},
\cite{Kong2002correlations}, \cite{Kong2003}, \cite{Kong2004} and 
\cite{Shi2005}.

The PMAS FOV covers an area of $1.2 \times 1.2$ kpc, with a spatial sampling
of 75 pc per spaxel. Two-dimensional maps of I~Zw~123 are shown in 
Fig.~\ref{Figure:izw123}.

In the continuum this galaxy is compact, with the intensity peak located in the
center of the outer circular isophotes.  All the emission-line maps have the
same pattern, with a single SF region slightly displaced (2 arcsec or 150 pc)
to the north of the continuum peak; the central isophotes are elongated in the
south direction, resembling a small tail. 

The three excitation maps display the same structure, tracing the SF region;
the values are typical of regions photoionized by stars, except in the 
[\ion{S}{ii}]/\Ha\ map, where the outer regions have relatively high values
(log[\ion{S}{ii}]/\Ha $\geq-0.4$); this could indicate that shocks are playing
a significant role.

The extinction map displays a maximum located in the same position as the SF
knot; the peak values of the \Ha/\Hb\ ratio implies an \EBV\ of about 0.3
mag.

The galaxy exhibits a blue spectrum, with prominent emission lines and no
evident absorption features. The oxygen abundance that we find by applying the
$T_\mathrm{e}$ method, $12+\log(\mathrm{O/H})=8.13\pm0.03$ agrees well 
with the value reported in \cite{Izotov1999}, 
$12+\log(\mathrm{O/H})=8.06\pm0.04$.

In both the \Ha\ and [\ion{O}{III}] velocity maps the velocity field appears
flat (the higher velocities seen in the outermost spaxels in the
[\ion{O}{III}] map are most likely due to noise).

\subsection{I~Zw~159 (=~UGCA~412, Mrk~1499)}
\label{SubSection:Izw159}

In \cite{Petrosian2007} I~Zw~159 is classified as a BCD/Irr. Optical surface
brightness photometry was published in
\cite{Doublier1997}, \cite{Doublier1999}, \cite{GildePaz2003} and 
\cite{GildePaz2005}. It is also
included in the spectroscopy study of BCDs by
\cite{Kong2002spectra}, \cite{Kong2002correlations}, \cite{Kong2003}, \cite{Kong2004} and 
\cite{Shi2005}, and is the only object
in our sample in common with the sample of galaxies studied by means of IFS by
\cite{Petrosian2002}.

The PMAS FOV covers a region of $3.4\times3.4$ kpc, with a spatial resolution
of 210 pc per spaxel. Two-dimensional maps of I~Zw~159 are shown in 
Fig.~\ref{Figure:izw159}. 

In the continuum the galaxy shows a central intensity peak and a boxy
morphology, with isophotes elongated around a northeast-southwest axis. In
all the emission lines we see the same pattern: a central peak, whose position
coincides with the continuum peak, surrounded by elongated isophotes.

All the excitation maps trace the SF knot; whereas in the
[\ion{O}{iii}]~$\lambda5007$/\Hb\ and [\ion{N}{ii}]~$\lambda6584$/\Ha\ maps
the values are consistent with  photoionization by young stars in the whole
galaxy, in the [\ion{S}{ii}]/\Ha\  and [\ion{O}{i}]/\Ha\  maps the outer
regions of the galaxy have higher values (log[\ion{O}{i}]/\Ha $\geq-1.0$,
log[\ion{S}{ii}]/\Ha $\geq -0.4$), indicating that an additional mechanism,
probably shocks, is acting there.

The galaxy shows an irregular extinction pattern, with a dust patch located in
the northern regions, where \EBV\ is as high as 0.45 mag.

The integrated spectra are blue, with some absorption wings around the Balmer
lines. With an oxygen abundance $12+\log(\mathrm{O/H})\simeq 8.3$, this galaxy 
also belongs to the high-metallicity BCDs branch. 

The velocity field in the \Ha\ and [\ion{O}{iii}] maps shows a clear overall
rotation along an axis roughly oriented south-north, with an amplitude of
about 70--80 km s$^{-1}$, in broad agreement with the velocity field
published by \cite{Petrosian2002}.

\section{Summary and conclusions}
\label{Section:Conclusions}

We present here what is to our knowledge the most extensive IFS analysis of
a sample of BCDs. This study is based on PMAS data, which cover a wavelength
range of 3590-6996 \AA, with a linear dispersion of 3.2 \AA per pixel, and map
an area $16\arcsec\times 16\arcsec$ with a spatial sampling of 
$1\arcsec\times 1\arcsec$. 

For all the sample galaxies we produced an atlas of two-dimensional
maps: two continuum bands, the brightest emission lines (i.e.
[\ion{O}{ii}]~$\lambda3727$,  \Hb\ , [\ion{O}{iii}]~$\lambda5007$,
[\ion{O}{i}]~$\lambda6300$, \Ha, [\ion{N}{ii}]~$\lambda6584$ and
[\ion{S}{ii}]~$\lambda\lambda6717,\;6731$) and the most relevant line ratios
(i.e. [\ion{O}{iii}]/\Hb, [\ion{O}{i}]/\Ha, [\ion{N}{ii}]/\Ha,
[\ion{S}{ii}]/\Ha\ and \Ha/\Hb) as well as the velocity field of
the ionized gas. Integrated spectroscopic properties of the most
prominent SF regions and of the whole galaxy have been also derived.

From this work we highlight the following results:

\begin{enumerate} 

\item All the objects except Mrk~750 and Tololo~1434+032 exhibit a mostly
regular morphology in the continuum, with one (or several for Mrk~32 and
Tololo 1434+032) central SF regions placed atop a more extended host galaxy.
The galaxy Mrk~750 reveals elongated outer isophotes, and Tololo~1434+032
displays a clumpy continuum morphology. 

All the galaxies show a similar morphology in the different mapped emission
lines, as expected for objects ionized by hot stars, and for most of the
galaxies the emission line morphology traces also the stellar component. 
\textnormal{Only for Mrk~32 and Tololo~1434+032 we found that the distribution
of the gaseous emission differs considerably from that of the stellar 
component.
Spatial discrepancies in the distribution of emission lines and continuum are
interpreted as signs of a spatial migration of the SF over the history of
the galaxies \citep{Petrosian2002}. However, small spatial offsets between
continuum and emission line peaks, as those seen in Tololo~1434+032, 
and which are indeed a common feature in compact starburst galaxies \citep{
HunterThronson1995,MaizApellaniz1998,Lagos2007}, are likely related to the
release of kinetic energy by massive stars and supernova explosions.}

\item   The different excitation maps produced for the same galaxies display
a similar pattern and trace the regions of star formation as expected
in objects ionized by hot stars. In three out of the eight sample galaxies,
namely Mrk~32, I~Zw~123 and I~Zw~159, higher values of [\ion{S}{ii}]/\Ha\  in
the outer galaxy regions suggest shocks. 


\item  

\textnormal{Six out of the eight objects display inhomogeneous 
extinction maps, with interstellar reddening values \EBV\ varying across 
the galaxy from $\leq0.1$ up to 0.7. 
This result stresses the importance of performing a bidimensional study of
the interstellar extinction even when dealing with the less luminous and more
compact BCDs as those studied here. Assuming a single, spatially constant
value for the extinction, as is usually done in long-slit or single-aperture
spectroscopic studies, can lead to large errors in the derivation of fluxes
and magnitudes in the different regions of the galaxy.}



\item  All SF regions in the sample galaxies have low electron densities,
ranging from $\leq100$ to 320 cm$^{-3}$, typical of classical \ion{H}{ii}
regions. 

\item The oxygen abundances in the present objects range from
$12+\log(\mathrm{O/H})=7.56$ to 8.44 ($Z=1/13Z_{\sun}$ to $Z=0.6Z_{\sun}$). 
We measured for the first time the oxygen abundances of Mrk~407 and Mrk~32.
The galaxy Mrk~407 is found to be a relatively high metallicity BCD, while the
oxygen abundance found for Mrk~32 \textnormal{from the
[\ion{O}{iii}]~$\lambda4363$ line flux would place} it in the list of
extremely metal-poor galaxies. These systems, with
$12+\log(\mathrm{O/H})\leq7.6$, are  excellent laboratories for galaxy
formation and evolution studies, as they allow us to study chemical
compositions and stellar populations in conditions approaching those of
distant protogalactic systems. However, they are also very difficult to find,
and at the present time only about ~30 extremely metal-deficient BCDs are
known \citep{Kunth2000,Kniazev2004,Papaderos2008}.


\item Wolf-Rayet features were measured in three out of the eight galaxies;
a marginal detection was reported for Mrk~32.

\item Three galaxies display a clear rotation pattern (Mrk~750, Mrk~206,
I~Zw~159); for Mrk~407 and Tololo~1434+032, although the maps are
noisier, both seem to indicate a low amplitude rotation around a preferred
axis. For Mrk~32, Mrk~475 and I~Zw~123 the velocity fields are nearly
flat.   

\end{enumerate}

This paper is part of a larger project that aims to map of the properties of
an externsive and representative sample of BCDs by means of IFS. Results for
five luminous BCDs were published in \cite{GarciaLorenzo2008}, and results for
the galaxies Mrk~409 and Mrk~1418, also observed with PMAS, have been shown in
\cite{Cairos2009Mrk409} and \cite{Cairos2009Mrk1418} respectively. The global
properties of the whole sample will be discussed in a forthcoming publication.

\begin{acknowledgements} 

L.~M.~Cair{\'o}s and C.~Kehrig acknowledge the Alexander von Humboldt
Foundation. N.~Caon and C.~Zurita are grateful for the hospitality of the
Astrophysikalisches Institut Potsdam. This research has made use of the
NASA/IPAC Extragalactic Database (NED), which is operated by the Jet
Propulsion Laboratory, Caltech, under contract with the National Aeronautics
and Space Administration. We acknowledge the usage of the HyperLeda database
(http://leda.univ-lyon1.fr). This work has been partially funded by the
spanish ``Ministerio de Ciencia y Innovaci{\'o}n'' through grants AYA 2007
67965 and HA2006-0032, and under the Consolider-Ingenio 2010 Program grant
CSD2006-00070: First Science with the GTC
(http://www.iac.es/consolider-ingenio-gtc/).

\end{acknowledgements}

\bibliographystyle{aa}
\bibliography{cairos-8feb}


\onecolumn
\begin{tiny}
\begin{landscape}
\begin{table*}
\begin{scriptsize}
\caption{Reddening-corrected line ratios, normalized to \Hb$^{(*)}$ for the
  sample of galaxies}
\label{Table:Linesflux1}
\begin{center}
\begin{tabular}{|ll|cccc|cccccc|}       
\hline\hline
$\lambda$  & Ion  & \multicolumn{4}{|c|}{Mrk 407} &                                   
                    \multicolumn{6}{|c|}{Mrk 32} \\    
           &      & \multicolumn{2}{|c}{Nuclear} & 
	            \multicolumn{2}{c|}{Integrated}  & 
                    \multicolumn{2}{|c}{Knot AB} & 
		    \multicolumn{2}{c}{Knot C} & 
		    \multicolumn{2}{c|}{Integrated} \\ 
           &      & $F_{\lambda}$  &  $-W_{\lambda}$  & 
	            $F_{\lambda}$  &  $-W_{\lambda}$  & 
		    $F_{\lambda}$  &  $-W_{\lambda}$  &
                    $F_{\lambda}$  &  $-W_{\lambda}$  & 
		    $F_{\lambda}$  &  $-W_{\lambda}$  \\  
\hline
3727 &  [\ion{O}{ii}]    &  
                      $4.53\pm0.54$  &   $ 43.8\pm 3.3$   &  
		      $4.95\pm1.22$  &   $ 49.9\pm10.3$   &  
                      $3.21\pm0.31$  &   $ 34.9\pm 1.8$   &  
		      $4.77\pm0.56$  &   $ 33.9\pm 2.6$   &  
		      $4.76\pm0.71$  &   $ 26.9\pm 2.4$   \\ 
3869 &  [\ion{Ne}{iii}]  & 
                      ---            &   ---              &  
		      ---            &   ---              &  
                      $0.44\pm0.08$  &   $  6.8\pm 1.2$   &  
		      ---            &   ---              &  
		      ---            &   ---              \\ 
3889 &  H8+HeI      & 
                      ---            &   ---              &  
		      ---            &   ---              &  
                      ---            &   ---              &  
		      ---            &   ---              &  
		      ---            &   ---              \\ 
3968 &  [\ion{Ne}{iii}]  & 
                      ---            &   ---              &  
		      ---            &   ---              &  
                      ---            &   ---              &  
		      ---            &   ---              &  
		      ---            &   ---              \\ 
4101 & H$\delta$         &  
                      ---            &   ---              &  
		      ---            &   ---              &  
                      $0.25\pm0.03$  &   $  4.1\pm0.4$    &  
		      ---            &   ---              &  
		      ---            &   ---              \\ 
4340 & H$\gamma$         &  
                      $0.27\pm0.05$  &   $  2.4\pm0.4$    &  
		      ---            &   ---              &  
                      $0.48\pm0.04$  &   $  9.3\pm0.6$    &  
		      $0.47\pm0.04$  &   $  4.6\pm0.3$    &  
		      $0.52\pm0.08$  &   $  4.0\pm0.5$    \\ 
4363 & [\ion{O}{iii}]    & 
                      ---            &   ---              &  
		      ---            &   ---              &  
                      $0.08\pm0.02$  &   $  1.5\pm0.5$    &  
		      ---            &   ---              &  
		      ---            &   ---              \\ 
4471 &  \ion{He}{i}      &
                      ---            &   ---              &  
		      ---            &   ---              &  
                      ---            &   ---              &  
		      ---            &   ---              &  
		      ---	     &   ---              \\ 
4861 & H$\beta$          &  
                      $1.00\pm0.00$  &   $ 10.5\pm0.3$    &  
		      $1.00\pm0.00$  &   $ 10.2\pm0.7$    &  
                      $1.00\pm0.00$  &   $ 24.2\pm0.7$    &  
		      $1.00\pm0.00$  &   $ 11.7\pm0.4$    &  
		      $1.00\pm0.00$  &   $  9.8\pm0.6$    \\ 
4959 & [\ion{O}{iii}]    & 
                      $1.17\pm0.06$  &   $ 12.8\pm0.4$    &  
		      $1.04\pm0.12$  &   $ 10.1\pm0.9$    &  
                      $0.99\pm0.04$  &   $ 25.8\pm0.7$    &  
		      $0.90\pm0.06$  &   $ 11.6\pm0.5$    &  
		      $0.87\pm0.09$  &   $  9.2\pm0.7$    \\ 
5007 &  [\ion{O}{iii}]   & 
                      $3.16\pm0.13$  &   $ 35.3\pm0.4$    &  
		      $2.70\pm0.23$  &   $ 26.2\pm0.9$    &  
                      $2.82\pm0.11$  &   $ 70.8\pm0.8$    &  
		      $2.53\pm0.12$  &   $ 33.0\pm0.6$    &  
		      $2.20\pm0.16$  &   $ 23.2\pm0.7$    \\ 
5876 &  \ion{He}{i}      & 
                      ---            &   ---              &  
		      ---            &   ---              &  
                      $0.12\pm0.02$  &   $  4.0\pm0.6$    &  
		      $0.16\pm0.03$  &   $  2.7\pm0.6$    &  
		      ---            &   ---              \\ 
6300 &  [\ion{O}{i}]    & 
                      ---            &   ---              &  
		      ---            &   ---              &  
                      $0.09\pm0.01$  &   $  3.4\pm0.5$    &  
		      $0.10\pm0.02$  &   $  1.9\pm0.4$    &  
		      $0.15\pm0.02$  &   $  2.4\pm0.3$    \\ 
6548 &  [\ion{N}{ii}]    & 
                      ---            &   ---              &  
		      ---            &   ---              &  
                      ---            &   ---              &  
		      ---            &   ---              &  
		      ---            &   ---              \\ 
6563 & H$\alpha$        & 
                      $2.86\pm0.16$  &   $ 53.0\pm0.4$    &  
		      $2.86\pm0.32$  &   $ 46.6\pm0.9$    &  
                      $2.87\pm0.12$  &   $107.4\pm1.4$    &  
		      $2.86\pm0.18$  &   $ 63.9\pm0.7$    &  
		      $2.86\pm0.26$  &   $ 46.4\pm0.7$    \\ 
6584 &  [\ion{N}{ii}]    & 
                      $0.26\pm0.02$  &   $  4.8\pm0.3$    &  
		      $0.23\pm0.04$  &   $  3.7\pm0.6$    &  
                      $0.14\pm0.02$  &   $  5.4\pm0.9$    &  
		      $0.23\pm0.03$  &   $  5.1\pm0.7$    &  
		      $0.21\pm0.04$  &   $  3.4\pm0.7$    \\ 
6678 &  \ion{He}{i}      &
                      $0.04\pm0.01$  &   $  0.8\pm0.2$    &  
		      ---            &   ---              &  
                      ---            &   ---              &  
		      ---            &   ---              &  
		      ---            &   ---              \\ 
6717 & [\ion{S}{ii}]     & 
                      $0.31\pm0.02$  &   $  6.0\pm0.3$    &  
		      $0.37\pm0.06$  &   $  6.2\pm0.7$    &  
                      $0.38\pm0.02$  &   $ 15.7\pm0.5$    &  
		      $0.33\pm0.03$  &   $  6.9\pm0.4$    &  
		      $0.49\pm0.06$  &   $  8.4\pm0.6$    \\ 
6731 & [\ion{S}{ii}]     & 
                      $0.25\pm0.02$  &   $  4.9\pm0.3$    &  
		      $0.30\pm0.05$  &   $  5.2\pm0.7$    &  
                      $0.29\pm0.02$  &   $ 11.7\pm0.5$    &  
		      $0.26\pm0.03$  &   $  5.4\pm0.5$    &  
		      $0.39\pm0.05$  &   $  6.5\pm0.6$    \\ 
      &          &  
                    \multicolumn{2}{|c}{$F(\Hb)=336.3\pm 45.8 $} &  
                    \multicolumn{2}{c|}{$F(\Hb)=408.4\pm110.9 $} &  
                    \multicolumn{2}{c}{ $F(\Hb)=113.4\pm  6.6 $} &  
                    \multicolumn{2}{c}{ $F(\Hb)= 24.4\pm  3.8 $} &  
                    \multicolumn{2}{c|}{$F(\Hb)=232.6\pm 52.6 $} \\ 
     &          &              
		    \multicolumn{2}{|c}{$\CHbeta=0.237\pm0.057 $} &  
                    \multicolumn{2}{c|}{$\CHbeta=0.165\pm0.113 $} &  
                    \multicolumn{2}{c}{ $\CHbeta=0.108\pm0.021 $} &  
                    \multicolumn{2}{c}{ $\CHbeta=0.167\pm0.065 $} &  
                    \multicolumn{2}{c|}{$\CHbeta=0.107\pm0.095 $} \\ 
     &          &	    
                    \multicolumn{2}{|c}{$\EBV=0.163\pm0.039$}  &  
                    \multicolumn{2}{c|}{$\EBV=0.114\pm0.078$}  &  
                    \multicolumn{2}{c}{ $\EBV=0.075\pm0.014$}  &  
                    \multicolumn{2}{c}{ $\EBV=0.115\pm0.045$}  &  
                    \multicolumn{2}{c|}{$\EBV=0.074\pm0.065$}  \\ 
\hline
\end{tabular}
\end{center}
(*) Notes. -- Reddening-corrected line fluxes, normalized to $F(\Hb)=1$. 
Equivalent widths of Balmer lines are corrected for underlying stellar 
absorption. The reddening coefficient, $\CHbeta$, \EBV\  
(derived as $0.69\times\CHbeta$) and the reddening-corrected \Hb\ flux, 
$F(\Hb) (\times 10^{-16}$ ergs cm$^{-2}$ s$^{-1}$) are listed for each 
region. 
The quoted uncertainties account for measurement, flux-calibration and 
reddening coefficient errors. 
\end{scriptsize}
\end{table*}
\end{landscape}
\end{tiny}

\onecolumn
\begin{tiny}
\begin{landscape}
\begin{table*}
\begin{scriptsize}
\caption{Reddening-corrected line ratios, normalized to \Hb$^{(*)}$}
\label{Table:Linesflux2}
\begin{center}
\begin{tabular}{|ll|cccc|cccc|}       
\hline\hline
$\lambda$  & Ion  & \multicolumn{4}{|c|}{Mrk~750} &             
                    \multicolumn{4}{|c|}{Mrk~206} \\    
           &      & \multicolumn{2}{|c}{Nuclear} & 
	            \multicolumn{2}{c|}{Integrated} &
                    \multicolumn{2}{|c}{Nuclear} & 
		    \multicolumn{2}{c|}{Integrated} \\
           &      &   $F_{\lambda}$  &  $-W_{\lambda}$ & 
	              $F_{\lambda}$  &  $-W_{\lambda}$ &
                      $F_{\lambda}$  &  $-W_{\lambda}$ & 
		      $F_{\lambda}$  &  $-W_{\lambda}$ \\  
\hline
3727 &  [\ion{O}{ii}]    &  
                      $2.04\pm0.17$  &  $106.8\pm 3.3$   &  
		      $2.41\pm0.23$  &  $ 86.6\pm 2.6$   &  
                      $3.85\pm0.33$  &  $ 97.3\pm 3.4$   &  
		      $4.96\pm1.27$  &  $219.7\pm50.7$   \\ 
3869 &  [\ion{Ne}{iii}]  &        
                      $0.43\pm0.03$  &  $ 28.0\pm 0.7$   &  
		      $0.49\pm0.05$  &  $ 24.4\pm 1.6$   &  
                      ---            &  ---              &  
		      ---            &  ---              \\ 
3889 &  H8+HeI      &  
                      $0.19\pm0.01$  &  $ 12.7\pm 0.6$   &  
		      $0.24\pm0.03$  &  $ 12.2\pm 1.2$   &  
                      ---            &  ---              &  
		      ---            &  ---              \\ 
3968 &  [\ion{Ne}{iii}]  &  
                      $0.27\pm0.02$  &  $ 18.8\pm 0.6$   &  
		      $0.30\pm0.03$  &  $ 17.4\pm 1.3$   &  
                      ---            &  ---              &  
		      ---            &  ---              \\ 
4101 & H$\delta$         &  
                      $0.26\pm0.01$  &  $ 19.0\pm 0.4$   &  
		      $0.22\pm0.02$  &  $ 19.3\pm 0.4$   &  
                      $0.21\pm0.03$  &  $  4.5\pm 0.7$   &  
		       ---           &  ---              \\ 
4340 & H$\gamma$         &  
                      $0.47\pm0.01$  &  $ 42.0\pm 0.5$   &  
		      $0.50\pm0.03$  &  $ 28.8\pm 1.2$   &  
                      $0.47\pm0.02$  &  $ 12.1\pm 0.5$   &  
		      $0.30\pm0.05$  &  $  5.6\pm 0.8$   \\ 
4363 & [\ion{O}{iii}]    &  
                      $0.06\pm0.01$  &  $  5.2\pm 0.3$   &  
		      $0.07\pm0.02$  &  $  4.1\pm 0.9$   &  
                      ---            &  ---              &  
		      ---            &  ---              \\ 
4471 &  \ion{He}{i}      &
                      $0.04\pm0.01$  &  $  3.5\pm 0.3$   &  
		      $0.06\pm0.01$  &  $  3.8\pm 0.8$   &  
                      ---            &  ---              &  
		      ---            &  ---              \\ 
4861 & H$\beta$          &  
                      $1.00\pm0.00$  &  $114.1\pm 0.3$   &  
		      $1.00\pm0.00$  &  $ 68.1\pm 0.9$   &  
                      $1.00\pm0.00$  &  $ 30.8\pm 0.3$   &  
		      $1.00\pm0.00$  &  $ 20.1\pm 1.0$   \\ 
4959 & [\ion{O}{iii}]    &  
                      $1.72\pm0.04$  &  $280.2\pm 0.7$   &  
		      $1.60\pm0.05$  &  $107.1\pm 1.1$   &  
                      $0.67\pm0.02$  &  $ 21.7\pm 0.3$   &  
		      $0.69\pm0.06$  &  $ 14.4\pm 1.0$   \\ 
5007 &  [\ion{O}{iii}]   &  
                      $5.00\pm0.12$  &  $801.9\pm 1.0$   &  
		      $4.75\pm0.14$  &  $323.4\pm 1.7$   &  
                      $1.94\pm0.05$  &  $ 64.5\pm 0.4$   &  
		      $2.03\pm0.13$  &  $ 43.6\pm 1.3$   \\ 
5876 &  \ion{He}{i}      &  
                      $0.10\pm0.01$  &  $ 17.2\pm 0.3$   &  
		      $0.12\pm0.01$  &  $ 11.4\pm 0.9$   &  
                      $0.13\pm0.01$  &  $  5.8\pm 0.2$   &  
		      $0.18\pm0.02$  &  $  4.9\pm 0.6$   \\ 
6300 &  [\ion{O}{i}]    &  
                      $0.03\pm0.01$  &  $  5.6\pm 0.3$   &  
		      ---            &  ---              &  
                      $0.07\pm0.01$  &  $  3.1\pm 0.3$   &  
		      ---            &  ---              \\ 
6548 &  [\ion{N}{ii}]    &  
                      ---            &  ---              &  
		      ---            &  ---              &  
                      $0.19\pm0.01$  &  $  9.2\pm 0.2$   &  
		      $0.21\pm0.03$  &  $  5.8\pm 0.6$   \\ 
6563 & H$\alpha$         &  
                      $2.86\pm0.07$  &  $984.6\pm 2.5$   &  
		      $2.86\pm0.14$  &  $325.5\pm 1.9$   &  
                      $2.86\pm0.08$  &  $140.4\pm 0.4$   &  
		      $2.86\pm0.23$  &  $ 78.4\pm 0.9$   \\ 
6584 &  [\ion{N}{ii}]    &  
                      $0.15\pm0.01$  &  $ 29.4\pm 0.4$   &  
		      $0.15\pm0.01$  &  $ 19.1\pm 0.8$   &  
                      $0.54\pm0.02$  &  $ 26.5\pm 0.2$   &  
		      $0.61\pm0.06$  &  $ 16.8\pm 0.8$   \\ 
6678 &  \ion{He}{i}      &
                      $0.03\pm0.01$  &  $  5.8\pm 0.3$   &  
		      $0.04\pm0.01$  &  $  5.3\pm 0.9$   &  
                      $0.03\pm0.01$  &  $  1.6\pm 0.2$   &  
		      ---            &  ---              \\ 
6717 & [\ion{S}{ii}]     & 
                      $0.16\pm0.01$  &  $ 37.9\pm 0.5$   &  
		      $0.20\pm0.01$  &  $ 23.2\pm 0.8$   &  
                      $0.33\pm0.01$  &  $ 16.4\pm 0.2$   &  
		      $0.40\pm0.04$  &  $ 10.7\pm 0.6$   \\ 
6731 & [\ion{S}{ii}]     & 
                      $0.12\pm0.01$  &  $ 29.4\pm 0.4$   &  
		      $0.15\pm0.01$  &  $ 17.9\pm 0.8$   &  
                      $0.26\pm0.01$  &  $ 12.8\pm 0.2$   &  
		      $0.29\pm0.03$  &  $  7.7\pm 0.6$   \\ 
      &          &  
                    \multicolumn{2}{|c}{$F(\Hb)=1784.0\pm 54.6 $} &  
                    \multicolumn{2}{c|}{$F(\Hb)=2454.2\pm340.5 $} &  %
                    \multicolumn{2}{c}{ $F(\Hb)=1353.0\pm 58.7 $} &  
                    \multicolumn{2}{c|}{$F(\Hb)=1973.8\pm383.2 $} \\ 
     &          &       
                    \multicolumn{2}{|c}{$\CHbeta=0.187\pm0.012 $} &   
                    \multicolumn{2}{c|}{$\CHbeta=0.140\pm0.060 $} &   
                    \multicolumn{2}{c}{ $\CHbeta=0.598\pm0.020 $} &   
                    \multicolumn{2}{c|}{$\CHbeta=0.649\pm0.081 $} \\  
&          &       
                    \multicolumn{2}{|c}{$\EBV=0.129\pm0.009$}  &   
                    \multicolumn{2}{c|}{$\EBV=0.097\pm0.041$}  &   
                    \multicolumn{2}{c}{ $\EBV=0.412\pm0.013$}  &   
                    \multicolumn{2}{c|}{$\EBV=0.448\pm0.056$}  \\  
\hline
\end{tabular}
\end{center}
\end{scriptsize}
\end{table*}
\end{landscape}
\end{tiny}

\onecolumn
\begin{tiny}
\begin{landscape}
\begin{table*}
\begin{scriptsize}
\caption{Reddening-corrected line ratios, normalized to \Hb$^{(*)}$}
\label{Table:Linesflux3}
\begin{center}
\begin{tabular}{|ll|cccccccc|cccc|}       
\hline\hline
$\lambda$  & Ion  & \multicolumn{8}{|c|}{Tololo~1434+032} &                                  
                    \multicolumn{4}{|c|}{Mrk~475} \\    
           &      & \multicolumn{2}{|c}{Knot A}  & 
	            \multicolumn{2}{c}{Knot B}   & 
		    \multicolumn{2}{c}{Knot CD}  & 
		    \multicolumn{2}{c|}{Integrated} &
                    \multicolumn{2}{|c}{Nuclear} & 
		    \multicolumn{2}{c|}{Integrated} \\
           &      &   $F_{\lambda}$  & $-W_{\lambda}$  & 
	              $F_{\lambda}$  & $-W_{\lambda}$  &   
		      $F_{\lambda}$  & $-W_{\lambda}$  & 
		      $F_{\lambda}$  & $-W_{\lambda}$  &
                      $F_{\lambda}$  & $-W_{\lambda}$  & 
		      $F_{\lambda}$  & $-W_{\lambda}$ \\  
\hline
3727 &  [\ion{O}{ii}]    &  
                      $2.19\pm0.22$  &  $153.4\pm  8.9$  &  
		      $2.87\pm0.43$  &  $123.5\pm 13.6$  &  
		      $3.94\pm1.28$  &  $490.1\pm145.9$  &  
		      $3.05\pm0.71$  &  $ 64.7\pm 13.1$  &  
                      $1.29\pm0.12$  &  $107.5\pm  5.6$  &  
		      $1.42\pm0.28$  &  $139.4\pm 24.4$  \\ 
 3869 &  [\ion{Ne}{iii}]    &  
                      $0.43\pm0.05$  &  $ 17.1\pm  1.5$  &  
		      $0.44\pm0.11$  &  $ 13.1\pm  3.1$  &  
		      ---            &  ---              &  
		      $0.68\pm0.19$  &  $ 11.3\pm  3.0$  &  
                      $0.41\pm0.03$  &  $ 44.0\pm  2.3$  &  
		      $0.47\pm0.09$  &  $ 51.3\pm  9.2$  \\ 
3889 &  H8+HeI      &  
                      $0.16\pm0.03$  &  $  7.4\pm  1.3$  &  
		      ---            &  ---              &  
		      ---            &  ---              &  
		      ---            &  ---              &  
                      $0.18\pm0.02$  &  $ 19.6\pm  1.8$  &  
		      ---            &  ---              \\ 
3968 &  [\ion{Ne}{iii}]  &  
                      $0.24\pm0.03$  &  $ 12.1\pm  1.4$  &  
		      $0.24\pm0.07$  &  $  9.6\pm  2.8$  &  
		      ---            &  ---              &  
		      ---            &  ---              &  
                      $0.27\pm0.02$  &  $ 30.2\pm  1.4$  &  
		      ---            &  ---              \\ 
4101 & H$\delta$         &  
                      $0.27\pm0.02$  &  $ 11.7\pm  0.9$  &  
		      $0.27\pm0.05$  &  $  8.1\pm  1.4$  &  
		      ---            &  ---              &  
		      ---            &  ---              &  
                      $0.27\pm0.01$  &  $ 28.6\pm  0.7$  &  
		      $0.23\pm0.03$  &  $ 18.7\pm  2.3$  \\ 
4340 & H$\gamma$         &  
                      $0.48\pm0.03$  &  $ 27.3\pm  1.2$  &  
		      $0.50\pm0.05$  &  $ 22.1\pm  2.1$  &  
		      $0.59\pm0.10$  &  $  8.1\pm  1.2$  &  
		      $0.60\pm0.10$  &  $  7.6\pm  1.8$  &  
                      $0.46\pm0.01$  &  $ 57.2\pm  0.8$  &  
		      $0.47\pm0.03$  &  $ 46.1\pm  1.8$  \\ 
4363 & [\ion{O}{iii}]    &  
                      $0.07\pm0.01$  &  $  2.1\pm  0.5$  &  
		      ---            &  ---              &  
		      ---            &  ---              &  
		      ---            &  ---              &  
                      $0.08\pm0.01$  &  $ 10.2\pm  0.6$  &  
		      $0.09\pm0.02$  &  $  7.7\pm  1.6$  \\ 
4471 &  \ion{He}{i}      &
                      ---            &  ---              &  
		      ---            &  ---              &  
		      ---            &  ---              &  
		      ---            &  ---              &  
                      $0.04\pm0.01$  &  $  4.6\pm  0.5$  &  
		      ---            &  ---              \\ 
4686 &  \ion{He}{ii}      &
                      ---            &  ---              &  
		      ---            &  ---              &  
		      ---            &  ---              &  
		      ---            &  ---              &  
                      $0.03\pm0.01$  &  $  4.5\pm  0.7$  &  
		      ---            &  ---              \\ 
4861 & H$\beta$          &  
                      $1.00\pm0.00$  &  $ 70.8\pm  0.9$  &  
		      $1.00\pm0.00$  &  $ 49.0\pm  2.0$  &  
		      $1.00\pm0.00$  &  $ 19.2\pm  1.3$  &  
		      $1.00\pm0.00$  &  $ 25.4\pm  1.4$  &  
                      $1.00\pm0.00$  &  $143.5\pm  0.9$  &  
		      $1.00\pm0.00$  &  $ 91.1\pm  2.3$  \\ 
4959 & [\ion{O}{iii}]    &  
                      $1.56\pm0.05$  &  $118.1\pm  1.3$  &  
		      $1.46\pm0.08$  &  $ 86.9\pm  2.5$  &  
		      $1.20 pm0.11$  &  $ 26.6\pm  1.6$  &  
		      $1.33\pm0.10$  &  $ 35.1\pm  1.5$  &  
                      $1.73\pm0.04$  &  $258.1\pm  1.2$  &  
		      $1.64\pm0.06$  &  $148.1\pm  2.6$  \\ 
5007 &  [\ion{O}{iii}]   &  
                      $4.58\pm0.13$  &  $307.3\pm  1.7$  &  
		      $4.32\pm0.21$  &  $254.6\pm  3.7$  &  
		      $3.37\pm0.27$  &  $ 73.1\pm  2.0$  &  
		      $3.98\pm0.25$  &  $104.4\pm  2.1$  &  
                      $5.16\pm0.13$  &  $799.4\pm  2.1$  &  
		      $4.93\pm0.18$  &  $411.5\pm  3.8$  \\ 
5876 &  \ion{He}{i}      &  
                      $0.12\pm0.01$  &  $  9.5\pm  0.9$  &  
		      ---            &  ---              &  
		      ---            &  ---              &  
		      ---            &  ---              &  
                      $0.10\pm0.01$  &  $ 19.1\pm  0.4$  &  
		      $0.10\pm0.01$  &  $  9.2\pm  0.9$  \\ 
6300 & [\ion{O}{i}]    &  
                      ---            &  ---              &  
		      ---            &  ---              &  
		      ---            &  ---              &  
		      ---            &  ---              &  
                      ---            &  ---              &  
		      ---            &  ---              \\ 
6548 &  [\ion{N}{ii}]    &  
                      $0.06\pm0.01$   &  $  7.4\pm  0.7$  &  
		      $0.05\pm0.01$   &  $  3.7\pm  1.1$  &  
		      ---             &  ---              &  
		      ---             &  ---              &  
                      ---             &  ---              &  
		      ---             &  ---              \\ 
6563 & H$\alpha$         &  
                      $2.86\pm0.11$   &  $340.0\pm  2.3$  &  
		      $2.86\pm0.20$   &  $207.1\pm  3.4$  &  
		      $2.86\pm0.31$   &  $ 94.2\pm  2.1$  &  
		      $2.86\pm0.26$   &  $102.5\pm  2.8$  &  
                      $2.86\pm0.08$   &  $636.3\pm  1.8$  &  
		      $2.83\pm0.12$   &  $347.1\pm  3.4$  \\ 
6584 &  [\ion{N}{ii}]    &  
                      $0.12\pm0.01$   & $ 14.1\pm  0.8$  &  
		      $0.13\pm0.02$   & $  9.3\pm  1.3$  &  
		      $0.22\pm0.05$   & $  6.9\pm  1.3$  &  
		      $0.14\pm0.04$   & $  4.9\pm  1.2$  &  
                      $0.09\pm0.01$   & $ 21.3\pm  0.6$  &  
		      $0.09\pm0.01$   & $  9.9\pm  1.3$  \\ 
6678 &  \ion{He}{i}      &
                      $0.04\pm0.01$   & $  4.7\pm  0.7$  &  
		      ---             &   ---            &  
		      ---             &   ---            &  
		      ---             &   ---            &  
                      $0.03\pm0.01$   & $  7.2\pm  0.6$  &  
		      $0.03\pm0.01$   & $  4.1\pm  1.0$  \\ 
6717 & [\ion{S}{ii}]     & 
                      $0.21\pm0.01$   & $ 23.8\pm  0.9$  &  
		      $0.27\pm0.04$   & $ 19.6\pm  2.2$  &  
		      $0.37\pm0.06$   & $ 11.0\pm  1.5$  &  
		      $0.28\pm0.05$   & $  9.7\pm  1.4$  &  
                      $0.14\pm0.01$   & $ 32.8\pm  0.6$  &  
		      $0.16\pm0.02$   & $ 24.0\pm  2.4$  \\ 
6731 & [\ion{S}{ii}]     & 
                      $0.15\pm0.01$   & $ 17.0\pm  0.8$  &  
		      $0.19\pm0.03$   & $ 13.9\pm  2.2$  &  
		      $0.24\pm0.05$   & $  7.2\pm  1.4$  &  
		      $0.22\pm0.05$   & $  7.9\pm  1.6$  &  
                      $0.11\pm0.01$   & $ 26.0\pm  0.6$  &  
		      $0.14\pm0.02$   & $ 21.2\pm  2.5$  \\ 
      &          &  
                    \multicolumn{2}{|c}{$F(\Hb)= 103.3\pm 10.1 $} &   
                    \multicolumn{2}{c}{ $F(\Hb)=  21.6\pm  3.7 $} &   
                    \multicolumn{2}{c}{ $F(\Hb)=  24.5\pm  6.5 $} &   
                    \multicolumn{2}{c|}{$F(\Hb)= 255.9\pm 56.9 $} &   
                    \multicolumn{2}{c}{ $F(\Hb)= 855.1\pm 40.6 $} &   
                    \multicolumn{2}{c|}{$F(\Hb)=1189.1\pm100.8 $} \\  
    &          &              
		    \multicolumn{2}{|c}{$\CHbeta= 0.119\pm0.041 $} &   
                    \multicolumn{2}{c}{ $\CHbeta= 0.040\pm0.072 $} &   
                    \multicolumn{2}{c}{ $\CHbeta= 0.131\pm0.111 $} &   
                    \multicolumn{2}{c|}{$\CHbeta= 0.099\pm0.093 $} &   
                    \multicolumn{2}{c}{ $\CHbeta= 0.019\pm0.018 $} &   
                    \multicolumn{2}{c|}{$\CHbeta= 0.018\pm0.034 $} \\  
		        &          &              
                    \multicolumn{2}{|c}{$\EBV=0.082\pm0.028$} &     
                    \multicolumn{2}{c}{ $\EBV=0.028\pm0.049$} &     
                    \multicolumn{2}{c}{ $\EBV=0.091\pm0.077$} &     
                    \multicolumn{2}{c|}{$\EBV=0.068\pm0.054$} &     
                    \multicolumn{2}{c}{ $\EBV=0.013\pm0.012$} &     
                    \multicolumn{2}{c|}{$\EBV=0.012\pm0.024$} \\    
\hline
\end{tabular}
\end{center}
\end{scriptsize}
\end{table*}
\end{landscape}
\end{tiny}

\onecolumn
\begin{tiny}
\begin{landscape}
\begin{table*}
\begin{scriptsize}
\caption{Reddening-corrected line ratios, normalized to \Hb$^{(*)}$}
\label{Table:Linesflux4}
\begin{center}
\begin{tabular}{|ll|cccc|cccc|}       
\hline\hline 
$\lambda$  & Ion  & \multicolumn{4}{|c|}{IZw~123} &
                    \multicolumn{4}{|c|}{IZw~159} \\    
           &      & \multicolumn{2}{|c}{Nuclear} & 
	            \multicolumn{2}{c|}{Integrated} &
                    \multicolumn{2}{|c}{Nuclear} & 
		    \multicolumn{2}{c|}{Integrated} \\
           &      &    $F_{\lambda}$  &  $-W_{\lambda}$ & 
	               $F_{\lambda}$  &  $-W_{\lambda}$ &
                       $F_{\lambda}$  &  $-W_{\lambda}$ & 
		       $F_{\lambda}$  &  $-W_{\lambda}$ \\  
\hline
3727 &  [\ion{O}{ii}]    &  
                      $1.75\pm0.17$  &  $ 55.3\pm2.9$  &  
		      $2.17\pm0.34$  &  $ 37.3\pm4.6$  &  
                      $3.10\pm0.27$  &  $ 79.7\pm3.7$  &  
		      $3.60\pm0.49$  &  $ 87.4\pm9.7$  \\ 
3869 &  [\ion{Ne}{iii}]    &  
                      $0.54\pm0.05$  &  $ 22.8\pm1.6$  &  
		      $0.47\pm0.15$  &  $ 13.6\pm4.3$  &  
                      $0.27\pm0.05$  &  $  6.5\pm1.1$  &  
		      ---            &    ---            \\ 
3889 &  H8+HeI      &  
                      $0.11\pm0.02$  &  $  4.9\pm1.0$  &  
		      ---            &  ---	       &  
                      ---            &  ---            &  
		      ---            &  ---            \\ 
3968 &  [\ion{Ne}{iii}]  &  
                      $0.26\pm0.02$  &  $ 11.5\pm0.8$  &  
		      ---            &   ---           &  
                      ---            &   ---           &  
		      ---            &   ---	       \\ 
4101 & H$\delta$         &  
                      $0.26\pm0.01$  &  $ 12.1\pm0.5$  &  
		      ---            &  ---	       &  
                      $0.26\pm0.02$  &  $  6.7\pm0.4$  &  
		      ---            &  ---	       \\ 
4340 & H$\gamma$         &  
                      $0.47\pm0.02$  &  $ 25.8\pm0.5$  &  
		      $0.39\pm0.04$  &  $ 14.6\pm1.3$  &  
                      $0.47\pm0.02$  &  $ 13.6\pm0.4$  &  
		      $0.47\pm0.03$  &  $  9.7\pm0.6$  \\ 
4363 & [\ion{O}{iii}]    &  
                      $0.07\pm0.01$  &  $  3.9\pm0.4$  &  
		      ---            &  ---            &  
                      ---            &  ---            &  
		      ---            &  ---            \\ 
4471 &  \ion{He}{i}      &
                      $0.06\pm0.01$  &  $  3.4\pm0.5$  &  
		      ---            &  ---            &  
                      ---            &  ---            &  
		      ---            &  ---            \\ 
4861 & H$\beta$          &  
                      $1.00\pm0.00$  &  $ 66.7\pm0.5$  &  
		      $1.00\pm0.00$  &  $ 44.8\pm1.6$  &  
                      $1.00\pm0.00$  &  $ 33.5\pm0.3$  &  
		      $1.00\pm0.00$  &  $ 24.4\pm0.6$  \\ 

4959 & [\ion{O}{iii}]    &  
                      $1.93\pm0.05$  &  $131.2\pm0.6$  &  
		      $1.69\pm0.09$  &  $ 73.7\pm1.8$  &  
                      $0.93\pm0.02$  &  $ 31.4\pm0.3$  &  
		      $0.87\pm0.03$  &  $ 19.3\pm0.4$  \\ 
5007 &  [\ion{O}{iii}]   &  
                      $5.78\pm0.14$  &  $406.7\pm1.0$  &  
		      $5.14\pm0.24$  &  $226.8\pm2.8$  &  
                      $2.78\pm0.07$  &  $ 95.0\pm0.4$  &  
		      $2.66\pm0.09$  &  $ 61.0\pm0.6$  \\ 
5876 &  \ion{He}{i}      &  
                      $0.09\pm0.01$  &  $  9.5\pm0.3$  &  
		      ---            &  ---            &  
                      $0.11\pm0.01$  &  $  4.7\pm0.2$  &  
		      $0.10\pm0.01$  &  $  3.2\pm0.5$  \\ 
6300 & [\ion{O}{i}]    &   
                      ---            &  ---            &  
		      ---            &  ---            &  
                      $0.07\pm0.01$  &  $  3.7\pm0.3$  &  
		      $0.09\pm0.02$  &  $  3.2\pm0.7$  \\ 
6548 &  [\ion{N}{ii}]    &  
                      ---            &  ---            &  
		      ---            &  ---            &  
                      $0.14\pm0.01$  &  $  7.8\pm0.2$  &  
		      ---            &  ---            \\ 
6563 & H$\alpha$         &  
                      $2.85\pm0.07$  &  $323.3\pm1.4$  &  
		      $2.86\pm0.19$  &  $204.1\pm3.4$  &  
                      $2.86\pm0.07$  &  $153.5\pm0.5$  &  
		      $2.85\pm0.11$  &  $104.5\pm1.0$  \\ 
6584 &  [\ion{N}{ii}]    &  
                      $0.14\pm0.01$  &  $ 15.9\pm0.6$  &  
		      $0.16\pm0.02$  &  $ 11.2\pm1.4$  &  
                      $0.35\pm0.01$  &  $ 18.8\pm0.3$  &  
		      $0.38\pm0.02$  &  $ 13.9\pm0.8$  \\ 
6678 &  \ion{He}{i}      &
                      $0.03\pm0.01$  &  $  4.2\pm0.5$  &  
		      ---            &  ---            &  
                      $0.03\pm0.01$  &  $  1.6\pm0.2$  &  
		      ---            &  ---            \\ 
6717 & [\ion{S}{ii}]     & 
                      $0.14\pm0.01$  &  $ 16.8\pm0.5$  &  
		      $0.23\pm0.03$  &  $ 17.3\pm1.8$  &  
                      $0.34\pm0.01$  &  $ 18.5\pm0.3$  &  
		      $0.41\pm0.02$  &  $ 15.3\pm0.6$  \\ 
6731 & [\ion{S}{ii}]     & 
                      $0.12\pm0.01$  &  $ 14.6\pm0.6$  &  
		      $0.17\pm0.03$  &  $ 12.7\pm1.7$  &  
                      $0.26\pm0.01$  &  $ 14.2\pm0.3$  &  
		      $0.30\pm0.02$  &  $ 11.3\pm0.6$  \\ 
      &          &  
                    \multicolumn{2}{c}{ $F(\Hb)=1263.3\pm 47.9 $}   &   
                    \multicolumn{2}{c|}{$F(\Hb)=1743.4\pm279.3 $}   &   
                    \multicolumn{2}{c}{ $F(\Hb)= 593.7\pm 20.3 $}   &   
                    \multicolumn{2}{c|}{$F(\Hb)= 751.6\pm 46.4$}    \\  
      &          &              
                    \multicolumn{2}{c}{ $\CHbeta=0.201\pm0.014$}     &   
                    \multicolumn{2}{c|}{$\CHbeta=0.188\pm0.067$}     &   
                    \multicolumn{2}{c}{ $\CHbeta=0.238\pm0.013$}     &   
                    \multicolumn{2}{c|}{$\CHbeta=0.248\pm0.024$}     \\  
     &          &              
                    \multicolumn{2}{c}{ $\EBV=0.139\pm0.010$}     &   
                    \multicolumn{2}{c|}{$\EBV=0.130\pm0.047$}     &   
                    \multicolumn{2}{c}{ $\EBV=0.164\pm0.010$}     &   
                    \multicolumn{2}{c|}{$\EBV=0.171\pm0.016$}     \\  
\hline
\end{tabular}
\end{center}
\end{scriptsize}
\end{table*}
\end{landscape}
\end{tiny}

\onecolumn
\begin{landscape}
\begin{table*}
\begin{scriptsize}
\caption{Physical parameters and chemical abundances}
\label{Table:Parameters}
\begin{center}
\begin{tabular}{|c|cc|ccc|cc|cc|}       
\hline\hline
Parameter         & \multicolumn{2}{|c|}{Mrk~407} &                                  
                    \multicolumn{3}{|c|}{Mrk~32}  &
                    \multicolumn{2}{|c|}{Mrk~750} &
                    \multicolumn{2}{|c|}{Mrk~206} \\    
                  & \multicolumn{1}{|c}{Nuclear}    & 
		    \multicolumn{1}{c|}{Integrated} &
                    \multicolumn{1}{|c}{Knot~AB}    & 
		    \multicolumn{1}{c}{Knot~C}      & 
		    \multicolumn{1}{c|}{Integrated} &
                    \multicolumn{1}{|c}{Nuclear}    & 
		    \multicolumn{1}{c|}{Integrated} &
                    \multicolumn{1}{|c}{Nuclear}    & 
		    \multicolumn{1}{c|}{Integrated} \\
\hline
$N_\mathrm{e}$([\ion{S}{ii}]) (cm$^{-3}$)    &  
                    188              &  
		    197              &                   
                    110              &  
		    154              &    
		    169              &  
                    $< 100$          &  
		    $< 100$          &                   
                    154              &  
		    $< 100$         \\  
$T_{\mathrm e}$([\ion{O}{ii}]) (10$^{4}$ K)   &        
                    ---              & 
		    ---              &                   
                    $1.54\pm0.22$    & 
		    ---              & 
		    ---              &  
                    $1.13\pm0.02$    &  
		    $1.23\pm0.09$    &                   
                    ---              &  
		    ---             \\  
$T_{\mathrm e}$([\ion{O}{iii}]) ($10^{4}$ K)  &  
                    ---              & 
		    ---              &                   
                    $1.77\pm0.31$    & 
		    ---              & 
		    ---              &  
                    $1.21\pm0.03$    &  
		    $1.34\pm0.13$    &                   
                    ---              & 
		    ---             \\  
$12+\log(\mathrm{O/H})$ -- ($T_\mathrm{e}$)        &  
                    ---              & 
		    ---              &                   
                    $7.56\pm0.10$    & 
		    ---              & 
		    ---              &  
                    $8.12\pm0.03$    &  
		    $8.00\pm0.07$    &                   
                    ---              & 
		    ---              \\  
$12+\log(\mathrm{Ne}^{++}/\mathrm{H}^{+})$        &  
                    ---              & 
		    ---              &                   
                    $6.88\pm0.18$    & 
		    ---              & 
		    ---              &  
                    $7.37\pm0.04$    &  
		    $7.27\pm0.12$    &                   
                    ---              & 
		    ---             \\  
$12+\log(\mathrm{S}^{+}/\mathrm{H}^{+})$        &  
                    ---              & 
		    ---              &                   
                    $5.80\pm0.09$    & 
		    ---              & 
		    ---              &  
                    $5.69\pm0.02$    &  
		    $5.70\pm0.06$    &                   
                    ---              & 
		    ---             \\  
$12+\log(\mathrm{N}^+/\mathrm{H}^+)$    &  
                      ---            & 
		      ---            &                   
                    $6.03\pm0.12$    & 
		    ---              & 
		    ---              &  
                    $6.34\pm0.02$    &  
		    $6.26\pm0.06$    &                   
                    ---              & 
		    ---             \\  
$\log(\mathrm{N/O})$      &
                    ---              & 
		    ---              &                   
                    $-1.19\pm0.20$   & 
		    ---              & 
		    ---              &  
                    $-1.29\pm0.05$   & 
		    $-1.26\pm0.12$   &                 
                    ---              & 
		    ---             \\  
$12+\log(\mathrm{O/H})$ -- (N2)                 &  
                    8.26             &  
		    8.24             &                   
                    8.16             &  
		    8.24             & 
		    8.22             &  
                    8.17             &  
		    8.17             &                   
                    8.44             &  
		    8.48            \\  
$12+\log(\mathrm{O/H})$ -- (O3N2)    &  
                    8.24             &  
		    8.24             &                   
                    8.17             &  
		    8.25             & 
		    8.25             &  
                    8.10             &  
		    8.11             &                   
                    8.41             &  
		    8.42            \\  
$\log([\ion{O}{iii}]5007/\Hb)$       &  
                    $0.50\pm0.02$    & 
		    $0.43\pm0.04$    &                   
                    $0.45\pm0.02$    & 
		    $0.40\pm0.02$    & 
		    $0.34\pm0.03$    &  
                    $0.70\pm0.01$    & 
		    $0.68\pm0.01$    &                   
                    $0.29\pm0.01$    & 
		    $0.31\pm0.03$   \\  
$\log([\ion{N}{ii}]6584/\Ha)$     &  
                    $-1.05\pm0.04$   & 
		    $-1.10\pm0.09$   &                   
                    $-1.30\pm0.08$   & 
		    $-1.09\pm0.07$   & 
		    $-1.14\pm0.10$   &  
                    $-1.28\pm0.02$   & 
		    $-1.27\pm0.04$   &                   
                    $-0.72\pm0.02$   & 
		    $-0.67\pm0.05$  \\  
$\log([\ion{S}{ii}]6717+6731/\Ha)$     & 
                    $-0.70\pm0.03$   & 
		    $-0.63\pm0.07$   &                   
                    $-0.63\pm0.02$   & 
		    $-0.69\pm0.04$   & 
		    $-0.51\pm0.05$   &  
                    $-1.00\pm0.01$   & 
		    $-0.91\pm0.03$   &                   
                    $-0.68\pm0.02$   & 
		    $-0.62\pm0.05$  \\  
$\log([\ion{O}{i}]6300/\Ha)$     & 
                    ---              &  
		    ---              &                   
                    $-1.50\pm0.07$   & 
		    $-1.47\pm0.10$   & 
		    $-1.27\pm0.07$   &  
                    $-2.04\pm0.03$   &  
		    ---              &                   
                    $-1.64\pm0.04$   &  
		    ---             \\ 
\hline
\end{tabular}
\end{center}
(*) Notes: 
$T_\mathrm{e}([\ion{O}{ii}])$ derived from the relation: 
$T_\mathrm{e}([\ion{O}{ii}])=0.72 \times T_\mathrm{e}([\ion{O}{iii}]) + 0.26$
found by \cite{Pilyugin2006}; 
$T_{\mathrm e}$([\ion{O}{iii}]): electron temperature measured from 
[\ion{O}{iii}]~$\lambda4363$; 
$12+\log(\mathrm{O/H})$ -- ($T_\mathrm{e}$): direct O/H abundance derived 
from $T_\mathrm{e}$([\ion{O}{iii}]); 
$12+\log(\mathrm{O/H})$ -- (N2): O/H derived from the N2 index 
\citep{PettiniPagel2004}; the associated uncertainty is $\pm0.38$; 
$12+\log(\mathrm{O/H})$ -- (O3N2): O/H derived from the O3N2 
index \citep{PettiniPagel2004}; the associated uncertainty is $\pm0.25$. 
\end{scriptsize}
\end{table*}
\end{landscape}

\onecolumn
\begin{landscape}
\begin{table*}
\begin{scriptsize}
\caption{Physical parameters and chemical abundances}
\label{Table:Parameters2}
\begin{center}
\begin{tabular}{|c|cccc|cc|cc|cc|}       
\hline\hline
Parameter  &       \multicolumn{4}{|c|}{Tololo~1434+032} &                                  
                   \multicolumn{2}{|c|}{Mrk~475}         &
                   \multicolumn{2}{|c|}{IZw~123}         &
                   \multicolumn{2}{|c|}{IZw~159}         \\    
           &       \multicolumn{1}{|c}{Knot~A}           & 
		   \multicolumn{1}{c}{Knot~B}            & 
		   \multicolumn{1}{c}{Knot~CD}           & 
		   \multicolumn{1}{c|}{Integrated}       &
                   \multicolumn{1}{|c}{Nuclear}          & 
		   \multicolumn{1}{c|}{Integrated}       &
                   \multicolumn{1}{|c}{Nuclear}          & 
		   \multicolumn{1}{c|}{Integrated}       &
                   \multicolumn{1}{|c}{Nuclear}          & 
		   \multicolumn{1}{c|}{Integrated}       \\
           &      &    &             &    &              &       & & &  &           \\
\hline
$N_\mathrm{e}$([\ion{S}{ii}]) (cm$^{-3}$)    &  
                    $< 100$          &  
		    $< 100$          &  
		    $< 100$          &  
		    150              &  
                    147              &  
		    323              &  
                    287              &  
		    $< 100$          &  
                    113              &  
		    $< 100$         \\  
$T_\mathrm{e}$([\ion{O}{ii}]) (10$^{4}$ K)   &        
                    $1.16\pm0.08$    &
		     ---             &   
		     ---             & 
		     ---             &  
                    $1.25\pm0.03$    & 
		    $1.29\pm0.09$    &  
                    $1.15\pm0.03$    &
		     ---             &  
                     ---             &  
		     ---            \\  
$T_\mathrm{e}$([\ion{O}{iii}]) ($10^{4}$ K)  &  
                    $1.25\pm0.11$    &  
		     ---             &  
		     ---             &  
		     ---             &  
                    $1.37\pm0.04$    &  
		    $1.43\pm0.13$    &  
                    $1.23\pm0.05$    &  
		    ---              &  
                    ---              &  
		    ---             \\  
$12+\log(\mathrm{O/H})$ -- ($T_\mathrm{e}$)        &  
                    $8.05\pm0.07$    &  
		     ---             &  
		     ---             & 
		     ---             &  
                    $7.92\pm0.02$    &  
		    $7.86\pm0.07$    &  
                    $8.13\pm0.03$    &  
		     ---             &  
                     ---             &  
		     ---            \\  
$12+\log(\mathrm{Ne}^{++}/\mathrm{H}^{+})$        &  
                    $7.31\pm0.12$    &  
		    ---              &  
		    ---	             &  
		    ---              &  
                    $7.17\pm0.05$    &  
		    $7.16\pm0.13$    &  
                    $7.43\pm0.06$    &  
		    ---              &  
                    ---              &  
		    ---             \\  
$12+\log(\mathrm{S}^{+}/\mathrm{H}^{+})$        &  
                    $5.76\pm0.06$    &  
		    ---              &   
		    ---              & 
		    ---              &  
                    $5.54\pm0.02$    &  
		    $5.60\pm0.06$    &  
                    $5.64\pm0.03$    &  
		    ---              &  
                    ---              &  
		    ---             \\  
$12+\log(\mathrm{N}^+/\mathrm{H}^+)$  &  
                    $6.26\pm0.06$    &  
		    ---              &   
		    ---              & 
		    ---              &  
                    $5.97\pm0.02$    &  
		    $5.97\pm0.08$    &  
                    $6.29\pm0.03$    &  
		    ---              &  
                    ---              &  
		    ---             \\  
$\log(\mathrm{N/O})$      &
                    $-1.32\pm0.12$   &  
		    ---              &   
		    ---              & 
		    ---              &  
                    $-1.25\pm0.05$   & 
		    $-1.25\pm0.15$   &  
                    $-1.25\pm0.07$   &  
		    ---              &  
                    ---              &  
		    ---             \\  
$12+\log(\mathrm{O/H})$ -- (N2)  &  
                    8.13             &  
		    8.14             & 
		    8.23             & 
		    8.15             &  
                    8.06             &  
		    8.07             &  
                    8.15             &  
		    8.17             &  
                    8.32             &  
		    8.34            \\  
$12+\log(\mathrm{O/H})$ -- (O3N2)    &  
                    8.08             &  
		    8.10             &  
		    8.21             & 
		    8.12             &  
                    8.01             &  
		    8.02             &  
                    8.07             &  
		    8.10             &  
                    8.30             &  
		    8.31            \\  
$\log([\ion{O}{iii}]5007/\Hb)$     &  
                    $0.66\pm0.01$    &  
		    $0.64\pm0.02$    &  
		    $0.53\pm0.03$    &  
		    $0.60\pm0.03$    &  
                    $0.71\pm0.01$    &  
		    $0.69\pm0.02$    &  
                    $0.76\pm0.01$    &  
		    $0.71\pm0.02$    &  
                    $0.44\pm0.01$    &  
		    $0.42\pm0.01$   \\  
$\log([\ion{N}{ii}]6584/\Ha)$     &  
                    $-1.38\pm0.04$   &  
		    $-1.34\pm0.07$   &  
		    $-1.11\pm0.11$   &  
		    $-1.32\pm0.12$   &  
                    $-1.54\pm0.02$   &  
		    $-1.51\pm0.06$   &  
                    $-1.31\pm0.02$   &  
		    $-1.26\pm0.07$   &  
                    $-0.91\pm0.02$   &  
		    $-0.88\pm0.03$  \\  
$\log([\ion{S}{ii}]6717+6731/\Ha)$     &  
                    $-0.89\pm0.03$   &  
		    $-0.79\pm0.05$   &  
		    $-0.66\pm0.08$   &  
		    $-0.75\pm0.07$   &  
                    $-1.06\pm0.02$   &  
		    $-0.98\pm0.04$   &  
                    $-1.04\pm0.02$   &  
		    $-0.86\pm0.05$   &  
                    $-0.68\pm0.01$   &  
		    $-0.60\pm0.02$  \\  
$\log([\ion{O}{i}]6300/\Ha)$    &  
                    ---              &  
		    ---              & 
		    ---              &  
		    ---              &  
                    ---              &  
		    ---              &                     
                    ---              &  
		    ---              &                     
                    $-1.60\pm0.04$   &  
		    $-1.49\pm0.10$  \\                    
\hline
\end{tabular}
\end{center}
\end{scriptsize}
\end{table*}
\end{landscape}

\end{document}